\pdfoutput=1
\documentclass[12pt,twoside,final]{huthesis}
\usepackage{bm,float}
\usepackage{graphics,amsmath}
\usepackage{tabularx,graphicx}
\usepackage{epsfig}
\usepackage{units}
\usepackage{amssymb}
\usepackage{cite}

\def\files{all}
\def\all{all}
\ifx\files\all  \else
\fi

\begin{document}

%

\newcommand{\deriv}[2]{\frac{d#1}{d#2}}
\newcommand{\derivc}[3]{\left. \frac{d#1}{d#2}\right|_{#3}}
\newcommand{\pd}[2]{\frac{\partial #1}{\partial #2}}
\newcommand{\pdc}[3]{\left. \frac{\partial #1}{\partial #2}\right|_{#3}}

\newcommand{\bra}[1]{\left\langle #1\right|}
\newcommand{\ket}[1]{\left|#1\right\rangle}
\newcommand{\braket}[2]{\left\langle #1 \left|#2\right.\right\rangle}
\newcommand{\braOket}[3]{\left\langle #1\left|#2\right|#3\right\rangle}

\def\cal#1{\mathcal{#1}}

\def\avg#1{\left< #1 \right>}
\def\abs#1{\left| #1 \right|}
\def\recip#1{\frac{1}{#1}}
\def\vhat#1{\hat{{\bf #1}}}
\def\smallfrac#1#2{{\textstyle\frac{#1}{#2}}}
\def\smallrecip#1{\smallfrac{1}{#1}}

\def\spshalf{{1\over{2}}}
\def\Orabi{\Omega_{\rm rabi}}
\def\btt#1{{\tt$\backslash$#1}}


\def\schrod{Schroedinger's Equation}
\def\helm{Helmholtz Equation}

\def\be{\begin{equation}}
\def\ee{\end{equation}}
\def\bea{\begin{eqnarray}}
\def\eea{\end{eqnarray}}
\def\bean{\begin{mathletters}\begin{eqnarray}}
\def\eean{\end{eqnarray}\end{mathletters}}

\newcommand{\tbox}[1]{\mbox{\tiny #1}}
\newcommand{\half}{\mbox{\small $\frac{1}{2}$}}
\newcommand{\pit}{\mbox{\small $\frac{\pi}{2}$}}
\newcommand{\sfrac}[1]{\mbox{\small $\frac{1}{#1}$}}
\newcommand{\mbf}[1]{{\mathbf #1}}
\def\text{\tbox}

\newcommand{\mV}{{\mathsf{V}}}
\newcommand{\mL}{{\mathsf{L}}}
\newcommand{\mA}{{\mathsf{A}}}
\newcommand{\lB}{\lambda_{\tbox{B}}}  
\newcommand{\ofr}{{(\mbf{r})}}       
\def\ofkr{(k;\mbf{r})}          
\def\ofks{(k;\mbf{s})}          
\newcommand{\ofs}{{(\mbf{s})}}       
\def\xt{\mbf{x}^{\tbox T}}      

\def\ce{\tilde{C}_{\tbox E}}        
\def\cew{\tilde{C}_{\tbox E}(\omega)}       
\def\ceqmw{\tilde{C}^{\tbox{qm}}_{\tbox E}(\omega)} 
\def\cewqm{\tilde{C}^{\tbox{qm}}_{\tbox E}} 
\def\ceqm{C^{\tbox{qm}}_{\tbox E}}  
\def\cw{\tilde{C}(\omega)}      
\def\cfw{\tilde{C}_{\cal F}(\omega)}        

\def\tcl{\tau_{\tbox{cl}}}      
\def\tcol{\tau_{\tbox{col}}}        
\def\terg{t_{\tbox{erg}}}       
\def\tbl{\tau_{\tbox{bl}}}      
\def\theis{t_{\tbox{H}}}        

\def\area{\mathsf{A}_D}         
\def\ve{\nu_{\tbox{E}}}         
\def\vewna{\nu_E^{\tbox{WNA}}}      

\def\dxcqm{\delta x^{\tbox{qm}}_{\tbox c}}  

\newcommand{\rop}{\hat{\mbf{r}}}    
\newcommand{\pop}{\hat{\mbf{p}}}

\newcommand{\sint}{\oint \! d\mbf{s} \,} 
\def\gint{\oint_\Gamma \!\! d\mbf{s} \,} 
\newcommand{\lint}{\oint \! ds \,}  
\def\infint{\int_{-\infty}^{\infty} \!\!}   
\def\dn{\partial_n}             
\def\aswapb{a^*\!{\leftrightarrow}b}        
\def\eps{\varepsilon}               

\def\dhdxt{\partial {\cal H} / \partial x}
\def\dhdx{\pd{\cal H}{x}}
\def\dhdxnm{\left( \pd{\cal H}{x} \right)_{\!nm}}
\def\dhdxnmsq{\left| \left( \pd{\cal H}{x} \right)_{\!nm} \right| ^2}

\def\bcs{\stackrel{\tbox{BCs}}{\longrightarrow}}    

\def\wx{\omega_x}
\def\wy{\omega_y}
\newcommand{\ofro}{({\bf r_0})}
\def\Eb{E_{\rm blue,rms}}
\def\Er{E_{\rm red,rms}}
\def\Es2{E_{0,{\rm sat}}^2}
\def\sb{s_{\rm blue}}
\def\sr{s_{\rm red}}

\def\ie{{\it i.e.\ }}
\def\eg{{\it e.g.\ }}
\newcommand{\etal}{{\it et al.\ }}
\newcommand{\ibid}{{\it ibid.\ }}

\def\gap{\hspace{0.2in}}

%

\newcounter{eqletter}
\def\mathletters{%
\setcounter{eqletter}{0}%
\addtocounter{equation}{1}
\edef\curreqno{\arabic{equation}}
\edef\@currentlabel{\theequation}
\def\theequation{%
\addtocounter{eqletter}{1}\thechapter.\curreqno\alph{eqletter}%
}%
}
\def\endmathletters{\setcounter{equation}{\curreqno}}


\def\kf{k_{\text F}}
\newcommand{\TL}{{\text{(L)}}}
\newcommand{\TR}{{\text{(R)}}}
\newcommand{\TLR}{{\text{L,R}}}
\newcommand{\VSD}{V_{\text{SD}}}
\newcommand{\GT}{\Gamma_{\text{T}}}
\newcommand{\DEL}{\mbox{\boldmath $\nabla$}}
\def\lf{\lambda_{\text F}}
\def\st{\sigma_{\text T}}
\def\stlr{\sigma_{\text T}^{\text{L$\rightarrow$R}}}
\def\strl{\sigma_{\text T}^{\text{R$\rightarrow$L}}}
\def\aeff{a_{\text{eff}}}
\def\aaeff{A_{\text{eff}}}
\def\gat{G_{\text{atom}}}
\newcommand{\LB}{Landauer-B\"{u}ttiker}

%
%

\newcommand{\nin}{\noindent}
\newcommand{\ka}{\kappa}
\newcommand{\alf}{\alpha_b}
\newcommand{\e}{\epsilon}
\newcommand{\s}{\sigma}
\newcommand{\alfe}{\alpha_e}
\newcommand{\alfo}{\alpha_{ph}}
\newcommand{\dlE}{\Delta E}
\newcommand{\dlt}{\Delta \tau}
\newcommand{\br}{{\bf r}}
\newcommand{\bq}{{\bf q}}
\newcommand{\bs}{{\bf s}}
\newcommand{\D}{\Delta}
\newcommand{\Db}{\Delta_0}
\newcommand{\Dr}{\Delta_{r}}
\newcommand{\Dren}{\Delta_{ren}}
\newcommand{\Dx}{\Delta_x}
\newcommand{\Dz}{\Delta_z}
\newcommand{\sx}{\sigma_x}
\newcommand{\sz}{\sigma_z}
\newcommand{\bk}{b_k}
\newcommand{\bdk}{b^\dag_k}
\newcommand{\ak}{a_k}
\newcommand{\adk}{a^\dag_k}
\newcommand{\om}{\omega}
\newcommand{\winf}{\omega_{inf}}
\newcommand{\wco}{\omega_{co}}
\newcommand{\wo}{\omega_{0}}
\newcommand{\wq}{\omega_{\bf\vec{q}}}
\newcommand{\Dco}{\Delta_{co}}
\newcommand{\Dox}{\Delta^x_0}
\newcommand{\Drx}{\Delta^x_r}
\newcommand{\Drenx}{\Delta^x_{ren}}
\newcommand{\Doz}{\Delta^z_0}
\newcommand{\coc}{\frac{\hbar\omega}{kT}}
\newcommand{\matel}{\left| \frac{\Dox}{\Db} \right|^2}
\newcommand{\Ddco}{\wco \left( \frac{\alpha_b}{\epsilon}
\right)^{1/\epsilon}}


\newcommand{\cg}{\chi^C}
\newcommand{\cgq}{\chi^C({\bf \vec{q}},\omega)}
\newcommand{\csi}{\chi^G}
\newcommand{\csiq}{\chi^G({\bf \vec{q}},\omega)}
\newcommand{\nusi}{\nu^G}
\newcommand{\nug}{\nu^C}
\newcommand{\vf}{v_{\rm F}}


\newcommand{\dle}{\Delta \epsilon}
\newcommand{\dlw}{\Delta \omega}
\newcommand{\dlec}{\Delta \epsilon_c}
\newcommand{\dlem}{\Delta \epsilon_{min}}
\newcommand{\eph}{E_{ph}}
\newcommand{\ma}{\frac{2ma^2}{\hbar^2}}
\newcommand{\map}{\Bigl(\frac{2ma^2}{\hbar^2}\Bigr)}
\newcommand{\al}{\alpha}
\newcommand{\rph}{\rho_{ph}}
\newcommand{\eikx}{e^{ikx}}
\newcommand{\ephc}{E_{\text{ph prox}}}
\newcommand{\li}{\text{lim}}
\newcommand{\ima}{\text{Im}}
\newcommand{\arctanh}{\text{ArcTanh}}
\newcommand{\re}{\text{Res}}

\ssp



\title{Dissipation in finite systems: Semiconductor NEMS, graphene NEMS, and metallic nanoparticles}
\author{C\'esar \'Oscar Seo\'anez Erkell}
\advisor{Dr. F. Guinea} 

\maketitle



\cleardoublepage
\dedication

\begin{quote}
\hsp \em \raggedleft

To my parents and grandparents

\end{quote}

\newpage
\cleardoublepage

\newpage
\thispagestyle{empty} \addcontentsline{toc}{section}{Resumen y
conclusiones}

\vspace{3.5cm}

{\bf {\LARGE Resumen}} \vspace{0.5cm}

Los procesos de decoherencia y disipación en sistemas mesoscópicos
han sido objeto en las últimas décadas de gran interés dado su papel
clave en el estudio y utilización de fenómenos cuánticos a escalas
cada vez mayores. Los espectaculares logros alcanzados en las
técnicas de fabricación y medición han permitido el diseño de
nanoestructuras móviles que hábilmente combinadas con otros
protagonistas de la historia física reciente, tales como puntos
cuánticos o gases electrónicos de baja dimensionalidad, otorgan la
fascinante oportunidad de estudiar la interacción electrón-fonón
cuanto a cuanto, con un potencial tecnológico prometedor. En la
mayoría de las aplicaciones vislumbradas, así como para la
observación de comportamiento cuántico en variables mecánicas
macroscópicas tales como el centro de masas de resonadores
micrométricos, es imprescindible una minimización de los procesos
disipativos que perturban la dinámica vibracional de estos sistemas.
En consecuencia se está llevando a cabo un gran esfuerzo actualmente
en el análisis y modelización de los mismos, tanto desde un punto de
vista experimental
como teórico.\\

Esta tesis forma parte de dicho esfuerzo. En concreto, dos de las
tres partes en que se divide están consagradas a la modelización de
algunos de los principales mecanismos de fricción presentes en dos
clases de sistemas nanoelectromecánicos (NEMS): nanoresonadores
construidos a partir de heteroestructuras de semiconductor, y
nanoresonadores en los que la parte móvil es un compuesto de carbono
de baja dimensionalidad, bien sea una lámina de grafeno o un
nanotubo.\\

En el primer caso nos hemos centrado en los procesos superficiales
de absorción de energía mecánica asociados al acabado imperfecto, la
presencia de impurezas y la estructura superficial parcialmente
desordenada, invocando las similitudes existentes con los procesos
de atenuamiento de ondas acústicas en materiales amorfos. El
resultado es un modelo de espines acoplados al conjunto de modos
vibracionales del resonador, analizado por medio de técnicas
desarrolladas para el estudio del \textit{Spin-Boson Model} y de
procesos de relajación. Asimismo se modelizan los posibles efectos
negativos que conlleva la deposición de electrodos metálicos sobre
las estructuras de
semiconductor.\\

Los osciladores de compuestos de carbono presentan una serie de
peculiaridades que los diferencian de sus homólogos semiconductores,
destacando su alto grado de cristalinidad y carácter metálico o
semimetálico. Ello implica la prevalencia de mecanismos de fricción
distintos, varios de los cuales se modelizan, destacando la
excitación de pares electrón-hueco en el resonador debido a cargas
distribuidas por la estructura, estudiada con
métodos perturbativos autoconsistentes.\\

La tercera y última parte se dedica a otro tipo de sistema
mesoscópico, las partículas nanometálicas, cuya respuesta óptica
viene dominada por la excitación colectiva denominada
\textit{plasmón de superficie}. La presencia en el espectro de
excitaciones electrónicas de pares electrón-hueco acoplados al
plasmón provoca su progresivo atenuamiento tras ser excitado
inicialmente por un campo eléctrico externo, por ejemplo un laser.
Este proceso se puede considerar como ejemplo de entorno disipativo
con un \textit{número finito} de grados de libertad en interacción
con el subsistema de interés, el plasmón. Los efectos que esta
finitud conlleva en la modelización teórica de la dinámica
electrónica son analizados en detalle, justificando la validez de
ciertas aproximaciones asumidas hasta ahora sin demostración previa.

\vspace{1.0cm}

{\bf {\LARGE Conclusiones.}} \vspace{0.5cm}

La fricción a bajas temperaturas asociada a los distintos procesos
superficiales en resonadores de semiconductor ha sido modelizada
adaptando el \textit{Standard Tunneling Model} utilizado para
explicar las propiedades acústicas de sólidos amorfos a
$\sim0.01\lesssim T\lesssim1$ K. El orden de magnitud así como la
débil dependencia con la temperatura observada en los experimentos
es cualitativamente reproducida. Sin embargo, llegar a un acuerdo
cuantitativo se antoja difícil hasta que no se posea un conocimiento
más detallado de los procesos y defectos estructurales presentes en
el resonador. \\

Diversos mecanismos disipativos presentes en nanoresonadores de
grafeno y nanotubos han sido analizados, obteniendo, al igual que en
el caso previo, dependencias paramétricas de los mismos en función
de las variables físicas relevantes del sistema, como la temperatura
y dimensiones características. La importancia relativa de los mismos
en función del régimen (amplitud de vibración, temperatura) ha sido
establecida, predominando a bajas temperaturas la disipación óhmica
debido a excitaciones electrónicas en el seno del resonador y a
altas temperaturas la atenuación por procesos termoelásticos. \\

Se ha presentado un modelo muy utilizado en el estudio de la
dinámica de plasmones superficiales y de la respuesta óptica de
clusters metálicos, analizando la coherencia interna de dicho
esquema teórico, justificando varios puntos clave del mismo que
hasta ahora se habían tomado como válidos sin justificación
rigurosa. De dicho análisis se obtiene asimismo información sobre
los tiempos característicos de evolución del plasmón y del resto de
excitaciones electrónicas acopladas al mismo, lo cual permite
fundamentar una aproximación Markoviana a la hora de estudiar la
dinámica del sistema acoplado plasmón - pares electrón-hueco.

\newpage
\cleardoublepage

\newpage
\thispagestyle{empty} \addcontentsline{toc}{section}{Abstract}

\vspace{3.5cm}

{\bf {\LARGE Abstract}} \vspace{0.5cm}

The physics of decoherence and dissipation in mesoscopic systems has
attracted great interest, given its key role in the study and use of
quantum phenomena at ever larger scales. The spectacular
achievements in fabrication and measurement techniques have opened
the doors to the design of mobile nanostructures which, combined in
smart ways with other recent cornerstones like quantum dots or
low-dimensional electron gases, grant the fascinating opportunity to
study electron-phonon interactions at an individual level, with a
promising technological potential. In most of the envisioned
applications, as well as to observe quantum behavior in macroscopic
mechanical variables like the center-of-mass of micrometric
resonators, a minimization of the dissipative processes perturbing
the vibrational dynamics of these systems is compulsory.
Consequently strong efforts are being made to analyze and model
them, both from the experimental as well as from the theoretical
sides.\\

This thesis exemplifies these efforts. Specifically, two out of the
three parts constituting it are consecrated to the modeling of some
of the main friction mechanisms found in two kinds of
nanoelectromechanical systems (NEMS): nanoresonators built from
semiconductor heterostructures, and nanoresonators whose mobile part
is a low-dimensional carbon compound, be it graphene or a
nanotube.\\

In the first case we have focussed in the surface processes
absorbing mechanical energy, related to the imperfect finish,
presence of impurities and partially disordered surface structure.
Similarities with acoustic wave damping in amorphous solids have
been invoked to build an adaptation of the \textit{Standard
Tunneling Model} which aims to provide a low temperature description
of such processes. The result is a model of spins coupled to the
ensemble of vibrational eigenmodes of the resonator, which is
analyzed by means of techniques developed for the study of the
\textit{Spin-Boson Model} and relaxation processes. Possible
negative effects due to metallic electrodes deposited on top of the
semiconductor structures are also modeled.\\

Oscillators made from carbon compounds, demonstrated for the first
time this year, display a series of peculiarities distinguishing
them from their semiconductor counterparts, specially their high
degree of crystallinity and semimetal or metal character. The
prevailing friction mechanisms thus differ, several out of which are
modeled, with the excitation of electron-hole pairs in the resonator
due to charges distributed
throughout the device dominating at low temperatures and small vibrational amplitudes. \\

The third and last part is devoted to another kind of mesoscopic
system, namely nanometallic clusters, whose optical response is
dominated by the so-called \textit{surface plasmon} collective
excitation. The presence in the electronic excitation spectrum of
one-body electron-hole pairs coupled to the plasmon causes its
progressive damping after its initial excitation by an external
electric field, e.g. a laser pulse. This process may be considered
as an example of finite dissipative environment interacting with the
subsystem subject of our interest, the plasmon. The consequences of
this finiteness on the theoretical modeling of the electronic
dynamics are analyzed in detail, justifying the validity of several
approximations assumed until now to hold without proof. This study
will also shed some light on the characteristic times of the
different electronic degrees of freedom, providing a basis for the
use of a Markovian approximation in the analysis of the plasmon
dynamics of nanometric clusters.

\begin{acknowledgments}
Looking back at these years of PhD, I really consider myself a very
fortunate person. All the people next to me have always given their
support and courage whenever I needed it. I have met many good,
generous, idealistic and hardworking people, whose invaluable
examples have been extremely stimulating, and a reason to reflect
and rejoice. I start with my advisor, Paco Guinea. His efficacy,
optimism and humility are simply impressive. Making use of his deep
physical intuition and thorough knowledge of numerous fields he
introduced me to fascinating research fields and also gave me the
possibility to assist to several schools and workshops, where I
could meet physicists and friends from many countries. Indeed, in
some of them I met for the first time some of the people with whom I
had the luck to collaborate, like Raj Mohanty and Rodolfo Jalabert.
\\
Two stages abroad have been fundamental to get to this point, one
with Antonio Castro-Neto in Boston and another with Rodolfo Jalabert
in Strasbourg and Augsburg. In both cases they, together with their
collaborators, made their best to help me in every issue to make me
feel like at home, and with them I could also learn different ways
of doing physics, a very enriching experience. Here go my thanks to
Antonio, Raj, Silvia, Marco, Guiti, Alexei and Johan (BU), and to
Rodolfo, Dietmar, Guillaume and Gert (IPCMS/U. Augsburg), and all
the rest of nice and helping people I met there. I appreciate as
well very much the effort made by Sebastián Vieira, Adrian Bachtold,
Enrique Louis, Rodolfo Jalabert and Pablo Esquinazi to be part of
the dissertation committee,
and the work done by Guillermo Gómez-Santos as tutor.\\
In any case, most of the time was spent here in Madrid, in a
wonderful group of the Condensed Matter Theory Department at the
ICMM. I doubt I will ever again find a working place where I feel so
comfortable. Moreover, there were many of us doing the PhD, so we
could share this experience, having often great fun. If you are
bored or feel depressed just knock on the door of offices 128 or
130, "mano de santo", as we say in spanish. Thank you so much for
everything, Juan Luis, Alberto, Debb, Suzana, Javi, Rafas, Félix,
Ramón, Geli, Luis, Virginia, Tobías, Leni, Belén, Berni, María,
David, Fernando, Carlos, Samuel, Juan, Ana and Mª José. In the case
of Geli, Ramón, Tobías, Pilar, Carlos, Alberto, Javi and Juan Luis
thank you very much for helping me with my computer and
physics-related doubts. Special thanks also to
Sonia, Ángel, Antonio, Cayetana and Puri, thank you for your sympathy and attentions. \\
Some words of gratitude cannot be skipped for my friends outside the
ICMM, your support is also very important to me: Javi, José, Héctor,
Juan, Teo, Alfredo, Fátima, Nuria, Ana, Abelardo, Juan, Rocío,
Izaskun, Ángela, Julio, Amadeo, Mario, Ful, Aurelio, Sabrina... I
also thank my spanish and swedish family, which I know is always
there in the good and bad circumstances: tío Mariano, tía Pili,
Pilar, Esther, Aniana, Titti, Lars-Johan..., and my grandparents,
who unfortunately are now gone.\\
And of course I am most indebted to my parents, your 28 years of
love and care are impossible to balance, I cannot ask for more,
tusen tack!

\end{acknowledgments}

\newpage
\cleardoublepage

\begin{citations}

\vspace{0.8in}

\ssp \noindent Parts of the contents of this work can be found in
the following publications:
\begin{itemize}
  \item C. Seoanez, F. Guinea and A. H. Castro-Neto (2007). \textit{Dissipation due to two-level systems in nano-mechanical
  devices}. Europhysics Letters \textbf{78}, 60002, preprint archive cond-mat/0611153.
  \item C. Seoanez, F. Guinea and A. H. Castro-Neto (2007). \textit{Surface dissipation in NEMS: Unified description with
  the Standard Tunneling Model, and effects of metallic electrodes.}.
  Sent to Physical Review B.
  \item C. Seoanez, G. Weick, R.A. Jalabert and D. Weinmann (2007). \textit{Friction of the surface plasmon by high-energy particle-hole pairs:
  Are memory effects important?}. The European Physical Journal D \textbf{44}, 351, preprint archive cond-mat/0703720.
  \item C. Seoanez, F. Guinea and A. H. Castro-Neto (2007). \textit{Dissipation in graphene and nanotube resonators}.
  Physical Review B \textbf{76}, 125427, preprint archive arXiv:07042225.
\end{itemize}

\end{citations}

\newpage
\addcontentsline{toc}{section}{Contents} \tableofcontents

\listoffigures

\listoftables


\newpage

\startarabicpagination


\include{ch1}

\part{Dissipation in semiconductor NEMS}

\chapter{NEMS}\label{chr1}
\section{Introduction: current research topics}
Nanoelectromechanical systems (NEMS) constitute a recent fascinating
subject of current research, both from fundamental and
application-oriented points of view \cite{C00,C02,B04,ER05,SR05}.
Inheritors of the bigger microelectromechanical systems (MEMS),
they are mesoscopic solid-state devices with a special ingredient:
one resonating component, with some of its characteristic dimensions
lying in the submicron regime, whose mechanical motion is integrated
into an electrical circuit. Typically this moving element can be a
bridge, a pillar, or a cantilever, one of whose mechanical
eigenmodes is excited and measured. Two features of these eigenmodes
are essential to explain the potential of NEMS: First, due
to its small dimensions, their modes vibrate at very high
frequencies $\wo$, currently from tens of MHz to GHz (microwave
frequencies). This is desirable if one wishes to study the quantum
regime where thermal fluctuations are smaller than the energy of a
phonon, $kT<\hbar\wo$ (for GHz frequencies, milliKelvin suffice),
and also for several high-speed signal processing components
\cite{Betal05,HFZMR05,BQKM07}.

Second, and perhaps more important, they exhibit extremely low damping, measured in terms of
the so-called \emph{quality factor}
\begin{equation}\label{Qdef1}
    Q(\wo)=\frac{\wo}{\Delta\wo}\,,
\end{equation}
where $\Delta\wo$ is the measured linewidth, reaching often $Q\sim10^5$ or higher.
This translates, for example in the case of NEMS-based single-electron transistors, into power dissipation rates several orders of magnitude below
the ones of common CMOS-based transistors found in standard microprocessors operating at similar frequencies \cite{BQKM07}. Thus, NEMS-based
computing technology constitutes a very attractive alternative for applications demanding GHz or lower clock-speeds, robustness and/or operation at
high temperatures ($T>200^{o}$C).

Apart from nanoscale versions of Babbage's 18th-century mechanical computers, these two features open the scope for many other technological
applications, among whom ultrasensitive sensors and actuators clearly outstand. Their working principle is easy: The interaction of the resonator
with other elements of the circuitry, like charges if the vibrating element is charged, or with
elements of the environment outside the device, cause measurable
shifts in the frequency of the eigenmode. The sensitivity depends crucially on the small dimensions of the device and the linewidth $\Delta\wo$ of the
mode, currently achieving sub-attonewton force detection \cite{MR01} and mass
sensing of individual molecules \cite{EHR04}. Other successes include single-spin detection
\cite{RBMC04}, study of Casimir forces at the nanometer scale
\cite{DLCFKJ05}, high-precision thermometry \cite{hopcroft:013505} or \textit{in-vitro}
single-molecule biomolecular recognition \cite{DKEM06}.\\

\begin{figure}[!t]
\begin{center}
\includegraphics[width=14cm]{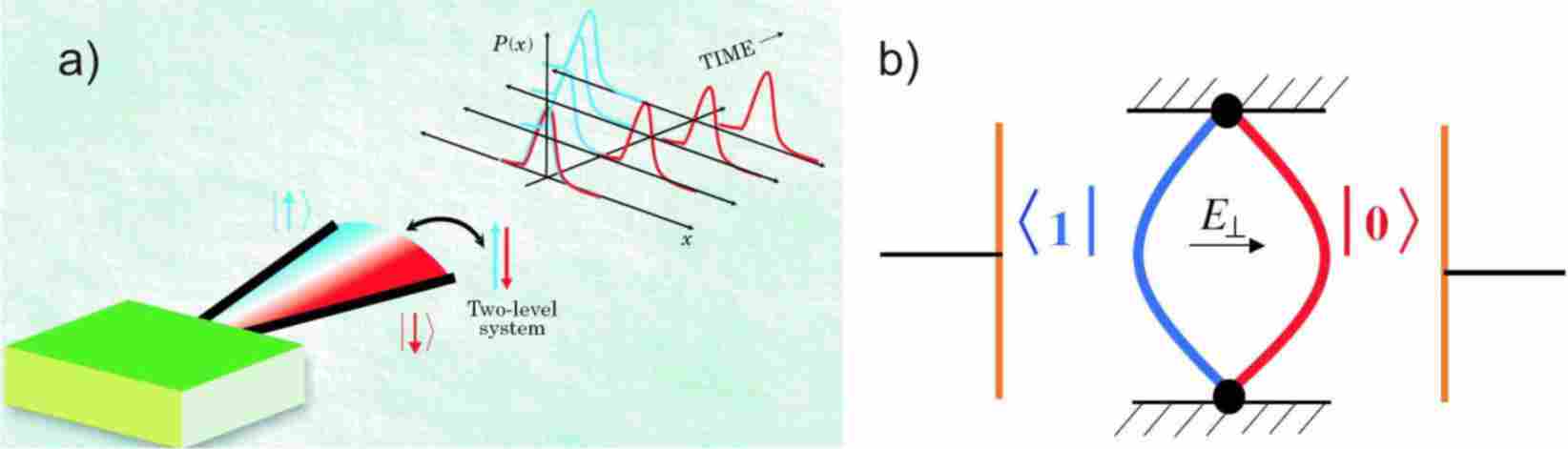}\\
\caption[Quantum superpositions with mechanical resonators]{\label{macro_super} Preparation of quantum superpositions of macroscopically distinct
 resonator states. a) From \cite{ER05}: the cantilever interacts with a two-level system (TLS) in such a way that the flexure depends on the TLS state,
 $|\uparrow\rangle$ or $|\downarrow\rangle$. If the TLS is prepared in a superposition state, the mechanical resonator state will be a linear combination
 of the two flexure states, with a probability distribution $P(x,t)$ as depicted to the right. b) When a doubly clamped bar is sufficiently compressed,
 it buckles, with two possible buckling positions. At low enough temperatures, tunneling through the energy barrier separating both states gives rise to
 eigenstates corresponding to superpositions of left and right buckled states, enabling its use as a qubit. From \cite{SHN06}}.
\end{center}
\end{figure}

\begin{figure}[!t]
\begin{center}
\includegraphics[width=13cm]{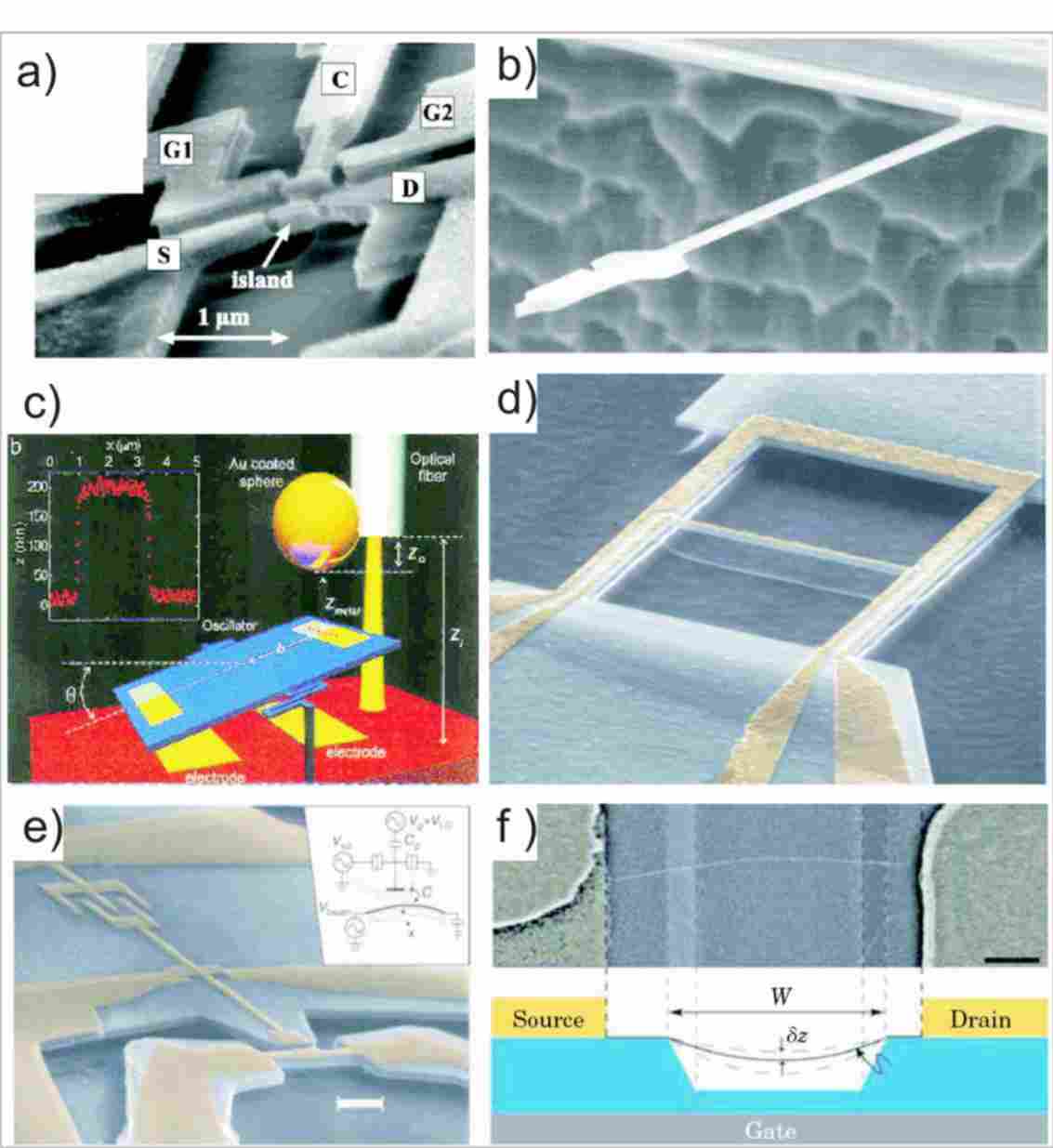}\\
\caption[NEMS examples]{\label{NEMS_examples} Several examples of
nanoelectromechanical devices (NEMS). a) Mechanical single electron
shuttle \cite{EWZB01}. b) Ultrasensitive magnetic force detector
used to detect a single electron spin \cite{RBMC04}. c) Setup using
a torsional oscillator used to study Casimir forces to check the
validity of Newtonian gravitation at the nanometer scale
\cite{DLCFKJ05}. d) Parametric radio-frequency amplifier providing a
thousandfold boost of signal displacement at 17 MHz \cite{SR05}. e)
A 116-MHz nanomechanical resonator coupled to a single-electron
transistor \cite{KC03}. f) Tunable carbon nanotube resonator
operating at 3-300 MHz \cite{Setal04}.}
\end{center}
\end{figure}

But leaving aside applications, they are also proving as very powerful tools in fundamental research:
\begin{itemize}
  \item They allow for a study of electron-phonon
interactions at a single quantum level, with setups prepared for the study of single electrons interacting with single phonons, or the analysis of
single electrons shuttled via mechanical motion \cite{EWZB01}.
\item At low enough temperatures several manifestations of a quantum behavior of the mechanical oscillator have been observed, or are expected to appear,
  like \cite{B04}:
  \begin{itemize}
    \item Zero-point motion detection and back-action forces due to the measurement process \cite{LBCS04,Netal06}. A displacement sensitivity of
    $10^{-4}$\AA  is needed, attainable with current cantilevers.
    \item Avoiding back-action forces, quantum non-demolition measurements of energy eigenstates
    \cite{santamore:144301,jacobs:147201,martin:120401,J07}.
    \item Quantization of thermal conductance \cite{B99,SHWR00}.
    \item Quantum tunneling between macroscopically distinct mechanical states, with several proposals for applications in quantum computing
    \cite{CLW01,GC05,SHN06,XWSOYS06}, see fig.(\ref{macro_super}).
    \item Preparation of minimum-uncertainty quantum squeezed states with a position uncertainty below that of zero-point fluctuations
    \cite{RSK05,RSK05b}.
  \end{itemize}
\end{itemize}
Figure (\ref{NEMS_examples}) shows some beautiful examples of setups for some of the applications mentioned above.

The study of quantum superpositions of mechanical
degrees of freedom in NEMS is also a straight way to assess two important issues: i) The validity of quantum mechanics for ever larger systems \cite{L02},
 and ii) The theoretical framework explaining environmentally-induced decoherence \cite{Letal87,W99}, a part of which we will use in this thesis to
 analyze the mechanisms limiting the quality factors $Q$ of the eigenmodes, and with them the observation of such quantum behavior.\\

As for most of these technological applications and fundamental
science works low-dissipation, high-$Q$ resonators are desirable,
several works have been devoted to the analysis of the different
sources of friction present in MEMS and NEMS
\cite{CR99,Yetal00,Eetal00,CR02,YOE02,Metal02,AM03,Hetal03,ZGSBM05,FZMR06,Mohetal07},
trying to determine the dominant damping mechanisms and ways to
minimize them. In the first two parts of this thesis we will analyze
them, both for the most common semiconductor-based resonators, with
vibrating structures made of Si, GaAs, SiN, SiC... (part I), and
also for carbon-materials based resonators, like nanotube
\cite{Setal04} or graphene \cite{Betal07} devices (part II). But
before we embark ourselves in such a task, it is convenient to learn
a few basic facts about NEMS.

\section{Semiconductor NEMS fabrication}
The performance of a given nanomechanical oscillator depends to a
great extent on its constituent materials and the manufacturing
steps they have gone through, not to speak about reproducibility and
scalability of results. A good review on MEMS and NEMS manufacturing
techniques can be found in \cite{C02}. We will focus on the most
widely used technique for building NEMS, namely Epitaxial
Heterostructure Nanomachining, applied to a (Si-SiO$_2$-Si)
heterostructure, chosen frequently by experimentalists due to the
high degree of crystallinity and perfection of the resulting
resonators, and more concretely in a study of dissipation with which
we will compare our predictions
\cite{ZGSBM05,SGN07,SGN07cond3,Mohetal07}.

The starting heterostructure is fabricated by oxygen ion implantation in a crystalline Si wafer, followed by a high-temperature annealing of several
hours, and is commonly known as SIMOX (Separation by IMplantation of OXigen) wafer. During the second step the previously formed mixed
buried layer Si-SiO$_2$ becomes purified through oxygen migration, constituting a separated
SiO$_2$ smooth buried layer, below a single-crystal recrystallized Si top layer. Typically, this single-crystal top layer, out of which
our resonator will be carved, contains a relatively large number of defects, mostly threading dislocations running from the SiO$_2$
layer to this one, with typical defect densities of about $10^3-10^5$/cm$^2$. The top layer presents variations in thickness $\sim10$ nm, as does
the silicon oxide underneath.

\begin{figure}[!t]
\begin{center}
\includegraphics[width=10cm]{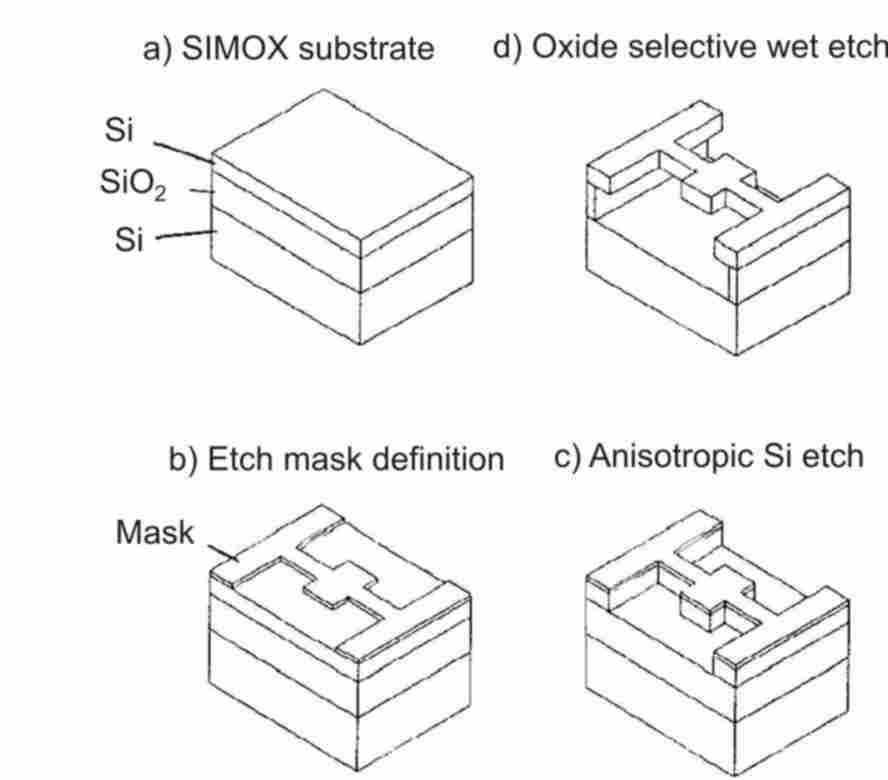}\\
\caption[NEM resonator fabrication: Nanomachining]{\label{NEM_fabrication} From \cite{C02}. Building a nanoresonator
using Epitaxial Heterostructure Nanomachining. (a) The process starts with a previously fabricated heterostructure, such as the
SIMOX (Separation by IMplantation of OXigen) Si-SiO$_2$-Si shown. (b) Using  electron beam lithography a mask is etched.
(c) Anisotropic dry etch such as Reactive Ion Etching. (d)
The oxide layer is removed using a selective SiO$_2$ wet etch, obtaining in this way the suspended structure. Metallization of
the top layer can be performed lithographically before the process.}
\end{center}
\end{figure}

To create the suspended structure, the wafer is subject to the nanomachining process depicted in fig.(\ref{NEM_fabrication}): first an etch mask
is defined using lithography: a PMMA photoresist is coated over the wafer and then patterned using a nm-resolution electron beam, constituting
a metal liftoff mask. Afterwards a layer resistant to Reactive Ion Etching (RIE) is deposited on top of it and lifted off, leaving behind
a RIE-resistant mask, fig.(\ref{NEM_fabrication}b). Then the RIE is applied on the sample enough time to remove the top Si layer except where the
protective mask lies, and remove also part of the underlying SiO$_2$ layer, fig.(\ref{NEM_fabrication}c). The suspended structure appears when
the device is submerged in a selective wet etch which removes the SiO$_2$, like hydrofluoric acid, fig.(\ref{NEM_fabrication}d).
Depending on the method to be used for the driving of the resonator,
a top metallic coating can be deposited at will for example using lithography, before the RIE mask is created.\\

Even though spectacular, the final result is far from perfect: the SIMOX wafer defects, together with surface roughness and imperfections
due to the resolution limitations of the different steps result in strong dissipation, as we will later see.

\section{Motion transduction of mechanical eigenmodes in semiconductor NEMS}
Well stablished transduction setups for the detection and driving of mechanical modes in MEMS, like
optical methods as fiber-optic interferometry and reflection, or piezoresistive displacement transduction,
cease to work properly for the submicrometer-sized NEMS. In the former case the light wavelengths exceed
the resonator transverse dimensions and diffraction effects spoil the scheme, while in the latter
dissipation heats the sample, precluding its use for low-temperature applications and studies.

For semiconductor NEMS, two other schemes have proved quite successful:
\subsection{Magnetomotive scheme}
As shown in fig.(\ref{NEM_magnetomotive}a), in this setup \cite{Metal02} the nanobridge is covered by a thin
metallic layer, connected in turn to an external circuit whose AC current is controlled
by the experimentalist, fixing its frequency $\om$ at will. The whole circuit is embedded in a strong static
magnetic field of several Tesla, perpendicular to the main axis of the bridge, as indicated in the figure.
When the electrons of the AC current flow back and forth at frequency $\om$, they experience a Lorentz force
due to the presence of the magnetic field perpendicular both to the field and the main axis, forcing the
bridge to oscillate up and down. The area enclosed by the circuit is thus a function of time, and so the
magnetic flux through it, therefore originating an induced electromotive force opposing the motion of the bridge,
which is detected by a network analyzer. When the frequency is tuned to the one of a given mechanical eigenmode,
the motion is strongly enhanced. With this scheme many modes can be excited, just by changing the circuit's frequency
to the one of the mode of interest. The presence of the top metallic layer adds extra dissipative sources which
will be analyzed later.

\subsection{Coupling to SETs and other sensitive electronic probes}
\begin{figure}[!t]
\begin{center}
\includegraphics[width=12cm]{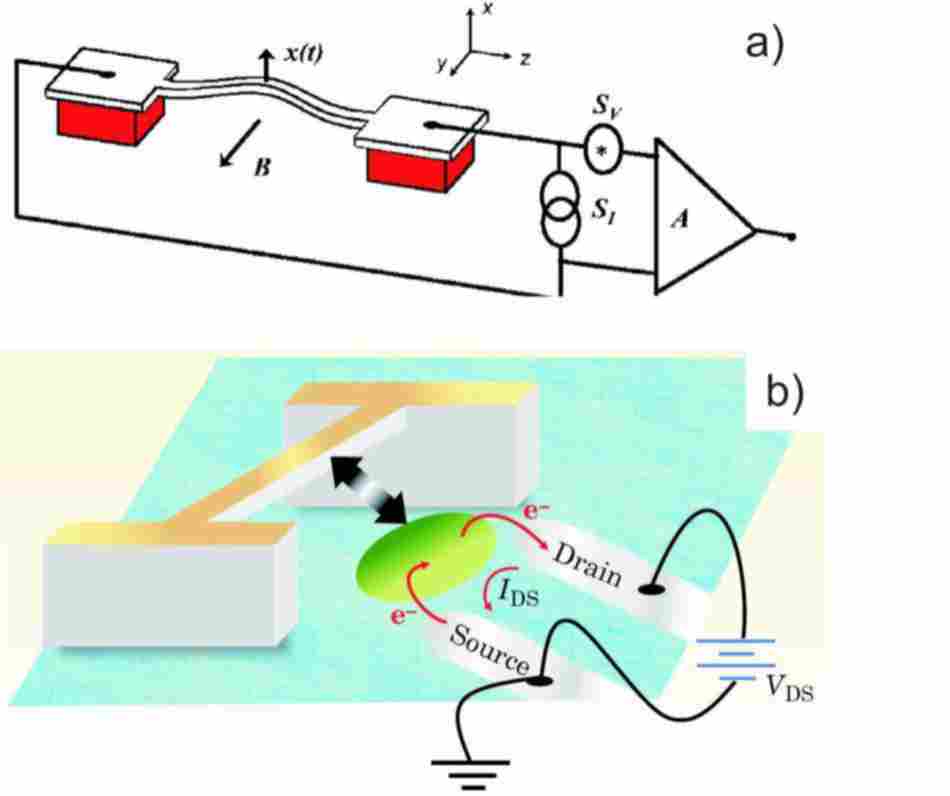}\\
\caption[NEMS detection and driving
schemes]{\label{NEM_magnetomotive} (a) Scheme of the magnetomotive
driving and detection setup with a nanobridge. The electrons flowing
back and forth through the bridge experience a Lorentz force which
drives the motion of the bridge. Due to the oscillation the area of
the loop formed by the circuit varies, and with it the magnetic
flux enclosed by it, inducing an electromotive force according to Faraday's law,
measured as a function of the AC electron current frequency by a
vector network analyzer \cite{Metal02}. (b) The vibrating bridge
interacts with the electrons flowing through a mesoscopic detector
such as a quantum point contact or quantum dot placed nearby, altering the current $I_{DS}$,
which is monitored with extremely high precision. Figures from \cite{ER05,SR05}.}
\end{center}
\end{figure}

The magnetomotive driving can be used in combination with more elements close to the resonator, to
study electron-phonon interactions in a previously unexplored regime, with a single-electron single-phonon
control. This can be achieved by coupling our magnetomotively driven charged resonator to a mesoscopic device
whose conductance can be controlled and measured with great accuracy, like quantum dots, quantum
point contacts or single-electron transistors (SETs) \cite{KC03}, as schematically
depicted in fig.(\ref{NEM_magnetomotive}b). Using similar ideas but different geometries, fascinating devices like
single-electron mechanical shuttles have been demonstrated \cite{EWZB01,SB04}.

\section{Nanoscale elasticity theory: Vibrational eigenmodes of long and thin rods}\label{vibmodesbeam}
Regarding the elastic properties of nanoresonators, it is clear that
as size shrinks a point is reached when classical continuum
mechanics ceases to be a good approximation. Nevertheless, for the
sizes of current semiconductor nanorods and wavelengths of the modes
excited ($\sim50$nm - $10\mu$m), much bigger than typical
interatomic spacing, continuum approximations continue to hold
\cite{SB98,LB99,SB02,SBNI02,LBTBK03,B04}, and classical elasticity
theory, using bulk concepts like Young's modulus $E$ or Poisson's
coefficient $\nu$, predicts with reasonable accuracy the eigenmodes
observed in experiments. In Appendix \ref{apmodeseqs} a brief review
of continuum elasticity concepts and the derivation of the equations
for the different modes present in quasi-1D resonators is given.
Here we state the main results. For a long and thin rod, considering
small deviations from equilibrium and wavelengths bigger than the
transversal dimensions, three kinds
of vibrational modes are found, namely longitudinal, flexural and torsional.\\

\begin{figure}[!t]
\begin{center}
\includegraphics[width=12cm]{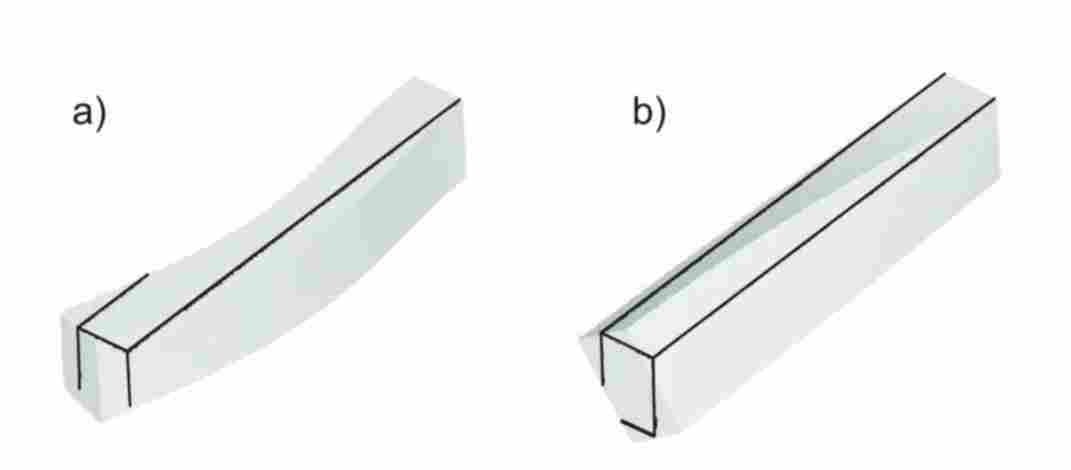}\\
\caption[Flexural and torsional modes of a
cantilever]{\label{Modes_rod} Examples of cantilever modes not
present in a bulk solid, with the black lines indicating the beam in
absence of movement serving as guide to the eye. a) Second flexural
(bending) mode. b) Fundamental (first) torsional (twisting) mode.}
\end{center}
\end{figure}

\textbf{Longitudinal} (or compressional) modes are as those found in bulk 3D solids but occur only along the direction
of the main axis, chosen to be the Z axis, with the Z-component $u_z$ of the local displacement field
$u_i=x_i'-x_{i,eq}$ obeying the wave equation
\begin{equation}\label{longitudinal}
    \frac{\partial^2u_z}{\partial z^2}=\frac{\rho}{E}\frac{\partial^2u_z}{\partial t^2}\,\,,
\end{equation}
$\rho$ being the mass density. \textbf{Torsional} vibrations, like the one of fig.(\ref{Modes_rod}b), are
parametrized in terms of the twisting angle $\phi(z)$, which obeys the wave equation
\begin{equation}\label{torsional}
    C\frac{\partial^2\phi}{\partial z^2}=\rho I \frac{\partial^2\phi}{\partial t^2}
\end{equation}
where $C$ is the so-called torsional rigidity. \textbf{Bending} (flexural) modes differ from the previous two
in that they do not obey a standard wave equation. They correspond to transversal displacements $X(z,t),Y(z,t)$
following the equations of motion
\begin{eqnarray}\label{bending}
    \nonumber EI_y\frac{\partial^4 X}{\partial z^4}&=&- \rho S \frac{\partial^2 X}{\partial t^2}\\
    EI_x\frac{\partial^4 Y}{\partial z^4}&=&- \rho S \frac{\partial^2 Y}{\partial t^2}\,\,.
\end{eqnarray}
Here $I_j$ are the inertia moments of the X and Y axes, and S is the cross-section area of the rod.
Even though these are not usual wave equations they admit plane wave-kind of solutions
$X(z,t),Y(z,t)\sim e^{i(kz-\omega t)}$, but with a \emph{quadratic}
dispersion relation, $\omega_{j}(k) = \sqrt{EI_j/(\rho S)}k^2$ . An example of bending mode for a cantilever can be seen
in fig.(\ref{Modes_rod}a).

\section{Dissipative mechanisms damping the vibrational eigenmodes of semiconductor nanoresonators}
Numerous dissipative sources limit the lifetime of an externally
excited mechanical eigenmode, absorbing its energy irreversibly.
Some of them, like surface impurities, clamping imperfections or gas
friction can be reduced devising solutions like working in high
vacuum conditions \cite{Metal02,MH07}, annealing the samples at the
right stage of the fabrication process
\cite{Yetal00,YOE02,WOE04,LVSLHP05} or engineering carefully the
clamping points to minimize the coupling to bulk vibrational modes
outside the resonator \cite{HFZMR05}. They are thus \emph{extrinsic}
mechanisms, as opposed to \emph{intrinsic} mechanisms, present even
in a theoretically perfect sample, such as anharmonic coupling with
other modes, electron-phonon or electron-electron interactions. A
brief description of the most important mechanisms follows.
\subsection{Extrinsic mechanisms}
\subsubsection{Gas or liquid friction}
Collisions with surrounding gas molecules can impose severe limitations on the quality factor, specially in those setups
designed to work in ambient conditions or in a liquid environment \cite{BW04,DKEM06,MH07}. Depending on the fluid density several
regimes appear which have to be analyzed with different frameworks, ranging from free molecular to continuum limits. $Q$ increases
with decreasing pressure $P$, and for $P<1$ mTorr it can be safely disregarded as the main frictional force for nanoresonators
with $Q\geq10^4$ \cite{Metal02,ER05}.
\subsubsection{Heating linked to the actuation or detection setup}
In the magnetomotive scheme the flow of electrons through the metallic top layer is not ballistic, and heating of the layer
is generated through their inelastic collisions, which in turn heats the semiconductor substrate underneath. Fortunately for the
usual current densities this effect seems to be irrelevant compared to others \cite{Metal02}. In schemes using actuation and detection
lasers illuminating the resonator, heating may play a more significant role, but we will not study it further in this thesis.
\subsubsection{Noise introduced by the external electrical circuit}
The external circuitry can be a source of extra damping and frequency shift, due to its finite impedance. This effect can
be used for applications requiring an external control of linewidth, like signal processing \cite{CR99,SCH02}.
\subsubsection{Clamping (attachment) losses}
This mechanism corresponds to the transfer of energy from the resonator mode to acoustic modes at the contacts and
beyond to the substrate through the anchorage areas \cite{JI68,PJ04}. The motion of the resonator forces as well a motion of the atoms linking it
to the rest of the device, irradiating in this way elastic energy to its surroundings.
This mechanism can be of special relevance for short and thick beams, while for long and thin ones it plays a secondary role \cite{HFZMR05,FZMR06}.
\subsubsection{Surface roughness and imperfections}
\begin{figure}
\begin{center}
\includegraphics[width=10cm]{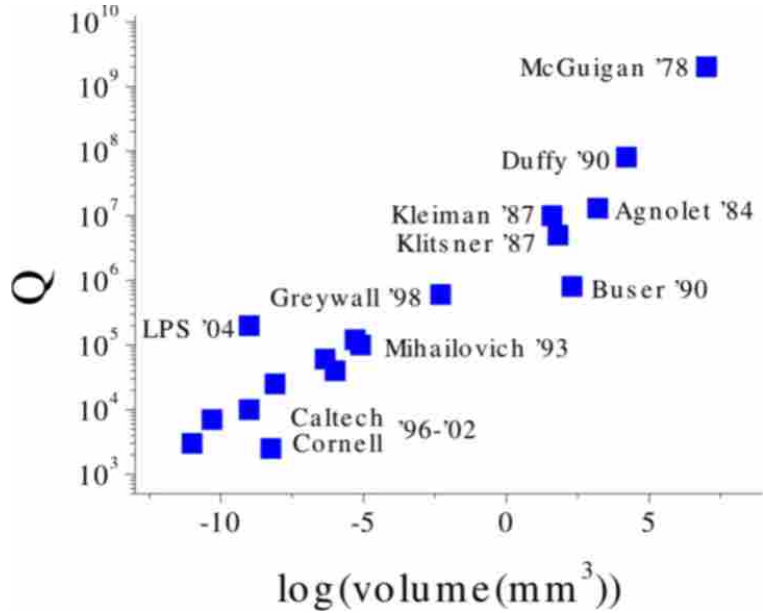}\\
\caption[Linear dependence of Q with the surface-to-volume ratio]{\label{Evolution_size} From ref.
\cite{ER05}. Evolution of reported quality factors in
monocrystalline mechanical resonators with size, showing a decrease
with linear dimension, i.e., with increasing surface-to-volume
ratio, indicating a dominant role of surface-related losses.}
\end{center}
\end{figure}

As mentioned previously, the structures obtained through current
methods present a certain degree of roughness of $\sim10$ nm. On top
of this, adsorbed impurities, dislocations and surface
reconstruction processes are all coupled to the local atomic
displacements linked to the resonator's motion, absorbing energy
from the vibrational eigenmode and transmitting it irreversibly to
the rest of degrees of freedom of the system. These imperfections
and impurities also exist in the core of the resonator structure,
but to a lesser degree as in the case of its surface, exposed to the
environment and with atoms less bound than in the bulk, more
susceptible to thermal rearrangements. Strong evidence has been
accumulated of the increasing role played by surfaces as the
resonator sizes shrink and surface-to-volume ratios grow, being the
dominating friction source in thin nanobeams at low temperatures: i)
The quality factor $Q$ roughly scales linearly with the former ratio
\cite{Yetal00,ER05}, as shown in fig.(\ref{Evolution_size}), ii) $Q$
has been observed to increase up to an order of magnitude when the
system went through a thermal treatment \cite{YOE00,WOE04,LVSLHP05}
or the surface was deoxidized \cite{YOE02}. In the next chapter we
will study in depth these surface-related losses, dominant for thin
beams at low temperatures, precisely the interesting regime where
quantum effects show up if a very high $Q$ is reached. Surface waves
can be also excited by the vibrational eigenmode due to the
roughness, but at low temperatures this process becomes strongly
suppressed \cite{Metal02}, and thus will be ignored in our study.

\subsubsection{Generation of e-h pairs in the metallic electrode due to electron-phonon coupling}
In the magnetomotive scheme, where the semiconductor bridge is covered with a thin metallic top layer, the flowing electrons
feel the Coulomb potential generated by static charges in the device and substrate. When the oscillating part is set into motion,
this potential is time-dependent, and gives the chance to the electrons to absorb part of the mechanical energy stored
in the eigenmode, creating e-h pairs. We will study this mechanism in the next chapter.

\subsection{Intrinsic mechanisms}
\subsubsection{Thermoelastic relaxation}
\begin{figure}
\begin{center}
\includegraphics[width=8cm]{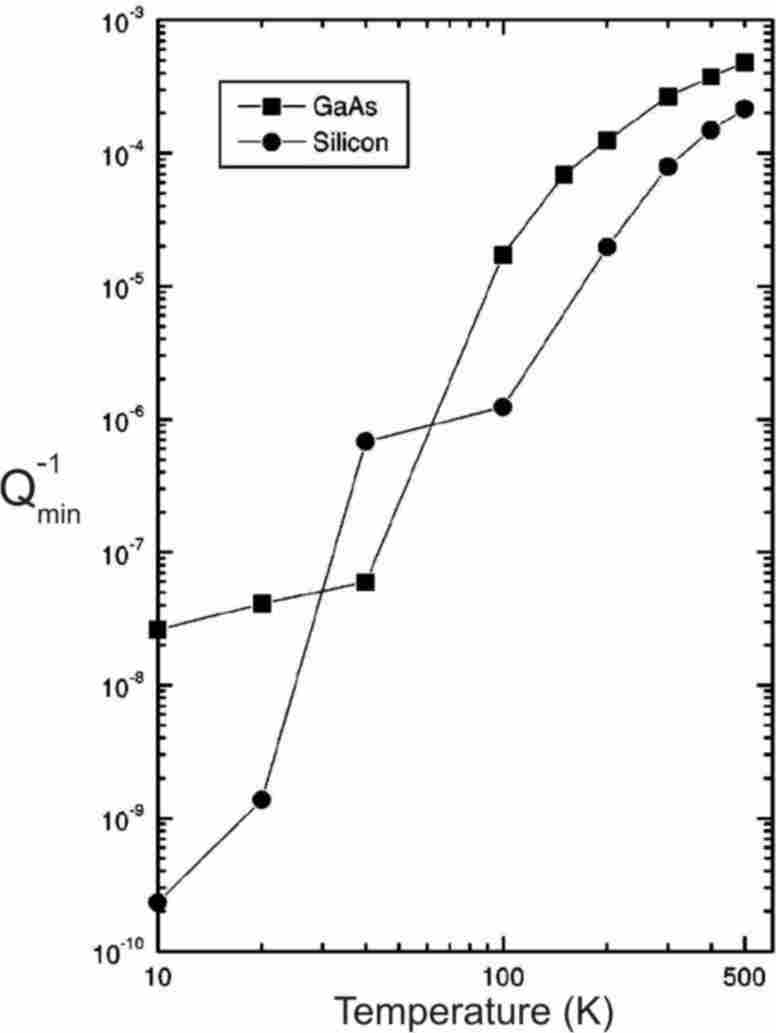}\\
\caption[Limits to Q imposed by the thermoelastic mechanism]{\label{thermodamp} From ref.
\cite{LR00}. Minimum dissipation $Q^{-1}_{min}$ imposed by the thermoelastic damping mechanism, as a
function of temperature, for any resonator made from GaAs or Si.}
\end{center}
\end{figure}

The strain field generated by the excited eigenmode modifies the local thermal equilibrium, with the coupling to the
local temperature field characterized by the linear expansion coefficient $\alpha=(1/L)dL/dT$. This translates into the local
creation of thermal phonons by the long-wavelength eigenmode, which then diffuse to points with different temperatures leading
to dissipation \cite{Z38,LR00,C02,U06}. This mechanism is strongly temperature dependent, being suppressed at low temperatures and dominating
at high ones. In \cite{LR00} a detailed study of this mechanism is carried out and the limits to the quality factor were established for
any resonator made with Si or GaAs, two of the commonest materials, see fig.(\ref{thermodamp}).

\subsubsection{Anharmonic mode coupling (phonon-phonon interactions)}
When the nanoresonator is strongly driven the harmonic approximation for the lattice potential breaks down, and higher order terms
which couple the former vibrational eigenmodes have to be included, giving a finite lifetime to a given externally excited mode.
At a temperature $T$, the externally excited mode will interact with thermal phonons, loosing its energy \cite{C02,U06}.
We will assume that the experiments are done at low excitations amplitudes, within the linear regime, where this effect can be
neglected. Nevertheless, it is worth mentioning the exciting physics and potential applications of strongly driven nanoresonators,
currently a subject of intensive research \cite{katz:040404,santamore:144301,URA04,BM05,SHN06,SIM07}.
\subsubsection{Surface reconstruction and amorphization}
Even if the fabrication processes and materials used were perfect, purely thermodynamic considerations imply that at
a temperature $T>0$ the sharp edges of a finite crystal begin to round off due to the increasing importance of entropy,
in a process called creation of vicinal surfaces \cite{W99}. The associated rearrangement of atoms and coupling of vibrational
modes limit the maximum quality factor attainable.\\

After this brief overview of semiconductor nanoresonators and friction mechanisms affecting them, we will focus our attention in Chapter \ref{chr2}
on the theoretical modeling of surface-related dissipation at low temperatures, borrowing ideas from sound attenuation in amorphous glasses.


\newpage
\cleardoublepage
\chapter{Surface dissipation in semiconductor NEMS at low temperatures}\label{chr2}
\section{Introduction}
\begin{figure}
\begin{center}
\includegraphics[width=10cm]{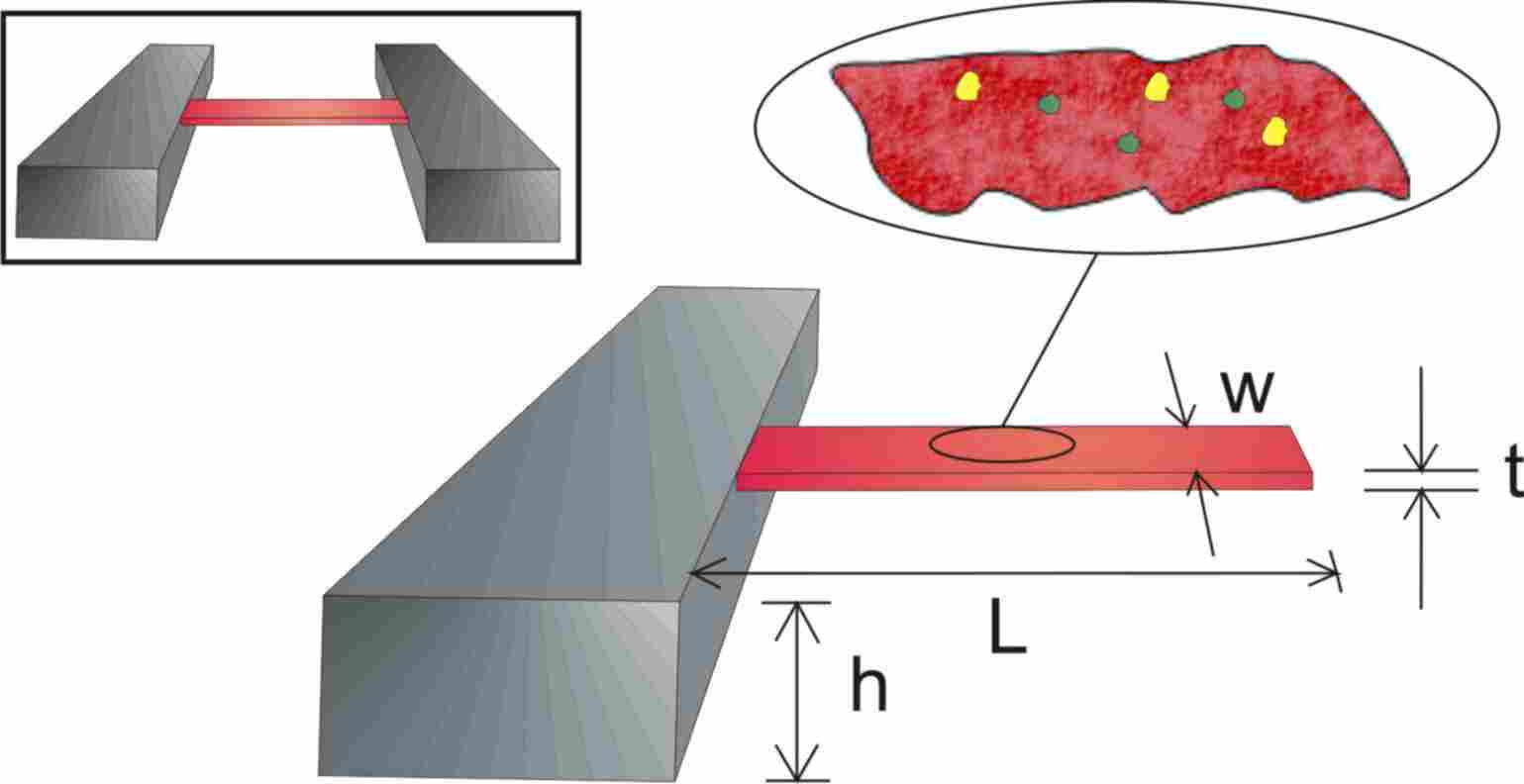}\\
\caption[Sketch of the nanoresonators with surface imperfections
limiting $Q$]{\label{sketch_resonator} Sketch of the systems whose
surface dissipation will be studied in this chapter. The inset shows
a doubly clamped beam, while the main figure shows a cantilever
characterized by its dimensions, width ($w$), thickness ($t$), and
length, ($L$), where $w \sim t \ll L$. The height above the
substrate is $h$. A schematic view of the surface is given,
highlighting imperfections like roughness and adsorbates, which
dominate dissipation at low temperatures.}
\end{center}
\end{figure}

In this chapter we will try to provide a unified theoretical
framework to describe the processes taking place at the surface of
nanoresonators at low temperatures, which have been observed to
dominate dissipation of their vibrational eigenmodes in this regime,
see fig.(\ref{sketch_resonator}). This is certainly a challenging
task, since as we have seen many different dynamical processes and
actors come into play (excitation of adsorbed molecules, movement of
lattice defects or configurational rearrangements), some of whom are
not yet well characterized, so simplifications need to be made. We
will provide such a scheme, based on the following considerations:
\begin{itemize}
  \item Experimental observations indicate that surfaces of otherwise
monocrystalline resonators acquire a certain degree of roughness,
impurities and disorder, resembling an amorphous structure
\cite{LTWP99,W99}.
  \item In amorphous solids the damping of acoustic waves
at low temperatures is successfully explained by the Standard
Tunneling Model \cite{AHV72,P72,P87,E98}, which couples the acoustic
phonons to a set of Two-Level Systems (TLSs) representing the
low-energy spectrum of all the degrees of freedom able to
exchange energy with the strain field associated to the vibration.
\end{itemize}

We will describe the attenuation of vibrations in nanoresonators due
to their amorphous-like surfaces in terms of an adequate adaptation
of the Standard Tunneling Model. We start reviewing such model, and
some concepts and techniques used to analyze the spin-boson
hamiltonian appearing in it. Then we proceed to apply in the proper
way the Standard Tunneling Model to the non-bulk nanoresonator
geometry. Expressions for the frequency shift and dissipation in
terms of the inverse quality factor $Q^{-1}$ for flexural and
torsional modes will be given, showing the scaling with dimensions,
temperature and other relevant parameters of these systems. We will
see that qualitative agreement with experimental observations is
obtained, discussing limitations and possible modifications of the
scheme to reach quantitative fitting to experiments. For NEMS
covered with metallic electrodes the friction due to electrostatic
interaction between the flowing electrons and static charges in the
device and substrate will be also studied.

\section{Damping of acoustic waves in amorphous solids: Standard Tunneling Model}
The description and understanding of the properties of amorphous
solids, where most of the simplifications linked to an infinite
periodic structure are absent, has been full of unexpected results
\cite{Z98}. In the low temperature regime, pioneering experiments by
Zeller and Pohl \cite{ZP71} revealed at $T<1$ K striking differences
between the thermal properties of dielectric amorphous solids as
compared to those of crystalline solids. In a Debye dielectric
crystalline solid it is well known that the heat capacity and the
thermal conductivity due to phonons behave as $C\sim T^3$ and
$\kappa\sim T^3$, but in the case of amorphous dielectrics none of
them is true, instead a $C\sim T^{1.2}$ (and much bigger in value as
compared to a crystalline form of the material) and $\kappa\sim
T^{1.8}$ behavior is observed. Apart from this, a heat release as a
function of time is observed on long time scales: after a sample is
heated and then rapidly cooled down to some temperature $T_0$, in
adiabatic conditions the sample begins gradually to heat itself
afterwards. Moreover, measuring their acoustic properties at low
temperatures, which a priori were expected to be similar to those
found in crystalline solids due to the long-wavelength vibrations involved, the following
dependence on the acoustic intensity was obtained \cite{HASND72}: at low
intensities, as the temperature is lowered the mean free path of the
phonon decreases, while at high intensities a monotonous increase is
observed.\\

Several theoretical scenarios were developed to explain these low-temperature features, the most successful of whom was the
so-called \emph{Standard Tunneling Model}, proposed simultaneously by Phillips and Anderson \textit{et al.} \cite{AHV72,P72}.

\begin{figure}
\begin{center}
\includegraphics[width=13cm]{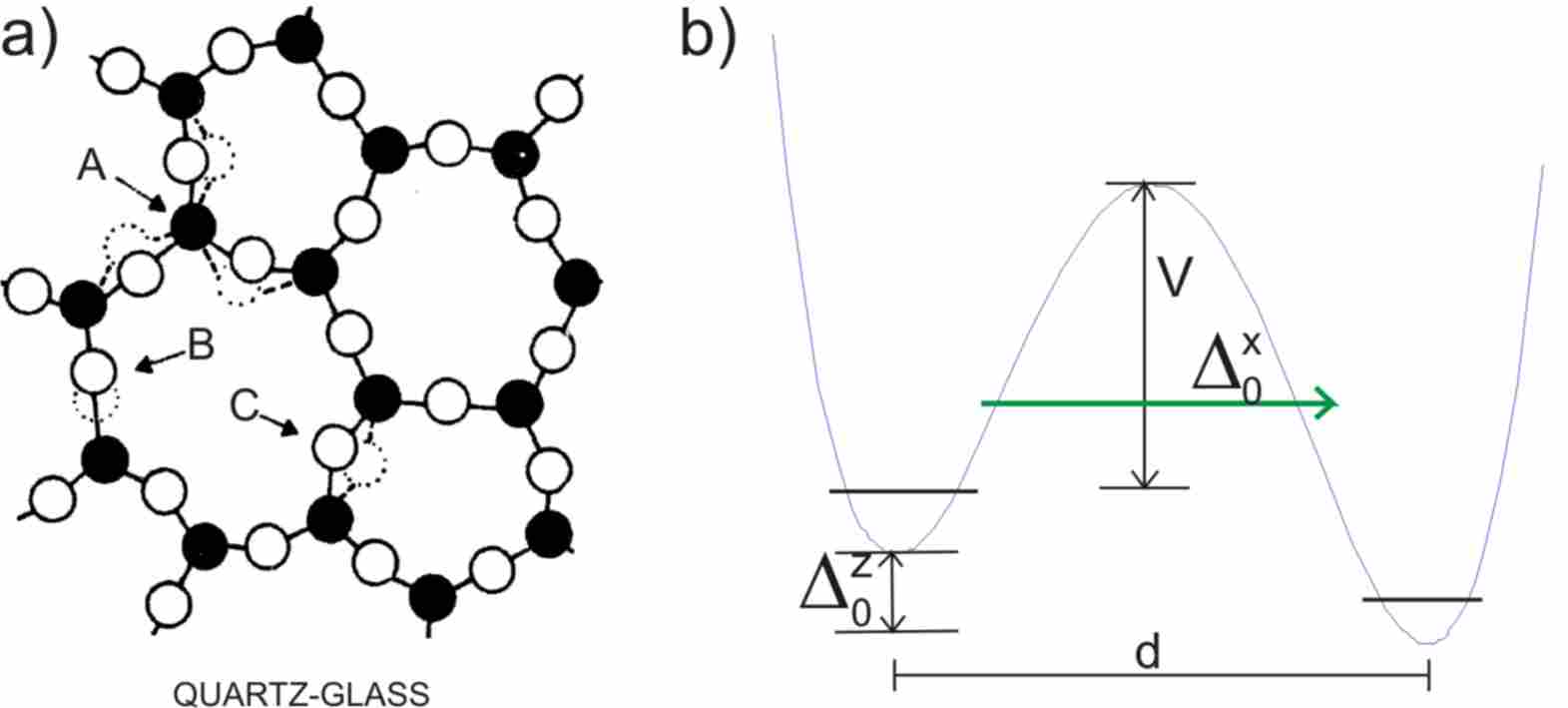}\\
\caption[Amorphous quartz structure with several Two-State candidates, and double well potential]{\label{Amorphous}
a) From \cite{JPAH76}. Schematic view of an amorphous quartz structure, showing three possible types of
two-state defects (A, B and C). b) Those two-states can be represented by a degree of freedom in a double well potential with an energy
barrier of height $V$, width $d$, and asymmetry between wells $\Doz$. At low temperatures only the ground states of both wells play a role in the
dynamics, becoming mixed by the tunneling through the barrier (green arrow), of amplitude $\Dox$. Thus the description is reduced to
that of a Two-Level System.}
\end{center}
\end{figure}

The tunneling model claims the existence of the following states, intrinsic to glasses: an impurity, an atom or cluster of
atoms within the disordered structure which present in their configurational space two energy
minima separated by an energy barrier (similarly to the dextro/levo
configurations of the ammonia molecule), as depicted schematically in fig.(\ref{Amorphous}a). These states can be modeled
as a degree of freedom tunneling between two potential wells. At low temperatures only the two lowest
eigenstates have to be considered, characterized by the bias $\Doz$
between the wells and the tunneling rate $\Dox$ through the barrier, see fig.(\ref{Amorphous}b).
This corresponds to the Two-Level System (TLS) hamiltonian:
\begin{equation}\label{TLShamiltonian}
    H_0=\Dox\sigma_x+\Doz\sigma_z\,\,,
\end{equation}
where $\sigma_i$ are Pauli matrices. Due to the disorder of the structure, a broad range of values for the TLSs parameters
$\Doz$ and $\Dox$ is expected to appear for the
ensemble of TLSs present in the amorphous solid. In order to be operative, the Standard Tunneling Model has to specify the properties
of this ensemble in terms of a probability distribution $P(\Dox,\Doz)$. The form of $P(\Dox,\Doz)$ can be
inferred from general considerations \cite{AHV72,P72}, as detailed in Appendices \ref{apDistribTLS} and \ref{apP0}, and is
furthermore supported by experiments \cite{E98}. The result is
\begin{equation}\label{TLSdistrib}
    P(\Dox,\Doz)=\frac{P_0}{\Dox}\,,
\end{equation}
with $P_0\sim10^{44}$J$^{-1}$m$^{-3}$, $\Dox>\Delta_{min}$
($\Delta_{min}$ fixed by the time needed to obtain a spectrum around
the resonance frequency of the excited eigenmode), and
$\varepsilon=\sqrt{(\Dox)^2+(\Doz)^{2}}<\varepsilon_{max}$, estimated to
be of the order of 5 K.

A last feature necessary to calculate the damping of acoustic waves caused by these low energy excitations
is the coupling TLSs - phonons. As justified in Appendix \ref{apCouplasymm}, the lattice distortions caused
by the propagation of the wave affect mainly the TLSs asymmetry $\Doz$, with the following final form for the hamiltonian
describing the coupled system TLSs + phonons:
\begin{equation}\label{H1}
    H=\sum_{k,j}\hbar\om_{k,j}b_{k,j}^\dag b_{k,j}+\sum_{\Dox,\Doz}\Bigl\{\Dox\sigma_x+\Bigl[\Doz+\sum_{k,j}
\lambda_{k,j} ( b_{k,j}^\dag + b_{k,j})\Bigr]\sigma_z\Bigr\}
\end{equation}
Here $b_{k,j}^\dag$ are the creation operators of the strain field waves $\textbf{u}(\textbf{r},t)$ whose components
$u_i(\textbf{r},t)$ obey the wave equations
\begin{equation}
    E\frac{\partial^2 u_i}{\partial x_{j}^{2}}=\rho\frac{\partial^2 u_i}{\partial t^2}
\end{equation}
The interaction hamiltonian is more specifically \cite{E98}
\begin{equation}\label{intham3D}
    H_{int}=\sigma_z \sum_{k_i,s} \lambda_{k_i,s}(b_{k_{i,s}}^{\dag}+b_{k_{i,s}})=
    \hbar \sigma_z \sum_{k_i,s} \Bigl [\gamma \sqrt{\frac{1}{2\rho L^3\hbar }} \frac{k_{i,s}}{\sqrt{\omega(k_{i,s})}}\Bigr ]
    (b_{k_{i,s}}^{\dag}+b_{k_{i,s}})\,,
\end{equation}
where $s$ is the polarization and $i$ the component of the wavevector. $L$ is the
lateral size of the sample, while $\gamma$ is the coupling constant, with an approximate
value $\gamma\sim1$ eV.

With the low energy TLS ensemble appearing in hamiltonian of eq.(\ref{H1}) characterized by a statistical distribution given by eq.(\ref{TLSdistrib}),
the strange thermal and acoustical properties of glasses at low temperatures mentioned in the beginning of this section
can be satisfactorily explained \cite{P87}: $C\sim T^{1}$ and $\kappa\sim T^{2}$ are obtained, close to the experimental
ones, whose discrepancies can be corrected by modifying slightly the
simplified probability distribution. The heat release experiments
are as well explained, because there is a broad distribution of
relaxation times for the initially thermally excited TLSs, and most
of them will relax very slowly, transferring gradually their energy to the
phonons of the solid, increasing its temperature. Concerning the
dependence of the phonon mean free path behavior on acoustic
intensity, it is explained by the saturation of the population of
excited TLSs, that eliminates the scattering mechanism of resonant
absorption and decay (to be explained later in this chapter) at high intensities. The scattering at higher
temperatures is dominated by non-resonant relaxational processes, also to be described later,
associated to the equilibration of the levels after a change in
population, which occurs at a rate resulting in a mean free path for
the phonon $l_{nres}\propto T^{-3}$.\\

The range of applicability of the Standard Tunneling Model is determined by the approximations involved.
The first one concerns the assumption that the TLSs interact only indirectly via their interaction with the phonons.
This is false below a certain temperature $T\sim50$ mK, when interactions of dipolar character among TLSs
cannot be disregarded \cite{E98,YL88,AM03,ERK04}. The high temperature limit of applicability of the model is
fixed by the moment when the TLS approximation, restricting the Hilbert space to the one spanned by the groundstates of the wells,
see fig.(\ref{Amorphous}b), is not valid anymore, at about $T\sim1$ K. The model also assumes weak intensities of the
acoustic waves excited, or otherwise anharmonic coupling among vibrational modes and higher order terms coupling
TLSs and phonons would become non-negligible.

\section{Theoretical analysis of the damping of phonons due to TLSs}
Given an externally pumped vibrational eigenmode of frequency $\wo$, we will divide its dissipation due to the presence of TLSs
into three parts: i) Resonant dissipation, ii) Friction caused by symmetric non-resonant TLSs, and iii) Relaxation processes due to
the bias $\Doz$.

For usual excitation amplitudes the first one can be ignored.
To obtain the energy dissipation rate corresponding to the other two mechanisms, we will proceed for each of them in two steps:
first we will study what is the effect of
a \emph{single} TLS, expressing the results in terms of its parameters $\Doz$, $\Dox$, and in a second step we will
\emph{sum} the individual contributions to dissipation \emph{over all the TLS ensemble}, characterized by the distribution (\ref{TLSdistrib}).

\subsection{Resonant dissipation}\label{secresdisip}
Those TLSs with their unperturbed excitation energies close to $\wo$
will resonate with the mode, exchanging energy quanta, with a rate
proportional to the mode's phonon population $n_{\wo}$ to first
order. For usual excitation amplitudes $\sim0.1-1${\AA} the vibrational
mode is so populated (as compared with the thermal population) that
the resonant TLSs become saturated, and their contribution to the
transverse (flexural) wave attenuation becomes negligible,
proportional to $n_{\wo}^{-1/2}\,\,\,\,$ \cite{Aetal74}.

\subsection{Dissipation of symmetric non-resonant TLSs}
\subsubsection{Dissipation caused by a single TLS}\label{secspinboson}
We will analyze here the dynamical behavior of one TLS coupled to the phonon ensemble:
\begin{equation}\label{H2}
    H=\Dox\sigma_x+\Doz\sigma_z+ \sum_{k,j}
\lambda_{k,j}\,\, \sigma_z( b_{k,j}^\dag +
b_{k,j})+\sum_{k,j}\hbar\om_{k,j}b_{k,j}^\dag b_{k,j}
\end{equation}
This hamiltonian is nothing but the well-known \textit{spin-boson model}, which applies to a great variety of
physical phenomena related to open dissipative systems, and has been subject of intensive
theoretical work \cite{Letal87,W99}. We will use some of this knowledge to describe the dynamics of the
single TLS in terms of the spectral function
\begin{equation}\label{spectral_TLS}
A ( \omega ) \equiv \sum_n \left| \left\langle 0 \left|\sigma_z \right| n  \right\rangle \right|^2 \delta ( \omega - \omega_n + \omega_0 )
\end{equation}
where $| n \rangle$ is an excited state of the total system TLS
plus vibrations. We will see how this function describes the absorption properties of a symmetric TLS
coupled to a vibrational ensemble through the operator $\sigma_z$.

The analysis of the spin-boson model for a TLS with asymmetry $\Doz\neq0$ is much more cumbersome than the symmetric TLS case,
if one tries to use the kind of techniques described in \cite{Letal87} that work for the symmetric case.
Thus, to learn about the effect of the asymmetry $\Doz$ in the absorption spectrum of the TLS we will
make a different kind of analysis, studying the delayed response of a TLS due to its bias $\Doz$, and
finally see its signatures in the absorption spectrum. A natural division of dissipative mechanisms
will arrive as a consequence of this analysis: one contribution, the relaxation mechanism, will correspond
to the presence of the bias, while the other will describe the dissipation occurring independently of the
presence or absence of bias.\\

The techniques we will use for the analysis of the symmetric case
are all described in \cite{Letal87}, mainly i) Simple
renormalization-group arguments to obtain effective or "dressed"
values $\Delta^x_{\rm{eff}}$, $\Delta^z_{\rm{eff}}$ of the TLS
parameters from $\Dox$, $\Doz$ due to the interaction with high
energy vibrational modes, ii) Characterization of the interaction
with the different kinds of vibrational modes based on the criteria
derived in the non-interacting blip approximation (NIBA, see
Appendix \ref{appath} for a brief review on the basic concepts and
results), iii) Second-order perturbation theory in the interaction
term of (\ref{H2}). These techniques are adequate for the regime
where the Standard Tunneling Model holds, described just before the
beginning of this section, and work better for very low vibration
amplitudes of the externally excited vibrational mode. Even within
the linear regime of the mode, above a certain amplitude other
approaches designed to study strongly driven TLSs become more suited
to describe the dynamics of at least part of the ensemble of TLSs,
namely numerical methods as Quantum Monte Carlo \cite{EW92,LWW98} or
Numerical Renormalization Group \cite{BTV03,BLTV05,ABV06,WD07}, or
theoretical studies of the driven spin-boson model
\cite{GSSW93,GH98}. We will assume that the TLSs in resonance with
the externally excited mode are saturated and thus do not contribute
significantly to its dissipation, while for the rest of TLSs the
vibration is assumed to be sufficiently weak for the techniques i),
ii) and iii) to hold.

\subsubsection{Spin-boson model for an almost symmetric TLS interacting with the nanoresonator modes}\label{sbonebath}
Now we take a TLS with $\Doz\ll\Dox$. In principle it is coupled to
three kinds of vibrational modes, those described in section
\ref{vibmodesbeam}.  The analysis of the resulting spectral function
(\ref{spectral_TLS}) would be greatly simplified if we could somehow
compare the effect of the different kinds of modes and conclude that
two of them have a negligible influence as compared with the third.
Fortunately the path integral NIBA approach developed by Leggett
\textit{et al.} \cite{Letal87} (see Appendix \ref{appath}) reveals
that the degree of influence of a given bosonic bath
$\sum_k\hbar\om_kb_k^\dagger b_k$ on the dynamics of a TLS coupled
to it through $H_{\rm{int}} \equiv \sigma_z \sum_k \lambda_k \left(
b_k^\dag + b_k \right)$ for low temperatures and small enough
coupling $\lambda_k$, is completely determined by the spectral
density of the bath
\begin{equation}\label{Specdens}
    J ( \omega ) \equiv \int_{-\infty}^{\infty}dt\,e^{i\om t}\langle H_{int}(t)H_{int}(0)\rangle=
    \sum_k\left| \lambda_k \right|^2 \delta (\omega - \omega_k )
\end{equation}
This function corresponds to the spectral decomposition of the operator $H_{\rm{int}}$, telling us the
rate at which transitions between the two eigenstates of the isolated TLS accompanied by excitation of a phonon happen,
as a function of $\om$. The three kinds of modes coupled to our TLS are characterized by a $J(\om)\sim\om^s$. As explained in the end of
Appendix \ref{appath}, this result, obtained for $T=0$, changes at $T>0$ because the thermal energy increases the
number of possible initial and final states, and $J(\om)$ changes to $J(\om,T\gg\om)\sim T^s$.
Then, if temperature is lowered the modes with a $J(\om)$ with the lowest power $s$ (weakest $T$ dependence) will affect the TLS
dynamics more strongly than the others, which to a first approximation can be ignored.

Focussing on the case of a TLS coupled to the vibrations present in a rod, the starting hamiltonian is
\begin{equation}\label{Hamiltonian}
    \textsl{H}=\Dox\sigma_x+\Doz\sigma_z+\sigma_z\textsl{F}(\partial_i  u_j)+H_{\rm{elastic}}
\end{equation}
where $ \partial_i u_j $ is a component of the deformation gradient
matrix, $\textsl{F}$ is an arbitrary function, and $H_{\rm{elastic}}$ represents the elastic energy stored due
to the atomic displacements, which will correspond in second quantization to the last term in eq.(\ref{H2}). Changing basis to
the energy eigenstates of the TLS, eq.(\ref{Hamiltonian}) becomes
$\textsl{H}= \varepsilon\sigma_z +
    [(\Delta_{0}^{x}/\varepsilon)\sigma_x+(\Delta_{0}^{z}/\varepsilon)\sigma_z]\textsl{F}(\partial_i u_j)+H_{\rm{elastic}}$.
We are considering for the moment only slightly biased TLSs for whom
$\Delta_{0}^{z}\ll\Delta_{0}$, so the last term can be ignored. A
further expansion of \textsl{F} to lowest order in the displacement,
together with a $\pi/4$ rotation of the eigenbasis, leads to:
\begin{equation}\label{Hamiltonian2}
    \textsl{H}=\varepsilon\sigma_x + \gamma (\Delta_{0}^{x}/\varepsilon)\sigma_z \partial_i u_j+H_{\rm{elastic}}\,,
\end{equation}
where $\gamma$ is the coupling constant, assumed to be as in amorphous 3D solids $\gamma\sim1$ eV.

The atomic displacements can be decomposed into the normal
vibrational modes, which for the cuasi-1D case of nanorods are the
ones obeying eqs.(\ref{longitudinal}-\ref{bending}). The vibrational
modes of a beam with fixed ends have a discrete spectrum, but we
will approximate them by a continuous distribution. This
approximation will hold as long as many vibrational modes become
thermally populated, $kT\gg\hbar\omega_{fund}$, where
$\omega_{fund}$ is the frequency of the lowest mode. The condition
is fulfilled in current experimental setups. The derivation of the
second quantized version of the sort of eq.(\ref{H2}) for each of
the modes and the corresponding spectral functions $J(\om)$ are
described in Appendix \ref{DerivJ}. Here we state the main results:

The compression and twisting modes lead to an ohmic spectral
function for $\omega \ll 2\pi c/R$ ($R$ being a typical transversal
dimension of the rod and $c$ the sound velocity), when the rod is
effectively 1D. In terms of the Young modulus of the material, $E$,
and the mass density, $\rho$, we get: $J_{{\rm comp}}( \omega ) =
\alpha_c |\omega|$, where,
\begin{equation}\label{alflongitudinal}
    \alpha_c= \gamma^2(2 \pi^2 \rho t w )^{-1} (E/\rho)^{-3/2} \, .
\end{equation}
A factor $(\Dox/\varepsilon)^2$
has been thrown away, as we consider almost symmetric TLS with $\Dox\sim\varepsilon$. We will proceed similarly with the other modes.
The twisting modes are defined by the torsional rigidity, $C = \mu
t^3w/3$ ($\mu$ is a Lande coefficient), and $I=\int dS x^2= t^3w/12$
(where $S$ is the cross-section). The corresponding spectral
function is given by: $J_{{\rm torsion}}(\omega) = \alpha_t
|\omega|$, where
\begin{eqnarray}\label{alftorsional}
\alpha_t = \gamma^2C (8\pi^2 \mu t w  \rho I)^{-1} (\rho I/C)^{3/2} \, .
\end{eqnarray}

The analysis of the flexural (bending) modes differs substantially
from the other ones, because they correspond to two fields that do not follow usual wave equations. The
normal modes have a quadratic dispersion,  $\omega_{j}(k) = \sqrt{E I_j/(\rho tw)} \, k^2$, leading to a \emph{sub-ohmic}
spectral function \cite{Letal87},
\begin{equation}\label{Jsub}
   \boxed{ J_{{\rm flex}}(\omega) = \alpha_b\sqrt{\wco} \sqrt{\omega}}\,\,,
\end{equation}
with,
\begin{eqnarray}\label{Jsubohmic}
 \alpha_b \sqrt{\wco} =
 0.3\frac{\gamma^2}{t^{3/2}w}\frac{(1+\nu)(1-2\nu)}{E(3-5\nu)}
   \Bigl(\frac{\rho}{E}\Bigr)^{1/4} \, ,
\end{eqnarray}
where $\nu$ is Poisson's ratio and $\wco\simeq \sqrt{EI_y/(\rho
tw)}(2\pi/t)^2$ is the high energy cut-off of the bending modes (frequency for whom the corresponding wavelength
is of the order of $t$, thus indicating the onset of 3D behavior). This example of sub-ohmic spin-boson model
is the first result of this thesis to be stressed, because the problem of a TLS
interacting with a sub-ohmic environment is interesting in its own
right \cite{Letal87,KM96,W99,S03,VTB05,GW88,STG06,ChT06,K04,IN92},
and the systems studied here provide a physical realization.

The bending modes prevail over the other, ohmic-like, modes as a
dissipative channel at low energies. One may ask at what frequency
do the torsional and compression modes begin to play a significant
role, and a rough way to estimate it is to see at what frequency do
the corresponding spectral functions have the same value, $J_{{\rm
flex}}(\om*)=J_{{\rm comp,tors}}(\om*)$. The results are
$\om*\sim30(1+\nu)^2(1-2\nu)^2(E/\rho)^{1/2}/[t(3-5\nu)^2]$ for the
case of compression modes and
$\om*\sim300(1-2\nu)^2(E/\rho)^{1/2}/[t(3-5\nu)^2(1+\nu)]$ for the
torsional. Comparing these frequencies to the one of the onset of 3D
behavior, $\wco$, they are similar, justifying a simplified model
where only flexural modes are considered.
\begin{figure}[t]
  \begin{center}
\includegraphics[width=7cm]{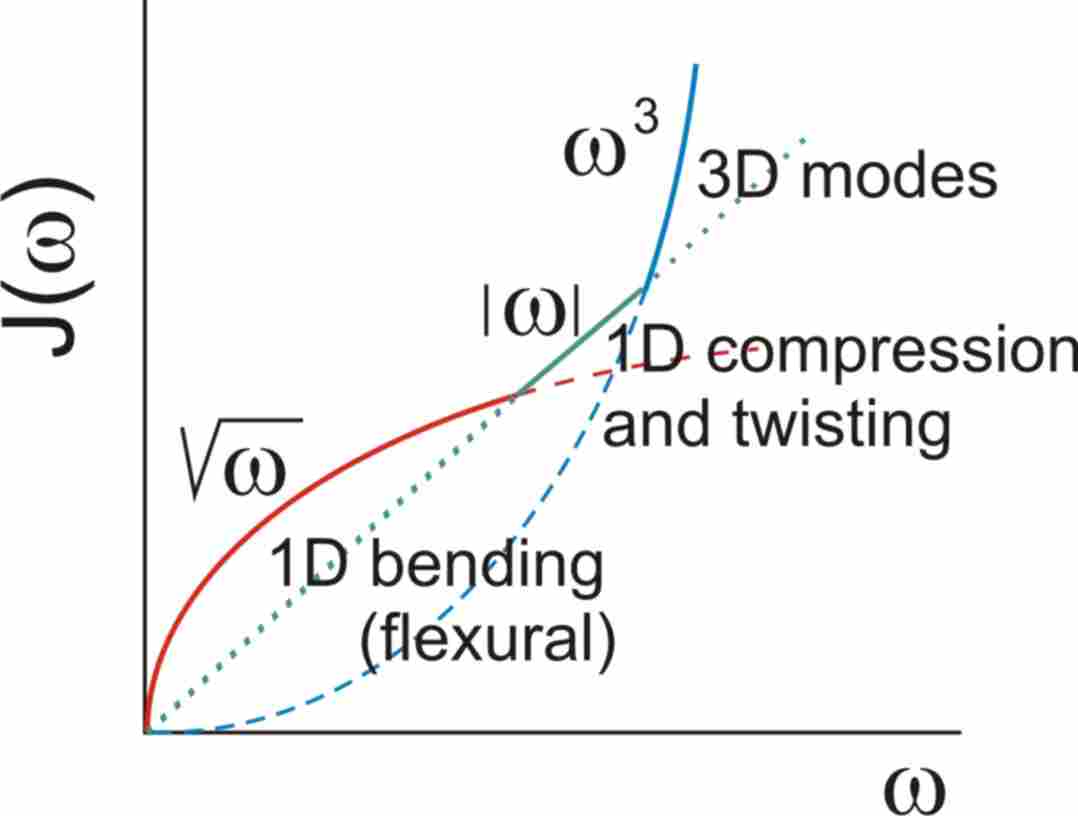}
\end{center}
    \caption{Sketch of the contributions to the spectral function which
      determines the dynamics of the TLS.}
    \label{spectral_function}
\end{figure}
Collecting the previous results, we find the spectral function
$J(\omega)$ plotted in Fig.[\ref{spectral_function}].

\subsubsection{TLS's dynamics}
\begin{figure}[!t]
\begin{center}
\includegraphics[width=11cm]{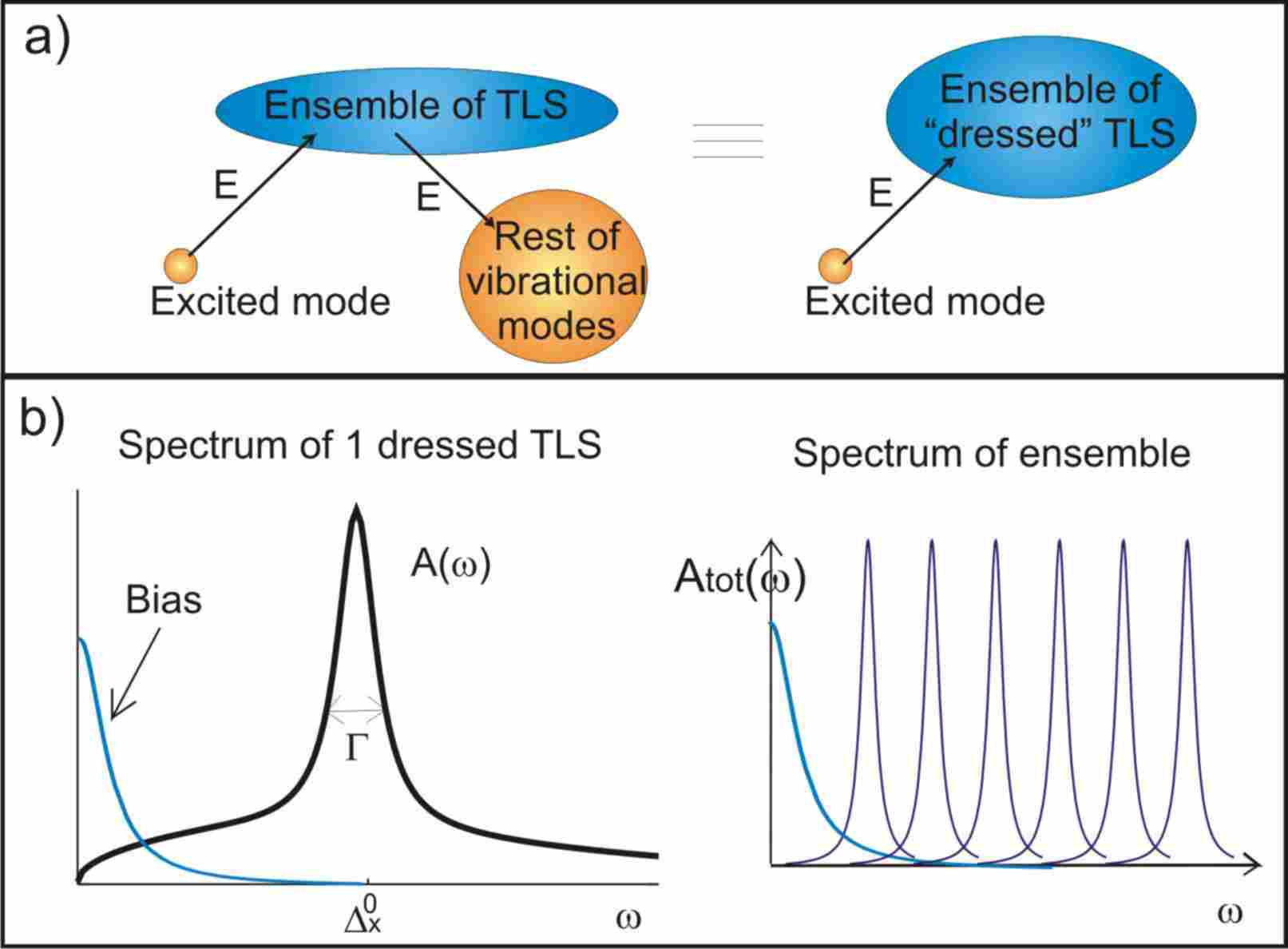}\\
\caption[Schematic view of the energy absorption due to TLSs, and representation of their spectral functions]
{\label{spectrum} a) Schematic
representation of the irreversible flow of energy from the
externally excited mode to the TLS ensemble, and from the ensemble
to the rest of vibrational modes. This process can be viewed as a
flow of energy from the excited mode to an ensemble of "dressed"
TLSs, with their dynamics modified by the presence of the
vibrational modes. b) Left: Spectral function $A(\om)$ of a single
dressed TLS, weakly damped ($\Gamma<\Dox$). A peak around $\om=0$
arises if the system is biased, corresponding to the relaxational
dissipation mechanism. Right: Total spectral function $A_{tot}(\om)$
of the ensemble of dressed TLSs.}
\end{center}
\end{figure}

The interaction between the bending modes and the TLSs affects both
of them. When a single mode is externally excited, as is done in
experiments, the coupling to the TLSs will cause an irreversible
energy flow, from this mode to the rest of the modes through the
TLSs, as depicted in fig.(\ref{spectrum}a). The dynamics of the TLSs
in presence of the vibrational bath determines the efficiency of the
energy flow and thus the quality factor of the excited mode. Taking
a given TLS plus the phonons, its dynamics is characterized by the
Fourier transform of the correlator
$\langle\sigma_z(t)\sigma_z(0)\rangle$, the spectral function
$A(\om)$, eq.(\ref{spectral_TLS}). We will see that, using second-order perturbation theory, the dissipation caused
by an ensemble of TLSs on a given externally excited vibrational mode of frequency $\wo$ is proportional
to $\sum_j A_j(\wo)$, with the index $j$ running over all the TLSs in the resonator, cf. eq.(\ref{FGRrate}).
We therefore study now $A(\om)$.\\

High-energy oscillators with $\om\gg\Dox$ adjust instantaneously to the value of $\sigma_z$, "dressing" the
wells' eigenfunctions and decreasing the overlap between them, reducing accordingly the tunneling amplitude $\Dox$ \cite{Letal87}.
This adiabatic renormalization of $\Dox$ can either reduce $\Dox$ to a finite value or suppress it completely.
For sub-ohmic coupling, eq.(\ref{Jsubohmic}), the self-consistent equation for the renormalized value $\Dren$ is:
\begin{equation}
\Dren = \Dox \exp\{- \alpha_b \sqrt{\wco}
\smallint^{\omega_{co}}_{\Dren} d \omega J (\omega)/\omega^2\}
\end{equation}
This equation has no solutions other than $\Dren = 0$ if
 $\Dox \ll \alpha_b^2 \wco$, so that the tunneling amplitude of the low energy TLSs
is strongly suppressed\cite{SD85,Letal87,W99,KM96}. The remaining
TLSs experience a shift and a broadening of the spectral function
function $A ( \omega )$. For a typical Si nanoresonator with $L\sim1\mu$m, $t\sim w\sim100$ nm, $\alf\ll1$,
and the shift is small, $\Dren\sim\Dox$, and will be neglected in the following.

In addition, $A ( \omega )$ acquires low and high energy tails, whose derivation can be found in
Appendix \ref{Aoffresonance}. At zero temperature the low energy tail is:
\begin{equation}
A ( \omega ) \propto
\alpha_b \frac{\sqrt{\wco \omega}}{ (\Dox)^2} \, \, \, \, \, \, \, \, \omega
\ll \Dox\,\,,
\label{off_res_TLS}
\end{equation}
while the high energy part takes the form $A ( \omega ) \propto \alpha_b
\sqrt{\wco} (\Dox)^2 \omega^{-7/2}$, for  $\omega \gg \Dox$. The
main features of $A ( \omega )$ are shown in fig.(\ref{spectrum}b). Finally, we obtain the width of
the resonant peak, $\Gamma ( \Dox )$, using Fermi's golden rule,
 $\Gamma(\Dox)= 16 \alpha_b\sqrt{\wco} \, \sqrt{\Dox}$.
This description is valid for wavelengths such that $1/L \ll k \ll
1/{\rm max}(w,t)$. For a biased TLS, the value of $\Doz$ is not renormalized by the phonons.

As mentioned before, the coupled system TLSs + vibrations can be
viewed, taking the coupling as a perturbation, from the point of
view of the excited mode $\wo$ as a set of TLSs with a modified
absorption spectrum. The TLSs, dressed perturbatively by the modes,
are entities capable of absorbing and emitting over a broad range of
frequencies, transferring energy from the excited mode $\wo$ to
other modes. The contribution to the value of the inverse of the
quality factor, $Q^{-1}(\wo)$, of all these non-resonant TLSs will
be proportional to
$A_{\rm{off-res}}^{\rm{tot}}(\wo)=\sum_{\Dox=0}^{\wo-\Gamma(\wo)}A(\Dox,\Doz,\wo)+
\sum_{\wo+\Gamma(\wo)}^{\e_{max}}A(\Dox,\Doz,\wo)\approx
2P\alf\sqrt{\wco/\wo}$ , a quantity measuring the density of states
which can be excited through $H_{\rm{int}}$ at frequency $\wo$, see
Appendix \ref{Aoffresonance} for details.

For an excited mode $\wo$ populated with $n_{\wo}$ phonons,
$Q^{-1}(\wo)$ is given by
\begin{equation}
    Q^{-1}(\wo)=\frac{\Delta E}{2\pi E_0}\,\,,
\end{equation}
where $E_0$ is the energy stored in the mode per unit volume, $E_0\simeq
n_{\wo}\hbar\wo / twL$, and $\Delta E$ is the energy fluctuations
per cycle and unit volume. $\Delta E$ can be obtained from Fermi's
Golden Rule:
\begin{equation}\label{FGRrate}
    \Delta E_{\rm{off-res}}^{\rm{tot}}\simeq \frac{2\pi}{\wo}\times\hbar\wo\times\frac{2\pi}{\hbar}n_{\wo}
    \left(\lambda \frac{k_{0}^{2}}{\sqrt{\wo}}\right)^2 A_{\rm{off-res}}^{\rm{tot}}(\wo)\,,
\end{equation}
and the inverse quality factor of the vibration follows. For finite
temperatures the calculation of $A_{\rm{off-res}}^{\rm{tot}}(\wo,T)$
is done in Appendix \ref{Aoffresonance}. It has to be noted that in experiments the \emph{observed linewidth}
is due to the total amount of \emph{fluctuations}, so at a finite temperature it corresponds to the addition of
emission and absorption processes. Thus in this context dissipation
means fluctuations, and not net loss of energy. The combined contribution of
these processes is
\begin{equation}\label{Mainresult}
 \boxed{(Q^{-1})_{\rm{off-res}}^{\rm{tot}}(\omega_0,T) \simeq 10 P_0t^{3/2} w\left(\frac{E}{\rho}\right)^{1/4} \frac{\alf^2\wco}{\wo}
 \coth\Bigl[\frac{\hbar\wo}{k_BT}\Bigr]}
\end{equation}
Before we continue with the third dissipative mechanism due to TLSs, a couple of words about the importance
of multi-phonon processes is in order: Until now prevalence of one-phonon processes in the interaction
among TLSs and vibrational modes has been assumed, but at
temperatures much higher than the frequencies of the relevant
phonons, multi-phonon processes should be taken into account. A useful indicator to estimate if one-phonon
processes suffice to describe the interaction between the TLS and the bath is the Fermi's Golden Rule result for the linewidth:
if the linewidth $\Gamma(\Dox,T)$ calculated in this way is much smaller than $\Dox$, the approximation can be taken as
good, while for $\Gamma(\Dox,T)\geq\Dox$ multiphonon processes have to be considered. The expression for the linewidth
at a finite temperature is
\begin{equation}\label{LinewidthT}
    \Gamma(\varepsilon,T)=16 \alpha_b\sqrt{\wco} \, \sqrt{\varepsilon}\coth\Bigl[\frac{\varepsilon}{2k_BT}\Bigr]
\end{equation}
One can estimate, using the probability distribution (\ref{TLSdistrib}), the
total number of overdamped TLSs, $\Gamma(\varepsilon,T)\geq\Dox$, in the volume fraction of the
resonator presenting amorphous features, $V_{amorph}$. With
$V_{amorph}\sim V_{tot}/10$, and using
$\Gamma(\varepsilon,T)\approx2T\Gamma(\varepsilon,T=0)/\varepsilon$, the
number of overdamped TLSs, $\varepsilon \leq \Gamma(\varepsilon,T)\rightarrow \varepsilon
\leq [30 \alf\sqrt{ \wco} T]^{2/3}$, is $N\approx P_0twL[30
\alf\sqrt{ \wco} T]^{2/3}$, which for typical resonator sizes
$L\sim1\mu$m, $t,w\sim0.1\mu$m is less than one for $T<1$ K.
Therefore, unless the resonator is bigger and/or $P_0$ too, the TLSs can be assumed to be underdamped.

\subsection{Contribution of biased TLSs to the linewidth: relaxation absorption}
This very general friction mechanism arises due to the phase delay
between stress and imposed strain rate. A detailed derivation for the case of
TLSs in amorphous solids can be found in Appendix \ref{aprelax}. In our context, for a given
TLS the populations of its levels take a finite time to readjust
when a perturbation changes the energy difference between its
eigenstates \cite{J72,E98}. This time $\tau$ corresponds to the
inverse of the linewidth (\ref{LinewidthT}). For a
coupling as the one described by $H_{int}$, a component $k$ of the
deformation gradient matrix, $\partial u_k\sim\langle k,
n_k|(k/\sqrt{\om_k})(\adk+a_k)|k, n_k\rangle$ associated to a
vibrational mode $|k,n_k\rangle$ modifies the asymmetry $\Doz$
between the energy levels
$\varepsilon_{1,2}=\mp\frac{1}{2}\sqrt{(\Dox)^2+(\Doz+\xi_k
\partial u_{k})^2}$ ($\xi_k$ the corresponding coupling constant with
the proper dimensions). So to first order the sensitivity of these
energies to an applied strain is proportional to the bias
\begin{equation}\label{Eofgamma}
    \frac{\partial \varepsilon^{i}}{\partial (\partial u_{k})}=\frac{(\Doz\mp\xi_k \partial u_{k})}{\varepsilon^{i}}\xi_k
    =\frac{\Delta_{z}^{tot}}{\varepsilon^{i}}\xi_k\,\,\,,
\end{equation}
and the response of the TLS will be $\propto
(\Delta_{z}^{tot}/\varepsilon)^2$. As derived in Appendix \ref{aprelax}, the imaginary part of the response, corresponding to
$Q^{-1}$ is also $\propto \tau/(1+\om^2\tau^2)$, which in terms of
$A(\om)$ is the lorentzian peak at $\om=0$ of fig.(\ref{spectrum}b).

\begin{figure}[!t]
\begin{center}
\includegraphics[width=10cm]{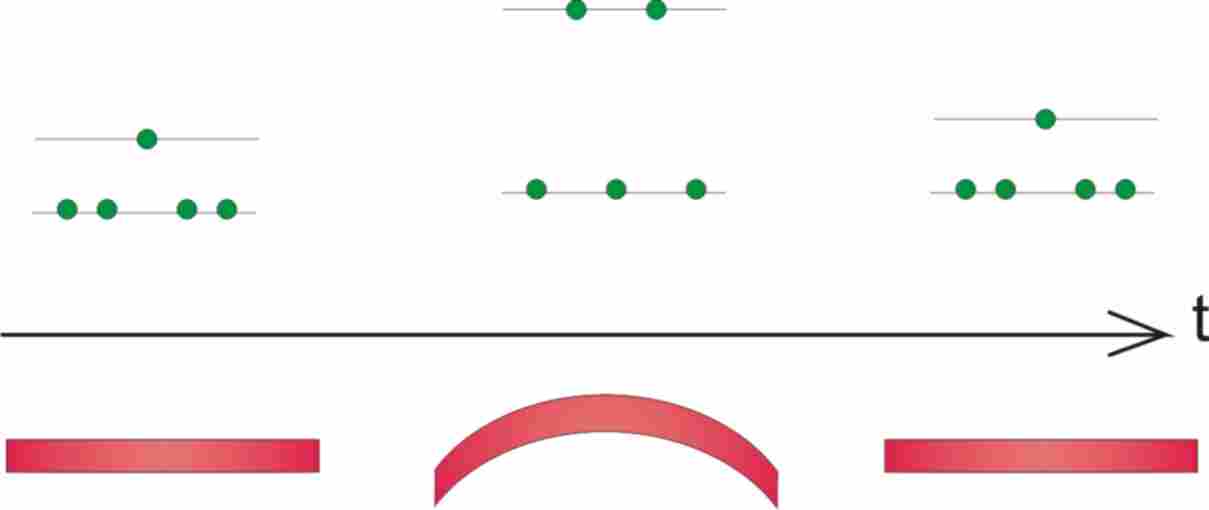}\\
\caption[Schematic representation of the effect of delay in the relaxation mechanism]{\label{relaxation_delay} Schematic
representation of the levels' population evolution of an ensemble of
5 identical TLSs with a response time $\tau\sim1/\om$, where $\om$
is the frequency of the bending mode excited, whose evolution is
also depicted in the lower part of the figure. The delay $\tau$ plus
the bias $\Doz$ give rise to the relaxational energy loss mechanism
of the mode, see text.}
\end{center}
\end{figure}

The mechanism is most effective when $\tau\sim1/\om$, where $\om$ is
the frequency of the vibrational mode; then, along a cycle of
vibration, the following happens (see fig.(\ref{relaxation_delay})):
When the TLS is under no stress the populations, due to the delay
$\tau$ in their response, are still being adjusted as if the levels
corresponded to a situation with maximum strain (and therefore of
maximum energy separation between them, cf.
$\varepsilon_{1,2}(u_k)$), so that the lower level becomes
overpopulated. As the strain is increased to its maximum value the
populations are still adjusting as if the levels corresponded to a
situation with minimum strain, thus overpopulating the upper energy
level. Therefore in each cycle there is a net absorption of energy
from the mechanical energy pumped into the vibrational mode.

If on the other hand $\tau\gg1/\om$ the TLSs levels' populations are
frozen with respect to that fast perturbation, while in the opposite
limit $\tau\ll1/\om$ the levels' populations follow adiabatically
the variations of $\varepsilon_{1,2}$ and there is neither a net absorption of
energy.

For an ensemble of TLSs, the contribution to $Q^{-1}$ is \cite{E98}
\begin{equation}\label{QEsq}
    Q^{-1}_{\rm{rel}}(\omega,T)=\frac{P_0\gamma^2}{E T}
    \smallint_{0}^{\varepsilon_{max}}d\varepsilon
\smallint_{u_{min}}^{1} du \frac{\sqrt{1-u^2}}{u}\frac{1}{\cosh^{2}\left(\varepsilon/2k_BT\right)}\frac{\omega\tau}{1+(\omega\tau)^2}
\end{equation}

The derivation of eq.(\ref{QEsq}) only relies in the assumptions of
an existence of well defined levels who need a finite time $\tau$ to
reach thermal equilibrium when a perturbation is applied, and the
existence of bias $|\Doz|>0$. This implies that such a scheme is
applicable also to our 1D vibrations, but is valid only if the
perturbation induced by the bath on the TLSs is weak, so that the
energy levels are still well defined. Therefore we will limit the
ensemble to underdamped TLSs for whom
$\hbar\Gamma(\varepsilon,T)<\varepsilon$. In eq.(\ref{QEsq}) the
factor $\cosh^{-2}\left(\varepsilon/2k_BT\right)$ imposes an
effective cutoff $\varepsilon<k_BT$, so that in
eq.(\ref{LinewidthT}) one can approximate
$\coth[\varepsilon/2k_BT]\sim2k_BT/\varepsilon$, resulting in
$\Gamma(\varepsilon)\sim1/\sqrt{\varepsilon}$, and thus the
underdamped TLSs will satisfy $\varepsilon \geq [30 \alf\sqrt{ \wco}
T]^{2/3}$. For $T\gg[32\alf\sqrt{\wco}]^2$, which is fulfilled for
typical sizes and temperatures (see Appendix \ref{Relanalysis} for
details):
\begin{equation}\label{QEsq2}
Q^{-1}_{\rm{rel}}(\wo,T)\approx\frac{20P_0\gamma^4}{t^{3/2}w}\frac{(1+\nu)(1-2\nu)}{E^2(3-5\nu)}\Bigl(\frac{\rho}{E}\Bigr)^{1/4}\frac{\sqrt{T}}{\wo}
\end{equation}
Here $V_{amorph}\sim V_{tot}/10$ was assumed.

\subsection{Comparison between contributions to $Q^{-1}$. Relaxation prevalence.}
\begin{figure}[!t]
\begin{center}
\includegraphics[width=14cm]{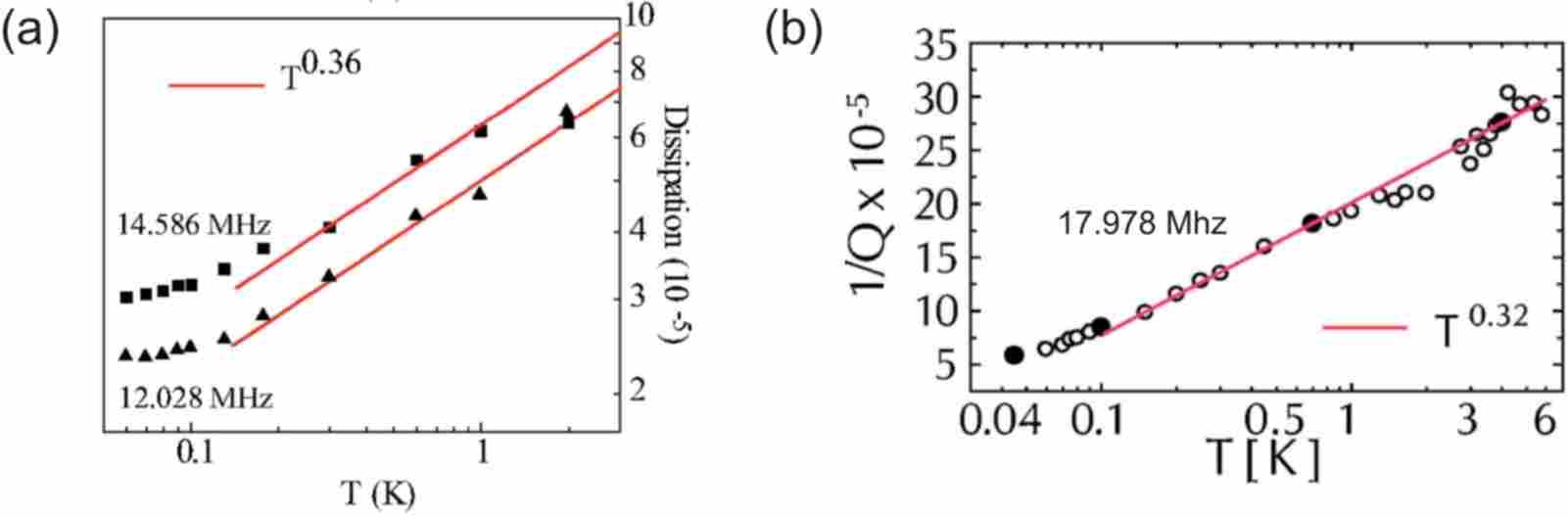}
    \caption[Low temperature dependence of Q for several doubly clamped beams]{Experimental results for the dissipation of
    the fundamental flexural (bending) mode of three different doubly-clamped resonators. The dissipation is expressed in terms of $Q^{-1}$, the
    inverse of the quality factor.(a) Results obtained in ref. \cite{ZGSBM05} for two silicon doubly clamped beams of dimensions
    $0.2\times0.3\times6\mu$m and $0.2\times0.3\times7\mu$m. Their fundamental bending modes, 14.586 and 12.028 MHz respectively,
    show a $Q^{-1}(T)\sim T^{0.36}$ optimum fit for $0.1<T<1$ K. (b) Results of ref. \cite{Mohetal07} for a similar type of beam
    made of GaAs, of dimensions
    $0.5\times0.5\times15\mu$m, exhibiting an power law dependence $Q^{-1}(T)\sim T^{0.32}$ for $0.08<T<3$ K.
    These behaviors have to be compared with our model, who predicts $Q^{-1}(T)\sim T^{1/2}$, see text for details.}
    \label{QMohanty}
\end{center}
\end{figure}

It is useful to compare the importance of the
contributions to $Q^{-1}$ coming from the last two mechanisms. For
that sake, we particularize the comparison to the case of the
fundamental flexural mode, which is the one usually excited and
studied, of a doubly clamped beam, with frequency
$\wo\approx6.5(E/\rho)^{1/2}t/L^2$ (for a cantilever these
considerations hold, with only a slight modification of the
numerical prefactors; the conclusions are the same). The result is:
\begin{equation}\label{Comparison1}
 \Bigl[\frac{Q^{-1}_{\rm{rel}}(T)}{Q^{-1}_{\rm{off-res}}(T)}\Bigr]_{\rm{fund}}\approx\frac{300t^{1/2}}{L}
 \frac{(3-5\nu)}{(1+\nu)(1-2\nu)}\frac{E}{\rho}\frac{1}{T^{1/2}}
\end{equation}
For a temperature $T=1$K, the result is as big as $10^6$ even for a
favorable case $t=1$nm, $L=1\mu$m, $E=50$GPa, $\nu=0.2$,
$\rho=3$g/cm$^3$. Then it can be concluded that for any reasonable temperature and dimensions
the dissipation is dominated by the relaxation mechanism, so that
the prediction of the limit that surfaces set on the quality factor
of nanoresonators is, within this model for the surface,
eq.(\ref{QEsq2}):
\begin{equation}\label{QEsq3}
 \boxed{Q^{-1}_{\rm{surface}}(\wo,T)\approx Q^{-1}_{\rm{rel}}(\wo,T)\sim\frac{T^{1/2}}{\wo}}
\end{equation}
For typical values $L\sim1\mu$m, $t,w\sim0.1\mu$m, $\gamma\sim5$ eV,
$P_0V_{amorph}/V_{tot}\sim10^{44}$J/m$^3$, and T in the range
1mK-0.5K the estimate for $Q^{-1}_{\rm{surface}}\sim10^{-4}$ gives
the observed order of magnitude in experiments like
\cite{ZGSBM05,Mohetal07}, and also predicts correctly a sublinear
dependence, but with a slightly higher exponent, $1/2$ versus the
experimental fit $0.36$ in \cite{ZGSBM05}, or $0.32$ in
\cite{Mohetal07}, see fig.(\ref{QMohanty}).

\section{Extensions to other devices}
\subsection{Cantilevers, nanopillars and torsional oscillators}
The extrapolation from doubly clamped beams to cantilevers
\cite{Metal05} and nanopillars \cite{SB04} is immediate, the only
difference between them being the allowed (k,$\om$(k)) values due to
the different boundary conditions at the free end (and even this
difference disappears as one considers high frequency modes, where
in both cases one has $k_n\approx(2n+1)\pi/2L$). All previous
results apply, and one has just to take care in the expressions
corresponding to the $Q^{-1}$ of the fundamental mode, for whom there
is more difference between the frequencies of both cases, the
cantilever one being $\wo^{cant}\approx(E/\rho)^{1/2}t/L^2$ as
compared to the doubly clamped case,
$\wo^{clamped}\approx6.5(E/\rho)^{1/2}t/L^2$.

\subsection{Effect of the flexural modes on the dissipation of torsional
  modes}
The contribution from the TLSs + subohmic bending mode environment
to the dissipation of a torsional mode of a given oscillator can be
also estimated. We will study the easiest (and experimentally
relevant\cite{SRCR04}) case of a cantilever. For paddle and double
paddle oscillators the geometry is more involved, modifying the
moment of inertia and other quantities. When these changes are
included, the analysis follows the same steps we will show.

\textbf{Relaxation absorption.} We assume, based on the previous
considerations on the predominant influence of the flexural modes on
the TLSs dynamics, as compared with the influence of the other
modes, that the lifetime $\tau=\Gamma^{-1}$ of the TLSs is given by
eq.(\ref{LinewidthT}). The change in the derivation of the
expression for $Q^{-1}$ comes in eq.(\ref{Eofgamma}), where the
coupling constant $\xi_k^{\rm{tors}}$ is different, which translates
simply, in eq.(\ref{QEsq}), into substituting
$\gamma^2\leftrightarrow\gamma_{\rm{tors}}^2$, and the corresponding
prediction for $Q^{-1}_{\rm{rel}}$
\begin{equation}\label{QEsqtors}
\boxed{Q^{-1}_{\rm{rel}}(\wo,T)\approx\frac{20P_0\gamma_{\rm{tors}}^2\gamma^2}{t^{3/2}w}\frac{(1+\nu)(1-2\nu)}{E^2(3-5\nu)}
\Bigl(\frac{\rho}{E}\Bigr)^{1/4}\frac{\sqrt{T}}{\wo}}\,,
\end{equation}
where now $\wo$ is the frequency of the corresponding flexural mode.
The range of temperatures and sizes for which this result applies is
the same as in the case of an excited bending mode.

\textbf{Dissipation of symmetric non-resonant TLSs} The modified
excitation spectrum of the TLS's ensemble,
$A_{\rm{off-res}}^{\rm{tot}}(\wo,T)$, remains the same, and the
change happens in the matrix element of the transition probability
of a mode $|k_0 , n_0\rangle$ appearing in eq.(\ref{FGRrate}),
$(\lambda k_{0}^{2}/\sqrt{\wo})^2$. The operator yielding the
coupling of the bath to the torsional mode which causes its
attenuation is the interaction term of the hamiltonian, which for
twisting modes is given by eq.(\ref{Hinttors}). In Appendix \ref{Tors} a detailed derivation of
eq.(\ref{E0tors}) is shown. The main results are stated here. Again, $Q^{-1}(\om_j)=\Delta E / 2\pi
E_0$, where the energy $E_0$ stored in a torsional mode
$\phi_j(z,t)=A\sin[(2j-1)\pi z/(2L)]\sin(\om_j t)$ per unit volume
is $E_0=A^2\om_j^2\rho (t^2+w^2)/48$ ($z$ is the coordinate along
the main axis of the rod). Expressing the amplitude $A$ in terms of
phonon number $n_j$, the energy stored in mode $|k_j , n_j\rangle$
is
\begin{equation}\label{E0tors}
    E_0(k_j,n)=\frac{1}{2}\frac{\hbar\om_j}{(t^3w+w^3t)L}(t^2+w^2)(2n_j+1)\,,
\end{equation}
the energy fluctuations in a cycle of such a mode is
\begin{equation}\label{DeltaEtors}
\Delta E
=\frac{2\pi}{\om_j}\times\hbar\om_j\times\frac{2\pi}{\hbar}\frac{\gamma^2\hbar\om_j}{16Ltw\mu}n_jA_{\rm{off-res}}^{\rm{tot}}(\om_j,T)\,,
\end{equation}
and the inverse quality factor
\begin{equation}\label{Qofftors}
    (Q^{-1})_{\rm{off-res}}^{\rm{tot}}(\wo,T)\approx0.04\frac{\gamma^4P}{t^{3/2}w}
  \frac{\rho^{1/4}(1+\nu)^2(1-2\nu)}{E^{9/4}(3-5\nu)}\frac{1}{\sqrt{\wo}}\coth\Bigl[\frac{\wo}{T}\Bigr]
\end{equation}
For sizes and temperatures as the ones used for previous estimates
the relaxation contribution dominates dissipation.
\section{Frequency shift}
Once the quality factor is known the relative frequency shift can be
obtained via a Kramers-Kronig relation (valid in the linear regime),
because both are related to the imaginary and real part,
respectively, of the acoustic susceptibility. First we will
demonstrate this, and afterwards expressions for beam and cantilever
will be derived and compared to experiments.
\subsection{Relation to the acoustic susceptibility}
In absence of dissipative mechanisms, the equation for the bending
modes is given by $-(12\rho/t^2)\partial^2 X/\partial
t^2=E\partial^4 X/\partial z^4$. The generalization in presence of
friction is
\begin{equation}\label{bendfriction}
    -\frac{12\rho}{t^2}\frac{\partial^2 X}{\partial t^2}=
(E+\chi)\frac{\partial^4 X}{\partial z^4}
\end{equation}
Where $\chi$ is a complex-valued susceptibility. Inserting a
solution of the form $X(z,t)=Ae^{i(kx-\om t)}$, where k is now a
complex number, one gets the dispersion relation
$\om=\sqrt{t^2(E+\chi)/(12\rho)}k^2$. Now, assuming that the
relative shift and dissipation are small, implying
$\rm{Re}(\chi)<<E$ , $\rm{Im}(k)<<\rm{Re}(k)$, the following
expressions for the frequency shift and inverse quality factor are
obtained in terms of $\chi$:
\begin{equation}\label{3dfreqshift}
    \left\{
      \begin{array}{ll}
        Q^{-1}&=\Delta\om/\om = -\rm{Im}(\chi)/E \\
        \delta\om/\om&=\rm{Re}(\chi)/2E
      \end{array}
    \right.
\end{equation}
Therefore a Kramers-Kronig relation for the susceptibility can be
used to obtain the relative frequency shift:
\begin{equation}\label{KK}
    \frac{\delta\om}{\om}(\om,T)=-\frac{1}{2\pi}P
\int_{-\infty}^{\infty}d\om'\frac{Q^{-1}(\om',T)}{\om'-\om}\,\,,
\end{equation}
where $P$ means here the principal value of the integral.
\subsection{Expressions for the frequency shift}
Relaxation processes of biased, underdamped TLSs dominate the
perturbations of the ideal response of the resonator, as we have
already shown for the inverse quality factor. For most of the
frequency range, $\om\geq [30 \alf\sqrt{ \wco} T]^{2/3}$,
$Q^{-1}(\om,T)\approx A\sqrt{T}/\om$ , with $A$ defined by
eq.(\ref{QEsq2}). The associated predicted contribution to the
frequency shift, using eq.(\ref{KK}), is
\begin{equation}\label{freqshift1}
    \boxed{\frac{\delta\om}{\om}(\om,T)\approx-\frac{A}{2\pi}\frac{\sqrt{T}}{\om}\log\Bigl[\Bigl|1-\frac{\om}{[30 \alf\sqrt{ \wco} T]^{2/3}}\Bigr|\Bigr]}
\end{equation}
For low temperatures, $\om\gg [30 \alf\sqrt{ \wco} T]^{2/3}$, the
negative shift grows towards zero as $\delta\om/\om
(\om,T)\sim-\sqrt{T}\log[\om/T^{2/3}]$, reaching at some point a
maximum value, and decreasing for high temperatures, $\om< [30
\alf\sqrt{ \wco} T]^{2/3}$, as $\delta\om/\om (\om,T)\sim
1/T^{1/6}$. Even though the prediction of a peak in $\delta\om/\om
(T)$ qualitatively matches the few experimental results currently
available \cite{ZGSBM05,Mohetal07}, it does not fit them
quantitatively.

\section{Applicability and further extensions of the model. Discussion}
As mentioned, the predictions obtained within this theoretical
framework do match qualitatively experimental results in terms of
observed orders of magnitude for $Q^{-1}(T)$, weak sublinear
temperature dependence, and presence of a peak in the frequency
shift temperature dependence. But quantitative fitting is still to
be reached, while on the experimental side more experiments need to
be done at low temperatures to confirm the, until now, few
results\cite{ZGSBM05,Mohetal07}.

\textbf{Applicability.} The several simplifications involved in the
model put certain constraints, some of which are susceptible of
improvement. We enumerate them first and discuss some of them
afterwards: i) The probability distribution $P(\Dox,\Doz)$, borrowed
from amorphous bulk systems, may be different for the case of the
resonator's surface, ii) The assumption of non-interacting TLSs,
only coupled among them in an indirect way through their coupling to
the vibrations, breaks down at low enough temperatures, where also
the discreteness of the vibrational spectrum affects our predictions
iii) When temperatures rise above a certain value, high energy
phonons with 3D character dominate dissipation, the two-state
description of the degrees of freedom coupled to the vibrations is
not a good approximation, and thermoelastic losses begin to play an
important role, iv) For strong driving, anharmonic coupling among
modes has to be considered, and some steps in the derivation of the
different mechanisms, which assumed small perturbations, must be
modified. This will be the case of resonators driven to the
nonlinear regime, where bistability and other phenomena take place.

The solution to issue i) is intimately related to a better knowledge
of the surface and the different physical processes taking place
there. Recent studies try to shed some light on this
question\cite{C07}, and from their results a more realistic
$P(\Dox,\Doz)$ could be derived, which remains for future work.
Before that point, it is easier to wonder about the consequences of
a dominant kind of dissipative process which corresponded to a set
of TLSs with a well defined value of $\Dox$ and a narrow
distribution of $\Doz$'s of width $\Delta_1$, as was suggested for
single-crystal silicon\cite{P88}. Following \cite{P88}, a
$Q^{-1}(T)\sim\sqrt{T}$ behavior is obtained for low temperatures
$T<\Delta_1$ if $\Gamma(\Dox,T)<\om$, and a
$Q^{-1}(T)\sim1/\sqrt{T}$ if $\Gamma(\Dox,T)>\om$, while at high
temperatures $T>\Delta_1$ a constant $Q^{-1}(T)\sim Q_0$ is
predicted for both cases. These predictions do not match better with
experiments than the results obtained with $P(\Dox,\Doz)\sim
P_0/\Dox$, so issue i) remains open.

We try now to have a first estimate of the temperature for which
interactions between TLSs cannot be ignored. Following the ideas
presented in \cite{E98}, we will estimate the temperature $T*$ at
which the dephasing time $\tau_{\rm{int}}$ due to interactions is
equal to the lifetime $\tau (T)=\Gamma^{-1}(T)$ defined in
eq.(\ref{LinewidthT}), for the TLSs that contribute most to
dissipation, which are those with $\e\sim\Dox\sim T$. For them
$\tau^{-1}(T)=\Gamma(T)\approx 40\alf\sqrt{\wco}\sqrt{T}$. The
interactions between the TLSs are dipolar, described by
$H_{\rm{int}}=\sum_{i,j}U_{1,2}\sigma_1^z\sigma_2^z$, with
$U_{1,2}=b_{12}/r_{12}^3$, $b_{12}$ verifying $\langle
b_{12}\rangle\approx0$, $\langle|b_{12}|\rangle\equiv
U_0\approx\gamma^2/E$ \cite{YL88,E98}. From the point of view of a
given TLS the interaction affects its bias,
$\Delta_j^z=(\Doz)_j+\sum_iU_{ij}\sigma_j^z$, causing fluctuations
of its phase $\delta \e_j(t)$, which have an associated
$\tau_{\rm{int}}$ defined by $\delta
\e_j(\tau_{\rm{int}})\tau_{\rm{int}}\sim1$. These fluctuations are
caused by those TLSs which, within the time $\tau_{\rm{int}}$, have
undergone a transition between their two eigenstates, affecting
through the interaction $H_{\rm{int}}$ the value of the bias of our
TLS. At a temperature $T$, the most fluctuating TLSs are those such
that $\e\sim\Dox\sim T$, and their density can be estimated, using
$P(\Dox,\Doz)$, as $n_T\approx P_0kT$. They will fluctuate with a
characteristic time
$\tau(T)\approx[40\alf\sqrt{\wco}\sqrt{T}]^{-1}$, so for a time
$t<\tau(T)$ the amount of these TLSs that have made a transition is
roughly $n(t)\approx P_0kTt/\tau(T)$. For a dipolar interaction like
the one described above, the average energy shift is related to
$n(t)$ by \cite{BH77} $\delta\e(t)\approx U_0 n(t)$. Substituting it
in the equation defining $\tau_{\rm{int}}$, and imposing
$\tau_{\rm{int}}(T*)=\tau(T*)$ gives the transition temperature
$T*\approx[6\alf\sqrt{\wco}/(U_0P_0)]^2$. For example, for a
resonator like the silicon ones studied in \cite{ZGSBM05},
$L=6\mu$m, $t=0.2\mu$m, $w=0.3\mu$m, the estimated onset of
interactions is at $T*\approx10$mK.

An upper limit $T_{\rm{high}}$ of applicability of the model due to
high energy 3D vibrational modes playing a significant role can be
easily derived by imposing $T_{\rm{high}}=\om_{\rm{min}}^{\rm{3D}}$.
The frequency $\om_{\rm{min}}^{\rm{3D}}$ corresponds to phonons with
wavelength comparable to the thickness of the sample,
$\om_{\rm{min}}^{\rm{3D}}=2\pi\sqrt{E/\rho}/t$. The condition is
very weak, as the value for example for silicon resonators reads
$T_{\rm{high}}\approx400/t$, with $t$ given in nm. At much lower
temperatures the two-state description of the degrees of freedom
coupled to the vibrations ceases to be realistic, with a high
temperature cutoff in the case of the model applied to amorphous
bulk systems of $T\sim5$K.

\begin{figure}[!t]
\begin{center}
\includegraphics[width=11cm]{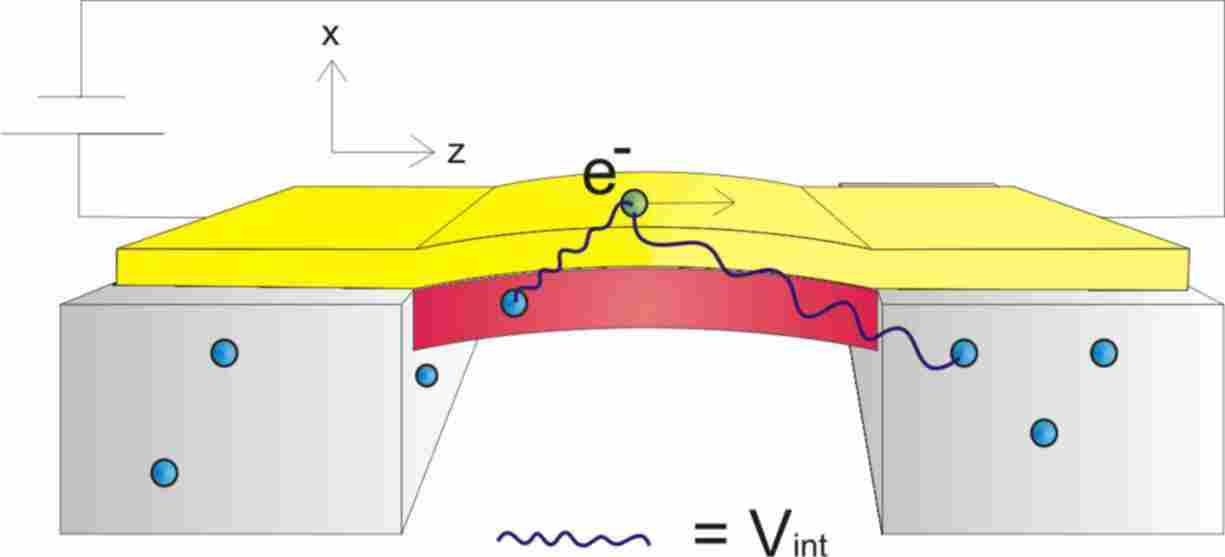}
    \caption[Sketch of the dissipation due to e-h excitations in the electrode layer]{Sketch of the distribution of charges in the device. When the
      system oscillates, these charges induce time dependent potentials which
      create electron-hole pairs in the metallic layer deposited on top of the beam, absorbing part of the mechanical energy of the flexural mode. See
      text for details.}
    \label{sketch_beam_charge}
\end{center}
\end{figure}
\section{Dissipation in a metallic conductor}
Many of the current realizations of nanomechanical devices monitor
the system by means of currents applied through metallic conductors
attached to the oscillators. The vibrations of the device couple to
the electrons in the metallic part. This coupling is useful in order
to drive and measure the oscillations, but it can also be a source
of dissipation. We will apply here the techniques described in
\cite{GJS04,G05} (see also \cite{WR95}) in order to analyze the
energy loss processes due to the excitations in the conductor.

We assume that the leading perturbation acting on the electrons in
the metal are offset charges randomly distributed throughout the
device. A charge $Q$ at position ${\bf \vec{R}}$ creates a
potential:
\begin{equation}
V ( {\bf \vec{r}} , t ) \equiv \frac{Q^2}{\epsilon_0 | {\bf \vec{R}}
( t ) -
  {\bf \vec{r}} ( t ) |}
\end{equation}
at a position ${\bf \vec{r}}$ inside the metal, see
Fig.[\ref{sketch_beam_charge}]. As the bulk of the device is an
insulator, this potential is only screened by a finite dielectric
constant, $\epsilon_0$. The oscillations of the system at frequency
$\wo$ modulate the relative distance $| {\bf \vec{R}} ( t ) - {\bf
  \vec{r}} ( t ) |$, leading to a time dependent potential acting on the
electrons of the metal.

The probability per unit time of absorbing a quantum of energy $\wo$
can be written, using second order perturbation theory, as
\cite{GJS04,G05}:
\begin{equation}
\Gamma = \int d {\bf \vec{r}}   d {\bf \vec{r}}' d t d t' V( {\bf
\vec{r}} , t )  V ( {\bf \vec{r}}' , t' )  \rm{Im} \chi [ {\bf
\vec{r}} - {\bf
    \vec{r}}' , t - t' ] e^{i \wo ( t - t' )}
\label{gamma_metal}
\end{equation}
where $\rm{Im} \chi [ {\bf \vec{r}} - {\bf
    \vec{r}}' , t - t' ]$ is the  imaginary part of the response function of
    the metal.

\textbf{Charges in the oscillating part of the device.} We write the
relative distance as ${\bf \vec{R}} ( t ) - {\bf
  \vec{r}} ( t ) = {\bf \vec{R}}_0 - {\bf \vec{r}}_0 + \delta {\bf \vec{R}}
  ( t ) - \delta {\bf \vec{r}} ( t ) $ and expand the potential, whose time-dependent part is approximately
\begin{equation}
V ( {\bf \vec{r}} , t ) \approx \frac{Q^2 \left[  {\bf \vec{R}}_0 -
{\bf
      \vec{r}}_0 \right] \cdot \left[ \delta {\bf \vec{R}}  ( t ) - \delta {\bf \vec{r}}
      ( t ) \right]}{\epsilon_0 | {\bf \vec{R}}_0 -  {\bf \vec{r}}_0 |^3}
\end{equation}
For a flexural mode, we have that $\left[ {\bf \vec{R}}_0 - {\bf
\vec{r}}_0 \right] \cdot[\delta {\bf \vec{R}}( t ) - \delta {\bf
\vec{r}}( t )]$ has turned into $[X_0-x_0]\cdot [\delta X(t) -
\delta x(t)]\sim t\cdot A\cdot\sin(\omega t)$, where $t=t_{\rm
ins}+t_{\rm metal}$ is the thickness of the beam and $A$ is the
amplitude of vibration of the mode. Thus the average estimate for
this case for the correction of the potential is
\begin{equation}
    \delta V ( {\bf \vec{r}} , t ) \sim \frac{Q^2}{\epsilon_0}\frac{t\cdot
A\cdot\sin(\omega t)}{L^3}\,\,,
\end{equation}
$L$ being the resonator's length.

The integral over the region occupied by the metal in
eq.(\ref{gamma_metal}) can be written as an integral over ${\bf
\vec{r}}_0$ and ${\bf
  \vec{r}}_0'$. The RPA dielectric constant of a dirty metal is:
\begin{equation}
{\rm Im} \chi ( {\bf \vec{r}} - {\bf \vec{r}}' , t - t' ) = \int
e^{i {\bf \vec{q}} (
  {\bf \vec{r}} - {\bf \vec{r}}' )} e^{i \om ( t - t' )} \frac{| \om |}{e^4 D \nu |
  {\bf \vec{q}} |^2}
\label{susc}
\end{equation}
where $e$ is the electronic charge, $D = \hbar v_{\rm F} l$ is the
diffusion constant, $ v_{\rm F}$ is the Fermi velocity, $l$ is the
mean free path, and $\nu \approx ( k_{\rm F} t_{\rm metal} )^2 / (
\hbar v_{\rm F} )$ is the one dimensional density of states.

Combining eq.(\ref{gamma_metal}) and eq.(\ref{susc}), and assuming
that the position of the charge, ${\bf \vec {R}}_0$ is in a generic
point inside the beam, and that the length scales are such that
$k_{\rm F}^{-1} , t_{\rm  metal} , t \ll L$, we can obtain the
leading dependence of $\Gamma$ in eq.(\ref{gamma_metal}) on $L$:
\begin{equation}
\Gamma \approx \frac{| \wo | A^2}{D \nu L} \left( \frac{t}{L}
\right)^2 \approx \frac{| \wo | A^2}{l k_{\rm F}^2 L^3} \left(
\frac{t}{t_{\rm metal}} \right)^2
\end{equation}
where we also assume that $| Q | = e$. The energy absorbed per cycle
of oscillation and unit volume will be $\Delta
E=(2\pi/\wo)\hbar\wo\Gamma_{\rm ph}/t^2L=2\pi\hbar\Gamma_{\rm
ph}/t^2L$, and the inverse quality factor $Q^{-1}_{\rm ph}(\wo)$
will correspond to
\begin{equation}\label{Qfactor1a}
    Q^{-1}_{\rm ph}(\wo)=\frac{1}{2\pi}\frac{\Delta E}{E_0}=\frac{\hbar\Gamma_{\rm ph}}{twL}\frac{1}{\frac{1}{2}\rho\wo^2A^2}\,\,,
\end{equation}
where $E_0$ is the elastic energy stored in the vibration and $A$
the amplitude of vibration. Substituting the result for $\Gamma$ one
obtains
\begin{equation}
Q^{-1} \approx \frac{2 \hbar}{l k_{\rm F}^2 L^4t^2\rho\wo} \left(
\frac{t}{t_{\rm metal}} \right)^2 \label{Q_metal}
\end{equation}
In a narrow metallic wire of width $t_{\rm metal}$, we expect that
$l \sim t_{\rm metal}$.

Typical values for the parameters in eq.(\ref{Q_metal}) are $k_{\rm
F}^{-1} \approx 1$\AA, $A \approx 1$\AA, $l \sim t_{\rm metal}
\approx 10$nm $\sim 10^2$\AA, $t \approx 100$nm $\approx 10^3$\AA
and $L \approx 1 \mu$m $\approx 10^4$\AA. Hence, each charge in the
device gives a contribution to $Q^{-1}$ of order $10^{-20}$. The
effect of all charges is obtained by summing over all charges in the
beam. If their density is $n_Q$, we obtain:
\begin{equation}
Q^{-1} \approx \frac{2 \hbar n_Q}{l k_{\rm F}^2 L^3\rho\wo} \left(
\frac{t}{t_{\rm metal}} \right)^2  \label{Q_metal_n}
\end{equation}
For reasonable values of the density of charges, $n_Q = l_Q^{-3} ,
l_Q \gtrsim 10$nm, this contribution is negligible, $Q^{-1} \lesssim
10^{-16}$.

\textbf{Charges in the substrate surrounding the device.}Many
resonators, however, are suspended, at distances much smaller than
$L$, over an insulating substrate, which can also contain unscreened
charges. As the Coulomb potential induced by these charges is long
range, the analysis described above can be applied to all charges
within a distance of order $L$ from the beam. Moreover, the motion
of these charges is not correlated with the vibrations of the beam,
so that now the value of $\left| \delta {\bf \vec{R}}  ( t ) -
\delta {\bf \vec{r}} ( t ) \right|$ has to be replaced by:
\begin{equation}
\left| \delta {\bf \vec{R}}  ( t ) - \delta {\bf \vec{r}} ( t )
\right| \approx A e^{i \wo t}\,\,, \label{rel_motion_2}
\end{equation}
and the value of $|{\bf \vec{R}}_0 - {\bf\vec{r}}_0|\sim L$.
Assuming, as before, a density of charges $n_Q = l_Q^3$, the effect
of all charges in the substrate leads to:
\begin{equation}
Q^{-1} \approx \frac{2 \hbar L}{l ( k_{\rm F} t_{\rm metal} )^2
l_Q^3 t^2\rho\wo}\approx 0.3\frac{\hbar L^3}{l ( k_{\rm F} t_{\rm
metal} )^2 l_Q^3 t^3\sqrt{E\rho}}\,\,,
\end{equation}
where the second result corresponds to the fundamental mode,
$\wo\approx 6.5(t/L^2)\sqrt{E/\rho}$. For values $L \approx 1 \mu$m,
$A \approx 1$\AA , $k_{\rm F}^{-1} \approx 1$\AA, $l \sim t_{\rm
metal} \approx 10$nm and $l_Q \sim 10$nm, we obtain $Q^{-1} \sim
10^{-9}$. Thus, given the values of $Q^{-1}$ reached experimentally
until now this mechanism can be disregarded, although it sets a
limit to $Q^{-1}$ at the lowest temperatures. It also has to be
noted that this estimate neglects cancelation effects between
charges of opposite signs.
\section{Conclusions}
Disorder and configurational rearrangements of atoms and adsorbed
impurities at surfaces of nanoresonators dominate dissipation of
their vibrational eigenmodes at low temperatures. We have given a
theoretical framework to describe in a unified way these processes. Based on the
good description of low temperature properties of disordered bulk
insulators provided by the Standard Tunneling Model
\cite{AHV72,P72,E98}, and in particular of acoustic phonon
attenuation in such systems, we adapt it to describe the damping of
1D flexural and torsional modes of NEMS associated to the
amorphous-like nature of their surfaces.

We have explained the damping of the modes by the presence of an
ensemble of independent Two-Level Systems (TLSs) coupled to the
local deformation gradient field $\partial_iu_j$ created by
vibrations. Flexural modes, with a high density of states at low
energies, lead to sub-ohmic damping, which can modify significantly
the distribution of TLSs. The problem of a TLS interacting with a
sub-ohmic environment is interesting in its own right
\cite{Letal87,KM96,W99,S03,VTB05,GW88,ChT06,K04,IN92}, and the
systems studied here provide a physical realization.\\

The different dissipation channels to which this ensemble gives rise
have been described, focussing the attention on the two most
important: relaxation dynamics of biased TLSs and dissipation due to
symmetric non-resonant TLSs. The first one is caused by the finite
time it takes for the TLSs to readjust their equilibrium populations
when their bias $\Doz$ is modified by local strains, with biased
TLSs playing the main role, as this effect is
$\propto[\Doz/\sqrt{(\Doz)^2+(\Dox)^2}]^2$. In terms of the
excitation spectrum of the TLSs, it corresponds to a lorentzian peak
around $\om=0$. The second effect is due to the modified absorption
spectrum of the TLSs caused by their coupling to all the vibrations,
specially the flexural modes. A broad incoherent spectral strength
is generated, enabling the "dressed-by-the-modes" TLSs to absorb
energy of an excited mode and deliver it to the rest of the modes
when they decay.

We have given analytical expressions for the contributions of these
mechanisms to the linewidth of flexural or torsional modes in terms
of the inverse quality factor $Q^{-1}(\wo)=\Delta\wo/\wo$, showing
the dependencies on the dimensions, temperature and other relevant
parameters characterizing the device. We have compared the two
mechanisms, concluding that relaxation dominates dissipation, with a
predicted $Q^{-1}(\wo,T)\sim T^{1/2}/\wo$ . Expressions have been
provided for damping of flexural modes in cantilevers and
doubly-clamped beams, as well as for damping of their torsional
modes.

Analytical predictions for associated frequency shifts have been
also calculated. Some important successes have been achieved, like
the qualitative agreement with a sublinear temperature dependence of
$Q^{-1}(T)$, the presence of a peak in the frequency shift
temperature dependence $\delta\om/\om(T)$, or the observed order of
magnitude of $Q^{-1}(T)$ in the existing experiments studying
flexural phonon attenuation at low
temperatures\cite{ZGSBM05,Mohetal07}. Nevertheless, the lack of full
quantitative agreement has led to a discussion on the assumptions of
the model, its links with the physical processes occurring at the
surfaces of NEMS, its range of applicability and improvements to
reach the desired quantitative fit.

Finally, we have also considered the contributions to the
dissipation due to the presence of metallic electrodes deposited on
top of the resonators, which can couple to the electrostatic
potential induced by random charges. We have shown that the coupling
to charges within the vibrating parts does not contribute
appreciably to the dissipation. Coupling to charges in the
substrate, although more significant, still leads to small
dissipation effects, imposing a limit at low temperatures
$Q^{-1}\sim10^{-9}$, very small compared to the values reached in
current experiments \cite{ZGSBM05,Mohetal07}.

\part{Friction mechanisms in graphene and carbon nanotube-based resonators}

\chapter{Graphene-based structures: promising materials}
\section{From 3D to 0D, 1D and 2D}
\begin{figure}[!b]
\begin{center}
\includegraphics[width=10cm]{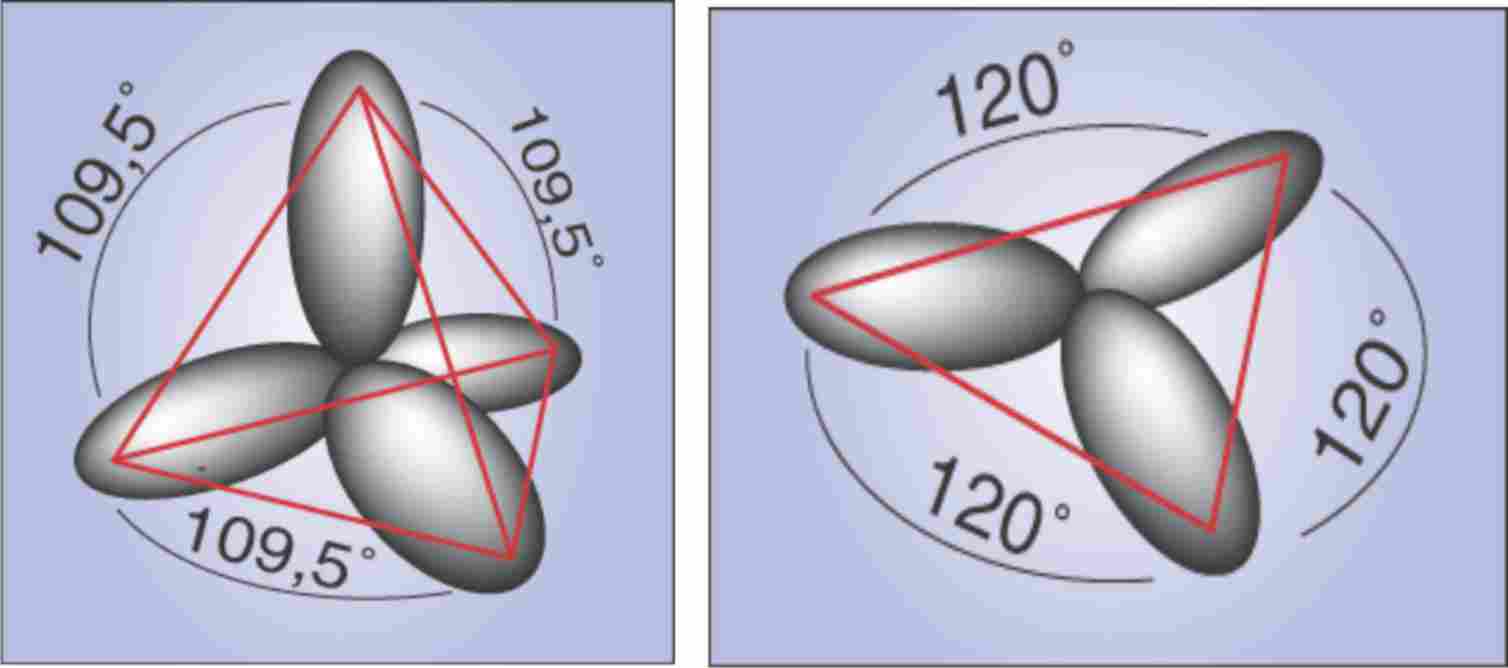}\\
\caption[Carbon orbitals]{\label{corbitals} The two main types of
hybrid orbitals formed by C atoms to create a solid. Left: sp$^3$
bonds. Right: sp$^2$, in plane, bonds. Image from
$\rm{http://en.wikipedia.org/wiki/Orbital}$\_hybridisation .}
\end{center}
\end{figure}

The versatility of carbon atoms to form all kinds of compounds with
completely different properties can be hardly overemphasized
\cite{P93}. When carbon atoms combine to constitute a solid, they
can form two types of bonds \cite{LGV06,GLV06}, see
fig.(\ref{corbitals}): the very strong sp$^3$ bonds present in
diamonds, where each atom has four first-neighbors located in the
vertices of a tetrahedron whose center is occupied by that atom; and
the more versatile sp$^2$ bonds, with three identical coplanar sp
bonds forming 120º angles between them, plus a fourth p-type orbital
perpendicular to that plane. As a result an hexagonal honeycomb
lattice with closest-neighbor distances of 1.4\AA$\,$ tends to form,
with in-plane $\sigma$ orbitals giving rigidity to the lattice and
out-of-plane $\pi$ orbitals providing delocalized charge carriers.

The stacking of these planes creates graphite, a highly anisotropic
material thanks to the differences between the intra and inter plane
bonds. The former are strong covalent bonds, whereas the latter are
weak van der Waals forces, explaining the easiness with which the
planes can slide one over the other, basis of the lubricant
properties of graphite. Indeed, a band-structure calculation renders
hopping amplitudes $t_{ab}=2.4$ eV between nearest neighbors within
a plane and $t_c=0.3$ eV between atoms of consecutive planes,
separated by 3.5\AA. This anisotropy led Wallace already in 1947
\cite{W47} to propose a simplified model where the interaction
between planes is neglected, and a single graphite plane is studied
to obtain approximately the electronic properties of graphite,
viewing it as a 3D solid formed by the stacking of 2D independent
hexagonal crystals.

\begin{figure}[!t]
\begin{center}
\includegraphics[width=12cm]{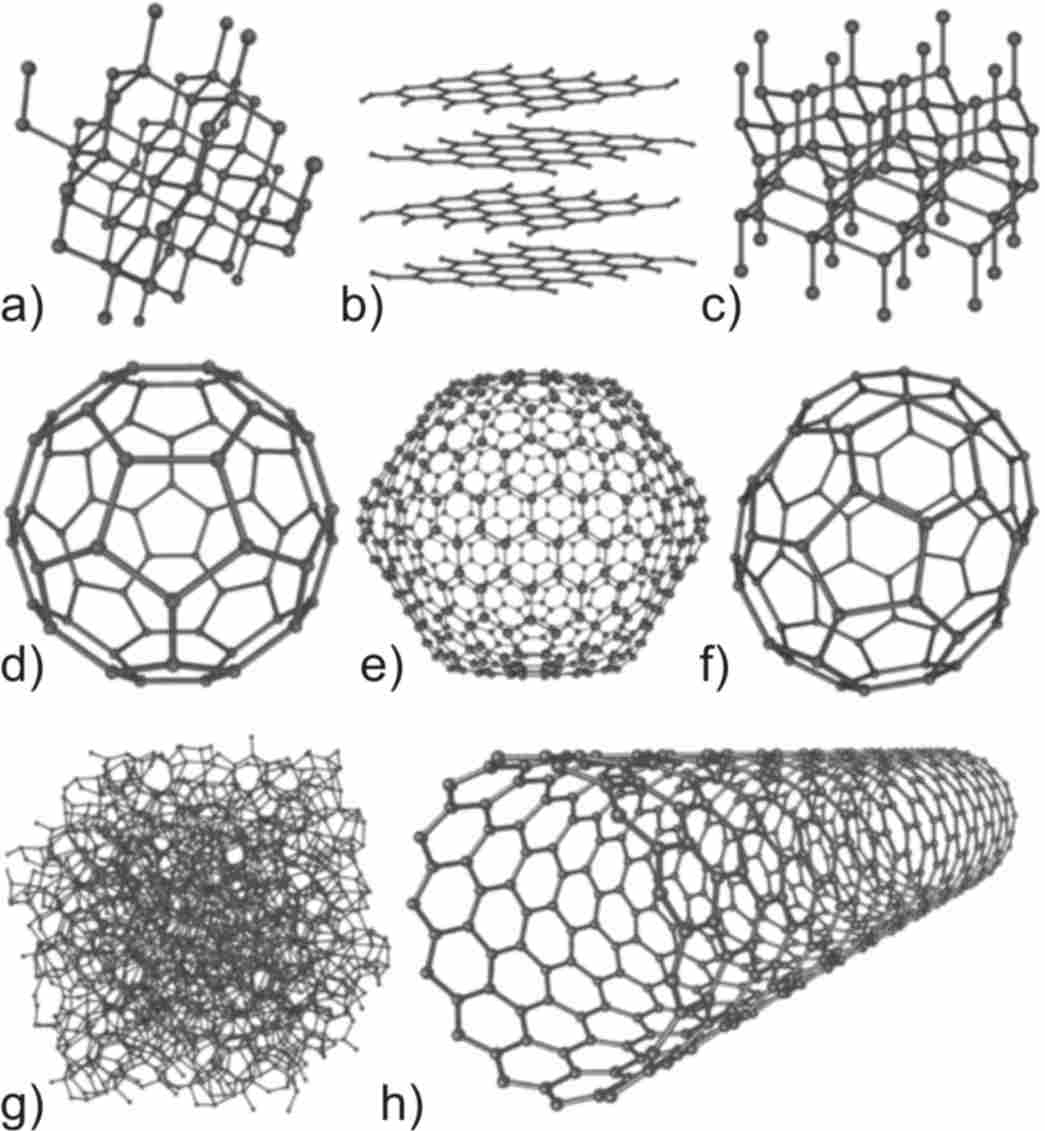}\\
\caption[Examples of carbon allotropes]{\label{callotropes} Eight
examples of allotropes of carbon: a) Diamond, b) Graphite, c)
Lonsdaleite, d) C$_{60}$ (Buckminsterfullerene or buckyball), e)
C$_{540}$, f) C$_{70}$, g) Amorphous carbon, and h) single-walled
carbon nanotube or buckytube. Image from
$\rm{http://en.wikipedia.org/wiki/Allotropes}$\_of\_carbon .}
\end{center}
\end{figure}

Fortunately for the mesoscopics community this is not the end of the
story of carbon compounds. The angles formed by the in-plane
$\sigma$ orbitals can be modified to a certain extent, so that not
only hexagons but also pentagons or heptagons of C atoms can be
formed. This automatically leads to non-planar structures, some of
whom may be energetically stable. This is the case of some closed
structures, carbon fullerenes \cite{DDE96}, whose main
representative is the C$_{60}$, a "football" $\sim$1 nm in diameter
composed of carbon hexagons and pentagons, called
buckminsterfullerene, see fig.(\ref{callotropes}d). The experimental
discovery of this new class of 0-dimensional carbon allotropes in
1985 by a team leaded by Kroto, Curl and Smalley \cite{KHOCS85} was
a breakthrough awarded with the Noble Prize in Chemistry in 1996,
opening several research lines in mesoscopics still active. Studies
have focussed on the physical and chemical properties of such
nano-objects, like heat conductance and high-temperature
superconductivity \cite{Hetal91}, and applications have been
proposed, for example in biomedics as vehicle for other molecules.
They have been even used in double-slit experiments to test the
validity of the wave-particle duality for
such big molecules \cite{ANVKZZ99}.\\

Some years later a second breakthrough took place: Ijima's paper on
the synthesis of carbon nanotubes \cite{I91} caught the attention of
the scientific community and boosted research on these carbon
structures, an example of whom is depicted in
fig.(\ref{callotropes}h). This figure shows a single-walled
nanotube, the result of the folding of a single graphite plane, but
there are as well multiple-walled nanotubes, with several concentric
folded graphite planes. Soon their amazing properties, with much
more technological potential than 0D fullerenes, became apparent:
depending on the way the carbon plane is cut and pasted when folding
it to make the tube, it can be semiconductor or metal, and by doping
them properly one can extend the scope to insulating and
superconductor states; they are also good candidates for
nanoelectronics, with extremely low scattering rates due to their
high degree of crystallinity, allowing electrons to travel
micrometers ballistically at room temperature \cite{DDA01,FS01}.
Coming to their mechanical properties, they are already used in the
fabrication of high strength composites thanks to their $\sim$1 TPa
(!) Young's Modulus and capacity to relax to their initial form
after severe deformations. Their sharp-ending geometry makes them
ideal field-emitters, with application in flat panel displays, and
good nanotools, serving as tips in several kinds of microscopes
(Atomic Force, Scanning Tunneling, Magnetic Resonance Force and
Scanning Nearfield Optical, Chemical/Biological Force Microscope
tips). Moreover, from the fundamental research point of view they
are also very attractive: the propagation of electrons only along
the axis makes them effective 1-dimensional systems where a
Luttinger liquid picture substitutes the conventional Fermi liquid
\cite{Etal01} applicable to bulk metals \cite{NP99}, due to the lack
of screening of the Coulomb interaction, as compared to the 3D case.
Quantum-mechanical interference effects, like the Aharonov-Bohm
effect in presence of magnetic fields, have been also observed and
studied \cite{Betal99}.\\

\begin{figure}[!t]
\begin{center}
\includegraphics[width=12cm]{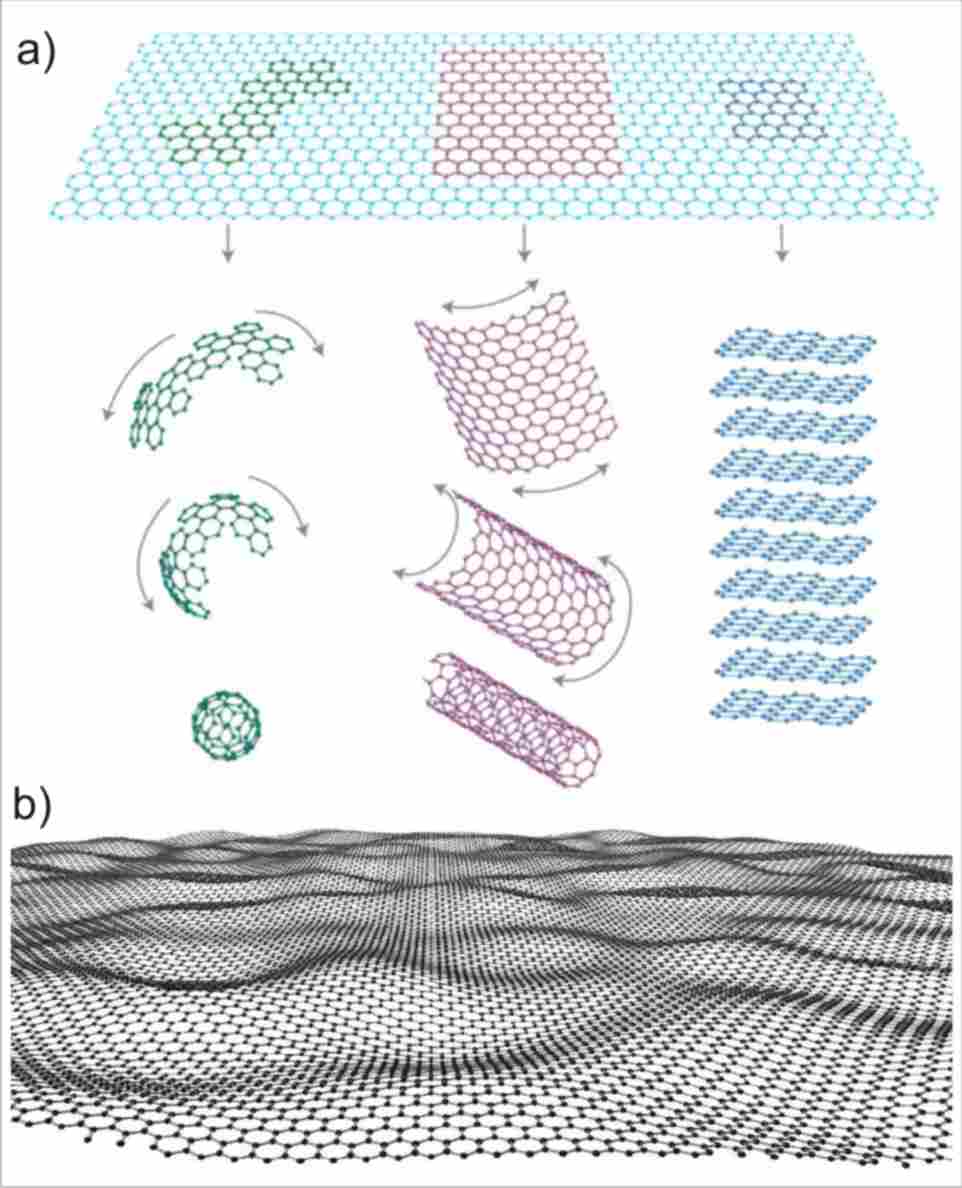}\\
\caption[Graphene and its derivates]{\label{graphene_mother} a)
Image from \cite{GN07}. The detailed understanding of graphene is
important for the comprehension of the other carbon materials
deriving from it: 0D buckyballs, 1D nanotubes and 3D graphite. b)
Pictorial representation of the observed crumpling of graphene
monolayers giving them thermodynamical stability. Extracted from
\cite{Metal07}.}
\end{center}
\end{figure}

The discovery of 0D fullerenes and 1D carbon nanotubes was believed
to put an end to the quest for carbon allotropes, as there were
convincing arguments and strong experimental evidence against the
thermodynamic stability of 2D crystals \cite{P35,LL80,M68,ETB06}.
The surprise came in 2004, when a single plane of graphite, the
so-called graphene, was identified over a SiO$_2$ substrate
\cite{Netal04}. Its existence over such an amorphous substrate or
even suspended in a different setup \cite{Metal07} can be reconciled
with the previous arguments by noting that the carbon monolayer
shows a certain amount of corrugation, with "mountains" ("valleys")
typically $\sim1$ nm high (deep) and with a basis of $\sim30$ nm
diameter, fig.(\ref{graphene_mother}b). The associated total free
energy is minimized, thanks to the strong reduction of thermal
vibrations linked to the loss of planarity, which compensates the
increase in elastic energy.

As depicted in fig.(\ref{graphene_mother}a), graphene can be
considered the "mother" of 3D graphite, 0D fullerenes and 1D carbon
nanotubes, increasing the attractiveness of its study. We will focus
in this second part of the thesis mainly in graphene-based devices,
so now a brief overview on its properties and virtues follows.

\section{Graphene}
Graphene is a remarkable material from many points of view. The
first one refers to its exceptional electronic quality: in spite of
the primitive methods employed to produce it, graphene crystals are
extremely pure, able to carry huge current densities $\sim10^8$
A/cm$^2$ (two orders of magnitude that of copper)\cite{Wi06}, with
electron mobilities $\mu$ exceeding $15.000$ cm$^2$V$^{-1}$s$^{-1}$
under ambient conditions \cite{Netal04,Netal05} and high carrier
concentrations $n>10^{12}$cm$^{2}$ in both electrically and
chemically doped devices. In terms of mean free paths ballistic
transport is observed on the submicrometer scale
($l_{\rm{mfp}}\approx0.3\mu$m at $T=300$K). For a recent
comprehensive review on the electronic properties of graphene see
\cite{CGPNG07}.

A second remarkable property: it is stable down to nanometer sizes,
where the confinement of electrons leads to the opening of
semiconductor gaps, changing completely its electrical properties.
Proceeding in this direction, Walt de Heer has proposed the use of
electron-beam lithography to pattern a graphene sheet the proper
way, carving in it nm ribbons serving as waveguides, and quantum
dots to create, from a single starting sheet, a transistor. A
semiconducting gap can be as well induced simply by growing
epitaxially graphene on a SiC substrate, according to recent studies
\cite{Zetal07}. Circuits using carbon nanotubes, however, need at
some points to have them connected to the circuit through
highly-resistive metal contacts, a major drawback also limiting the
downscaling of silicon microchips, absent in graphene-taylored
circuits. These features raise hopes on graphene as a candidate for
future electronics.

\begin{figure}[!t]
\begin{center}
\includegraphics[width=12cm]{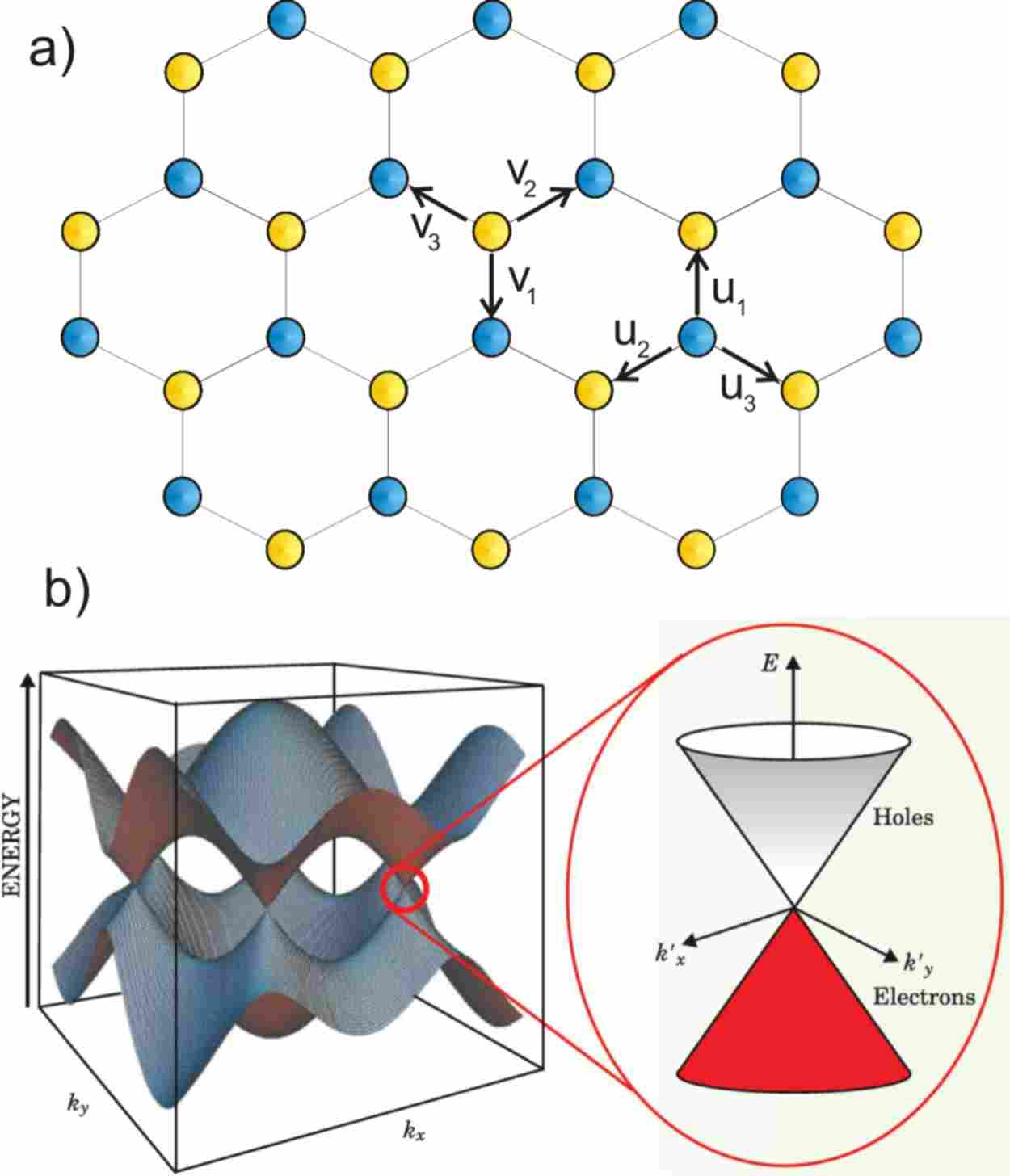}\\
\caption[Graphene's relativistic dispersion
relation]{\label{Dirac_dispersion}a) Graphene lattice can be seen as
composed of two interpenetrated triangular sublattices A and B, with
sublattice vectors $\textbf{u}_i$ and $\textbf{v}_i$. b) Plot of
graphene's bandstructure as derived with a simple band-model for the
honeycomb lattice \cite{W47}, eq.(\ref{Disprelgraphene}). When the
graphene layer is neutral, the Fermi level is situated just at the 6
so-called Fermi points, one of whom is zoomed at the right. At low
energies, the dispersion relation around those Fermi points is
linear, relativistic-like, $E=\hbar v_F|\textbf{k}|$ . Image from
\cite{Wi06}.}
\end{center}
\end{figure}

But even if those hopes were not fulfilled there is another aspect
which justifies by itself research on graphene: Using the band-model
of Wallace \cite{W47} one can see that the electrons in a honeycomb
lattice can be described at low energies in terms of quasiparticles
obeying the (2+1)-dimensional Dirac equation. This means that at low
energies the charge carriers in graphene behave as though they were
massless Dirac fermions, but with a "speed of light" reduced with
respect to vacuum photons by a factor $300$. This is clearly
reflected in fig.(\ref{Dirac_dispersion}b).

The main steps for its derivation go as follows: In the hexagonal
lattice there are two inequivalent atoms per unit cell, so the
lattice can be seen as composed of two interpenetrated triangular
sublattices A and B, see fig.(\ref{Dirac_dispersion}a). An atom of a
given sublattice is surrounded by three nearest neighbors belonging
to the other sublattice. Once the filled $\sigma$ orbitals determine
the honeycomb geometry, the properties of graphene can be calculated
considering only the $\pi$ orbitals with energies close to
$\varepsilon_F$. Starting with a tight-binding hamiltonian with
nearest-neighbor interactions
\begin{equation}\label{Hgraphene}
    H=-t\sum_{<i,j>}a_i^\dagger a_j\,,
\end{equation}
where $a_i^\dagger$ is the creation operator for site $i$, so that a
generic wave function can be written in this localized basis as
\begin{equation}\label{Psigraphene}
    \Psi = \sum_{i\in A}c_Ae^{i\textbf{k}\cdot\textbf{r}_i}a_i^\dagger |0\rangle+\sum_{i\in B}c_Be^{i\textbf{k}\cdot\textbf{r}_i}a_i^\dagger |0\rangle\,,
\end{equation}
one can find the eigenfunctions and eigenenergies of $H$ by solving
the matrix equation for the coefficients $c_A$ and $c_B$
\begin{equation}\label{Hgraphene2}
    \left(
      \begin{array}{cc}
        0 & -t\sum_{i}e^{i\textbf{k}\cdot\textbf{u}_i} \\
        -t\sum_{i}e^{i\textbf{k}\cdot\textbf{v}_i} & 0 \\
      \end{array}
    \right)\left(
             \begin{array}{c}
               c_A \\
               c_B \\
             \end{array}
           \right)=E(\textbf{k})\left(
             \begin{array}{c}
               c_A \\
               c_B \\
             \end{array}
           \right)
\end{equation}
The vectors $\textbf{u}_i$ and $\textbf{v}_i$ are defined in
fig.(\ref{Dirac_dispersion}a). Solving the equation one obtains the
dispersion relation represented in fig.(\ref{Dirac_dispersion}b),
namely
\begin{equation}\label{Disprelgraphene}
    E(\textbf{k})=\pm
    t\sqrt{1+4\cos^2\Bigl[\frac{\sqrt{3}}{2}ak_x\Bigr]+4\cos\Bigl[\frac{\sqrt{3}}{2}ak_x\Bigr]\cos\Bigl[\frac{\sqrt{3}}{2}ak_y\Bigr]}\,,
\end{equation}
with a lattice constant $a=1.42$\AA . With one electron per $\pi$
orbital (half-filling) the Fermi level is located just where the
Fermi surface collapses into 6 points, two of whom inequivalent.
Expanding eq.(\ref{Disprelgraphene}) for low energies around these
so-called Dirac points the dispersion relation is linear in
$\textbf{k}$
\begin{equation}\label{Dispgraphene}
    E(\textbf{k})=\hbar v_F|\textbf{k}|
\end{equation}
This is in stark contrast with the usual quadratic dispersion
relation found in bulk metals or semiconductors. The fermi velocity
$v_F$ is given by $\hbar v_F=(3/2)ta$. Also within this low energy
subspace around the Dirac points, one can Taylor-expand and derive a
low-energy effective model:
\begin{equation}\label{HDirac}
    H\approx\frac{3}{2}ta\left(
                           \begin{array}{cc}
                             0 & \delta k_x+i\delta k_y \\
                             \delta k_x-i\delta k_y & 0 \\
                           \end{array}
                         \right)=\hbar v_F
                         (k_x\sigma_x+k_y\sigma_y)\,\,,
\end{equation}
arriving as anticipated at the Dirac hamiltonian in (2+1)
dimensions, $\sigma_j$ being the Pauli matrices. Experiments confirm
this relativistic behavior of pseudoparticles close to the Fermi
point, most clearly seen in recent ARPES results \cite{BOSHR07},
which reproduce fig.(\ref{Dirac_dispersion}b).

The vanishing density of states at the Fermi points means that in
neutral graphene there are few quasiparticles to screen the Coulomb
interaction. The resulting strong Coulomb interaction results,
according to some works \cite{GGV99,SHT07}, into a marginal Fermi
liquid behavior, with a quasiparticle lifetime
$\tau(|\varepsilon-\varepsilon_F|)\sim |\varepsilon-\varepsilon_F|$,
as compared to the quadratic dependence present in Fermi liquids.
Nevertheless it has to be noted that this question is not yet fully
settled. When the graphene layer is doped with carriers the Fermi
level separates from the Dirac points and locates in a region where
the density of states is no more vanishing, so that now the Coulomb
interaction can be efficiently screened, and a standard Fermi liquid
picture applies.

When one does not have a monolayer but a stack of them, the
interlayer interaction modifies the dispersion relation, displacing
slightly the conduction and valence bands creating the so-called
electron and hole pockets, keeping a very low density of states
(though no more vanishing) at neutrality. Numerous studies address
the evolution of the physical properties as the number of layers of
the system $N$ is increased, starting from a single 2D graphene
monolayer \cite{Netal04,Netal06b,T06,PP06}. It turns out that the 3D
graphite electronic structure is very soon recovered, with about
just 10 layers. Before that, the evolution of the electronic
spectrum can be roughly classified into single layer, bilayer and
$3\leq N<10$ cases, becoming increasingly complicated as $N$ grows.
We will not move further into this issue, and limit ourselves as
well to mention that many other interesting phenomena occur in
graphene as a consequence of the interplay of its peculiar
bandstructure and finite size effects, like the observation of an
anomalous quantum Hall effect \cite{Netal05,ZTSK05,Netal06b}, a
finite value of the electric conductivity at zero field
\cite{Netal05,GN07}, or unclear results about its magnetic response \cite{E06}.\\

\begin{figure}[!t]
\begin{center}
\includegraphics[width=12cm]{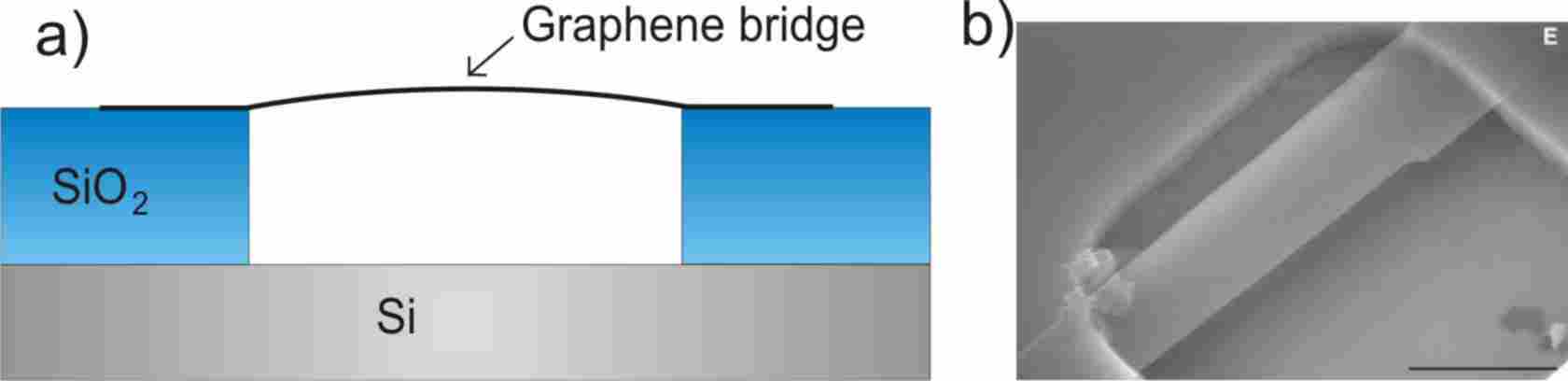}\\
\caption[Scheme of a graphene-based
resonator]{\label{graphene_bridge}a) Schematic side view of the
graphene-based resonator setup of ref.\cite{Betal07}. b) Scanning
electron microscope image from \cite{Betal07}, showing a few-layer
($N\sim$2) graphene resonator. Scale bar, 1 $\mu$m.}
\end{center}
\end{figure}

The extraordinary mechanical and electrical properties of graphene
mono and multilayers have not passed unnoticed to the NEMS
community. Recently Bunch \textit{et al} \cite{Betal07} demonstrated
the first example of electromechanical nanobridge where the
vibrating bridge is a single or multilayer graphene sheet, see
fig.(\ref{graphene_bridge}). The advantages of the use of this
material as compared with semiconductors like silicon or GaAs are
clear: the strength of the $\sigma$ bonds provides stability even to
a single suspended carbon plane, as opposed to crystal Si or GaAs
devices, allowing for much lighter resonators (the lightest one can
think of) with thus higher mass or force sensitivity. Moreover,
those bonds result in an extremely high Young's Modulus, which
together with the large surface area of these devices enhances the
detector's sensitivity as well. As to the friction mechanisms
limiting the quality factor of the resonator's modes, the high
degree of crystallinity of graphene minimizes the dissipative
processes coming from impurities and imperfections, which as we saw
dominate damping in semiconducting nanoresonators at low
temperatures.

Graphene resonators of course present some drawbacks too, in the
form of new dissipative mechanisms absent in semiconductor-based
devices, or increased importance of some of the decoherence
processes already described. The detailed description of these
processes is the subject of the following chapter. As a final
remark, we mention another interesting proposal for graphene as the
basis for ultrasensitive gas sensors, based on the detection of
local changes in the carrier concentration when a molecule is
adsorbed \cite{SGMJHBN07}. This is only possible thanks to the
extremely low intrinsic electronic noise in graphene.


\newpage
\cleardoublepage

\chapter{Dissipative processes in graphene resonators}\label{chg2}
\section{Introduction}
In this chapter we will analyze several dissipative mechanisms
affecting graphene resonators, both mono- and multilayer, like the
ones fabricated by Bunch \textit{et al}\cite{Betal07}, comparing
them with the semiconductor-based NEMS we have studied in the first
part of the thesis. We start with a brief description of these
devices.
\subsection{Experimental setup}
The heart of the setup consists of a mono- or multilayer graphene
sheet exfoliated on top of a previously patterned trench etched in a
SiO$_2$ surface, figs.(\ref{graphene_bridge}) and (\ref{optical1}a).
This SiO$_2$ layer, $\sim$300 nm thick, is thermally grown from a
starting underlying crystalline Si wafer. The movement of the sheet
is detected with a fast photodiode by looking at variations in the
intensity of the reflected light from a 632.8 nm He-Ne laser
focussed on the resonator, which are due to the size variation of
the interferometer formed by the suspended graphene sheet and the
silicon back plane, see fig.(\ref{optical1}b). Bunch \textit{et al}
use two different driving schemes to induce motion:

\subsubsection{Electrical drive}
As shown in Fig.(\ref{optical1}a), a gate voltage $V_g$ is applied
between the graphene and the silicon layers, with both a static (DC)
component and a small time-varying (AC) one, inducing a charge
$q=C_g V_g$, where $C_g$ is the capacitance between the two layers.
The attraction between the charges of the two layers causes an
electrostatic force downward on the graphene. The total
electrostatic force on the sheet is
\begin{equation}\label{Electro_force1}
    F_{el}=\frac{1}{2}C'_gV_g^2\simeq\frac{1}{2}C'_gV_g^{DC}\Bigl(V_g^{DC}+2\delta
 V_g\Bigr)=\frac{1}{2}C'_g\Bigl(V_g^{DC}\Bigr)^2+C'_gV_g^{DC}\delta V_g
\end{equation}
where $C'_g=dC_g/dz$ is the derivative of the capacitance with
respect to the distance between the plates of the capacitor. The
second term of the electrostatic force leads to the motion of the
oscillator, with increased amplitude if the driving frequency $\om$
is close to a resonance $\wo$.

\subsubsection{Optical drive}
\begin{figure}[]
\begin{center}
$\begin{array}{cc}
  \textbf{(a)} & \textbf{(b)} \\
  \includegraphics[width=7cm]{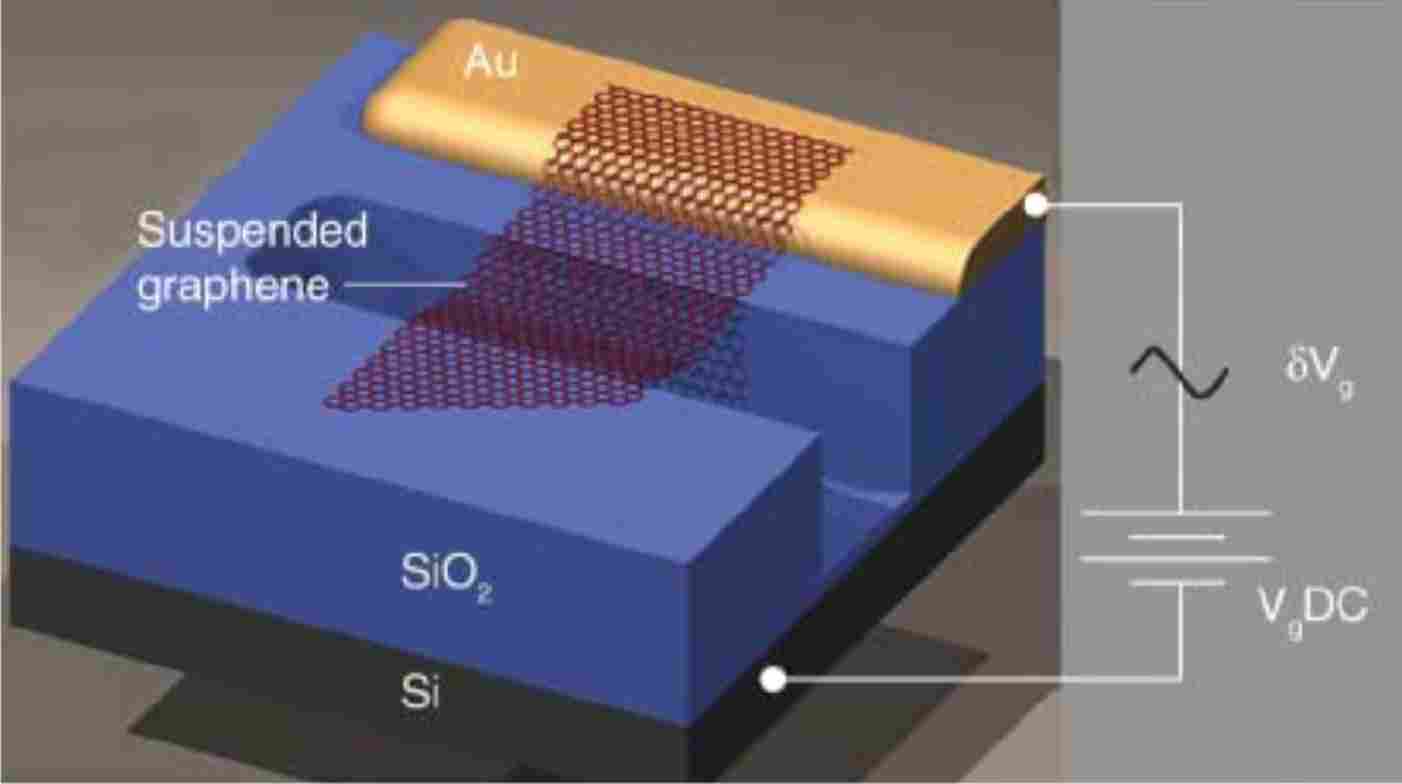} & \includegraphics[width=7cm]{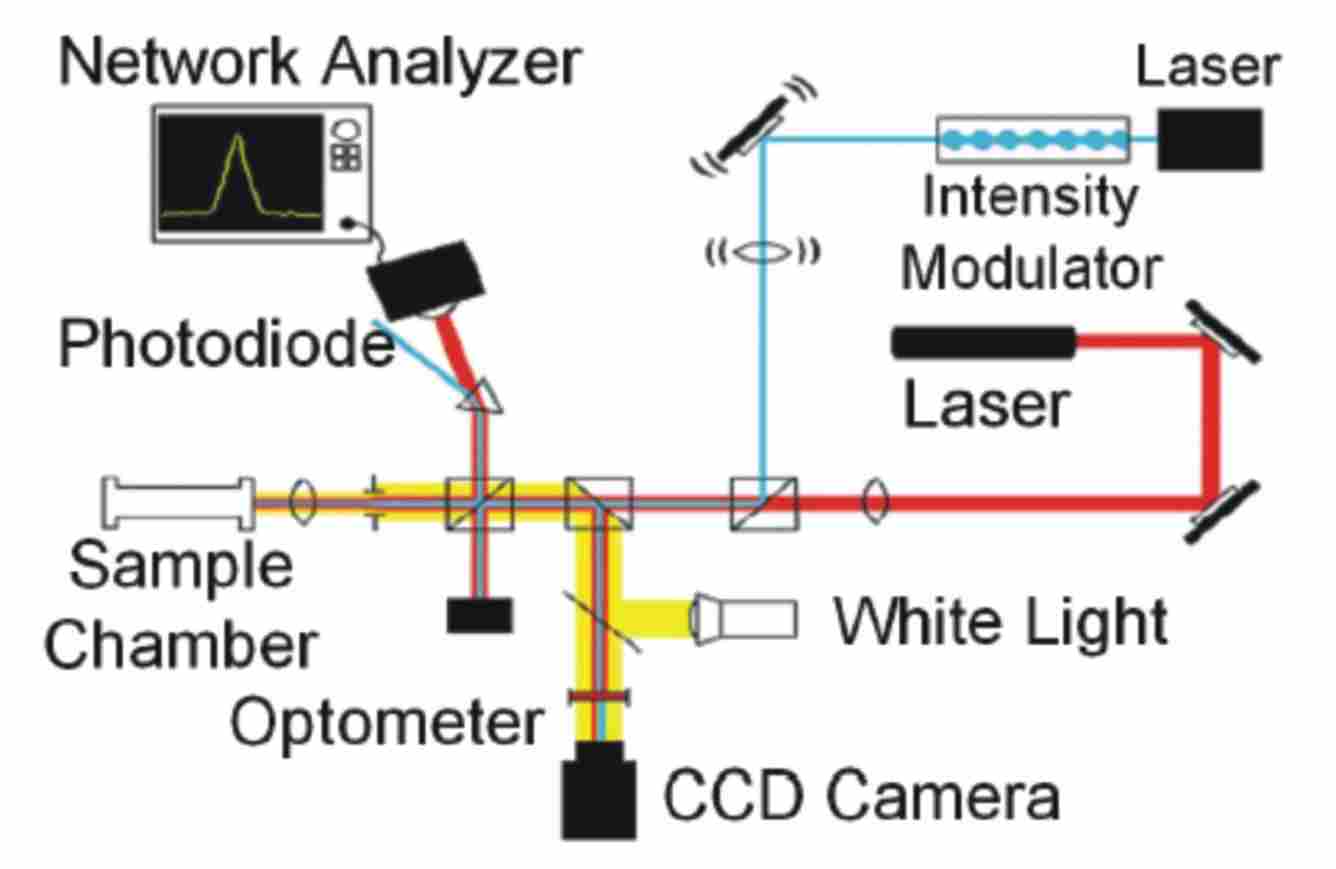}
\end{array}$\\
\caption[Graphene resonator's setups]{\label{optical1} (a) Scheme of
the resonator and the electrical actuation setup. (b) Optical
actuation and detection scheme. From \cite{Betal07}.}
\end{center}
\end{figure}

A 432 nm diode laser is intensity modulated at a frequency defined
by the network analyzer and focussed on the suspended graphene,
causing a periodic contraction/expansion of the layer that leads to
motion (see Fig.(\ref{optical1}b)).

\subsection{Quality factors measured}
\begin{figure}[!t]
\begin{center}
\includegraphics[width=14cm]{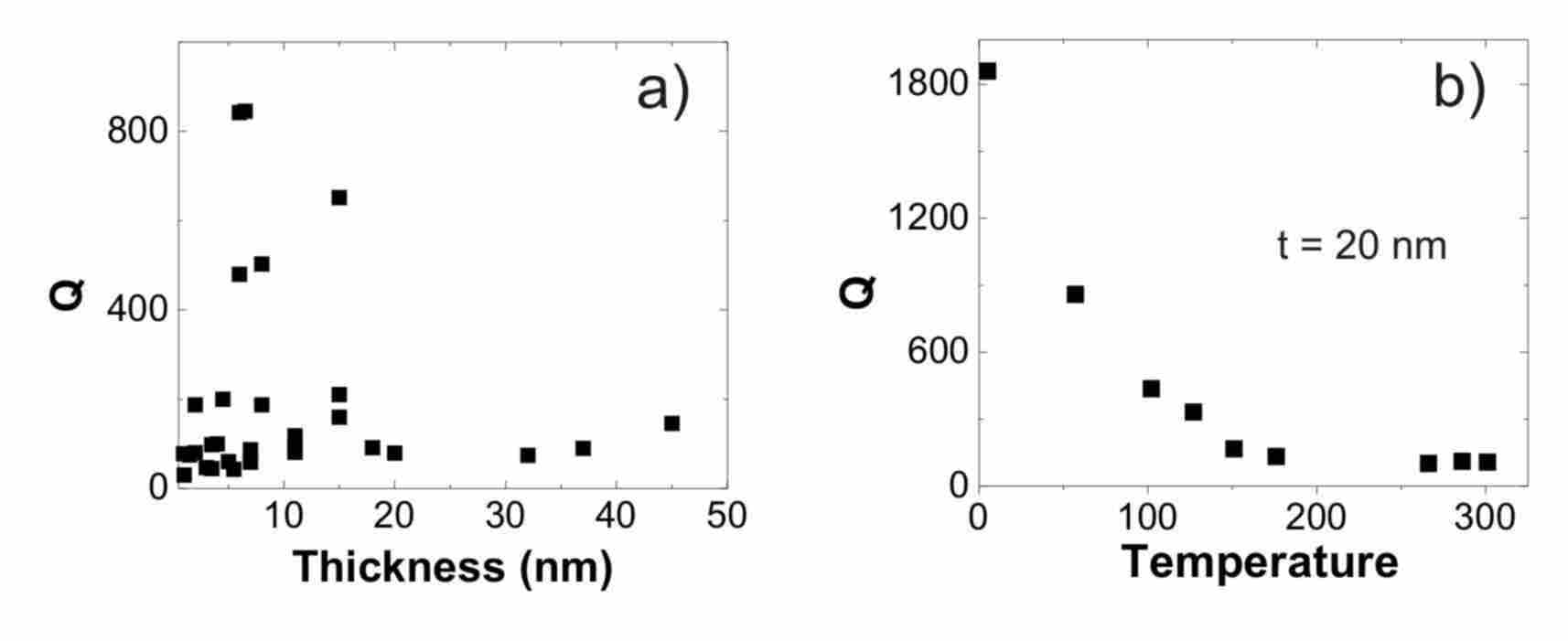}\\
\caption[Q of graphene-based resonators as a function of their
thickness]{\label{Qgraphene} (a) The quality factor $Q$ of the
fundamental mode for all resonators studied in \cite{Betal07} does
not show a clear increase with thickness, contrary to semiconductor
NEMS. Thus surface processes like the ones described in previous
chapters do not seem to dominate damping in these devices, in
agreement with graphene's  well-known high degree of crystallinity.
(b) Temperature dependence of $Q$ for a resonator 20 nm thick,
showing a strong increase of the quality as the temperature is
lowered. Courtesy from A. van der Zande (unpublished).}
\end{center}
\end{figure}

The quality factors $Q$ measured at room temperature for the
fundamental mode of several resonators, with thicknesses ranging
from the monolayer case to $\sim100$ layers, are shown in
fig.(\ref{Qgraphene}a). The most prominent feature noted by the
authors is a surprising lack of dependence of $Q$ on the thickness
of the resonators, leading them to suggest that the dominant
dissipation mechanism is different from that of standard
semiconductor NEMS. Comparing with results obtained for carbon
nanotubes, they mention that the similarity of the structure and
observed magnitude of $Q$ perhaps points to a common dissipation
mechanism, unknown until now for CNT resonators. They suggest also
that an extrinsic mechanism such as clamping loss or fluctuating
charge noise may dominate dissipation in graphene resonators. In a
preliminary study of the temperature dependence of $Q$, they observe
a sharp increase of the quality as T is lowered, as shown in
fig.(\ref{Qgraphene}b).

\subsection{Damping mechanisms present in graphene resonators}
Prior to analyzing in detail several of the mechanisms limiting the
quality of the bending modes, we will briefly enumerate and describe
the (known) most prominent ones, comparing them with the case of
semiconductor NEMS:
\subsubsection{Novel or enhanced mechanisms with respect to semiconductor resonators}
The ability of graphene to sustain large carrier densities when an
external voltage is applied is a feature distinguishing it from the
NEMS studied in the first part of this thesis, and as we have seen
is used in one driving scheme. When the oscillating graphene layer
is charged, the presence of charge in the Si electrode gives rise to
friction associated to the \textbf{excitation of electron-hole pairs
in the graphene layer}. This is due to the time-dependent Coulomb
potential that the charges in the graphene layer feel, created by
the charges in the electrode.

And viceversa, when the Si electrode is charged, the time-dependent
Coulomb potential due to the oscillating charged graphene layer
causes \textbf{excitation of electron-hole pairs in the electrode},
absorbing mechanical energy. These processes are present in
semiconductor devices only when the setup includes a metallic layer
deposited on top of the resonator (magnetomotive technique), and
play usually a secondary role due to the low charge density in the
non-mobile part of the device. Graphene resonators, on the contrary,
have a higher charge density both in the graphene bridge and in the
Si electrode in the setup with electrical drive, increasing the
importance of this dissipative channel. Moreover, this mechanism is
also present in the optical driving setup, as graphene is known to
be, even in absence of applied voltages, locally charged, with
so-called electron and hole puddles corresponding to the sheet
corrugation \cite{Metal07b}, though its importance will be smaller
as compared to the electrical drive setup case.\\

Apart from their presence in the graphene bridge and the Si
electrode, \textbf{trapped charge} in the Si-SiO$_2$ interface, as
well as charge distributed throughout the volume of the oxide is
known to exist \cite{S81,N82}, limiting for example the performance
of MOS-based devices. In our case it will add an extra Coulomb
potential to the one generated by the charge in the Si electrode,
enhancing the \textbf{creation of e-h pairs
in the graphene layer} with the corresponding damping of the vibration.\\

A process also missing in current semiconducting resonators which
can damp the oscillation is the breaking and healing of bonds gluing
the graphene sheet to the SiO$_2$ substrate, conveniently
denominated \textbf{Velcro effect} \cite{VZ07}. We will do some
estimates based on the known chemical properties of SiO$_2$ surfaces
and conclude that it may become a strong source of energy loss if
the resonator is strongly driven into the non-linear regime, being
otherwise negligible.

\subsubsection{Other mechanisms a priori non-negligible in both kinds of resonators}
Some of the trapped charged impurities throughout the device are not
static entities, but have their own dynamics, coming frequently
associated to the same defects causing surface dissipation in
semiconductor NEMS. A clear example corresponds to charges switching
between metastable trapping sites, which can be modeled by a set of
two-level systems (TLSs). They generate the previously mentioned
\textbf{fluctuating charge noise}, a real headache for the quantum
computing community in need for long coherence qubit times
\cite{DH81,We88,PFFF02,BRSCHDC04,SSMM07}. We will model the ones of
the amorphous SiO$_2$ substrate with the usual TLS distribution
employed in studies with amorphous materials \cite{P88}, concluding
that this mechanism can be neglected.\\

\textbf{Attachment losses} due to the coupling of the bridge's
oscillation to the phonons of the rest of the device through the
anchorage areas will as well be estimated. Another source of
dissipation which can be of importance, specially at high
temperatures, is the \textbf{thermoelastic coupling} of the strain
field of the excited resonator's bending mode to the rest of its
modes, causing local variations in the temperature field. Its
contribution will prove to be as well of secondary
importance.\\

For all of the aforementioned mechanisms we will give a numerical
estimate of their contribution to the quality factor at $300$ K, and
determine its temperature dependence, using a prototypical resonator
characterized by the parameters given in table
\ref{table_parameters}. The results are summarized in table
\ref{table_results}.

\subsubsection{Heating by the lasers}
In the optical drive setup, the intense actuation laser induces a
\textbf{local heating} of the resonator, reducing the quality
factor. To a lesser extent this mechanism is also present in the
electrical drive setup due to the weak He-Ne monitoring laser. We
will not analyze in detail this effect, concentrating in the
electric drive setup. The interesting study of this heating process
is yet a task to be performed. Nevertheless, a first observation can
be made, based on fig.(\ref{Qgraphene}b): in that experiment the
intensity of the lasers used is not varied as T is lowered, so its
effect on $Q$ should be roughly independent of T, but as a drastic
increase of $Q$ is observed, it leads us to think that the heating
by the lasers is not the dominant source of dissipation, in that
temperature range at least.

\begin{table}[!t]
\begin{center}
 \begin{tabular}[c]{||c|c||}
  \hline \hline
  \multicolumn{2}{||c||}{System properties}  \\ \hline
  Dimensions & \\ Thickness $t$ & $10\cdot10^{-9}$ m\\
  Width $w$ & $10^{-6}$ m \\
  Length $L$ & $10^{-6}$ m \\
   Height above substrate
$d$ & $300\cdot10^{-9}$ m \\ \hline Frequency  $f_0$
   & 100 MHz \\
Amplitude $A$ &0.5 nm \\ \hline Carrier density $\rho_C$ &$10^{12}
{\rm cm}^{-2}$ \\ \hline \hline \multicolumn{2}{||c||}{Properties of
graphite} \\ \hline Mass density $\rho_M^C$  & 2200 kg/m$^{3}$ \\
\hline
  Elastic constants & \\
$E$ & $10^{12}$ Pa \\
  $\nu$ & 0.16 \\ \hline
Debye temperature $\theta_{D}$ & $\sim
  570\,K$ \\ \hline
 Specific heat $C_p$ & 700 J / Kg. K \\ \hline
Thermal conductivity $\kappa$ & 390 W / m . K \\ \hline \hline
\end{tabular}
\end{center}
\caption{Parameters used in the estimates presented in the chapter,
adapted to the systems studied in\protect{\cite{Betal07}}. Bulk data
taken from\protect{\cite{P93}}.} \label{table_parameters}
\end{table}


\section{Estimates for the damping caused by the different
mechanisms}\label{mech_graph}

\begin{table}[!t]
\begin{center}
\begin{tabular}{||l|c|c||}
\hline \hline
& $Q^{-1} ( T = 300 K )$ &Temperature \\ & &dependence \\
\hline Charges in the SiO$_2$ &$10^{-7} - 10^{-6}$ &$T$ \\
  Charges in graphene sheet &$10^{-2}$ &$T$ \\
 and metallic gate & & \\
 Velcro effect &\rm{Absent} &$T^0$ \\
  Two-level systems &$10^{-22}$ &$A+BT$ \\
 Attachment losses& $10^{-6} - 10^{-5}$ & $T^0$ \\
 Thermoelastic losses &$10^{-7}$ &$T$
\\ \hline \hline
\end{tabular}
\end{center}
\caption{Contribution of the mechanisms considered in section
\ref{mech_graph} to the inverse quality factor $Q^{-1}(T)$ of the
systems studied in\protect{\cite{Betal07}}.} \label{table_results}
\end{table}

\subsection{Coupling to fixed charges in the ${\rm SiO}_2$ substrate}
\label{app_charge}

We first calculate the dissipation induced in the graphene layer by
a single fixed charge located in the SiO$_2$ substrate, and
afterwards sum over all fixed charges.

The time dependent component of the unscreened potential induced by
a charge separated by a distance $d$ in the vertical direction from
the graphene layer, acting on an electron at position ${\bf
\vec{r}}$ in the graphene layer is given, approximately, by:
\begin{equation}
V ( {\bf \vec{r}} , t ) \approx \frac{e^2 d A e^{i \omega_0 t}}{( |
{\bf \vec{r}} |^2 + d^2 )^{3/2}} \label{potential_charge}
\end{equation}
where $A$ is the amplitude of the flexural mode, and $\omega_0$ its
frequency. Its Fourier transform is:
\begin{eqnarray}\label{potential_transform}
\nonumber V ( {\bf \vec{q}} , \omega ) &=&\int d{\bf
\vec{r}}e^{i{\bf \vec{q}}{\bf \vec{r}}}V ( {\bf \vec{r}} , \omega )=
-e^2 A \delta (\omega - \wo )\frac{\partial}{\partial
d}\Bigl\{\int_0^{2\pi}d\theta\int_0^{\infty}dr\frac{r
e^{iqr\cos\theta}}{\sqrt{r^2+d^2}}\Bigr\}\\
&=&2\pi e^2A e^{-qd}\delta (\omega - \wo )
\end{eqnarray}
This potential is screened by the polarizability of the graphene
layer\cite{WSSG06}. In an RPA self-consistent calculation (see
Appendix \ref{apclean_metal} and Section \ref{linear_response}) the
screening is expressed in terms of a renormalization of the charge,
replacing $e^2$ as follows:
\begin{equation}\label{charge_renorm}
e^2 \rightarrow {e^*}^2 = \frac{e^2}{1 + (e^2 / | {\bf \vec{q}} |)
{\rm Re} [\chi_0 ( | {\bf\vec{q}} | , \omega ) ]} \approx \frac{|
{\bf \vec{q}} |}{{\rm Re} [\chi_0 ( | {\bf \vec{q}} | , \omega ) ]}
\end{equation}
where $\chi_0$ is the susceptibility of the graphene layer.
Rigorously this result is obtained only when 2D translational
invariance is assumed, something we will do, ignoring border
effects. At low energy and momenta the value of the susceptibility
tends to the compressibility (density of states) of the electrons in
the layer. For the monolayer and multilayer cases we have:
\begin{equation}
\lim_{| {\bf \vec{q}} | \rightarrow 0 , \omega \rightarrow 0} {\rm
Re} [ \chi_0 ( | {\bf\vec{q}} | , \omega ) ] = \left\{
\begin{array}{lr} \frac{\kf}{\hbar\vf}
&N=1 \\ \frac{N \gamma}{\hbar^2\vf^2} &N \ne 1 \end{array} \right.
\end{equation}
where $N$ is the number of layers and $\gamma$ is the interlayer
hopping element. The result for the monolayer can be immediately
obtained assuming a linear dispersion relation $\varepsilon
(k)=\hbar\vf k$, while for a stack with $N$ layers, the model with
one interlayer hopping element\cite{GNP06} has been used, which
gives rise to $2 N$ low energy bands, most of which show a quadratic
dispersion.

Using Fermi's golden rule applied to the perturbation potential of
eq.(\ref{potential_charge}) screened as described in
eq.(\ref{charge_renorm}) one arrives at eq.(\ref{transprob2}) of
Appendix \ref{apChi1} for the width of the graphene mode. Fourier
transforming to momentum space ($v({\bf \vec{q}},\om)$ = $v(-{\bf
\vec{q}},\om)$):
\begin{equation}
\Gamma_{\rm ph} \approx \int d^2 {\bf \vec{q}} | v ( {\bf \vec{q}} )
|^2 {\rm Im} \chi_0 ( {\bf \vec{q}} , \wo )\,. \label{Fermi_charge}
\end{equation}
The low energy and momentum limit of graphene layers ${\rm Im}
\chi_0 ( {\bf \vec{q}} , \wo )$ is given by \cite{WSSG06,GNP06}
\begin{equation}
{\rm Im} \chi_0 ( {\bf \vec{q}} , \wo ) \approx \left\{
\begin{array}{lr}
    \frac{| \omega | \kf}{\vf^2 | {\bf \vec{q}} |} &N=1 \\
\frac{| \omega | \gamma^2 N^{3/2}}{\vf^2 | {\bf \vec{q}} |
\sqrt{\rho}} &N \ne 1
    \end{array} \right.
\label{susc_charge}
\end{equation}
where, for $N \ne 1$, $\rho$ is the total carrier density. This last
expressions are valid for lengths bigger than the mean free path,
$l\gg l_{mfp}$.

The energy absorbed per cycle of oscillation and unit volume will be
$\Delta E=(2\pi/\wo)\hbar\wo\Gamma_{\rm ph}/twL=2\pi\hbar\Gamma_{\rm
ph}/twL$, and the inverse quality factor $Q^{-1}_{\rm ph}(\wo)$ will
correspond to
\begin{equation}\label{Qfactor1}
    Q^{-1}_{\rm ph}(\wo)=\frac{1}{2\pi}\frac{\Delta E}{E_0}=\frac{\hbar\Gamma_{\rm ph}}{twL}\frac{1}{\frac{1}{2}\rho\wo^2A^2}=
    \frac{2\hbar\Gamma_{\rm ph}}{M\wo^2A^2}\,\,,
\end{equation}
where $E_0$ is the elastic energy stored in the vibration and $M$ is
the total mass of the resonator. Calculating eq.(\ref{Fermi_charge})
and substituting in eq.(\ref{Qfactor1}) for a single graphene layer,
a single charge turns out to give a contribution to the inverse
quality factor of:
\begin{equation}
\boxed{Q^{-1} \sim \frac{1}{\kf
d}\frac{2\hbar}{M\omega_0d^2}}\label{charge}
\end{equation}
where $\kf = \pi \sqrt{\rho_C}$, and $\rho_C$ is the density of
carriers in the graphene sheet. We can estimate this density as a
function of the gate voltage $V_g$ treating our system as a
capacitor \cite{Netal04},
\begin{equation}\label{capacitorformula}
  \rho_{C}=\frac{\e_0\,\e}{e}\frac{V_g}{d}
\end{equation}
where $\e_0$ is the permittivity of free space, $\e$ the relative
permittivity of the material between the charged plates of the
capacitor (in this case the graphene layer and the Si), and $d$ the
distance between the plates. In the region where the layer is
suspended, $\e=1$, and for typical voltages applied $V_g\sim100$ V
the resulting carrier density turns out to be
$\rho_{C}\sim2\cdot10^{12}$ cm$^{-2}$. Thus $\kf d \sim 10^2 - 10^3
\gg 1$. Eq.(\ref{charge}) can be generalized to a graphene sheet
with $N$ layers using eq.(\ref{susc_charge}):
\begin{equation}
\boxed{Q^{-1} \sim \frac{1}{\sqrt{N \rho} d}
\frac{2\hbar}{M\omega_0d^2}} \label{charge_N}
\end{equation}
The suppression with the number of layers is due to the increased
screening in this system.

The total contribution to the inverse quality factor is obtained by
multiplying eqs.(\ref{charge}) or (\ref{charge_N}) by the total
number of charges $N_{\rm ch}$. An upper bound to the effective
surface density of local charges, deduced from some models for the
electric conductivity of graphene\cite{NM07,HAS06}, is $\rho_{\rm
ch} \sim 10^{12} {\rm cm}^{-2}$. In Appendix \ref{apcharge_imp} some
facts about these charges are collected from the literature, and an
independent estimate for $\rho_{\rm ch}$ is made, obtaining a
smaller value, eq.(\ref{oxchargedens2}). The reason for this
discrepancy is still unclear to us. Using the first estimate for
$\rho_{\rm ch}$ and the parameters in Table[\ref{table_parameters}],
we find $N_{\rm ch} \sim 10^4$ and $Q^{-1} \sim 10^{-11}$ at low
temperatures.

This mechanism leads to ohmic dissipation, as the energy is
dissipated into electron-hole pairs in the metallic graphene layer.
Hence, as deduced in Appendix \ref{apTdepohm}, the temperature
dependence of this mechanism is given by $Q^{-1} ( T ) \sim Q^{-1} (
0 ) \times ( kT / \hbar\wo )\,,$ and $Q^{-1} \sim 10^{-6}$ at 300 K.

The analysis presented here does not consider the additional
screening due to the presence of a metallic gate. In that case, one
needs to add to the potential from a static charge,
eq.(\ref{potential_charge}), a contribution from the image charge
induced by the gate. This effect will reduce the coupling between
the graphene layer and charges in the vicinity of the gate.

\subsection{Ohmic losses at the graphene sheet and the metallic
gate.}

The electrons in the vibrating graphene layer induce a time
dependent potential on the metallic gate which is sometimes part of
the experimental setup. The energy is transferred to electron-hole
pairs created at the gate or at the graphene layer. These processes
contribute to the energy loss and decoherence of electrons in
metallic conductors near gates\cite{GJS04,G05}.

The coupling between charge fluctuations in the two metallic systems
is due to long range electrostatic interactions. The corresponding
hamiltonian is
\begin{eqnarray}\label{Hwithint}
   H = \frac{1}{2}\Bigl\{\int_Cv_{scr}(z,{\bf
    \vec{r}},t)\rho^C(z,{\bf \vec{r}},t) &+&\int_Gv_{scr}(0,{\bf
    \vec{r}}',t)\rho^G({\bf \vec{r}}')\Bigl\}\\
   \nonumber&+&\int_C\frac{1}{2\rho_M tw}\Pi^2+\frac{1}{2}\frac{Et^3w}{12}\Bigl[ \Bigl(\frac{\partial^2 \phi}
{\partial x^2}\Bigr)^2 +\Bigl(\frac{\partial^2 \phi} {\partial
y^2}\Bigr)^2\Bigr]
\end{eqnarray}
where the indices $G$ and $C$ stand for the gate and graphene layer,
respectively. $\rho_M$ is the mass density of the graphene sheet,
and $t,w,E$ its thickness, width and Young modulus, whereas
$\phi({\bf\vec{r}},t)$ represents the vibrating amplitude field of
bending modes and $\Pi=\partial\textsl{L} / \partial\dot{\phi}$ is
its conjugate momentum ($\textsl{L}$ is the Lagrangian). The
self-consistent screened potentials $v_{scr}(z,{\bf \vec{r}},t)$,
$v_{scr}(0,{\bf \vec{r}},t)$ are calculated as a function of the
bare potentials $v_0(z,{\bf \vec{r}},t)$, $v_0(0,{\bf \vec{r}},t)$
in Appendix \ref{metallic_layers}.

As in the case of eq.(\ref{potential_charge}), the time-dependent
part of the bare potentials couples the electronic degrees of
freedom and the mechanical ones through the charge $\rho({\bf
\vec{r}})$ and amplitude of the vibrational mode, $A_{\bf \vec{q}}$,
and would give rise to a term in the quantized hamiltonian of the
form
\begin{equation}\label{intelectrmech}
    H_{int}\propto \rho({\bf \vec{r}})A_{\bf \vec{q}}\propto(b^\dagger_{{\bf
    \vec{q}}}+b_{{\bf
    \vec{q}}})\sum_{{\bf \vec{k}},{\bf \vec{k}'}}[c^\dagger_{{\bf \vec{k}}+{\bf
    \vec{k}'}}c_{{\bf \vec{k}}}+\rm{h.c}]
\end{equation}
where $A_{\bf \vec{q}}$ and $\rho({\bf \vec{r}})$ have been
expressed in terms of creation and annihilation operators of phonons
$({\bf \vec{q}},\wq)$ and electrons of a 2D Fermi gas, respectively.

But a realistic model requires taking into account the screening of
the potential associated to these charge fluctuations. In terms of
the screened potentials, the induced broadening of the mode $({\bf
\vec{q}},\wq)$ of the graphene layer is, extrapolating
eq.(\ref{transprob2}) of Appendix \ref{apChi1} to this case, as
\cite{GJS04}
\begin{equation}\label{gamma_gate}
    \Gamma(\wq)=\sum_{\alpha = {\rm G,C}} \int d^3
  {\bf \vec{r}} \int d^3{\bf\vec{r}}' \Bigl\{ \rm{Re}V_{\rm  scr}^\alpha ( {\bf \vec{r}} , \wq )\times\rm{Re} V_{\rm scr}^\alpha ( {\bf \vec{r}}' , \wq )
  \times\rm{Im} \chi^\alpha [{\bf \vec{r}}- {\bf \vec{r}}' , \wq ]\Bigr\}
\end{equation}
The static screening properties, $\lim_{{\bf \vec{q}} \rightarrow 0}
{\rm Re} \chi^{\alpha}  ( {\bf \vec{q}} , 0 )$,  of the graphene
layer and the gate are determined by their electronic
compressibilities (densities of states), $\nug$ and $\nusi$
respectively. We will assume that the distance between the graphene
and the gate is much larger than the electronic elastic mean free
path in either material, so that their polarizability is well
approximated by the one of a diffusive dirty metal, see Appendices
\ref{apdirty_metal} and \ref{apdirty_metal2}:
\begin{equation}\label{susclayers}
\chi^\alpha ( {\bf \vec{q}} , \omega ) \approx \frac{\nu^\alpha
D^\alpha |
  {\bf \vec{q}} |^2}{D^\alpha | {\bf \vec{q}} |^2 + i \omega}
\end{equation}
where $D^\alpha = \vf^\alpha l^\alpha$ is the diffusion constant,
and $l^\alpha$ is the elastic mean free path. The two dimensional
conductivity is $g^\alpha = \kf^\alpha l^\alpha$.

We have assumed the gate to be quasi two dimensional. This
approximation is justified when the distance between the gate and
the graphene layer is much larger than the width of the gate.
Appendix \ref{apthicknessSi} shows how for typical oxide thicknesses
and voltages applied it is fulfilled. In this situation, the
broadening of the mode, eq.(\ref{gamma_gate}), can be expressed,
analogously to eq.(\ref{Fermi_charge}), as
\begin{equation}\label{gammamode2}
    \Gamma(\wq)\approx\int d^2{\bf \vec{k}}\Bigl\{|v_{scr}(d,{\bf \vec{k}},\wq)|^2\rm{Im}\cg+|v_{scr}(0,{\bf
    \vec{k}},\wq)|^2\rm{Im}\csi\Bigr\}
\end{equation}
The screened potentials for a graphene layer oscillating in an
eigenmode $({\bf \vec{q}},\wq)$ of amplitude $A_{\bf \vec{q}}$, have
in a first approximation only one momentum component, $v_{scr}({\bf
\vec{k}},\wq)=v_{scr}({\bf \vec{q}},\wq)\delta({\bf \vec{k}}-{\bf
\vec{q}})$, and these components are (see Appendix
\ref{metallic_layers})
\begin{equation}\label{vscr0}
    \left\{
      \begin{array}{l}
         v_{scr}(d,{\bf \vec{q}},\wq)=\frac{q\Bigl[\cg\Bigl(e^{qd}+e^{-qd}\Bigr)-2\csi e^{qd}\Bigr]\rho_CA_{\bf \vec{q}}e^{-qd}}
          {2\cg\csi\Bigl(1- e^{-2qd}\Bigr)} \\
         v_{scr}(0,{\bf \vec{q}},\wq)=\frac{|{\bf \vec{q}}|\Bigl[-\nug\Bigl(e^{2qd}+1\Bigr)+2\nusi \Bigr]\rho_0A_{\bf \vec{q}}e^{-qd}}
         {2\nug\nusi\Bigl(1- e^{-2qd}\Bigr)}
      \end{array}
    \right.\,\,
\end{equation}
where $q=|{\bf \vec{q}}|$ and $\rho_C$ is the charge density in the
graphene layer. The results for $\Gamma(\wq)$ and $Q^{-1}(\wq)$ can
be formulated in terms of the total charge in the graphene layer,
$Q_C = \int d^2 {\bf \vec{r}} \rho_C \approx L \times w \times
\rho_C$.

In the limit of short separation between the layers, $d\ll L$, which
is the situation present in current experimental setups,
substitution of eq.(\ref{vscr0}) into eq.(\ref{gammamode2}) leads to
\begin{equation}\label{gammamode5}
    \Gamma(\wo)\approx\frac{\wo A^2 Q_C^2}{4d^2}\Bigl(\frac{1}{\nug D^C}+\Bigl(\frac{\nusi}{\nug}\Bigr)^2\frac{1}{\nusi D^G}\Bigr)
\end{equation}
The limit $D|{\bf\vec{q}}|^2\gg\om$ for the imaginary part of the
susceptibility of a dirty metal,
$\rm{Im}\chi({\bf\vec{q}},\om)\approx\om\nu/D|{\bf\vec{q}}|^2$, has
been used, and is justified in Appendix \ref{apdirty_metal}. The
first term in the summation describes losses at the graphene sheet,
and the second at the gate. The associated inverse quality factor,
according to eq.(\ref{Qfactor1}), is given by
\begin{equation}\label{gammamode6}
    \boxed{Q^{-1}(\wo)\approx\frac{\hbar Q_C^2}{2M\wo d^2}\Bigl(\frac{1}{\nug D^C}+\Bigl(\frac{\nusi}{\nug}\Bigr)^2\frac{1}{\nusi
    D^G}\Bigr)}
\end{equation}
To make numerical estimates, we use the parameters in Table
[\ref{table_parameters}], with $\nu^C(E)= E/2 \pi \hbar^2 v_F^2$,
$v_F\approx10^6$ m/s for a single layer of graphene, and $\nu^C(E)=
( N \gamma ) /\hbar^2 \vf^2$ for a stack of $N$ layers\cite{GNP06}.
Carriers in graphene stacks have large mobilities\cite{Netal04}, and
we take $D^C\nu^C\approx10^3$. Typical charge densities for the
graphene layer are, as discussed in the previous subsection, $\rho_C
\sim 10^{12} {\rm cm}^{-2}$, leading to a total charge $Q_C \sim
10^4$. For these parameters, the contribution of the graphene sheet
is $Q^{-1} \sim 10^{-8}$. The relative contribution from the gate
depends on the distance to the graphene sheet. For a Si layer with
$D^G\nu^G\approx 10^3$ and at short distances, the contribution to
the damping from the gate is of the same order as that of the
graphene sheet.

Damping is associated to the creation of e-h pairs in a metal, which
implies that this mechanism is ohmic. The inverse quality factor
should thus increase linearly with temperature as described in
Appendix \ref{apTdepohm}, leading to $Q^{-1} \sim 10^{-2}$ at 300 K.

\subsection{Breaking and healing of surface bonds: Velcro effect.}
In the fabrication process of the device, the graphene flake is
deposited on the silica (SiO$_2$) substrate. The surface properties
of silica are largely determined by the nature of silanol (SiOH)
groups present there \cite{MM90}, due to the hydroxyl functional
group that has chemical reactivity, as compared to the inert
siloxane (Si-O-Si) surface \cite{SG95}. In Appendix
\ref{apsilica_surface} more information relevant to us is given on
silica and its surface.

Graphene becomes linked to the silica surface through hydrogen bonds
created by those silanol groups. When the flake is set into motion,
some of this bonds may repeatedly break and heal (the Velcro
effect\cite{VZ07}), causing dissipation of the energy stored in the
vibration. Numerical estimates are difficult to make, but
nevertheless two qualitative arguments showing that its role in the
damping is probably negligible can be presented:

i) This mechanism is expected to be temperature independent, in
contrast with the strong decrease of friction observed as
temperature is lowered \cite{Betal07}, fig.(\ref{Qgraphene}b).

ii) The elastic energy stored in a typical graphene oscillator of
lateral dimensions $w\sim1\mu$m is $E_0=(1/2)\rho w^2 t \wo^2
A^2\sim 10$eV, when the amplitude is $A\sim 1$nm. This means about
$\sim 10^{-5}$eV per nm$^2$. On the other hand, the energy per
hydrogen bond is about $10^{-1}$eV, and typical radical densities at
SiO$_2$ surfaces are \cite{DPX98} $\sim 1 {\rm nm}^{-2}$. Hence the
elastic energy available on average for each hydrogen bond is much
less than the energy stored in the bond. Only rare fluctuations,
where a significant amount of energy is concentrated in a small area
will be able to break bonds, and in this way to induce energy
dissipation. Note, however, that this argument ceases to be valid
for very large amplitudes $\gtrsim 30$nm. For higher amplitudes,
this mechanism can induce significant losses.

\subsection{Fluctuating charge noise: Dissipation due to two-level systems.}
As we have seen in Part I of this thesis, in an amorphous material
an atom or a few atoms can have two nearly degenerate
configurations. A vibration modifies the energy difference between
these situations, and this coupling leads to the damping of acoustic
phonons in amorphous SiO$_2$\cite{E98}. We have seen this mechanism
is expected to dominate friction in many semiconductor
NEMS\cite{Metal02}. But with graphene resonators, we expect the
graphene sheet to show a high degree of crystallinity, so we will
only consider two-level systems (TLSs) in the rest of the structure.

The TLSs can only dissipate energy if they are coupled in some way
to the vibrating graphene sheet. A possible mechanism is the
existence of charge impurities associated to these defects
(fluctuating charges), which are electrostatically coupled to the
conducting electrons in the graphene.

We expect this mechanism to be less effective in the device
considered here than in NEMs made of semiconducting materials, as
now the TLSs reside in the SiO$_2$ substrate, not in the vibrating
structure. The coupling, arising from long range forces, will be in
comparison accordingly suppressed, by a factor of order $( a / d
)^n$, where $a$ is a length comparable to the interatomic
separation, and $n$ describes the decay of the coupling ($n=1$ for
the Coulomb potential between charged systems).

The temperature dependence of the contribution of TLSs to $Q^{-1}$
is determined by the density of states of the modes coupled to the
TLSs and the distribution of TLSs in terms of their paramenters
(tunneling amplitude $\Dox$ and bias $\Doz$)\cite{SGN07}.The
hamiltonian describing the coupling of the effective TLS's and the
oscillating graphene sheet is given by \cite{SGN07}
\begin{equation}\label{Ham_TLS}
    H=\e\,\sigma_x+\gamma\frac{\Dox}{\e}\sigma_z\sum_{\textbf{k}}\lambda_{\textbf{k}}(b_{\textbf{k}}+b_{\textbf{-k}}^{\dagger})+
\sum_{\textbf{k}}\hbar\om_{\textbf{k}}b_{\textbf{k}}^{\dagger}b_{\textbf{k}}
\end{equation}
where $\e=\sqrt{(\Dox)^2+(\Doz)^2}$, $\gamma$ is the coupling
constant, which will be strongly suppressed in these devices as
compared to attenuation of acoustic waves in amorphous materials,
$\gamma\sim1\rm{eV}\times( a / d )^n$, $b_{\textbf{k}}^{\dagger}$
represent the phonon creation operators associated to the different
vibrational modes of a sheet, and
$\sum_{\textbf{k}}\lambda_{\textbf{k}}(b_{\textbf{k}}+b_{\textbf{-k}}^{\dagger})$
represents the coupling to the strain tensor $u_{ik}$. There are two
types, compression modes (longitudinal waves) and bending modes. The
damping is due to the initial transfer of energy from the
vibrational mode studied by the experimentalists to the TLSs, which
in a second step transfer this energy to the rest of the modes. The
properties of the spin-boson model, eq.(\ref{Ham_TLS}), are fully
determined by the power-law $s$ of the spectral
function\cite{Letal87}, $ J ( \omega ) \equiv \sum_k
\left|\gamma\lambda_k\Dox /\e \right|^2 \delta ( \omega - \omega_k
)\sim \alpha\wco^{1-s}\om^s $, where $\omega_k$ is the frequency of
mode $k$, $\alpha$ is an adimensional constant and $\wco$ is the
upper cutoff of the phonon bath. For this system, compression modes
gives rise to a superohmic, $s=2$, bath, while the bending modes
constitute an ohmic bath, $s=1$, and thus will prevail as a source
of dissipation at low temperatures\cite{Letal87}. We will therefore
restrict our analysis to the dissipation caused by the ohmic
component of the vibrational spectrum.\\
A \textbf{detailed calculation} of the dissipation due to this
mechanism is carried out in \textbf{Appendix \ref{apTLSgraphene}}.
Here we state the main results. One arrives at $J(\om)=\alpha\om$,
with
\begin{equation}\label{bending_sheet5}
\alpha\approx
4\Bigl(\gamma\frac{\Dox}{\e}\Bigr)^2\frac{\rho_M^{1/2}(1+\nu)^{3/2}
 (1-\nu)^{1/2}}{\hbar t^2E^{3/2}(9+\frac{3\nu}{1-2\nu})}
\end{equation}
Here $\nu$ is the Poisson ratio of graphene. Choosing fairly
symmetrical TLSs, $\Dox/\e\sim1$, for the parameters in table
\ref{table_parameters}, $\alpha\sim10^{-5}\times (a/d)^{2n}$, very
small. The relaxation of the TLSs dissipates the energy of the
vibration, giving rise to the following expression for the inverse
quality factor \cite{E98},
\begin{equation}\label{Qgeneral}
    Q^{-1}(\om,T)=\frac{P\gamma^2}{EkT}\int_0^{\e_{max}} d\e\int_{u_{min}}^{1}
du\frac{\om }{u\sqrt{1-u^2}}\,C(\om,T)
\end{equation}
where $u=\Dr/\e$, $\e_{max}\sim5$ K, and $(u\sqrt{1-u^2})^{-1}$
comes from the probability density of TLS's in an amorphous solid,
like SiO$_2$. $Q^{-1}(\om,T)$ is a function of $C(\om,T)$, the
Fourier transform of the correlation function
$C(t,T)=\langle\sigma_z(t)\sigma_z(0)\rangle_T$. For biased TLSs and
$\alpha\ll1$ an extensive analysis of $C(\om,T)$ is performed
in\cite{W99}, where several expressions are provided in different
limits. Using them, the estimate for $Q^{-1}(\om,T)$ follows:
\begin{equation}\label{limitsTLSs}
   \boxed{ \left\{
      \begin{array}{ll}
        Q^{-1}(\om,T)\approx\frac{P\gamma^2}{E\hbar\om}\Bigl\{\frac{4\pi}{3}\alpha\e_{max}+\frac{\pi^2}{3}\alpha^2kT\Bigr\}, & \hbox{$kT>\e_{max}$} \\
        Q^{-1}(\om,T)\approx\frac{P\gamma^2\alpha}{E\hbar\om}\frac{4\pi}{3}kT, & \hbox{$kT<\e_{max}$}
      \end{array}
    \right.}
\end{equation}

In the range of temperatures of current experiments (5K$<$T$<$300K),
the dependence of dissipation with T is weak, and $Q^{-1}\sim
10^{-6}\times (a/d)^4\sim 10^{-22}$. The main uncertainty of the
calculation has been the use of the TLSs' distribution assumed for
amorphous solids \cite{P87}, but due to the small value of $\alpha$
a weak dissipation is expected also with a modified distribution.
Thus the conclusion is that the relative importance of TLSs damping
is much smaller for graphene than for other NEMs
devices\cite{CR02,B04,ER05}.

\subsection{Other friction mechanisms}
\subsubsection{Attachment losses.}
The main expressions needed are given in\cite{PJ04}. When $d\gg t$,
and $d$ is much smaller than the wavelength of the radiated elastic
waves in the SiO$_2$ substrate, the contribution to the inverse
quality factor is given by
\begin{equation}\label{atloss1}
    Q^{-1}\approx \frac{w}{L}\Bigl(\frac{t}{d}\Bigr)^2
    \sqrt{\frac{\rho_M^CE^C(1-(\nu^O)^2)}{\rho_M^OE^O}}
\end{equation}
where the superscript $O$ applies to the silicon oxide, and $\nu^O$
stands for Poisson's ratio. The range of values of the quality
factor varies from $Q^{-1}\approx5\cdot10^{-6}$ for a graphene
monolayer, to $Q^{-1}\approx5\cdot10^{-3}$ for a stack with 30
layers and $t = 10$nm. These quantities probably overestimate the
attachment losses, as they do not include the impedance at the
SiO$_2$-graphene interface.

This damping process due to energy irradiated away from the
resonator should not depend on temperature.

\subsubsection{Thermoelastic effects.} When the phonon mean free path
of the acoustic phonons is shorter than the wavelength of the mode
under study, the acoustic phonons can be considered a dissipative
environment coupled to the mode by anharmonic terms in the ionic
potential\cite{Z38,LR00,U06}. These anharmonic effects are described
by the expansion coefficient, $\alpha$, and the thermal
conductivity, $\kappa$. We follow the analysis in\cite{Z48}. For a
rectangular beam vibrating at a frequency $\omega$ the inverse
quality factor is
\begin{equation}\label{Zenerdiss}
    Q^{-1}_Z(T)=\frac{E\alpha^2T}{C_p}\frac{\omega\tau_Z}{1+(\omega\tau_Z)^2}
\end{equation}
where $E$ is the Young Modulus, $C_p$ is the specific heat at a
constant pressure, and $\tau_Z$ is the thermal relaxation time
associated with the mode, which in the case of a flexural vibration
is given by $\tau_Z=t^2C_p/(\pi^2\kappa)$. This estimate assumes
that the graphene sheet is weakly deformed, and that the typical
relaxation time is associated to the diffusion of phonons over
distances comparable to the thickness of the sheet.

Although better approximations are available in the
literature\cite{LR00}, eq.(\ref{Zenerdiss}) is enough for an
estimate of the order of magnitude of $Q^{-1}$. Using the parameters
from table[\ref{table_parameters}], for $t=10$ nm and $f\sim100$
MHz, we find that
 $\omega\tau_Z\ll 1$, and
\begin{equation}\label{Zenerdiss2}
    Q^{-1}_Z(T=300 \rm{K})\approx\frac{E\alpha^2T\omega
t^2}{\pi^2\kappa}\sim5\cdot10^{-7}
\end{equation}

\section{Extension to nanotube oscillators.}
The analysis presented here can be extended, in a straightforward
way, to systems where the oscillating part is a nanotube.

We expect in these devices a larger impedance between the modes of
the nanotube and those of the substrate, so that attachment losses
will be suppressed with respect to the estimate presented here for
graphene.

The damping mechanisms which require long range forces between the
moving charges in the nanotube and degrees of freedom of the
substrate (fluctuating and static charges) will not be significantly
changed. A nanotube of length $L$ at distance $d$ from the substrate
will interact with a substrate area of order $( L + d ) \times d$. A
similar estimate for a graphene sheet of length $L$ and width $w$
gives an area $\sim ( L + d ) \times ( w + d )$. As $L \sim w \sim d
\sim 1 \mu$m, the two areas are comparable.

On the other hand, ohmic losses induced in the nanotube will be
reduced with respect to the two dimensional graphene sheet, as the
number of carriers is lower in the nanotube.

Finally, we expect a longer phonon mean free path in the nanotube,
which implies that thermoelastic effects will be reduced .

\section{Conclusions.}
We have considered six possible dissipation mechanisms which may
lead to damping in a graphene mesoscopic oscillator. The main
results are summarized in Table[\ref{table_results}]. We expect that
the calculations give the correct order of magnitude and dependence
on external parameters.

We find that at high temperatures the leading damping mechanism is
the ohmic losses in the metallic gate and the graphene sheet. This
effect depends quadratically with the total charge at the graphene
sheet, which can be controlled by the gate voltage.

When speaking about the room temperature prevailing damping source,
however, a disturbing fact obtained in ref. (\cite{Betal07}) has to
be noted: the independence of the magnitude of $Q$ at room
temperatures for both the optical and electrical driving setups.
Unless a mechanism like optical heating were of the order of the
predicted dissipation due to ohmic losses, we are missing something.
In principle optical heating does not seem to be an important
friction source, as it should lead to an almost temperature
independent $Q(T)$, contradicting results like the one shown in
fig.(\ref{Qgraphene}b).

At low temperatures attachment losses limit the quality of the
vibration. If the resonator is strongly driven, a new damping
mechanism may come into play, the Velcro effect, which may limit
substantially the quality factor as compared with the slightly
driven case. The high crystallinity of the resonators eliminates the
main source of dissipation in semiconducting resonators, namely
surface-related effective TLSs coupled to the local strain field.

These conclusions apply with only slight modifications to carbon
nanotube-based resonators.

A deeper knowledge on the charge distribution throughout the device
as a function of the parameters characterizing the fabrication
process and the operation of the resonator for the different setups
would prove of great interest, as it may lead to substantial
improvements in the quality factor through the design of charge
minimization strategies, and shed some light on the prevailing
damping mechanisms for these devices. More measurements of the
quality factor as a function of temperature would be also desirable
for these purposes.


\part{Dissipation of collective excitations in metallic nanoparticles}

\chapter{Metallic clusters: optical response and collective excitations}
\section{From atoms to bulk}
\begin{figure}[t]
\begin{center}
\includegraphics[width=7cm]{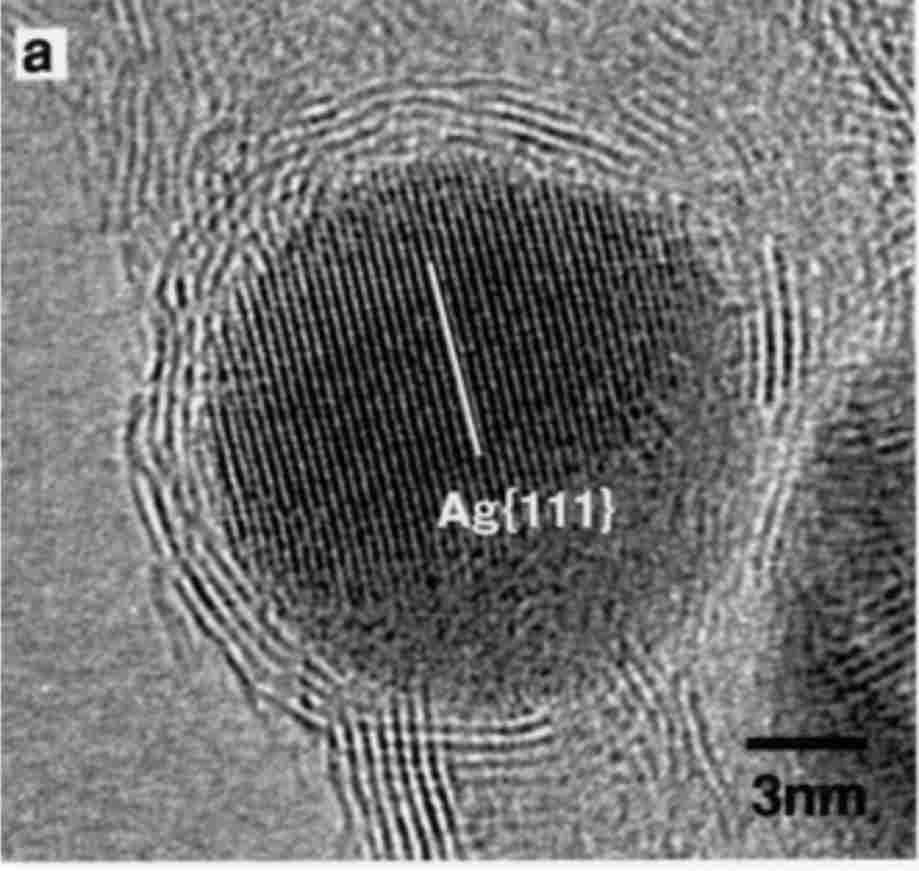}\\
\caption[HREM image of an Ag nanoparticle]{\label{nanop1} High
resolution electron microscopy image of an Ag nanoparticle
encapsulated within a boron nitride nanocage (from ref.
\cite{OKNS00}).}
\end{center}
\end{figure}

\footnote{For a very good and detailed review on the issues covered
in this chapter see \cite{W06}. It has been the main bibliographic
source of this chapter and the beginning of the next.}
Nanometer-sized systems are the natural playground where the
crossover from atomic-like to bulk behavior is observed. Among the
simplest and most studied ones are clusters of alkaline or noble
metal atoms, with radii $a$ between 0.5 and several nm (few atoms to
hundreds of thousands, respectively, see Fig. \ref{nanop1}). The
simpler electronic structure of alkaline atoms renders the
theoretical modeling of alkaline clusters easier as compared to the
case of noble metals (with both s and d electrons), but their high
reactivity imposes stringent conditions on their production:
clusters cannot be formed embedded in a solid matrix, so they are
created, starting from an alkaline vapor at a high temperature, by
expansion into vacuum within an inert carrier gas flux, a method
called \textit{seeded beam expansion} \cite{KKS82}. Noble metals
circumvent this problem, allowing for a more controlled fabrication
and study by ion implantation in an inert matrix and subsequent high
temperature annealing \cite{DYNMN99,OKNS00}. Moreover, their
chemical stability is required for their use as biological markers,
one of their most promising applications \cite{BTMLO02,Cetal03}. We
will comment
further on this issue later.\\

The crossover size to observe bulk behavior depends on the physical
property under study, and is moreover never clearcut. Two
ingredients are the main responsibles of several new and interesting
phenomena occurring in this \textit{interregnum}. On the one hand,
the spectrum of the electronic states remains discrete, with the
energy levels broadened by the interactions with the environment. On
the other, interferences and coherent behavior of electrons happen
due to their coherent motion inside the system: inelastic or
phase-breaking scattering lengths are larger than the size of the cluster.\\

The strongest evidence of the quantization of the electronic
spectrum in metallic nanoparticles is the so-called
\textit{electronic shell structure} observed for the first time in
1984 by Knight \textit{et al.} \cite{Ketal84}. In the nuclear
shell-model \cite{BM75,RS80} a given nucleon of a nucleus can be
considered to a first approximation as confined by a mean-field
potential created by the rest of nucleons. Analogously for clusters,
the valence electrons originating metallic bands in the bulk can be
thought of as dwelling in a mean-field potential created by the rest
of valence electrons and all the ionic cores comprising the cluster.
An immediate consequence of such a model is the existence of certain
"magic numbers" of valence electrons (or correspondingly, atoms) for
whom the system is energetically more stable, namely those filling
angular-momentum shells, as in the nuclear shell-model. In
\cite{Ketal84}, the histogram of clusters as a function of their
size detected in a molecular beam of sodium seeded in argon showed a
clear preference for $N=8,20,40,58$ and 92, which was explained with
the use of such a shell model with a spherically symmetric potential
$V(r)$ of the Woods-Saxon form \cite{WS54}. It turns out that a
sphere is a reasonable approximation for electronically closed-shell
nanoparticles \cite{H93,B93}, while for open-shell structures
distortions occur due to the Jahn-Teller effect \cite{H93,JT37}. We
will focus our attention in this thesis on the first, simpler
case.\\

Bulk-like behavior is frequently observed for surprisingly small
clusters. Furthermore, the size dependence of quantitative features
can already be smooth for very small systems. For instance, the
binding energy of ${\rm Na}_9$ clusters is quite close to that of
${\rm Na}_8$, even if these two systems are very different from the
molecular point of view \cite{H93}. Such a continuity points,
together with the aforementioned evidence of electronic shells
\cite{Ketal84,Ketal85}, towards the relatively minor importance of
the ionic cores and supports the descriptions based on the jellium
model, where the conduction electrons are subject to a uniform
neutralizing background \cite{B93}.\\

Electronic properties of metallic nanoclusters owe their uniqueness
to two facts: first the already mentioned size range, where an
evolution from discrete to continuum spectrum takes place, and
second, the existence of a surface, a finite boundary that, due to
the small size of these systems, has profound implications, the most
relevant of which is the appearance of the so-called \textit{surface
plasmon}, a collective excitation that dominates the optical
response for all clusters except the smallest ones, for whom
transitions between single particle levels dictate the optical
response. Indeed, the question of how large the size of the
nanoparticle has to be in order to observe this collective
excitation in such a finite system has the surprising answer that a
very small cluster, like for instance ${\rm Na}_6$, may already be
enough \cite{Setal91,YB91}.\\

As we will see in next subsection, the surface plasmon corresponds
to the excitation of a periodic displacement of the electronic
center of mass with respect to the positive ionic background. The
jellium approximation for this background allows to decompose the
electronic Hamiltonian into a part associated with the electronic
center of mass, a part describing the relative coordinates (treated
in the mean-field approximation), and finally a coupling between the
two subsystems. The coupling between the center of mass and the
relative coordinates causes decoherence and dissipation of this
collective state once it is excited.\\

The treatment of the surface plasmon as a quantum particle (the
electronic center of mass degree of freedom) therefore provides a
model system for the study of decoherence and quantum dissipation in
confined nanoscopic systems, where the role of the electronic
correlations is preponderant and the excitation spectrum arises from
a \textit{finite} number of particles. The finite number of
electrons makes the term "environment" not completely justified in
this situation, and leads us to the following question, which will
be addressed in this thesis: How large in size do we need to go to
be allowed to describe the relative coordinates of the electron gas
as an environment damping the collective excitation? The existence
of several experimental techniques to observe the surface plasmon
dynamics, and the use of such excitation in biomedical applications
add even more interest to the subject.\\

Based on the mentioned decomposition and mean-field approach an
extensive and powerful analytical study of the plasmon dynamics is
feasible, incorporating the effects of finite size, a dielectric
material, or the finite temperature of the electron gas. With the
use of semiclassical expansions it allows to i) obtain the
parametric dependence (in size, temperature, Fermi energy, etc) of
the addressed quantities like the surface plasmon lifetime or its
resonance frequency, ii) compute the optical properties of
arbitrarily large clusters, and with some extensions also describe
smaller ones, iii) be extended to nonlinear processes such as those
resulting from the interaction with strong laser fields iv) address
coherence and dissipation of the surface plasmon in both weak and
strong laser field driving of the nanoparticle
\cite{YB91,YB92,MWJ02,MWJ03,WMWJ05,WIJW06,WIWJ07,WWIJ07}.\\

Such a wealth of analytical results and gain of knowledge demands
the use of certain approximations, some of which have been assumed
until now to be valid without proof. In the context of this thesis
we will first present the theoretical scheme and then justify its
approximations, confirming the validity of previous results and
putting the approach on a more firm basis. In particular, we will
determine which are the energies of the electronic excitations that
are active in the damping of the surface plasmon, and what is the
characteristic response time of the large enough electronic
environment. The latter is important in justifying the Markovian
approximation that is assumed to describe the dynamics of the
surface plasmon coupled to particle-hole excitations \cite{WIWJ07}.
Other complimentary techniques, like the time-dependent local
density approximation (TDLDA), will not be analyzed. We will start
by giving a first physical image of the surface plasmon, and
summarizing available experimental techniques for its study and use
in biomedical applications.

\section{Surface plasmon excitation} 
\begin{figure}[t]
\begin{center}
\includegraphics[width=10cm]{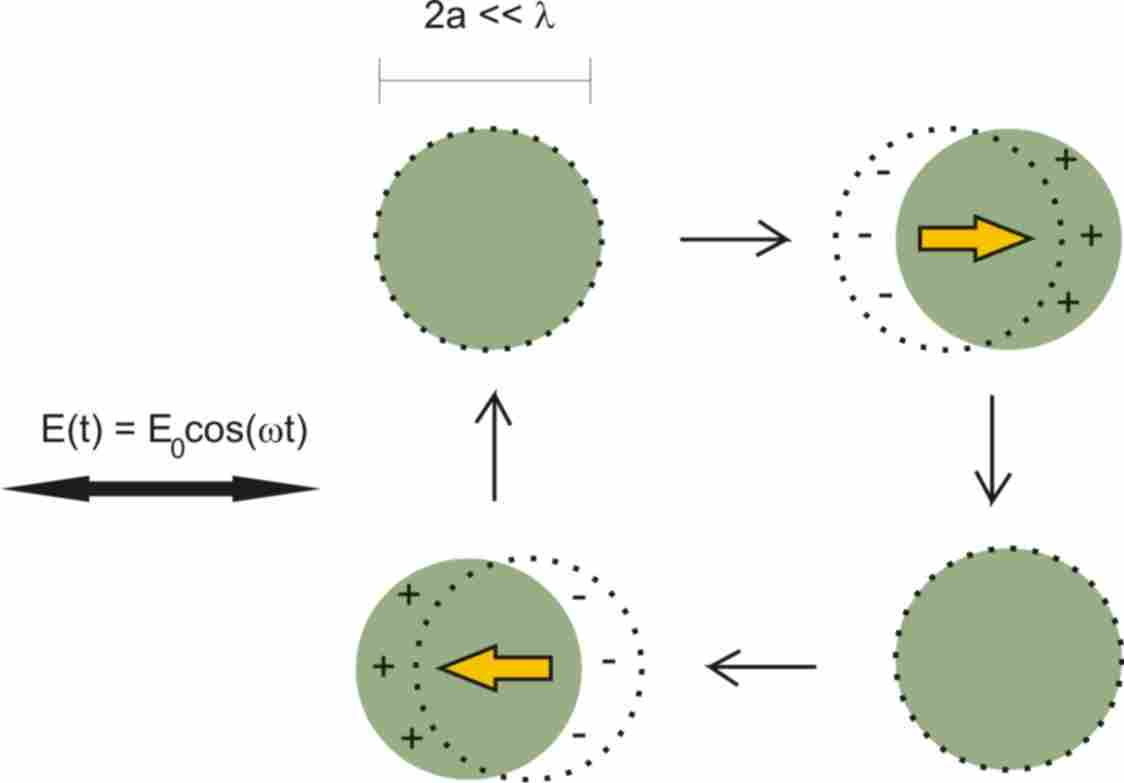}\\
\caption[Schematic representation of the surface plasmon
mode]{\label{oscplasmon} Schematic representation of the surface
plasmon resonance in small metallic clusters. The wavelength
$\lambda$ of the exciting electrical field E(t) is much larger than
the diameter $2a$ of the cluster. Thus, the electrical field
throughout the nanoparticle is nearly uniform. The shaded regions
symbolize the ionic background, while the dashed circles represent
the electronic cloud. The "-" and "+" signs account for the
uncompensated negative and positive charges. The electrical field
exerts a force on the positively charged ions and an equal force on
the electrons in the opposite direction, separating their centers of
mass mainly due to the motion of the light electrons. The ions,
being much more massive, remain almost at rest. The resulting
surface charges create a dipolar restoring field (thick arrow)
opposing E(t).}
\end{center}
\end{figure}

The study of the response of
metallic nanoparticles to externally applied electromagnetic fields
reveals the main features of the electronic spectrum just outlined,
with the optical response dominated by a collective resonance, the
surface plasmon. What happens to the system when such a collective
state is excited? For optical fields, the wavelength is bigger than
the diameter $2a$ of the system and the frequency is such that the
light electronic cloud is able to react to the variations of the
field, whereas the ionic background, three orders of magnitude more
massive, cannot follow the perturbation and remains approximately
static. The result is the uniform relative displacement of electron
and ionic clouds, creating unbalanced charges at the surface of the
nanoparticle (a dipole) which try to restore the equilibrium
situation in absence of an external field. The dipole is forced to
oscillate with the frequency of the applied field (depicted in fig.
\ref{oscplasmon}), and its magnitude peaks at the so-called
\textit{surface plasmon frequency}, the natural frequency of the
dipole, fixed by the electronic density, the spherical geometry, and
the charge and mass of the electrons. So the surface plasmon is
nothing but the self-sustained state of motion for which the center
of mass of the negative charges oscillates with respect to the
center of mass of the more massive ionic cores.

Applying classical Maxwell's equations to the case of a dielectric
metal sphere with appropiate boundary conditions, Mie obtained in
1908 \cite{KV95,M08} a first estimate for this frequency, the
classical Mie frequency $\omega_{\rm M}=\omega_\textrm{p}/\sqrt{3}$,
where $\omega_\textrm{p}=(4\pi n_{\rm e} e^2/m_{\rm e})^{1/2}$ is
the bulk plasma frequency and $e$, $m_{\rm e}$, and $n_{\rm e}$
denote the electronic charge, mass, and density, respectively. In a
real cluster surface effects lead to a small reduction of the
surface plasmon frequency with respect to the Mie value $\omega_{\rm
M}$ \cite{B93,GGI02,WIJW06}. Typical values for the surface plasmon
energy are about several eV, belonging thus to the energy range
corresponding to bound single-particle states \cite{BB94}.\\

\subsection{Probing and using the surface plasmon}
Several techniques show and use the existence of the surface plasmon
to study electronic properties of metallic clusters. We will mention
three of them:
\subsubsection{Photoabsorption experiments}
The cluster is illuminated by a light of a certain frequency, and
the amount of absorbed photons is determined and given in terms of
the photoabsorption cross section $\sigma(\om)$, which is the
probability that the system absorbs a photon times the area
illuminated by the photon beam. The presence of the single-particle
states derived in the electronic shell model should be signalled by
several peaks in $\sigma(\om)$, but they are usually overwhelmed by
the peak of the collective excitation \cite{BCLS93}, thus indicating
a substantial transfer of spectral weight to build the surface
plasmon.

\subsubsection{Noble metal nanoclusters as biological markers}
\begin{figure}[!h]
\begin{center}
\includegraphics[width=10cm]{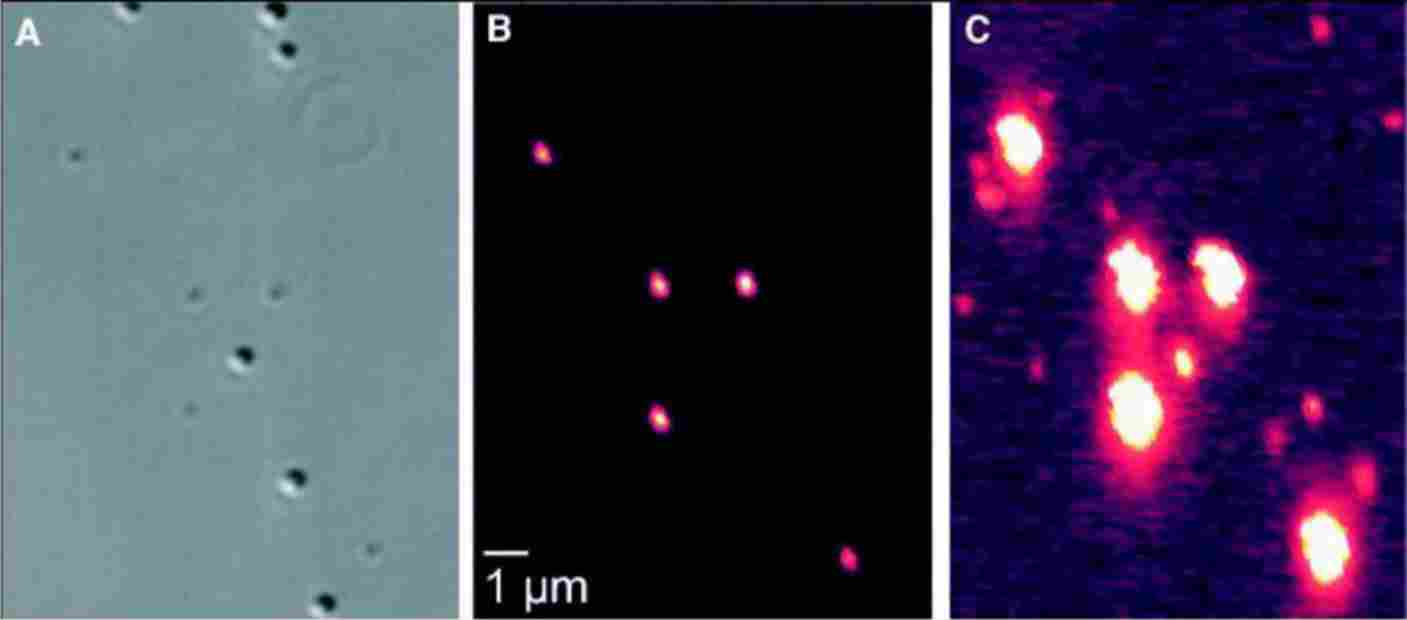}\\
\caption[Application of noble metal nanoparticles as biological
tracers]{\label{photothermal} Noble metal nanoparticles can be used
as biological tracers, thanks to their strong light scattering and
absorption in the optical range due to the existence of the surface
plasmon collective excitation. From ref. \cite{BTMLO02}. (A)
Differential interference contrast (DIC) based on Rayleigh
scattering of light and (B and C) photothermal images measuring
local heating due to strong light absorption, of a sample containing
300-nm-diameter latex spheres, 80-nm-diameter gold spheres, and
10-nm-diameter gold spheres. The heating intensity was 30 kW/cm$^2$
(B) or 1.5 MW/cm$^2$ (C). The 80-nm gold spheres appear on all three
images and allowed direct comparisons of the same imaged sample
area. In the DIC image, the 10-nm particles are not visible, whereas
the 300-nm latex spheres give very strong signals and the 80-nm gold
spheres give weaker ones. In (B), a photothermal image at low power
shows only the large metal particles. In (C), at high heating power,
the 10-nm particles are clearly visible, whereas the 80-nm gold
spheres saturate the detection capability. The strongly scattering
latex spheres are completely absent in the photothermal images, as
they do not absorb light as efficiently as a metallic cluster.}
\end{center}
\end{figure}

An ideal optical label for large molecules should generate an
intense optical signal; it should also be small, durable, chemically
inert, and apt to bind to the molecule of interest in a controlled
manner. Noble metal clusters meet all of these requirements when
illuminated with a frequency close to the one of their surface
plasmons. They overcome the main shortages present in other
candidates like fluorescent dyes or nanocrystals of II-VI
semiconductors, as they do not photobleach like the former and do
neither optically saturate at reasonable exciting intensities like
the latter. Several techniques take advantage of the excitation of
the surface plasmon as a means to track the nanoparticle adhered to
the molecule of interest: nanoparticles bigger than 40 nm in
diameter can be imaged, based on Rayleigh scattering, in an optical
microscope by illuminating in dark-field at the plasmon frequency
\cite{SSMS00}, with differential interference contrast (DIC) and
video enhancement \cite{GSS88}, or with total internal reflection
\cite{Setal00}. For smaller nanoparticles, whose Rayleigh scattering
is too small to be detected, local heating associated to its strong
light absorption causes a temperature change which can be monitored
by a sensitive interference method similar to DIC \cite{BTMLO02},
see fig. \ref{photothermal}.

\subsubsection{Pump-probe experiments: plasmon dynamics}
The use of femtosecond pulsed lasers in time-resolved pump-probe
spectroscopy has rendered possible to experimentally address the
surface plasmon dynamics \cite{BHMD00,FVFHN00,LKLA99,LUCS01}. An
initial ultrashort laser pulse lasting around a hundred femtoseconds
pumps energy into the nanoparticle by exciting the surface plasmon
mode. After a certain time delay a second, much weaker, pulse, the
probe laser, tests the absorption spectrum of the hot electron gas
in the cluster. By repeating the experiment for several values of
the delay an image of the time evolution of the relaxation process
is obtained.\\

\begin{figure}[!h]
\begin{center}
\includegraphics[width=10cm]{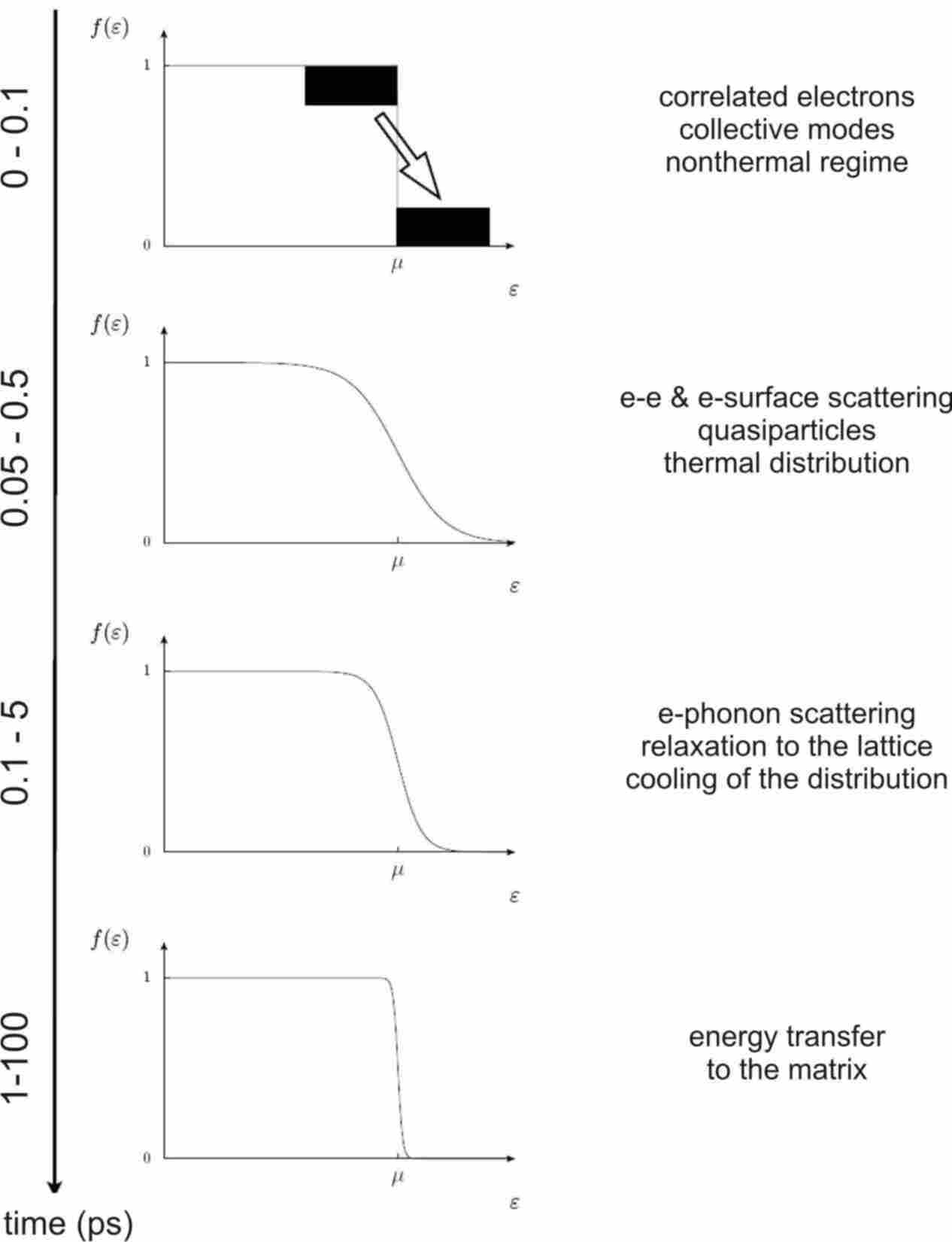}\\
\caption[Relaxation processes after excitation of the surface
plasmon by a laser]{\label{relaxplasmon} From ref. \cite{W06}.
Sketch of the relaxation processes in a metallic nanoparticle after
an excitation by a strong and ultrashort femtosecond laser pump
pulse. $f(\e)$ is a sketch of the electronic distribution, while
$\mu$ is the chemical potential. The arrow in the uppermost image
symbolizes the effect of the pump laser on the electronic system.}
\end{center}
\end{figure}

Usually the experiments are performed on an ensemble of noble metal
clusters embedded in an inert matrix, taking care that typical
separations between clusters guarantee independence of their
response. Several timescales are involved in the relaxation, linked
to the different characteristic times of the degrees of freedom
participating: surface plasmon, single-particle electron
excitations, nanoparticle phonons, and embedding matrix phonons. As
schematically depicted in fig. \ref{relaxplasmon}, the energy is
initially stored in the surface plasmon collective mode, causing a
non-thermal, highly-correlated, electronic distribution $f(\e)$. In
a femtosecond scale the surface plasmon decays through
electron-electron and electron-surface scattering, transferring its
energy to the rest of (single-particle) electronic degrees of
freedom, which accordingly heat and reach a thermal equilibrium
distribution with temperatures up to hundreds of degrees, depending
on the intensity of the pump laser. Phonons of the nanoparticle,
with typical frequencies $\sim10^{12}$ Hz, come into play on a much
longer timescale, of the order of a picosecond, and absorb part of
the energy stored in the electronic distribution, cooling it. The
relaxation process ends with the transfer of energy to the phonons
of the matrix where the nanoparticle is embedded, on the timescale
of several picoseconds, reaching the system as a whole thermal
equilibrium.\\

\begin{figure}[!t]
\begin{center}
\includegraphics[width=10cm]{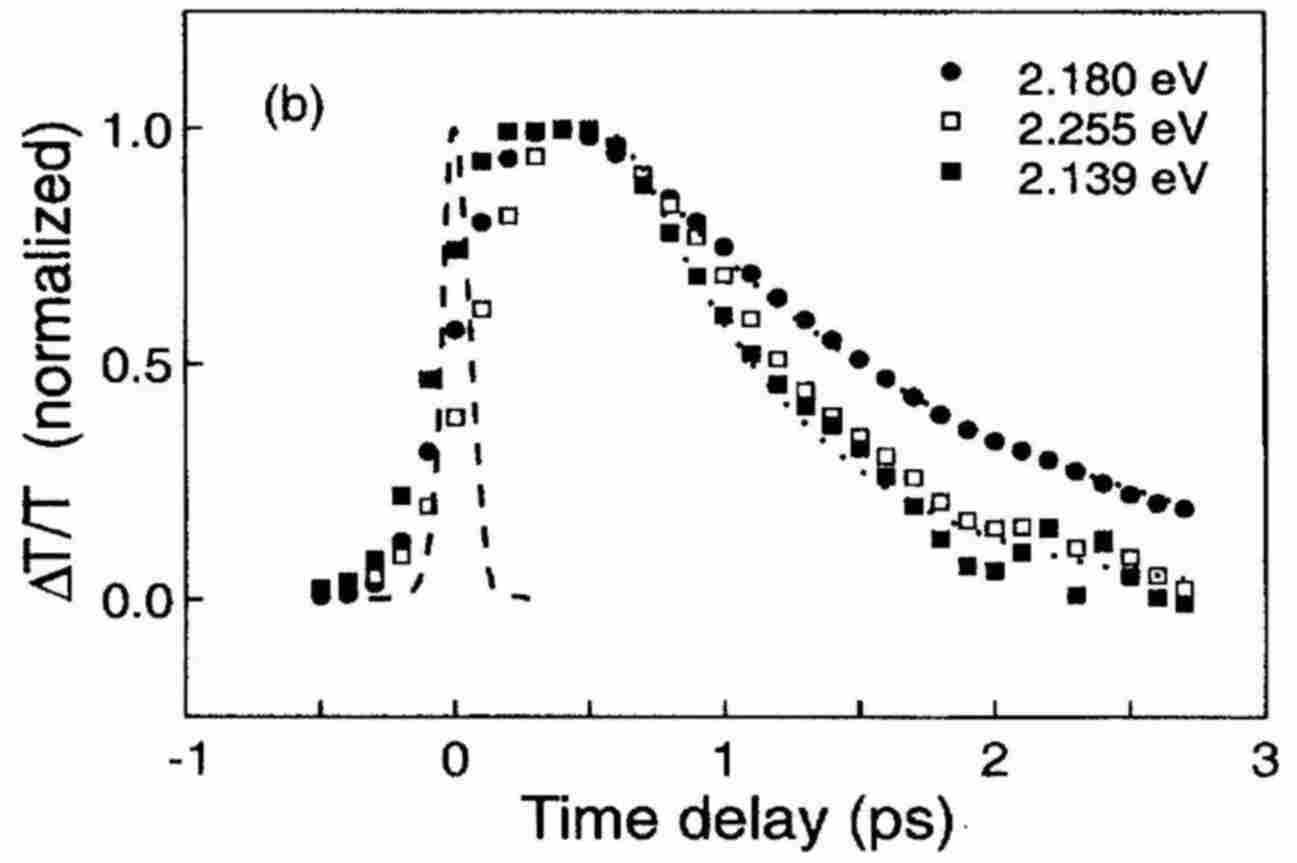}\\
\caption[Time evolution of differential transmission
$\Delta\mathcal{T}/\mathcal{T}$ for Cu
nanoparticles]{\label{relaxCu} Experimental differential
transmission $\Delta\mathcal{T}/\mathcal{T}$ as a function of the
delay time between the pump and probe pulses, for different probe
photon energies. The nanoparticles in this experiment are made of
copper and embedded in a glass matrix. They have an average diameter
of 10 nm. The plasmon resonance is located close to
$\hbar\om_{sp}\simeq 2.2$ eV. (Reproduced from Ref. \cite{BMCD95})}
\end{center}
\end{figure}

In experiments, the normalized difference of transmission of the
sample with and without the pump laser field, called differential
transmission $\Delta\mathcal{T}/\mathcal{T}$, is the quantity
measured to study the relaxation process:
$\Delta\mathcal{T}/\mathcal{T}=(\mathcal{T}_{on}-\mathcal{T}_{off})/\mathcal{T}_{off}$.
An example is given in fig. \ref{relaxCu}, showing the differential
transmission as a function of the time delay between the pump and
probe pulses, for different probe energies in the vicinity of the
surface plasmon resonance. The decrease of
$\Delta\mathcal{T}/\mathcal{T}$ as a function of time reflects the
relaxation of the electronic energy to the lattice.

\newpage
\cleardoublepage
\chapter{Electronic dynamics in metallic nanoparticles: theoretical
model}\label{ch3}
This chapter describes a very successful theoretical framework for
the analytical description of the electron dynamics in metallic
clusters, expressed in terms of two types of excited states, the
single-particle e-h excitations, and the many-body surface plasmon
collective state. In a basis of relative + center of mass electronic
coordinates, e-h and surface plasmon excitations will correspond to
the relative and center of mass coordinates, respectively. Some key
assumptions needed for the conceptual soundness of the model will be
underlined, whose justification will be provided in the final part
of the chapter.
\section{Electronic hamiltonian: relative and collective coordinates}
The systems we will analyze are neutral, spherical, metallic
nanoparticles of radius $a$ containing $N$ valence electrons. They
thus have, as mentioned in the previous chapter, closed angular
momentum shells ($N=8,20,40,58...$, the so-called "magic numbers"
\cite{B93}). We will work within the adiabatic Born-Oppenheimer
approximation \cite{AM76}, treating the ions as static, compared
with the lighter and faster valence electrons. Moreover, we will
work within the jellium approximation, which neglects effects of the
ionic structure and treats the ions as a continuous, homogeneous
positively charged background of total charge $+Ne$, producing a
charge density in the case of a spherical nanoparticle
$\rho_i(r)=(Ne/V)\Theta(a-r)$, with $V=4\pi a^3/3$ the volume of the
cluster and $\Theta(x)$ the Heaviside step function. The subsequent
model is simple enough to be applied for the description of big
clusters containing up to several thousands of atoms, and the
effects linked to the ionic structure turn out to be important only
for the smallest cases, $N<10$ \cite{B93}.\\

The electronic hamiltonian for an alkaline-metal nanoparticle in
vacuum is in this context given by
\begin{equation}\label{Helectronic}
    H=\sum_{i=1}^{N}\Bigl[\frac{p_i^2}{2m_e}+U(r_i)\Bigr]+\frac{e^2}{2}\sum_{\substack {i,j=1 \\ (i\neq j)}}\frac{1}{|\textbf{r}_i-\textbf{r}_j|}\,,
\end{equation}
where $e$ and $m_e$ are the electronic charge and mass,
$\textbf{r}_i$ is the position of the $i^{th}$ conduction electron,
$r_i=|\textbf{r}_i|$, $\textbf{p}_i$ is the associated momentum, and
$U(r)$ represents the electrostatic interaction between the
electrons and the ionic background.\\

For a noble metal cluster embedded in a dielectric medium, the
electron-electron interaction in (\ref{Helectronic}) is modified by
the presence of the d-electrons and the matrix in terms of two
dielectric constants $\e_d$ and $\e_m$, and the expression for the
Coulomb potential is much more cumbersome due to the loss of
translational invariance. For a thorough study of this case see
\cite{W06,WMWJ05}.\\

The shape of the confining potential $U(r)$ for a charged sphere of
radius $a$ and total charge $+Ne$, harmonic inside the nanoparticle
with the Mie frequency $\om_M = \sqrt{Ne^2/m_ea^3}$ and Coulomb-like
outside, can be readily obtained using Gauss's theorem,
\begin{equation}\label{Harlomb}
    U(r)=\frac{Ne^2}{2a^3}(r^2-3a^2)\Theta(a-r)-\frac{Ne^2}{r}\Theta(r-a)
\end{equation}
If the many-body eigenstates $|f\rangle$ and eigenenergies $E_f$ of
the hamiltonian (\ref{Helectronic}) are known, the photoabsorption
cross section $\sigma(\om)$ can be calculated using Fermi's golden
rule \cite{M70}
\begin{equation}\label{Photoabs}
    \sigma(\om)=\frac{4\pi e^2\om}{3c}\sum_f|\langle
    f|z|0\rangle|^2\delta(\hbar\om-E_f+E_0)
\end{equation}
The ground state is $|0\rangle$, and the electric field is assumed
to be along the $z$ axis. From $\sigma(\om)$ the frequency
$\om_{pl}$ and linewidth $\gamma_{pl}$ of the surface plasmon are
immediately obtained. But the calculation of the many-body
eigenstates becomes exceedingly difficult for clusters bigger than a
few electrons, and further approximation schemes have to be
developed.

The most useful one starts by changing the electronic coordinate
basis to the center of mass $\textbf{R}=\sum_i \textbf{r}_i/N$ and
relative coordinates $\textbf{r}'_i=\textbf{r}_i-\textbf{R}$. The
corresponding conjugated momenta are $\textbf{P}=\sum_i\textbf{p}_i$
and $\textbf{p}'_i=\textbf{p}_i-\textbf{P}/N$, respectively, and
(\ref{Helectronic}) becomes
\begin{equation}\label{Helectronic2}
    H=\frac{\textbf{P}^2}{2Nm_e}+H_{rel}+\sum_{i=1}^N [U(|\textbf{r}'_i+\textbf{R}|)-U(r'_i)]
\end{equation}
where the hamiltonian for the relative-coordinate system reads
\begin{equation}\label{Hrelative}
H_{rel}=\sum_{i=1}^N\Bigl[\frac{p'_i\,^2}{2m_e}+U(r'_i)\Bigr]+\frac{e^2}{2}\sum_{\substack
{i,j=1 \\ (i\neq j)}}\frac{1}{|\textbf{r'}_i-\textbf{r'}_j|}
\end{equation}
A first glance comparing eqs. (\ref{Helectronic}) and
(\ref{Hrelative}) may make us wonder where is the difference between
them, and what is our motivation to go from (\ref{Helectronic}) to
(\ref{Helectronic2}). Regarding the difference between hamiltonians,
note that one of the relative coordinates is a function of the rest,
as $\sum_i\textbf{r}'_i=0$. The change of variables, on the other
hand, is justified as it is the right one if we are interested in
the study of the surface plasmon collective excitation, which was
shown in the previous chapter to correspond to electronic
center-of-mass oscillations.

Assuming small displacements for the center of mass,
$|\textbf{R}|\ll a$, the dependence on the relative coordinates in
the last term of (\ref{Helectronic2}) drops out,
$U(|\textbf{r}'+\textbf{R}|)-U(r')\simeq \textbf{R}\cdot\nabla
U(r')+(1/2)(\textbf{R}\cdot\nabla)^2U(r')$, where derivatives are
taken at $\textbf{r}'=\textbf{r}$. Choosing the oscillation axis of
the center of mass in the $z$-direction, $\textbf{R}=Z\textbf{e}_z$,
one obtains with (\ref{Harlomb})
\begin{eqnarray}\label{Hrelative2}
  \textbf{R}\cdot\nabla U(r')&=& Zm_e\om^2_M\Bigl[z'\Theta(a-r')+\frac{z'a^3}{r'^3}\Theta(r'-a)\Bigr] \\
  (\textbf{R}\cdot\nabla)^2U(r') &=&
   \nonumber Z^2Ne^2\Bigl[\frac{1}{a^3}\Theta(a-r')+\frac{1-3\cos^2\theta'}{r'^3}\Theta(r'-a)\Bigr]\simeq
  Z^2\frac{Ne^2}{a^3}\Theta(a-r')
\end{eqnarray}
where in the last line a second-order coupling in $Z$ between the
center-of-mass and the relative-coordinate system for $r'>a$ has
been neglected as compared with the first-order one expressed in the
second term of the right-hand side of the first line. Inserting
(\ref{Hrelative2}) into (\ref{Helectronic2}),
\begin{equation}\label{Helectronic3}
    H=\frac{\textbf{P}^2}{2Nm_e}+\frac{1}{2}\frac{Ne^2}{a^3}\textbf{R}^2\sum_{i=1}^N\Theta(a-r_i)+H_{rel}+H_c=H_{cm}+H_{rel}+H_c\,,
\end{equation}
where the coupling $H_c$ between the center-of-mass and the relative
coordinates is, to first order in the displacement $\textbf{R}$ of
the center of mass
\begin{equation}\label{Helectronic4}
    H_c=\sum_{i=1}^N\textbf{R}\cdot\nabla U(r'_i)\Bigr|_{\textbf{R}=0}\,,
\end{equation}
with $\textbf{R}\cdot\nabla U(r'_i)\Bigr|_{\textbf{R}=0}$ given in
(\ref{Hrelative2}). Notice that due to the spread of the electrons'
wavefunctions outside the geometrical boundary of the cluster,
$r=a$, there is a finite probability for them to have $r_i>a$, which
due to the sum over $\Theta(a-r_i)$ in (\ref{Helectronic3}), results
in a reduction of the center-of-mass oscillation frequency from the
classical Mie value, $\tilde{\om}_M=\om_M\sqrt{1-N_{out}/N}<\om_M$.
This quantum correction is the so-called \textit{spill-out effect}
\cite{H93,B93,KV95}.

The structure of (\ref{Helectronic3}) is typical for quantum
dissipative systems \cite{W99}: The system under study ($H_{cm}$) is
coupled via $H_c$ to an environment or "heat bath" described by
$H_{rel}$. The system-environment coupling results in dissipation
and decoherence of the collective excitation.

\section{Mean-field approximation and second quantization of the
hamiltonian}\label{secMeanfield} The presence in $H_{rel}$ of
electronic interactions makes the problem still a formidable one and
demands further simplifications. A mean-field approximation for the
relative-coordinates' degrees of freedom, together with a second
quantization procedure, will render a tractable hamiltonian suited
for the study of collective
excitations in metallic clusters.\\

\begin{figure}[t]
\begin{center}
\includegraphics[width=10cm]{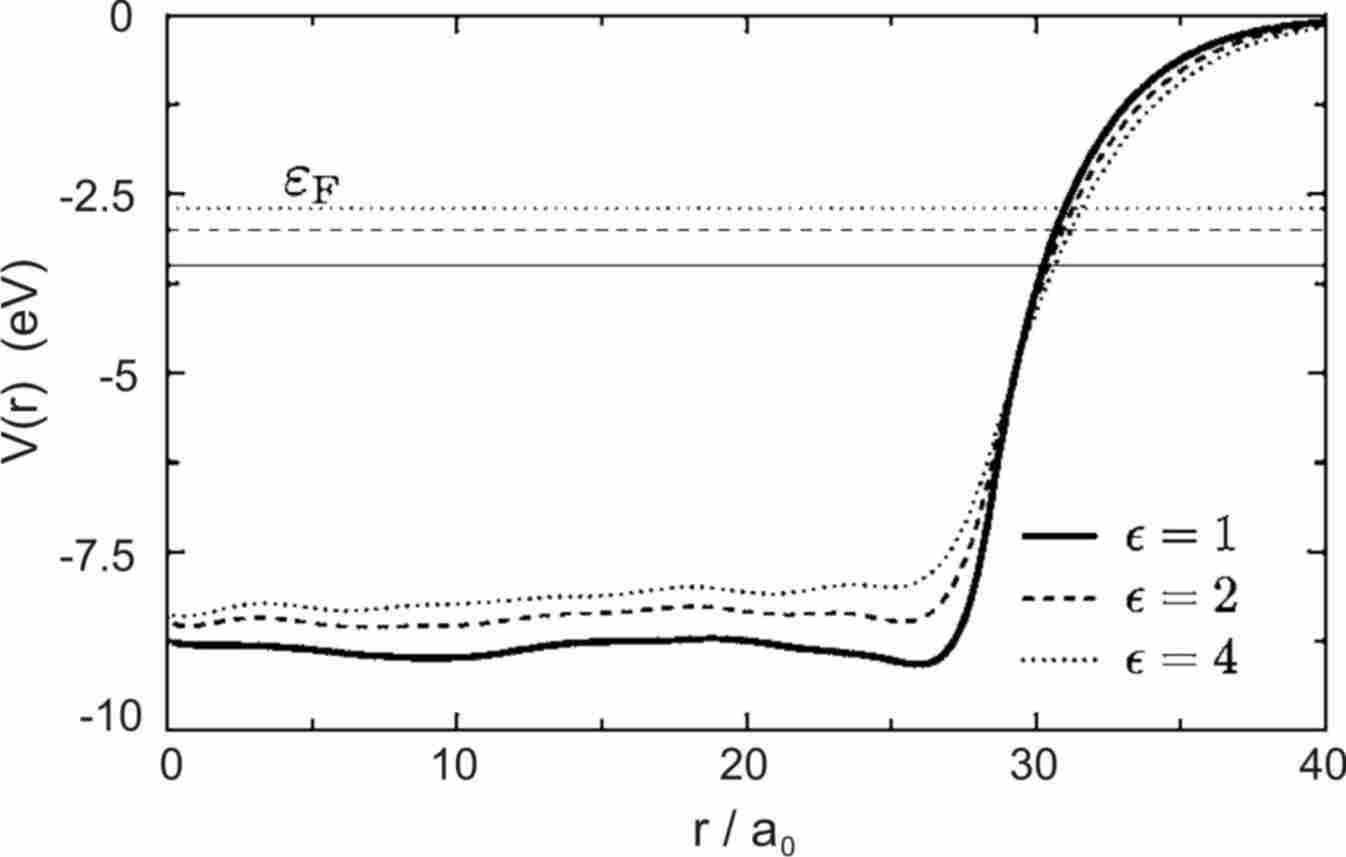}\\
\caption[Self-consistent mean-field potential $V$ from TDLDA
numerical calculations]{\label{TDLDA} From \cite{WMWJ05}.
Self-consistent mean-field potential $V$ as a function of the radial
coordinate (in units of the Bohr radius $a_0 = 0.53{\AA}$) from
TDLDA numerical calculations (at $T = 0$) for a nanoparticle with $N
= 832$ valence electrons. The mean distance between electrons $r_s =
(3/4\pi n_e)^{1/3} = 3.02 a_0$ (silver), corresponding to
$a\simeq28.5 a_0$; for Na $r_s=3.93a_0$ \cite{AM76}, yielding
similar results for $V(r)$, as the $\epsilon=1$ case is equivalent
to the one of an alkaline-metal cluster in vacuum. For most purposes
in analytical calculations $V$ can be approximated by a step
function with discontinuity at $r=a$.}
\end{center}
\end{figure}

The harmonic oscillator hamiltonian $H_{cm}$ for the center-of-mass
is quantized as usual with bosonic creation and annihilation
operators \cite{BF04},
\begin{equation}\label{hcm}
    \left\{
      \begin{array}{l}
        b = \sqrt{\frac{Nm_e\tilde{\om}_M}{2\hbar}}Z+\frac{i}{\sqrt{2Nm_e\hbar\tilde{\om}_M}}P_Z \\
        b^\dagger=\sqrt{\frac{Nm_e\tilde{\om}_M}{2\hbar}}Z-\frac{i}{\sqrt{2Nm_e\hbar\tilde{\om}_M}}P_Z
      \end{array}
    \right.
\end{equation}
where $P_Z$ is the conjugated momentum to $Z$. Eigenstates
$|n\rangle$ of $H_{cm}$ satisfy $b|n\rangle=\sqrt{n}|n-1\rangle$ and
$b^\dagger|n\rangle=\sqrt{n+1}|n+1\rangle$, and $H_{cm}$ is
expressed in terms of $b$ and $b^\dagger$, subtracting the
zero-point energy $\hbar\tilde{\om}_M/2$, as
 \begin{equation}\label{hcm2}
    H_{cm}=\hbar\tilde{\om}_Mb^\dagger b
 \end{equation}
In the mean-field approximation for the hamiltonian $H_{rel}$,
eq.(\ref{Hrelative}), we forget the fact that one of the relative
coordinates is not independent of the rest (hopefully causing $1/N$
errors which become negligible as the number of electrons $N$
grows), and substitute $H_{rel}$ by the mean-field approximation of
eq.(\ref{Helectronic}), mimicking the effect of the rest of
electrons on a given one with an effective one-body potential
$V(\textbf{r})$ felt by the latter:
\begin{equation}\label{Hrelative3}
    H_{rel}\approx\sum_{i=1}^N H_{sc}[r'_i,p'_i]=\sum_{i=1}^N\Bigl[\frac{p'_i\,^2}{2m_e}+V(r'_i)\Bigr]
\end{equation}
We have used $V(\textbf{r})=V(r)$, valid for a spherical
nanocluster. Numerical calculations using time-dependent local
density approximation (TDLDA) within the spherical jellium model at
zero temperature provide information about the shape of the
potential $V(r)$ \cite{E84,E84b,GK85,GDP96,CRSU00}. In this approach
the original hamiltonian (\ref{Helectronic}) is replaced by a
self-consistent hamiltonian like (\ref{Hrelative3}) where $V(r)$ is
a local function containing exchange and correlation terms. Results
show that for most purposes $V(r)$ can be very well approximated in
analytical calculations by a step function, with the jump located at
$r=a$. An example of TDLDA result for $V(r)$ can be found in fig.
\ref{TDLDA}.

From the self-consistent single-particle hamiltonian
(\ref{Hrelative3}) a second-quantized version immediately follows in
terms of fermionic creation and annihilation operators
$c^\dagger_\alpha$ and $c_\alpha$,
\begin{equation}\label{Hrelative4}
    H_{rel}=\sum_\alpha \e_\alpha c^\dagger_\alpha c_\alpha\,\,,
\end{equation}
where $c^\dagger_\alpha$ creates the one-body state $|\alpha\rangle$
whose wavefunction in the coordinate representation satisfies
$H_{sc}\psi_\alpha(\textbf{r})=\e_\alpha\psi_\alpha(\textbf{r})$.\\

The coupling term $H_c$ between the center-of-mass and the relative
coordinates becomes in terms of the bosonic and fermionic operators
$b^\dagger$ and $c^\dagger_\alpha$
\begin{equation}\label{Hcoupl}
    H_c=\Lambda(b^\dagger+b)\sum_{\alpha\,\beta}d_{\alpha\beta}c^\dagger_\alpha
    c_\beta\,\,,
\end{equation}
where $\Lambda=\sqrt{\hbar m_e\om_M^3/2N}$, and
\begin{equation}\label{dab}
    d_{\alpha\beta}=\langle\alpha|\Bigl[z\Theta(a-r)+\frac{za^3}{r^3}\Theta(r-a)\Bigr]|\beta\rangle
\end{equation}
is the matrix element between two eigenstates of the unperturbed
mean-field problem. In the constant $\Lambda$ the approximation
$\tilde{\om}_M\approx\om_M$ has been included, as it leads to
$1/a^2$ negligible corrections in the quantities one wants to
compute, like the lifetime of the surface plasmon, using the second
quantized approximation for (\ref{Helectronic}),
\begin{equation}\label{Helectronic5}
    H=\hbar\tilde{\om}_Mb^\dagger b+\Lambda(b^\dagger+b)\sum_{\alpha\,\beta}d_{\alpha\beta}c^\dagger_\alpha
    c_\beta+\sum_\alpha \e_\alpha c^\dagger_\alpha c_\alpha
\end{equation}

\section{Surface plasmon linewidth and finite size effects}
\subsection{Mechanisms causing the decay of the surface plasmon}
The finite lifetime of the surface plasmon excitation can be related
to several processes. When Mie performed his study of the optical
response of metal clusters using Maxwell's equations, he used a
simple free-electron Drude model \cite{AM76} for the dielectric
constant describing the metal. Within this approximation,
\begin{equation}\label{Drudee}
    \e(\om)=\e_d-\frac{\om_p^2}{\om(\om+i\gamma_i)}\,,
\end{equation}
where $\om_p=\sqrt{4\pi n_ee^2/m_e}$ is the plasma frequency,
$\gamma_i^{-1}$ the relaxation or collision time, while $e$, $m_e$
and $n_e$ stand for the electron charge, mass and bulk density,
respectively. $\e_d$ is the dielectric constant of the metal without
the free-carrier contribution. Inserting (\ref{Drudee}) in the
expression he had derived for the photoabsorption cross section, he
obtained
\begin{equation}\label{Drudee2}
    \sigma(\om)=\frac{9\om_p^2\mathcal{V}}{2c}\frac{\e_m^{3/2}}{(\e_d+2\e_m)^2}\frac{\gamma_i/2}{(\om-\om_M)^2+(\gamma_i/2)^2}\,,
\end{equation}
where $\e_m$ is the dielectric constant of the medium in which the
particle is embedded, and $\mathcal{V}$ the volume of the cluster.
Expression (\ref{Drudee2}) predicts the existence of the surface
plasmon at the Mie frequency $\om_M$, with a linewidth $\gamma_i$.
This value of $\gamma_i$ accounts for the various processes leading
to elastic or inelastic scattering in \emph{bulk} metals, such as
interactions with phonons, electrons, impurities, etc. Experiments
give, however, a much larger value for the linewidth, indicating
that the plasmon lifetime is due mainly to quantum mechanical
effects arising in such a confined system, absent in Mie's
treatment. For example, for Sodium $\hbar\gamma_i\simeq20$ meV,
while the observed linewidth for Na clusters with $a=2.5$ nm is more
than 10 times this value.

\begin{figure}[t]
\begin{center}
\includegraphics[width=10cm]{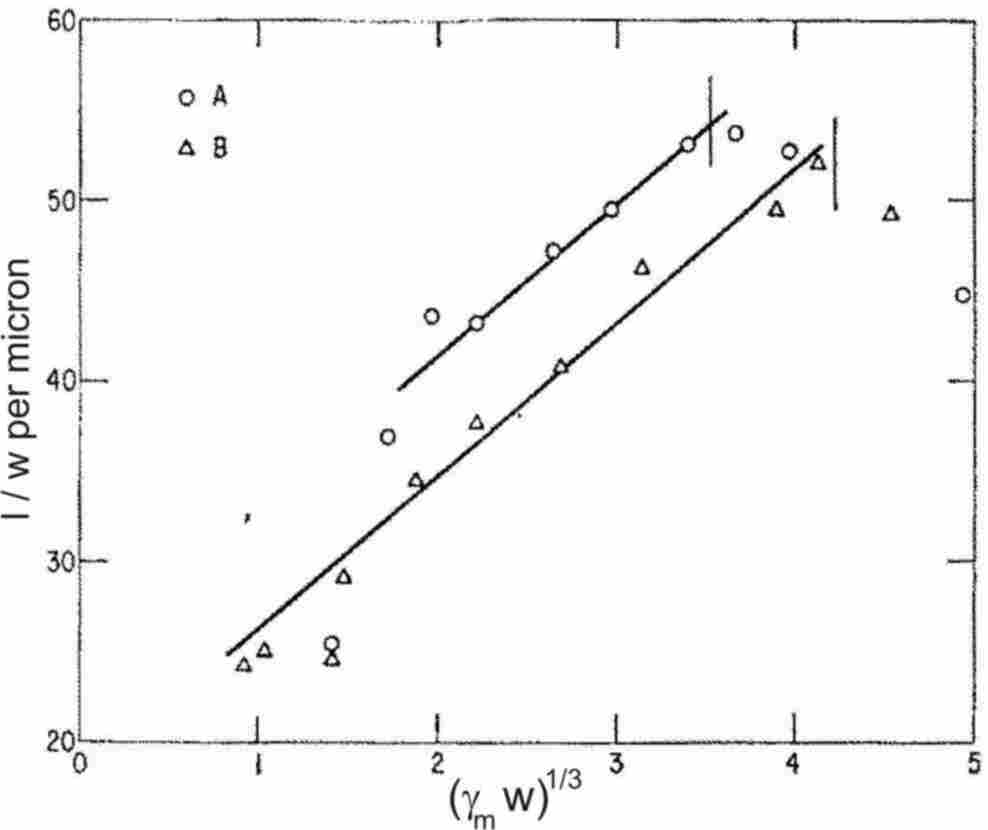}\\
\caption[Linear scaling with size of the surface plasmon's
lifetime]{\label{landaudamping} From \cite{D65}. Experimental
results for the inverse surface plasmon linewidth of silver
nanoparticles embedded in a glass matrix. The quantity represented
on the abscissa is proportional to the radius $a$ of the
nanoparticles (between $\sim$4 and 10 nm), while the vertical axis
shows the inverse of the linewidth, $\tau=\gamma^{-1}$. It clearly
demonstrates the linear size-scaling of the surface plasmon
lifetime. The circles and triangles correspond to two different
samples.}
\end{center}
\end{figure}

We have already described one such finite-size effect, see
eq.(\ref{Helectronic5}) and fig.(\ref{oscplasmon}): the motion of
the electronic center of mass linked to the excitation of the
surface plasmon results in a net surface charge creating a
time-varying dipolar electric field inside the cluster, which
couples to the mean-field one-particle states calculated in the
absence of such a field. Through this coupling, represented by the
second term of the right-hand side of (\ref{Helectronic5}), the
one-particle states become excited, absorbing energy from the field,
and thus damping the surface plasmon. This decay of the surface
plasmon into one-particle one-hole (1p-1h) uncorrelated
single-particle excitations which have the same energy as the
collective mode is called \textit{Landau damping}, just as the decay
in bulk metals of the bulk plasmon into 1p-1h pairs
\cite{BB94,NP99}. The associated linewidth was first calculated
using linear response theory \cite{K57} and the
fluctuation-dissipation theorem \cite{CL00} in \cite{KK66,BS89}, and
can also be obtained applying Fermi's Golden Rule with the coupling
hamiltonian in (\ref{Helectronic5}), \cite{WMWJ05}
\begin{equation}\label{landaudamp}
    \gamma = 2\pi\Lambda^2\sum_{ph}|d_{ph}|^2\delta(\hbar\om_M-\varepsilon_p+\varepsilon_h)\,,
\end{equation}
where $|p\rangle$ and $|h\rangle$ represent, respectively, particle
and hole states of the mean-field problem. The result renders a
plasmon linewidth with a characteristic $\gamma(a)\sim 1/a$
dependence, observed in experiments as shown in fig.
(\ref{landaudamping}):
\begin{equation}
\label{gamma_intro} \gamma(a)=\frac{3v_{\rm F}}{4a}
g_0\left(\frac{\varepsilon_{\rm F}}{\hbar\omega_{\rm M}}\right)\,,
\end{equation}
with $\varepsilon_{\rm F}=\hbar^2k_{\rm F}^2/2m_{\rm e}$ and $v_{\rm
F}$ the Fermi energy and velocity, respectively. The monotonically
increasing function $g_0$ is defined as
\begin{align}
\label{g_0} g_0(x)=&\;\frac{1}{12x^2} \Big\{
\sqrt{x(x+1)} \big(4x(x+1)+3\big) -3(2x+1) \ln{\left(\sqrt{x}+\sqrt{x+1}\right)} \nonumber\\
&-\Big[ \sqrt{x(x-1)} \big(4x(x-1)+3\big) -3(2x-1)
\ln{\left(\sqrt{x}+\sqrt{x-1}\right)}\Big]\Theta(x-1) \Big\}\,,
\end{align}
$\Theta(x)$ being the Heaviside step function. It is very well
approximated by
\begin{equation}\label{approxg0}
    \tilde{g}_0(x)=(1+(225x/64)^2)^{-1/4}
\end{equation}
In the most interesting size range $0.5$ nm$\leq a\leq10$ nm, where
the transition from atomic to bulk behavior of physical properties
takes place, the Landau damping mechanism dominates, limiting the
lifetime of the collective excitation.

For large clusters, $a\geq10$ nm, \textit{radiation damping} due to
the interaction of the surface plasmon with the external
electromagnetic field dominates our mode's attenuation \cite{KV95}.
The lifetime associated to this mechanism can be shown to scale with
size as $\gamma_{rad}\sim a^3$ \cite{KK66}, explaining why it is
dominant for large clusters, becoming negligible for smaller ones in
comparison with the Landau damping, who presents a weaker dependence
of $\gamma$ on $a$. For radii smaller than \unit[0.5]{nm} the
interaction with the ionic background becomes dominant, and the
jellium model no longer provides a useful description. We will focus
our attention in this thesis in the size range where Landau damping
dominates.

\subsection{Landau damping and the double-counting problem}
\label{doublecounting} The final form of the hamiltonian
(\ref{Helectronic5}) is an \emph{almost} perfect starting point for
the description of the surface plasmon dynamics, in particular to
address the Landau damping mechanism, with the plasmon
(center-of-mass) degree of freedom coupled through $H_c$ to the
single-particle excitations, which can be viewed as a dissipative
environment for the plasmon. If the bosonic operator $b^\dagger$ and
the fermionic operators $c^\dagger_\alpha$ corresponded to
well-defined, independent degrees of freedom, one could proceed
without further considerations to calculate for example
perturbatively the effect of the relative coordinates on the plasmon
lifetime, but as we have already noted, the final form of $H_{rel}$
is obtained forgetting the fact that there are not $N$ but $N-1$
independent relative coordinates. Thus, it seems that from $N$
original degrees of freedom in (\ref{Helectronic}) we have arrived
at $N+1$ after the mean-field approximation and second quantization
procedure, double-counting in some way the plasmon degree of freedom
in $H_{rel}$. From the point of view of the study of the plasmon
dissipative dynamics, this spoils the virtues of our approximate
hamiltonian (\ref{Helectronic5}), as we have to guarantee that the
environment to which we couple our system does not include our
system.

Fortunately, previous knowledge coming from a related problem of
nuclear physics, the description of the \textit{giant dipolar
resonance} of nuclei \cite{RS80,YDG82}, gives a hint on how to
circumvent this inconsistency. This nuclear resonance corresponds to
a collective motion of the (charged) protons with respect to the
(neutral) neutrons, when an external electromagnetic field is
applied, analogously to what happens in metallic clusters. Using a
slightly different approach to obtain the approximate eigenstates of
the nuclei, it was shown that the collective state is to a very good
approximation a linear superposition of low-energy 1p-1h excitations
of the Hartree-Fock ground state. The restriction to this kind of
excitation, neglecting multiple p-h states, is a good approximation
to describe the low energy physics of the system. The 1p-1h
excitations are not true eigenstates due to the non-Hartree-Fock
part of the Coulomb interaction, named residual interaction, which
mixes them giving rise to more realistic eigenstates of the system,
most of which are very similar to the ones obtained at the
Hartree-Fock level, except for one linear combination, whose high
energy lies \emph{outside} the low-energy sector where its 1p-1h
constituents are located, and which corresponds to the collective
resonance. We will give a detailed derivation of these results in
the following section.

The surface plasmon is thus basically originated from a combination
of low-energy one-particle mean-field states, showing up when the
Coulomb interaction is included in its full complexity. Coming back
to the description of its dynamics, if we want to calculate its
Landau damping, keeping a clear separation between the surface
plasmon and the degrees of freedom constituting its dissipative
environment, the previous results suggest the following two-step
procedure \cite{YDG82,YB92}:
\begin{enumerate}
  \item Divide the Hilbert space spanned by the 1p-1h Hartree-Fock states into a
  low-energy (called restricted subspace, $S_R$) and a high energy (called additional subspace, $S_A$)
  subspaces. Introduce now the residual interaction in $S_R$, and obtain the eigenstates, one of whom will
  be the surface plasmon, having an energy lying in $S_A$.
  \item To calculate the lifetime of the surface plasmon due to
  Landau damping, just apply Fermi's Golden Rule using the residual
  interaction coupling the plasmon to the quasi-continuum of states
  in $S_A$ degenerate with it.
\end{enumerate}
In the context of the calculation of Landau damping using
(\ref{Helectronic}), the residual interaction is given by $H_c$, and
to avoid the mixing of the surface plasmon and what we consider its
environment, we just have to determine a safe energy cutoff $\dlec$
below the surface plasmon frequency $\om_{pl}$ such that (i) the
value of $\om_{pl}$ is insensitive to the increase of $\dlec$,
showing that the relevant low-energy Hartree-Fock 1p-1h excitations
building up the surface plasmon lie below it, and (ii) only 1p-1h
excitations above $\dlec$, which can be thus regarded as belonging
to a subspace independent of the surface plasmon, are included in
Fermi's Golden Rule sum. The latter excitations are by definition
$S_A$, so that the linewidth has an expression
\begin{equation}\label{Landamping}
    \gamma\propto\sum_{ph\in S_A}|\langle p|z|h\rangle|^2\delta(\hbar\om_{pl}-\e_p+\e_h)
\end{equation}
Here $ph$ is a particle-hole excitation composed of a particle with
energy $\e_p$ and a hole with energy $\e_h$. In this calculation of
$\gamma$ the concrete estimate for the value of $\dlec$ is
irrelevant as long as $\dlec<\om_{pl}$ \cite{YB92}, but the
approximate theoretical description of other physical quantities
might depend on the cutoff. For example, if one uses
(\ref{Helectronic5}) to study the decay and decoherence of the
surface plasmon, the timescales that characterize the dynamics of
the electronic environment in $S_A$ \cite{WIWJ07} depend on the
cutoff energy. Moreover, it has been shown in \cite{WIJW06} that the
environment-induced redshift of the resonance frequency depends
logarithmically on the cutoff.

We will in the following section derive an estimate for this cutoff
based on sound physical arguments.

\section{The surface plasmon as a superposition of low-energy particle-hole excitations}
To show this fact and find an estimate for $\dlec$ we will forget in
this section about the model based in a separation of center-of-mass
plus relative coordinates described in the beginning of previous
section, whose main result was eq. (\ref{Helectronic5}), and use
instead the approach borrowed from the description of giant dipolar
resonances in nuclei which has already been briefly introduced.
\subsection{Separable interaction ansatz}
We now obtain the simplest model which starting from a mean-field
approach includes the essential features of the Coulomb interaction
originating the surface plasmon. The starting point is the simplest,
mean-field approximation of the hamiltonian (\ref{Helectronic}),
where the one-body electronic states obtained come from a
self-consistent calculation like the TDLDA one (fig. \ref{TDLDA}).
The problem with this naive approximation is that it fails to
describe the ability of the Coulomb interaction to create
self-sustained collective excitations, like the surface plasmon we
want to study. We have to introduce an additional component $\delta
V(r)$ to the self-consistent potential $V(r)$ which captures this
feature of the Coulomb interaction. To obtain $\delta V(r)$, we
recall the fact that the occurrence of collective vibrational modes
in a system governed by independent-particle motion can be
understood in terms of the variations in the average one-body
potential produced by an oscillation in the nanoparticle electronic
density. Such variations in the one-particle potential give rise to
excitations in the electronic motion, and a self-sustained
collective motion is obtained if the induced density variations are
equal to those needed to generate the oscillating
potential \cite{BM75}.\\

In our case, from the several collective modes susceptible of
appearing in our system we focus our attention on the simplest,
dipolar in character, surface plasmon. Starting from the knowledge
of the existence of such a mode, the procedure to obtain the
associated $\delta V(r)$ is then straightforward: calculate the
variation in the average one-body potential $V(r)$ produced by an
electronic density variation $\delta n(\textbf{r})$ linked to the
excitation of the collective mode. This was done in \cite{WMWJ05},
as follows:

Assuming that at equilibrium the electron density is uniform within
a sphere of radius $a$, $n_e(r) = n_e\Theta(a - r)$, a rigid
displacement with a magnitude $Z$ along the $z$-direction changes
the density at \textbf{r} from $n_e(\textbf{r})$ to
$n_e(\textbf{r}-\textbf{R})=n_e(\textbf{r})+\delta n_e(\textbf{r})$.
To first order in the field $\textbf{R}=Z\textbf{e}_z$, $\delta
n_e(\textbf{r})=-\textbf{R}\cdot\nabla
n_e(\textbf{r})=Zn_e\cos(\theta)\delta(r-a)$, where
$\theta=\arccos(\textbf{r}\cdot\textbf{e}_z/|\textbf{r}|)$. The
related change in the mean-field potential seen by the electrons is
thus
\begin{equation}\label{separable1}
    \delta V(\textbf{r},\textbf{R})=\int d^3\textbf{r}'\delta
    n_e(\textbf{r}')V_C (\textbf{r},\textbf{r}')
\end{equation}
where $V_C (\textbf{r},\textbf{r}')$ denotes the Coulomb interaction
between electrons. The term $\delta H$ which has to be added to the
mean-field hamiltonian (\ref{Hrelative3}) is then
\begin{equation}\label{separable2}
    \delta H=\sum_{i=1}^N\delta V(\textbf{R},\textbf{r}'_i)
\end{equation}
A simple expression for $\delta V(\textbf{r},\textbf{R})$ can be
obtained, given the spherical symmetry of the problem, using a
multipolar decomposition for the Coulomb interaction \cite{J75},
namely
\begin{equation}\label{separable3}
     \delta V(\textbf{r},\textbf{R})=Z\frac{4\pi
     n_ee^2}{3}\Bigl[z\Theta(a-r)+\frac{za^3}{r^3}\Theta(r-a)\Bigr]\,,
\end{equation}
so $\delta V(\textbf{r},\textbf{R})$ is actually the same as
$\textbf{R}\cdot\nabla U(r)$, eq.(\ref{Helectronic2}). A further
simplification is obtained if we remember that, even though the
electronic wavefunctions spread outside the geometrical boundaries
of the nanocluster, giving rise to small quantum corrections like
the spill-out effect, the expected influence of the exact form of
$\delta V(\textbf{r},\textbf{R})$ for $r>a$ is very small. We can
then simplify $\delta V$ by extending its dipolar character also for
$r>a$, $\delta V(\textbf{r},\textbf{R})\approx(4\pi n_ee^2/3)Z\cdot
z$, with a resulting \textit{separable form} for $\delta H$
\begin{equation}\label{separable4}
    \delta H\approx\frac{1}{2N}\frac{4\pi n_ee^2}{3}\sum_{i\,j}^Nz_iz_j
\end{equation}
where $Z=\sum_i^Nz_i/N$ has been used. The term separable applies to
those interactions satisfying $V(x_i,x_j)\sim f(x_i)\cdot f(x_j)$,
$f(x)$ being a generic function of $x$, like (\ref{separable4}). The
results that would be obtained without the use of this approximation
would include $1/N$ corrections to the ones we will present, like
the estimate for $\dlec$, so for our purposes we can work using the
separable ansatz. Expressed in terms of the mean-field hamiltonian
basis $|\alpha\rangle$, the second-quantized version of
(\ref{separable4}) reads
\begin{equation}
\label{eq:V_res} \delta H=\sum_{\alpha\beta\gamma\delta} \lambda
d_{\alpha\gamma} d_{\delta\beta}^{*} c_\alpha^\dagger
c_\beta^\dagger c_\delta c_\gamma \,,
\end{equation}
where $d_{\alpha\gamma}=\langle \alpha|z|\gamma\rangle$ are dipole
matrix elements and $\lambda$ is a positive constant.

\subsection{Linear response theory and Random Phase Approximation for the surface
plasmon}\label{linear_response} We will analyze the response of the
nanocluster to a time-dependent weak external perturbation, and
obtain this response in a self-consistent way, leading to the RPA
theory of response. The result will be written in terms of a
response function, yet to be defined, whose poles determine the
energies of the excited states. We will particularize to the
separable potential of the dipolar field (\ref{eq:V_res}), and see
what are its effects on the energies of the self-consistent
mean-field states. We will obtain a secular equation for those RPA
eigenenergies, and show how to solve it
graphically. \\

We follow \cite{BB94} and \cite{RS80}. Suppose we know the
stationary one-particle eigenstates $|i\rangle$ and eigenenergies
$e_i$ of the mean-field, time-independent hamiltonian
(\ref{Hrelative3}), and now introduce a time-dependent external
perturbation $V_x({\bf \vec{r}},t)=V_x({\bf \vec{r}})e^{i\om t}$ of
a given frequency $\om$, interacting with the system through its
particle density $n({\bf \vec{r}})$,
\begin{equation}\label{pertham}
    H_{int}(t)=\int d{\bf \vec{r}}V_x({\bf \vec{r}},t)n({\bf \vec{r}})
\end{equation}
Then, the correction to the eigenstates of the system is, to first
order in $H_{int}(t)$,
\begin{equation}\label{eigenf}
    \phi_i'(t)=\Bigl[\phi_i+\frac{1}{2}\sum_j\langle j|V_x({\bf \vec{r}})|i\rangle\Bigl(\frac{e^{i\om t}}{e_i-e_j-\hbar\om}+
    \frac{e^{-i\om t}}{e_i-e_j+\hbar\om}\Bigr)\Bigr]e^{-ie_it/\hbar}\,,
\end{equation}
where $\phi_i$ is the single-particle wave function of state
$|i\rangle$. The particle density becomes now $n({\bf
\vec{r}},t)=\sum_i^{occ}|\phi_i'({\bf \vec{r}},t)|^2=n_0({\bf
\vec{r}})+\delta n({\bf \vec{r}})e^{i\om t}$, with
\begin{equation}\label{partdens1}
    \delta n({\bf \vec{r}})=\sum_i^{occ}\sum_j\langle j|V_x({\bf \vec{r}})|i\rangle\langle
    i|n({\bf \vec{r}})|j\rangle\Bigl(\frac{1}{e_i-e_j-\hbar\om}+
    \frac{1}{e_i-e_j+\hbar\om}\Bigr).
\end{equation}
We now define the independent particle retarded response function
$R^0$ for the density change at ${\bf \vec{r}}$ induced by a
potential field at ${\bf \vec{r}}'$, as
\begin{equation}\label{response1}
    R^0 ({\bf \vec{r}},{\bf \vec{r}}',\om)=\sum_i^{occ}\sum_j\langle j|n({\bf \vec{r}})|i\rangle\langle
    i|n({\bf \vec{r}}')|j\rangle\Bigl(\frac{1}{e_i-e_j-\hbar\om}+ \frac{1}{e_i-e_j+\hbar\om}\Bigr),
\end{equation}
In terms of $R^0$, the response to an arbitrary perturbation
potential can be calculated from the integral,
\begin{equation}\label{partdens2}
    \delta n({\bf \vec{r}},\om)=\int d{\bf \vec{r}}'R^0({\bf \vec{r}},{\bf \vec{r}}',\om)V_x({\bf \vec{r}}').
\end{equation}
This is the independent particle response, but one more ingredient
has to be included to construct the RPA theory of response. So far
the only perturbation was the external field, but it is clear that
the induced density oscillation $\delta n({\bf \vec{r}},\om)$ will
cause the self-consistent field to oscillate at the same frequency
as well. We can define the change of the mean-field potential with
respect to small density changes in terms of the functional
derivative $\delta V({\bf \vec{r}})/\delta n({\bf \vec{r}}')$.In the
case of an external long-wavelength field we have seen that the main
modification to the mean-field potential is the presence of a
dipolar electric field, described in (\ref{eq:V_res}). In terms of
$\delta V({\bf \vec{r}})/\delta n({\bf \vec{r}}')$ the time-varying
mean field is then
\begin{equation}\label{response1b}
   \delta V({\bf \vec{r}})=\int d{\bf \vec{r}}'\frac{\delta V({\bf \vec{r}})}{\delta n({\bf \vec{r}}')}\delta n({\bf \vec{r}}')
\end{equation}
Adding this potential to the external potential, we obtain the
implicit equation for the self-consistent density,
\begin{equation}\label{response1c}
   \delta n_{RPA}({\bf \vec{r}},\om)=\int d{\bf \vec{r}}R^0({\bf \vec{r}},{\bf \vec{r}}_2,\om)\Bigl\{V_x({\bf \vec{r}}_2)+
   \int d{\bf \vec{r}}'\frac{\delta V({\bf \vec{r}}_2)}{\delta n({\bf \vec{r}}')}\delta n_{RPA}({\bf \vec{r}}')\Bigr\}
\end{equation}
It is more useful to express the self-consistent density in terms of
the external field, defining in this way the RPA response function
$R^{RPA}$ (we will always work with retarded response functions, so
 it will be assumed in the following, and not mentioned explicitly as "retarded"),
\begin{equation}\label{partdens2b}
    \delta n_{RPA}({\bf \vec{r}},\om)=\int d{\bf \vec{r}}'R^{RPA}({\bf \vec{r}},{\bf \vec{r}}',\om)V_x({\bf
    \vec{r}}')\,.
\end{equation}
From (\ref{response1c}) and (\ref{partdens2b}) follows the implicit
equation that the RPA response function must satisfy:
\begin{equation}\label{response1d}
   R^{RPA}({\bf \vec{r}},{\bf \vec{r}}',\om)=R^0({\bf \vec{r}},{\bf \vec{r}}',\om)+\int d{\bf \vec{r}}_2d{\bf \vec{r}}_3
   R^0({\bf \vec{r}},{\bf \vec{r}}_2,\om) \frac{\delta V({\bf \vec{r}}_2)}{\delta n({\bf \vec{r}}_3)}R^{RPA}({\bf \vec{r}}_3,{\bf \vec{r}}',\om)
\end{equation}
If the external perturbation were static, $\om=0$, equations
(\ref{response1c}) and (\ref{response1d}) reduce simply to the
Hartree-Fock self-consistent solution to the problem, where the
occupied eigenstates contribute to the total effective one-body
potential in the hamiltonian, which in turn changes the occupied
wavefunctions, starting a cycle that ends when selfconsistency
between eigenfunctions and total potential is reached. The RPA
approximation is therefore nothing but the small amplitude limit of
the time-dependent mean field approach.

The next step is to repeat the linear response calculations,
imposing again (i)a self-consistent mean field approach for the
total hamiltonian and (ii)small amplitude limit for the
perturbation, but working in the basis and language in which
(\ref{eq:V_res}) is written. The derivation of the response becomes
much more tedious and lengthy (see e.g. \cite{RS80,BB94}), so we
will state here the main points and results, the latter being very
similar in structure to those we have obtained:\\

\begin{figure}[!h]
\begin{center}
\includegraphics[width=15cm]{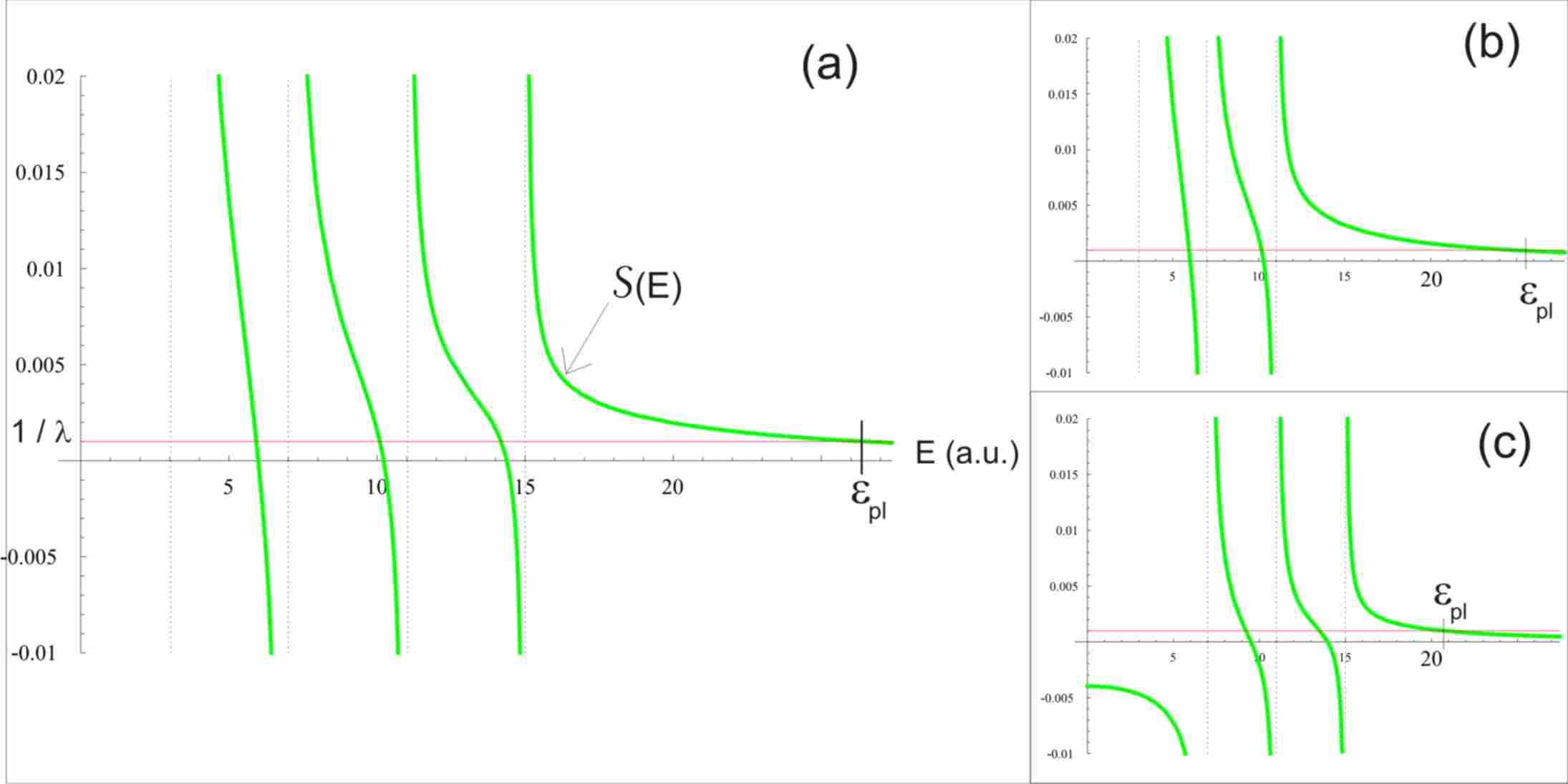}
\caption[Graphical solution of a secular equation of the kind (\ref{RPAsum_secular})]{\label{RPAsumfig}%
(a) Schematic example of the graphical solution of a secular
equation of the kind (\ref{RPAsum_secular}). In this case the dipole
matrix elements are assumed to have a dependence with $\Delta
\varepsilon_{ph}$, $d_{ph}\sim1/\Delta \varepsilon_{ph}^2$, which
corresponds to the realistic case of metallic clusters. $1/\lambda$
is chosen to be $1/\lambda=0.002$ (arbitrary units), while the sum
includes only 4 1p-1h excitations, whose excitations energies in
absence of the dipolar residual interaction (\ref{eq:V_res}) are
$\Delta \varepsilon_{ph}=3,7,11,15$ (arbitrary units): ${\cal
S}(E)=(1/3)/(E^2-3^2)+(1/7)/(E^2-7^2)+(1/11)/(E^2-11^2)+(1/15)/(E^2-15^2)$.
Due to the residual interaction the eigenstates and eigenenergies
change, with the new excitation eigenenergies $E$ given by the
crossing points where ${\cal S}(E)=1/\lambda$. The highest
excitation energy corresponds to the surface plasmon collective
mode, $\varepsilon_{pl}$. (b) and (c) reflect the influence of high
and low 1p-1h excitations in the value of $\varepsilon_{pl}$,
respectively; in (b) $\Delta \varepsilon_{ph}=15$ is absent, causing
a slight shift of the surface plasmon energy $\varepsilon_{pl}$,
while in (c) $\Delta \varepsilon_{ph}=3$ is absent, resulting in a
strong shift of $\varepsilon_{pl}$. This points to the predominant
role of low-energy p-h excitations in the formation of the
collective mode.}
\end{center}
\end{figure}

Take the creation and destruction operators $c_\alpha^\dagger$ and
$c_\alpha$ of the mean-field states of the unperturbed hamiltonian.
The external perturbation is chosen to be a one-body operator,
$V_x(t)=V_xe^{-i\om t}+V^\dagger_xe^{i\om
t}=\sum_{kl}v_{kl}(t)c_k^\dagger c_l$. As also the rest of operators
in the hamiltonian will be approximated to self-consistent one-body
operators, one assumes
\begin{itemize}
  \item First, that the ground state wavefunction $|\rm{RPA}(t)\rangle$ at
any time is a Slater determinant, imposing in this way
selfconsistency.
  \item Second, one assumes that the external field
$V_x(t)$ is weak, introducing only small perturbations around the
stationary Hartree-Fock ground state $|\rm{HF}\rangle$ of the
unperturbed hamiltonian,
$|\rm{RPA}(t)\rangle=|\rm{HF}\rangle+\delta|\psi(t)\rangle$.
\end{itemize}
The combination of Slater-determinant sort of solution plus small
perturbations with respect to $|\rm{HF}\rangle$ limits automatically
the Hilbert space out of which $|\rm{RPA}(t)\rangle$ is constructed,
to the stationary Hartree-Fock ground state $|\rm{HF}\rangle$ plus
the set of all its 1p-1h excitations $c_p^\dagger
c_q|\rm{HF}\rangle$ ($c_q$ annihilates an occupied one-particle
state of $|\rm{HF}\rangle$, while $c_p^\dagger$ populates a
previously empty one-particle state). This shows up in the so-called
\textit{spectral representation} of the RPA response function
obtained
\begin{equation}\label{Polarization1}
    R_{pqp'q'}^{RPA}(\om)=\sum_{i> 0}\Bigl(\frac{\langle0|c_q^\dagger c_p|i\rangle\langle i|c_{p'}^\dagger c_{q'}|0\rangle}
    {\hbar\om-\e_i+\e_0}-\frac{\langle0|c_{p'}^\dagger c_{q'}|i\rangle\langle i|c_q^\dagger
    c_p|0\rangle}{\hbar\om+\e_i-\e_0}\Bigr)\,\,,
\end{equation}
where the index pairs $pq$ and $p'q'$ run only over particle-hole
($ph$) and hole-particle ($hp$) pairs, while all other matrix
elements vanish in RPA order. Unsurprisingly, $R_{pqp'q'}^{RPA}$
satisfies an implicit equation analogous to (\ref{response1d}),
\begin{equation}\label{Polarization2}
   R_{pqp'q'}^{RPA}=R_{pqp'q'}^{0}+\lambda\sum_{\substack {p_1q_1 \\ p_2q_2}}R_{pqp_1q_1}^{0}d_{p_1q_2} d_{p_2q_1}^{*}R_{p_2q_2p'q'}^{RPA}
\end{equation}
In (\ref{Polarization2}) we have introduced the spectral
representation $R_{pqp'q'}^{0}$ of the response function of the free
system,
\begin{equation}\label{Polarization1b}
    R_{pqp'q'}^{0}(\om)=\delta_{pp'}\delta_{qq'}\frac{\sum_{i> 0}|\langle i|c_{p}^\dagger c_{q}|0\rangle|^2-|\langle i|c_q^\dagger c_p|0\rangle|^2}
    {\hbar\om-\e_p+\e_q}\,,
\end{equation}
and we have also particularized to our case, where $V_x(t)$ is the
dipolar field represented by the corresponding operator
(\ref{eq:V_res}). Thanks to the separable form of the interaction,
we can proceed further with (\ref{Polarization2}) to get
\begin{equation}\label{Polarization3}
    R^{RPA}(\om)=R^0(\om)[1+\lambda R^{RPA}(\om)]\,\,\,\rightarrow\,\,\,R^{RPA}(\om)=\frac{R^0(\om)}{1-\lambda R^0(\om)}
\end{equation}
with
\begin{equation}\label{Polarization4}
    R^0(\om)=\sum_{pqp'q'}d_{pq}^{*}R_{pqp'q'}^{0}d_{p'q'}=\sum_{ph}|d_{ph}|^2\Bigl(\frac{1}{\hbar\om-\e_p+\e_h}-\frac{1}{\hbar\om+\e_p-\e_h}\Bigr)
\end{equation}
The last sum runs over 1p-1h states, $|ph\rangle$, whose energy with
respect to the RPA ground state is
$\Delta\varepsilon_{ph}=\e_p-\e_h$. The poles of $R^{RPA}(\om)$ give
the excitation energies $E$ we are looking for, which are thus
defined by the secular equation
\begin{equation}
\label{RPAsum_secular} \boxed{\frac{1}{\lambda}={\cal S}(E)}
\end{equation}
Here we have defined the RPA sum
\begin{equation}
\label{RPAsum} {\cal S}(E)= \sum_{ph}\frac{2
\Delta\varepsilon_{ph}|d_{ph}|^2} {E^2-\Delta \varepsilon_{ph}^2}\,.
\end{equation}
Nuclear physics textbooks (see, e.g., Ref.~\cite{RS80}) show how to
solve the secular equation \eqref{RPAsum_secular} graphically under
the implicit assumption that the p-h spectrum is bounded in energy.
Basically one plots ${\cal S}(E)$ and a horizontal line at the
ordinate value $y=1/\lambda$. The intersections of ${\cal S}(E)$
with the horizontal line give the solutions $E$, which turn out to
be merely small renormalizations of the p-h energies $\Delta
\varepsilon_{ph}$ except for the largest excitation energy
$\varepsilon_{pl}$ which corresponds to the collective mode, see
fig.(\ref{RPAsumfig}) for a schematic example. This state is nothing
but the surface plasmon collective state. In
fig.(\ref{RPAsumfig}b-c) it shows how $\varepsilon_{pl}$ is a
function mostly of the initial low-energy 1p-1h excitations,
depending only slightly on the high-energy ones, for a decaying
dipole matrix element $d_{ph}\sim1/\Delta \varepsilon_{ph}^2$. This
decaying behavior will be demonstrated in next subsection, ensuring
also the applicability of this schematic picture even though in a
metallic cluster the p-h excitations are not bounded from above on
the scale of the plasma energy.

\subsection{Fast decay of $\textsl{\textbf{d}}_{ph}$. Plasmon state
built from low-energy p-h excitations} \label{sec_spslephe}
\begin{figure}[!t]
\begin{center}
\includegraphics[width=12cm]{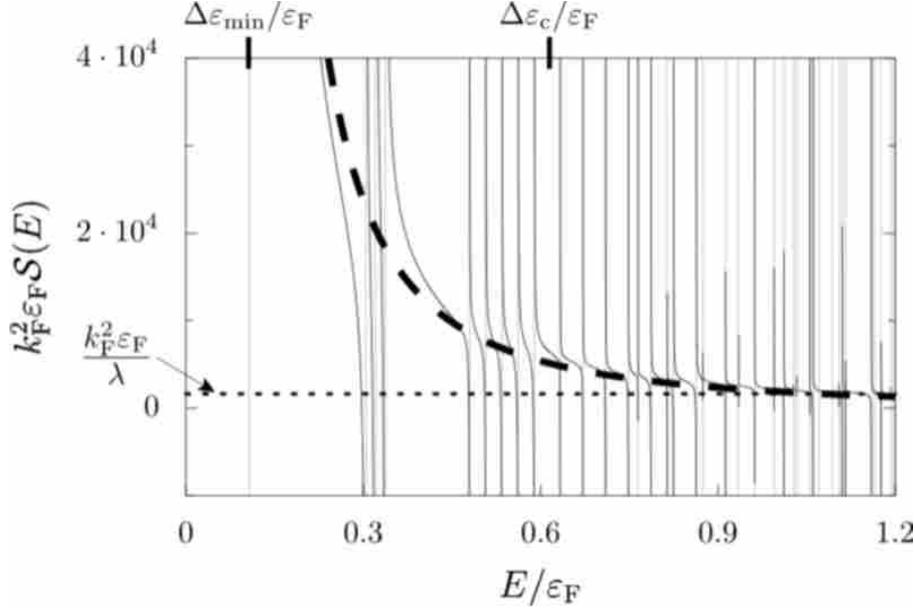}
\caption[RPA sum showing how the plasmon is built from low-energy p-h excitations]{\label{sum}%
RPA sum \eqref{RPAsum_inter} for a ${\rm Na}$ nanoparticle with
$k_{\rm F}a=30$ (solid black line). The p-h excitation energies
$\Delta\varepsilon_{ph}$ (represented by vertical grey lines) have
been obtained from the semiclassical spectrum \eqref{Ediff}. There
are about 15000 degenerate excitations in the interval shown, and
the smallest excitation
$\Delta\varepsilon_\textrm{min}\simeq\varepsilon_\textrm{F}/10$ has
a degeneracy ${\cal N}=380$. The dashed line takes only into account
the contribution of $\Delta\varepsilon_\textrm{min}$ (see
Eq.~\eqref{RPA_mini}). The cutoff energy
$\Delta\varepsilon_\textrm{c}$ separating the two RPA subspaces is
also shown in the figure (see Eq.~\eqref{estimatedlec3}). Above
$\Delta\varepsilon_\textrm{c}$ the two sums are close to each other,
showing that for high energies the RPA sum is essentially given by
the contributions coming from the low-energy p-h excitations. The
horizontal dotted line indicates the position of the coupling
constant $\lambda$ entering the RPA secular equation
\eqref{RPAsum_secular}, according to the estimate \eqref{lambda}.}
\end{center}
\end{figure}

The
secular equation (\ref{RPAsum_secular}) depends crucially on the
form of the dipole matrix element $d_{ph}$. We will show now more
rigorously that due to its fast decay
$d_{ph}(\Delta\varepsilon)\sim1/\Delta\varepsilon^2$, cf. eq.
(\ref{D1}), the plasmon excitation energy depends only on the lowest
1p-1h excitations. Assuming that the p-h states are confined within
a hard-wall sphere, one can calculate $d_{ph}$, see Appendix
\ref{apDipole} for a detailed derivation. $d_{ph}$ is factorized
into radial and angular parts \cite{YB92}
\begin{equation}
\label{d_ph} d_{ph} = {\cal A}_{l_p l_h}^{m_p m_h} {\cal R}_{l_p
l_h}(\varepsilon_p , \varepsilon_h)\,.
\end{equation}
The angular part is expressed in terms of Wigner-$3j$ symbols as
\begin{equation}\label{sp_angular}
    {\cal A}_{l_p l_h}^{m_p m_h} =(-1)^{m_p} \sqrt{(2l_p+1)(2l_h+1)}\times
\begin{pmatrix}
l_p & l_h & 1 \\
0 & 0 & 0
\end{pmatrix}
\begin{pmatrix}
l_p & l_h & 1 \\
-m_p & m_h & 0
\end{pmatrix}
\end{equation}
and sets the selection rules for the total and azimuthal angular
momenta, respectively:
\begin{equation}\label{selrules}
    l_p=l_h\pm 1\,\,\,\,\,\,,\,\,\,\,\,\,m_p=m_h
\end{equation}
The radial part depends on the energies of the p-h states as
\begin{equation}
\label{R_ph} {\cal R}_{l_p l_h}(\varepsilon_p ,
\varepsilon_h)=\frac{2\hbar^2}{m_{\rm e} a}
\frac{\sqrt{\varepsilon_p
\varepsilon_h}}{\Delta\varepsilon_{ph}^2}\,.
\end{equation}
With the help of equations \eqref{d_ph}--\eqref{R_ph} and using the
aforementioned dipole selection rules (\ref{selrules}), we can write
the RPA sum \eqref{RPAsum} as
\begin{equation}
\label{RPAsum_inter} {\cal S}(E)= \frac{32k_{\rm
F}^{-2}}{3}\left(\frac{\varepsilon_{\rm F}}{k_{\rm F}a}\right)^2
\sum_{\substack{n_h, l_h\\n_p, l_p=l_h\pm1}}
f_{l_p}\frac{\varepsilon_p\varepsilon_h}
{(E^2-\Delta\varepsilon_{ph}^2)\Delta\varepsilon_{ph}^3}\,,
\end{equation}
where $n_h$ and $n_p$ are radial quantum numbers. We have defined
$f_{l_p}=l_h+1$ if $l_p=l_h+1$ and $f_{l_p}=l_h$ if $l_p=l_h-1$.
Using the semiclassical quantizations \eqref{qcondition2} and
\eqref{Ediff} of Appendix~\ref{app_Emin}, we finally obtain the
result shown in Figure~\ref{sum}.

In the figure we show (solid black line) the resulting
$E$-dependence of \eqref{RPAsum_inter}. The crossings of this curve
with the horizontal dotted one with height $1/\lambda$ yield the
excitation spectrum within the RPA. The lowest p-h energy estimated
in Appendix~\ref{app_Emin} is
\begin{equation}
\label{SpacingGuillaume}
\Delta\varepsilon_\textrm{min}\simeq\frac{\varepsilon_{\rm
F}}{k_{\rm F}a/\pi}\,.
\end{equation}
The existence of a gap in the excitation spectrum of the dipole
operator stems from the quantization of the states in a hard-wall
sphere and the dipole selection rules. Whenever the energy $E$
coincides with a p-h excitation energy $\Delta\varepsilon_{ph}$, we
have a divergence in ${\cal S}(E)$ (see vertical grey lines in
Fig.~\ref{sum}). For the lowest energies $E$ of the interval
considered the sum is dominated by the term associated with the
divergence closest to $E$. On the other hand, for the largest
energies $E$, the fast decay of $d_{ph}$ with
$\Delta\varepsilon_{ph}$ means that the divergences are only
relevant for energies extremely close to them. Away from the
divergences the sum is dominated by the contributions arising from
the low-energy p-h excitations.

To gain physical insight and proceed further in the analysis of the
electron dynamics in metallic nanoparticles, we introduce the
typical dipole matrix element for states separated by a given energy
difference $\Delta\varepsilon$,
\begin{equation}\label{Dmieq}
    d^\textrm{p-h}(\Delta\varepsilon)=\Bigg[\frac{1}{\rho^\textrm{p-h}(\Delta\varepsilon)}
\sum_{ph}{|d_{ph}|}^2\delta(\Delta\varepsilon-\Delta\varepsilon_{ph})\Bigg]^{1/2}=\Bigg[\frac{1}{\rho^\textrm{p-h}(\Delta\varepsilon)}
\int_{\varepsilon_{\rm F}-\Delta\varepsilon}^{\varepsilon_{\rm F}}
{\rm d}\varepsilon\, {\cal
C}(\varepsilon,\Delta\varepsilon)\Bigg]^{1/2}\,.
\end{equation}
Here, we have introduced two quantities,
$\rho^\textrm{p-h}(\Delta\varepsilon)$ and ${\cal
C}(\varepsilon,\Delta\varepsilon)$. ${\cal
C}(\varepsilon,\Delta\varepsilon)$ is the local density of dipole
matrix elements
\begin{equation}
\label{lddme} {\cal C}(\varepsilon,\Delta\varepsilon)=
\sum_{ph}|d_{ph}|^2\delta(\varepsilon-\varepsilon_h)
\delta(\varepsilon+\Delta\varepsilon-\varepsilon_p)\,.
\end{equation}
As shown in Appendix~\ref{app_A}, this can be expressed as
\begin{equation}
\label{C} {\cal C}(\varepsilon,\Delta\varepsilon)=
\frac{1}{3\pi^2}\frac{a^2}{\Delta\varepsilon^2}
F\left(\frac{\varepsilon}{\Delta\varepsilon}\right)\,,
\end{equation}
where
\begin{equation}
\label{F}
F(x)=(2x+1)\sqrt{x(x+1)}-\ln{\left(\sqrt{x}+\sqrt{x+1}\right)}\,.
\end{equation}
For $\Delta\varepsilon\ll\varepsilon$, equation \eqref{C} simplifies
to
\begin{equation}
\label{Csmallde} {\cal
C}(\varepsilon,\Delta\varepsilon)\simeq\frac{2a^2}{3\pi^2}
\frac{\varepsilon^2}{\Delta\varepsilon^4}\,.
\end{equation}

In expression \eqref{Dmieq},
\begin{equation}
\label{rph1} \rho^\textrm{p-h}(\Delta\varepsilon)=\sum_{ph}
\delta(\Delta\varepsilon_{ph}-\Delta\varepsilon)
\delta_{l_h,l_p\pm1}\delta_{m_h,m_p}
\end{equation}
is the density of p-h excitations with energy $\Delta\varepsilon$
respecting the dipole selection rules. In order to simplify the
presentation, we do not consider spin degeneracy factors. In
Appendix~\ref{app_B} we show that for $\Delta\varepsilon\ll
\varepsilon_\textrm{F}$ we have
\begin{equation}
\label{rph3}
\rho^\textrm{p-h}(\Delta\varepsilon)\simeq\frac{\left(k_{\rm
F}a\right)^4}{4\pi^2} \frac{\Delta\varepsilon}{\varepsilon_{\rm
F}^2}\,,
\end{equation}
and therefore in such a limit the typical matrix element
\eqref{Dmieq} can be approximated by
\begin{equation}
\label{D1} \boxed{d^\textrm{p-h}(\Delta\varepsilon)\simeq
\frac{2\sqrt{2}k_{\rm F}^{-1}}{\sqrt{3}k_{\rm F}a}
\left(\frac{\varepsilon_{\rm F}}{\Delta\varepsilon}\right)^2}\,.
\end{equation}
To check the validity of our estimate of the typical dipole matrix
element, there is an \emph{exact} sum rule
$\sum_{ph}\Delta\varepsilon_{ph}|d^\textrm{p-h}|^2=(3/4\pi)\hbar^2N/2m_{\rm
e}$ \cite{LS91,YB92}, satisfied also in the presence of Coulomb
interactions between electrons. Performing the sum using \eqref{D1}
    \begin{equation}\label{sumruleapprox}
\sum_{ph}^{E_F}\Delta\varepsilon_{ph}|d^\textrm{p-h}(\Delta\varepsilon)|^2\simeq\int_{\dlem}^{E_F}d\Delta\varepsilon
 \Delta\varepsilon\Bigl[\int_{E_F-\Delta\varepsilon}^{E_F}d\varepsilon C(\varepsilon,\Delta\varepsilon)\Bigr]
\end{equation}
one obtains about 70\% of the exact result $(3/4\pi)\hbar^2N/2m_{\rm
e}$ \cite{B93}. This is quite reasonable regarding all the
approximations we made to obtain \eqref{D1}.

In order to emphasize once more the importance of the low-energy p-h
excitations, we present in Figure~\ref{sum} (dashed line) the
contribution to the RPA sum coming only from the infrared p-h
excitation energy with the appropriate degeneracy factor $\cal N$
(see Eq.~\eqref{deg1} in App.~\ref{app_Emin}). Indeed, we can
estimate \eqref{RPAsum} as
\begin{equation}
{\cal S}(E)\approx{\cal N}\times\frac{2\Delta\varepsilon_{\rm min}
\left[d^\textrm{p-h}(\Delta\varepsilon_{\rm min})\right]^2}
{E^2-\Delta\varepsilon_{\rm min}^2}\,.
\end{equation}
With the results \eqref{SpacingGuillaume}, \eqref{D1}, and
\eqref{deg1}, we obtain
\begin{equation}
\label{RPA_mini} {\cal S}(E)\approx\frac{64k_{\rm
F}^{-2}}{3\pi^5}(k_{\rm F}a)^3\frac{\varepsilon_{\rm
F}}{E^2-(\varepsilon_{\rm F}\pi/k_{\rm F}a)^2}\,.
\end{equation}
While in the lower part of the energy interval in Figure~\ref{sum}
the two curves exhibit considerable discrepancies, in the second
half of the interval they are very close (except of course at the
divergences). Since the collective excitation is found in this last
interval, we see that it is mainly the low-energy p-h excitations
that are relevant for the definition of the collective excitation.

Since the resonance energy is known from experiments, we can obtain
the value of the coupling constant $\lambda$. Indeed, using the
estimate \eqref{RPA_mini} evaluated at $E=\hbar\omega_{\rm M}$ in
the secular equation \eqref{RPAsum_secular}, we obtain
\begin{equation}
\label{lambda} \frac{1}{\lambda}\simeq \frac{64k_{\rm
F}^{-2}}{3\pi^5}(k_{\rm F}a)^3 \frac{\varepsilon_{\rm
F}}{(\hbar\omega_{\rm M})^2}
\end{equation}
to leading order in $k_{\rm F}a\gg1$. Notice that this result is
consistent with the estimate obtained from the energy-weighted sum
rule given in \cite{YB92}. For the case studied in Figure \ref{sum},
our estimate \eqref{lambda} yields $k_{\rm F}^2\varepsilon_{\rm
F}/\lambda\approx1600$. As the radius $a$ increases the lowest p-h
energy decreases like $1/a$, the degeneracy at this value increases
as $a^2$, and the density of p-h excitations contributing to
(\ref{RPAsum}), $\rho^\textrm{p-h}(\Delta\varepsilon)$, grows as
$a^4$. This increase in the number of excitations contributing to
the sum is partially cancelled by the $1/a$ behaviour of the typical
dipole matrix element, resulting in a decrease of the coupling
constant $\lambda$ proportional to the number of particles in the
cluster, $\lambda\sim 1/a^3$, and a value of the plasmon frequency
which remains almost unaffected. Therefore, for a larger
nanoparticle size, the divergences shown in Figure~\ref{sum} would
be more dense in energy, starting at a lower energy, and the
vertical scale would increase as $a^3$, thus obtaining a similar
value for the plasmon excitation energy $\hbar\omega_{\rm M}$.
\subsection{Separation of the reduced and additional particle-hole
subspaces} \label{sec_sraphs} As we have shown in the last section,
the high-energy part of the p-h spectrum is not crucial for the
determination of the energy of the collective excitation. In what
follows we make this statement more quantitative and estimate the
upper-bound cutoff $\Delta\varepsilon_\textrm{c}$ of the low-energy
excitations that we need in order to obtain a stable position of the
surface plasmon.

In order to obtain a quantitative estimate of the cutoff, we require
that by changing it from $\Delta\varepsilon_\textrm{c}$ to
$(3/2)\Delta\varepsilon_\textrm{c}$, the position of the plasmon
changes only by a fraction of its linewidth $\gamma$, the smallest
energy scale with experimental significance. Our criterion leads to
the condition
\begin{align}
\label{estimatedlec2} {\cal S}_{\Delta\varepsilon_{\rm
c}}(\hbar\omega_{\rm M})= {\cal S}_{\frac32\Delta\varepsilon_{\rm
c}}(\hbar\omega_{\rm M}+\hbar\gamma)\,,
\end{align}
with the RPA sum $\cal S$ that has been defined in \eqref{RPAsum}.
The additional subscript refers to the upper bound of the p-h
energies.

The left-hand side of \eqref{estimatedlec2} can be estimated
according to
\begin{equation}
{\cal S}_{\Delta\varepsilon_{\rm c}}(\hbar\omega_{\rm M})
\simeq\int_{\Delta\varepsilon_\textrm{min}}^{\Delta\varepsilon_\textrm{c}}
{\rm d}\Delta\varepsilon\,
\rho^\textrm{p-h}(\Delta\varepsilon)\frac{2\Delta\varepsilon
\left[d^\textrm{p-h}(\Delta\varepsilon)\right]^2}{(\hbar\omega_{\rm
M})^2-\Delta\varepsilon^2}\,,
\end{equation}
with $d^\textrm{p-h}$ and $\rho^\textrm{p-h}$ as defined in
\eqref{Dmieq} and \eqref{rph1}, respectively. Using \eqref{rph3} and
\eqref{D1}, we obtain
\begin{equation}
{\cal S}_{\Delta\varepsilon_{\rm c}}(\hbar\omega_{\rm
M})\simeq\frac{4k_{\rm F}^{-2}}{3\pi^2} \left(k_{\rm
F}a\frac{\varepsilon_{\rm F}}{\hbar\omega_{\rm M}}\right)^2
\left(\frac{1}{\Delta\varepsilon_{\rm min}}
-\frac{1}{\Delta\varepsilon_{\rm c}} \right)
\end{equation}
to leading order in $k_{\rm F}a$. Similarly,
\begin{align}
{\cal S}_{\frac32\Delta\varepsilon_{\rm c}}(\hbar\omega_{\rm
M}+\hbar\gamma) \simeq&\;\frac{4k_{\rm F}^{-2}}{3\pi^2} \left(k_{\rm
F}a\frac{\varepsilon_{\rm F}}{\hbar\omega_{\rm M}}\right)^2
\left(1-\frac{2\gamma}{\omega_{\rm
M}}\right)\left(\frac{1}{\Delta\varepsilon_{\rm min}}
-\frac{2}{3\Delta\varepsilon_{\rm c}}\right)\,.
\end{align}
Using the expressions \eqref{gamma_intro} and
\eqref{SpacingGuillaume} for $\gamma$ and $\Delta\varepsilon_{\rm
min}$, respectively, finally yields according to the criterion
\eqref{estimatedlec2} the main result of this chapter, an estimate
for the cutoff energy based on sound physical arguments
\begin{equation}
\label{estimatedlec3}\boxed{
\Delta\varepsilon_\textrm{c}\simeq\frac{\pi}{9g_0(\varepsilon_{\rm
F}/\hbar\omega_{\rm M})} \hbar\omega_\textrm{M}}\,,
\end{equation}
with the function $g_0$ defined in \eqref{g_0}. For Na clusters, we
have $\varepsilon_\textrm{F}/\hbar\omega_\textrm{M}=0.93$ and our
criterion yields a value $\Delta\varepsilon_{\rm c} \simeq
(3/5)\varepsilon_{\rm F}$.

We have verified the robustness of our criterion
\eqref{estimatedlec2} by exploring different physical parameters,
like the size of the cluster. Indeed, it can be seen in
Figure~\ref{sum} that above $\Delta\varepsilon_\textrm{c}$ the RPA
sum evaluated from \eqref{RPAsum_inter} (solid line) and the
estimate \eqref{RPA_mini} (dashed line) are close to each other,
showing that for high energies the RPA sum is essentially given by
the contributions coming from the low-energy p-h excitations. We
have checked that this feature of $\Delta\varepsilon_{\rm c}$ is
independent of the size $a$ of the nanoparticle.

In order to have a well-defined collective excitation, the cutoff
energy $\Delta\varepsilon_{\rm c}$ must obviously be larger than the
minimal p-h excitation energy \eqref{SpacingGuillaume}. For Na, we
find that this condition is already verified with only $N=20$
conduction electrons in the nanoparticle, in agreement with the
experiments of reference \cite{Setal91} and the numerical
calculations of reference \cite{YB91}.

Eq. (\ref{estimatedlec3}) is the key ingredient we needed to avoid
the problem of the clear separation between the plasmon and its
environment, as discussed in section \ref{doublecounting}. The
splitting in low
$\Delta\varepsilon_{ph}<\Delta\varepsilon_\textrm{c}$ and
high-energy $\Delta\varepsilon_{ph}>\Delta\varepsilon_\textrm{c}$
p-h excitations gives physical soundness to the method based on the
separation of the electronic degrees of freedom into centre-of-mass
and relative coordinates \cite{GGI02, WIJW06, WMWJ05, WWIJ07}
described in the first two sections of the chapter. Now that the
Fermi's Golden Rule calculation of the Landau damping of the plasmon
(\ref{landaudamp}) is justified, provided one sums over p-h pairs
with $\Delta\varepsilon_{ph}>\Delta\varepsilon_\textrm{c}$, we will
use the estimate (\ref{estimatedlec3}) to compare in next section
the characteristic time scales of the surface plasmon and its
high-energy p-h environment. We will show that the memory of this
environment is very short compared to the timescale characterizing
the evolution of the plasmon, its decay time $\gamma(a)^{-1}$,
justifying in this way a Markovian approach for the analysis of the
plasmon dynamics in presence of such a dissipative bath, cf.
\cite{WIWJ07}.

\section{Dynamics of the relative-coordinate system}
\label{sec_tdrcs} 
\begin{figure}[!t]
\begin{center}
\includegraphics[width=11cm]{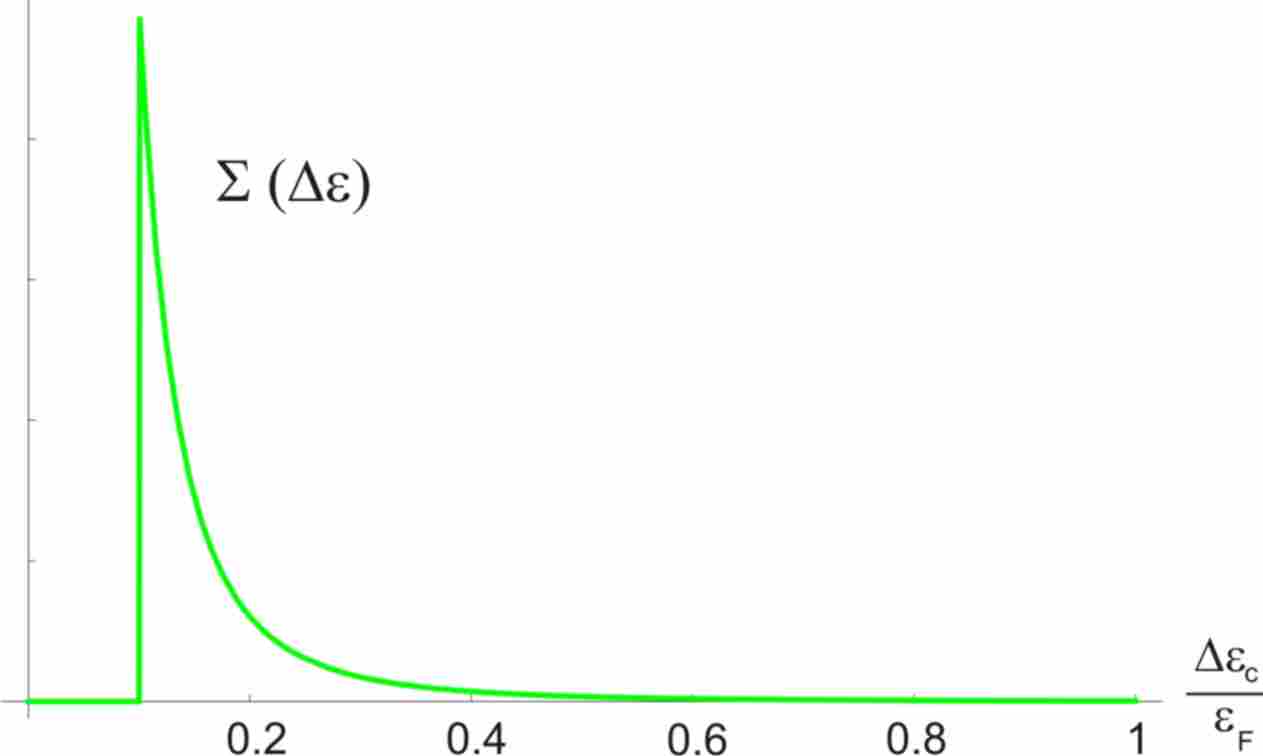}\\
\caption[Plot of the Fourier transform of $C(t)$,
$\Sigma(\Delta\varepsilon)$]{\label{sigmafig} Schematic plot of the
spectral function $\Sigma(\Delta\varepsilon)$, Fourier transform of
$C(t)$, with $\Delta\varepsilon_\textrm{c}$ taken to be
$\Delta\varepsilon_\textrm{c}=0.1\varepsilon_F$. The sharp peak is a
consequence of the fast decrease of
$d^\textrm{p-h}(\Delta\varepsilon)$ with $\Delta\varepsilon$, eq.
\eqref{D1}, which overwhelms the increase in the density of
particle-hole excitations contributing to the damping of the
plasmon, eq.(\ref{rph3}), giving in this way a preponderant role in
the dissipative bath to the particle-hole excitations with energies
close to $\Delta\varepsilon_\textrm{c}$.}
\end{center}
\end{figure}

The decomposition \eqref{Helectronic5} of $H$
suggests to treat the collective coordinate as a simple system of
one degree of freedom which is coupled to an environment with many
degrees of freedom. The latter are the relative coordinates
described by $H_{\rm rel}$. This is the approach taken in
\cite{WIWJ07}. In this picture, the time evolution of the
center-of-mass system (i.e., the surface plasmon) strongly depends
on the dynamics of the relative-coordinate system. Such a dynamics
is characterized by a correlation function which can be written at
zero temperature as\cite{WIWJ07}
\begin{equation}
\label{Cenv1} C(t)=\Lambda^2\sum_{ph} |d_{ph}|^2 {\rm e}^{{\rm
i}\Delta\varepsilon_{ph}t/\hbar}
\Theta(\Delta\varepsilon_{ph}-\Delta\varepsilon_\textrm{c})\,.
\end{equation}
$C(t)$ is nothing but the Fourier transform of the spectral function
of this dissipative bath \cite{Letal87,W99}
\begin{equation}
\label{Sigma} \Sigma(\Delta\varepsilon)=\frac{2\pi}{\hbar}\Lambda^2
\sum_{ph}|d_{ph}|^2\delta(\Delta\varepsilon-\Delta\varepsilon_{ph})
\Theta(\Delta\varepsilon_{ph}-\Delta\varepsilon_\textrm{c})
\end{equation}
which has been calculated in \cite{WIJW06}. Finite temperatures were
shown to result in a small quadratic correction. Consistently with
the results of the preceding sections, we employ our low energy
estimates \eqref{rph3} and \eqref{D1} to obtain
\begin{equation}
\label{Sigma_result} \Sigma(\Delta\varepsilon)\simeq\frac{3v_{\rm
F}}{4a} \left(\frac{\hbar\omega_{\rm M}}{\Delta\varepsilon}\right)^3
\Theta(\Delta\varepsilon-\Delta\varepsilon_\textrm{c})\,.
\end{equation}
This result is consistent with the one of reference \cite{WIJW06}
(see Eq.~(34) in there) in the limit
$\Delta\varepsilon\ll\hbar\omega_{\rm M}$ and for zero temperature.
A schematic plot is given in fig.(\ref{sigmafig}), showing the
dominant role that the p-h excitations with energies close to
$\Delta\varepsilon_\textrm{c}$ have in the properties of the
environment, pointing to a correlation time $\tau_\textrm{cor}$ of
the bath of order $\sim \hbar/\Delta\varepsilon_\textrm{c}$.

In principle we could calculate $C(t)$ by taking the inverse Fourier
transform of \eqref{Sigma_result}. The decay of $C(t)$ \emph{for
very long times} is dominated by the discontinuity of $\Sigma$ at
$\Delta\varepsilon_\textrm{c}$. This is somehow problematic since
$\Delta\varepsilon_\textrm{c}$ can only be estimated as we did in
Section~\ref{sec_sraphs}, and since the functional form of the long
time decay depends on how sharply the cutoff is implemented.
However, it is important to realize that \emph{it is not the very
long time behavior that determines the relevant decay of the
correlation function}, but rather the typical values at which $C(t)$
is reduced by an important factor from its initial value $C(0)$. We
then estimate the correlation time as the mean decay time of $C(t)$,
\begin{align}
\label{estimatortau} \langle\tau_{\rm
cor}\rangle&=\left|\int_{0}^{\infty}{\rm d}t\,t\frac{\rm d}{{\rm
d}t} \left(\frac{C(t)}{C(0)}\right)\right|=\frac{1}{C(0)} \left|
\int_{0}^{\infty}{\rm d}t\, C(t) \right|\,.
\end{align}
If $C(t)$ were an exponentially decreasing function,
$\langle\tau_{\rm cor}\rangle$ would simply reduce to the inverse of
the decay rate. Using the definition \eqref{Sigma}, we get
\begin{equation}
\label{estimatortau2} \langle\tau_\textrm{cor}\rangle=
\frac{\hbar\displaystyle
\int_{\Delta\varepsilon_\textrm{c}}^{\infty}\textrm{d}\Delta\varepsilon
\Sigma(\Delta\varepsilon)/\Delta\varepsilon} {\displaystyle
\int_{\Delta\varepsilon_\textrm{c}}^{\infty}\textrm{d}\Delta\varepsilon
\Sigma(\Delta\varepsilon)}\,.
\end{equation}
Given the fast decay of the function
$\Sigma\sim1/\Delta\varepsilon^3$, the above integrals are dominated
by their lower limit $\Delta\varepsilon_\textrm{c}$ and we have
\begin{equation}
\label{estimatortau3} \langle\tau_{\rm
cor}\rangle\simeq\frac{2}{3}\frac{\hbar}{\Delta\varepsilon_\textrm{c}}\,.
\end{equation}
Since $\Delta\varepsilon_\textrm{c}$ is of the order of
$(3/5)\varepsilon_\textrm{F}$, we see that the response time (or
correlation time) of the electronic environment is of the order of
its inverse Fermi energy.

The estimate of the characteristic response time of the electronic
environment is crucial in justifying the Markovian approximation
used in \cite{WIWJ07}. In that work, the degrees of freedom
corresponding to the relative coordinates were integrated out and
treated as an incoherent heat bath that acts on the collective
coordinate. Such an approach relies on the fast response of the
environment as compared to the time evolution of the surface
plasmon. The typical scale for the latter is the inverse of the
decay rate, $\tau_{\rm sp}=1/\gamma$. Using \eqref{gamma_intro} and
\eqref{estimatedlec3}, we therefore have
\begin{equation}
\label{tratio1} \boxed{\frac{\tau_{\rm sp}}{\langle\tau_{\rm
cor}\rangle}= \frac{\pi}{9[g_0(\varepsilon_{\rm F}/\hbar\omega_{\rm
M})]^2} \frac{\hbar\omega_{\rm M}}{\varepsilon_{\rm F}}k_\textrm{F}
a}\,.
\end{equation}
For the example of Na nanoparticles worked in
Section~\ref{sec_spslephe}, we have
$\tau_\textrm{sp}/\langle\tau_{\rm cor}\rangle\simeq k_{\rm F}a$.
This is a safe limit since in not too small nanoparticles, $k_{\rm
F}a\gg1$. As the size of the cluster increases, the applicability of
the Markovian approximation is more justified. This is expected
since the electronic bath has more and more degrees of freedom,
approaching an ``environment" in the sense of quantum dissipation.
Since the physical parameters of alkaline nanoparticles entering
\eqref{tratio1} are close to that of noble-metal clusters, the
dynamics of the surface plasmon can be expected to be Markovian in
that case too.

\section{Conclusions}
\label{sec_conclusions} We have studied the role of particle-hole
excitations on the dynamics of the surface plasmon. A key concept in
this analysis is the separation into low-energy excitations which
lead to the collective excitation once they are mixed by the
residual dipolar interaction, and high-energy excitations that act
as an environment damping the resonance. Using the random phase
approximation and assuming the separability of the residual
interaction, we have established a criterion for estimating the
cutoff energy separating the low- and high-energy subspaces. The
resulting cutoff energy is approximately $(3/5)\varepsilon_{\rm F}$
for the case of Na nanoparticles.

Since the number of electrons in the cluster is finite, the
assumption that the high-energy particle-hole excitations act on the
collective excitation as an environment, introducing friction in its
dynamics, may be questionable. What settles this issue is the ratio
between the typical evolution time of the collective excitation and
the one of the high-energy particle-hole excitations. The former is
given by the inverse of the plasmon linewidth, while the latter is
obtained from the decay of the correlation function of the
environment. We have found that this ratio improves with increasing
cluster size. Even for a small cluster with $a=\unit[1]{nm}$, the
ratio is approximately 10, justifying the use of the Markovian
approximation which assumes a fast time evolution of the environment
with respect to the one of the collective excitation.

The relevance of memory effects in the electronic dynamics of small
clusters is of current interest, due to the advance in time-resolved
experimental techniques \cite{BHMD00,FVFHN00,LKLA99,LUCS01}.
First-principle calculations have recently addressed this issue by
comparing time-dependent density functional theories with and
without memory effects for small gold clusters \cite{KB06}. For very
small clusters ($N<8$) memory effects were shown to be important. It
has to be noted, however, that the memory considered in \cite{KB06}
is that of the electron gas as a whole, while we are concerned in
this work with the memory arising from the dynamical evolution of
the relative-coordinate subsystem. It would be interesting to
consider cluster sizes intermediate between the ones considered in
the present work and those of reference \cite{KB06} in order to
study the emergence of memory effects.


\newpage
\cleardoublepage


\appendix


\chapter{Appendix to Chapter \ref{chr1}}
\label{apchr1}

\section{Modes' equations of motion of quasi-1D
resonators.}\label{apmodeseqs}
\subsection{Elasticity basics.}

We will follow ref.\cite{LL59}. The starting point is to define the
displacement vector of a given point $x_i$ in the solid,
$u_i=x_i'-x_i$, representing the difference between the coordinates
of the same point before and after the deformation. The related
magnitude that appears everywhere in the theory of elasticity is the
so-called \textit{strain tensor} (symmetric by definition):
\begin{equation}\label{strain}
    u_{ik}=\frac{1}{2}\Bigl(\frac{\partial u_i}{\partial x_k}+\frac{\partial u_k}{\partial x_i}+
    \frac{\partial u_l}{\partial x_i}\frac{\partial u_l}{\partial
    x_k}\Bigr)\simeq \frac{1}{2}\Bigl(\frac{\partial u_i}{\partial x_k}+\frac{\partial u_k}{\partial x_i}\Bigr)
\end{equation}
Then the \textit{stress tensor} $\sigma_{ik}$ is defined,
considering the forces $F_i$ exerted on a small volume $dV$ due to
the deformation of a body:
\begin{equation}\label{stress}
    \int F_i dV = \int \frac{\partial \sigma_{ik}}{\partial
    x_k}dV=\oint\sigma_{ik}dS_k\phantom{,}\rightarrow\phantom{,}F_i=\frac{\partial \sigma_{ik}}{\partial x_k}
\end{equation}
With these two quantities one can study the thermodynamics of the
deformation: take a deformed body and suppose there is a small
variation in the deformation, so that $u_i\rightarrow u_i +\delta
u_i$, then the internal work done by the internal forces is given by
\begin{equation}\label{work}
    \int\frac{\partial \sigma_{ik}}{\partial x_k}\delta u_idV
\end{equation}
We will consider bodies at a fixed temperature $T$, so the
thermodynamic potential we will use is the free energy $F=U-TS$,
with a variation of $F$ in a small deformation at a constant $T$
given by $dF=\sigma_{ik}du_{ik}$, implying $\sigma_{ik}=\partial
F/\partial u_{ik}\bigr|_T$. Now, for small deformations, choosing as
non-deformed state the one in absence of external forces, we can
expand $F$ as a function of $u_{ik}$, obtaining
$F=\frac{1}{2}\lambda_{iklm}u_{ik}u_{lm}$, so that
$\sigma_{ik}=\lambda_{iklm}u_{lm}$ (Hooke's law). For an isotropic
body the expression reduces to $F=F_0 + (\lambda/2)u_{ii}^{2}+\mu
u_{ik}^{2}$ ($\lambda$ and $\mu$ are the Lam\'e coefficients), which
can be reexpressed as
\begin{equation}\label{apFrigidity}
    F=\mu
    \Bigl(u_{ik}-\frac{1}{3}\delta_{ik}u_{ll}\Bigr)^2+\frac{K}{2}u_{ll}^{2}\phantom{XXX},\phantom{XXX}K=\lambda + \frac{2}{3}\mu
\end{equation}
defining the rigidity modulus $K$. If the deformation is homogeneous
(the strain tensor is constant throughout the body), and more
specifically a simple extension or compression along the z axis, one
has
\begin{equation}\label{apYoung}
    \sigma_{zz}=E\times u_{zz}\,\,,
\end{equation}
defining the Young modulus,
\begin{equation}\label{apYoung2}
    E=\frac{9K\mu}{3K+\mu}
\end{equation}
The ratio between transversal contraction $u_{xx}$ and longitudinal
elongation $u_{zz}$ is Poisson's coefficient
\begin{equation}\label{apPoisson}
    \nu=-\frac{u_{xx}}{u_{zz}}=\frac{1}{2}\frac{3K-2\mu}{3K+\mu}\,.
\end{equation}

\subsection{The case of a rod.} We start defining a local
coordinate system $(\xi,\eta,\zeta)$, attached to the point, that
coincides with $(x,y,z)$ in the absence of deformation (the rod is
then along the Z axis). Next we define
$\vec{\Omega}=d\vec{\phi}/dl$, the rate of variation of the relative
rotation angle between the local coordinate systems of two adjacent
points ($dl$ is taken along the bar's main axis). For example, a
pure bending has $\vec{\Omega}=\Omega_{\eta}\vec{\eta}$. Now we
define a coordinate system fixed in space, independent of the rod,
with respect to whom a point in the rod will be given by a vector
$\vec{r}$. The tangent vector at that point is given by
$\vec{t}=d\vec{r}/dl$, and the curvature by $d\vec{t}/dl$, with
modulus $1/R$, defining the curvature radius. For a slight bending
one has
\begin{equation}\label{omega}
    \vec{\Omega}=\vec{t}\wedge\frac{d\vec{t}}{dl}\simeq\vec{t}
\wedge\frac{d^2\vec{r}}{dz^2}=\Bigr(-\frac{d^2Y}{dz^2},\frac{d^2X}{dz^2},0\Bigl)
\end{equation}
In terms of $\vec{\Omega}$ the free energy of a rod is deduced to be
\begin{equation}\label{rodenergy}
    F_{rod}=\frac{1}{2}\int
    \Bigl\{I_1E\Omega_{\xi}^{2}+I_2E\Omega_{\eta}^{2}+C\Omega_{\zeta}^{2}\Bigr \}dl
\end{equation}
where $I_1$ and $I_2$ are the principal inertia moments of the
section of the rod, and C is the so-called torsional rigidity. In
the case of a slight bending the expression reduces to
\begin{equation}\label{energybending}
    F_{bending}=\frac{E}{2}\int
    \Bigl\{I_1\Bigl(\frac{d^2X}{dz^2}\Bigr)^{2}+I_2\Bigl(\frac{d^2Y}{dz^2}\Bigr)^{2}\Bigr\}dz
\end{equation}
Now we will deduce the equations for the modes as slight deviations
from the equilibrium situation, and the first step towards our goal
is the analysis of the equilibrium equations.

\subsubsection{Equilibrium conditions.} If there are external forces
$\vec{F}_{ext}$ applied to the rod, and external moments
$\vec{M}_{ext}$ associated with them, the rod will suffer a
deformation until the internal forces and moments compensate the
external ones, so in equilibrium one must have
\begin{equation}\label{apeqcond}
    \left\{
      \begin{array}{l}
        \nonumber \vec{F}_{ext}+\vec{F}_{int}=0 \\
        \vec{M}_{ext}+\vec{M}_{int}=0
      \end{array}
    \right.
\end{equation}
Considering the case of slight bending, for which $\vec{\Omega}$ is
given by eq.(\ref{omega}), the internal moment $\vec{M}_{int}$
associated to the internal stresses is
$\vec{M}_{int\phantom{,}bending}=(-EI_1Y'',EI_2X'',0)$, where the
double prime represents $\partial^2/\partial z^2$. Taking a small
volume of length $dl$ and supposing an external force per unit
length along the rod $\vec{F}_{ext}=\vec{K}=(K_x,K_y,K_z)$, the
first of the equilibrium equations results in
$d\vec{F}/dl=-\vec{K}$, and the second in
$d\vec{M}_{int}+[d\vec{l}\wedge\vec{K}]=0\rightarrow
d\vec{M_{int}}/dl=\vec{F}\wedge\vec{t}$. Deriving the second
equation with respect to $l$, and taking into account the small
value of $d\vec{t}/dl$ (slight bending), one arrives at
$d^2\vec{M}_{int}/dl^2=\vec{t}\wedge\vec{K}$, so that
\begin{equation}\label{apeqcond2}
    \left\{
      \begin{array}{l}
        EI_2\,\partial^4 X/\partial z^4= K_x \\
        EI_1\,\partial^4 Y/\partial z^4 = K_y
      \end{array}
    \right.
\end{equation}
and the internal forces will be given by
\begin{equation}\label{apeqcond3}
    \left\{
      \begin{array}{l}
        \nonumber F_x= -EI_2X''' \\
        F_y= -EI_1Y'''
      \end{array}
    \right.
\end{equation}
\subsubsection{Mode's equations.} To obtain the equations of motion
of an elastic medium, we start from Newton's equation equating the
product of mass and acceleration to the internal forces the mass is
subject to
\begin{equation}\label{newton}
    \rho \ddot{u}_i =\frac{\partial \sigma_{ik}}{\partial x_k}
\end{equation}
In the simplest case of \textbf{longitudinal} vibrations we have,
using eq.(\ref{apYoung})
\begin{equation}\label{aplongitudinal}
    \rho \ddot{u}_z =\frac{\partial \sigma_{zk}}{\partial x_k}\rightarrow \frac{\partial^2u_z}{\partial
    z^2}=\frac{\rho}{E}\frac{\partial^2u_z}{\partial t^2}
\end{equation}
This is the usual wave equation with linear dispersion relation. For
the \textbf{bending} (flexural) case, the equations of motion are
obtained just by substituting $-K_x$ and $-K_y$ in the equilibrium
equations (\ref{apeqcond2}) by $\rho S\ddot{X}$ and $\rho S\ddot{Y}$
(S = transversal section of the rod):
\begin{eqnarray}\label{apbending}
    \nonumber EI_y\frac{\partial^4 X}{\partial z^4}&=&- \rho S \frac{\partial^2 X}{\partial t^2}\\
    EI_x\frac{\partial^4 Y}{\partial z^4}&=&- \rho S \frac{\partial^2 Y}{\partial t^2}\,\,.
\end{eqnarray}
These equations admit plane wave-kind of solutions
$X(z,t),Y(z,t)\sim e^{i(kz-\omega t)}$, but with a quadratic
dispersion relation, $\omega_{j}(k) = \sqrt{EI_j/(\rho S)}k^2$. In
the case of \textbf{torsional} vibrations, the equations follow from
equating the moment $C\,\partial \Omega_{\zeta}/\partial z$ to the
temporal derivative of the angular momentum $\mathcal{L}$ per unit
length, $\mathcal{L}=\rho I \partial \phi/\partial t$. In the last
expression $\phi$ is the rotation angle of the considered section,
so that $\partial \phi/\partial z=\Omega_{\zeta}$, and $I$ is the
inertia moment of the transversal section with respect to the center
of mass, $I=\int (x^2+y^2)dS$, leading finally to
\begin{equation}\label{aptorsional}
    C\frac{\partial^2\phi}{\partial z^2}=\rho I \frac{\partial^2\phi}{\partial t^2}
\end{equation}
Again a wave equation with linear dispersion relation for the
variable $\phi$.

\subsubsection{Shape and frequency of the bending eigenmodes}
Take a rod of length $L$, thickness $t$, width $w$, and mass density
$\rho$. There are several possible configurations of a rod; we
choose a doubly clamped beam, with both ends fixed. In the
experiment the fundamental mode of transversal vibration in one of
the directions, lets say X, is excited, and its frequency $\wo$ is
measured. Now we will solve the equation for $X(z,t)$ following
\cite{Landau} with fixed ends for the rod, and obtain a relation
between $\wo$ and the Young modulus E, from which we will calculate its value.\\
We try a solution of the form $X(z,t)=X_0(z)\times \cos(\omega t +
\delta)$. Inserting it in the eq.(\ref{apbending}) for $X(z,t)$, we
obtain an equation that $X_0(z)$ has to verify:
 \begin{equation}\label{X0}
    \frac{d^4X_0(z)}{dz}=\omega^2\frac{\rho S}{EI_y}X_0(z)=\ka^4X_0(z)
 \end{equation}
The solution to this equation is of the form
\begin{equation}\label{X02}
    X_0(z)=A \cos(\ka z)+ B\sin(\ka z)+ C\cosh(\ka z)+ D\sinh(\ka z)
\end{equation}
Now, the boundary conditions we have to impose in this solution are,
for fixed ends, $X_0=\frac{dX_0}{dz}=0$ in $z = 0,L$. The final form
of $X_0(z)$ in terms of $\ka$, and the equation for the possible
values of $\ka$, are
\begin{eqnarray}
  \nonumber X_0(z) &=& A \Bigl \{ (\sin\ka L - \sinh\ka L)(
  \cos\ka z-\cosh\ka z)-(\cos\ka L - \cosh\ka L)(\sin\ka z-\sinh\ka z) \Bigr \}\\
  1 &=& \cos\ka L\cosh\ka L
\end{eqnarray}
The possible values for $\ka$ can be estimated to be, for $\ka L>1$,
 $\ka \simeq \frac{2n+1}{2}\frac{\pi}{L}$, because
the $\cosh\ka L$ takes big values that have to be "compensated" by
the $\cos\ka L$, for their product to be 1. Therefore the density of
states in the k-space is approximately constant,
$\frac{dN}{dk}\simeq \frac{L}{\pi}$. The first solution for $\ka$ is
$\ka_{0}\simeq\frac{\sqrt{22.4}}{L}$, already close to
$\frac{3}{2}\frac{\pi}{L}$, and from the relation between $\ka$ and
$\omega$ ($\omega(\ka) = \sqrt{\frac{EI_y}{\rho S}}\times \ka^2$) we
get
\begin{equation}\label{omegamin}
    \wo=\frac{22.4}{L^2}\sqrt{\frac{EI_y}{\rho S}}
\end{equation}




\chapter{Appendix to Chapter \ref{chr2}}
\label{apchr2}

\section{Some details about the Standard Tunneling Model.}\label{apSTM}
\subsection{Distribution function of the TLSs} \label{apDistribTLS}
From ref.\cite{P87}. The tunneling model proposes the existence of
an ensemble of TLSs, each one characterized by a hamiltonian:
\begin{equation}\label{apTLShamiltonian}
    H_0=\Dox\sigma_x+\Doz\sigma_z
\end{equation}
The task is to determine the probability density distribution
$P(\Dox,\Doz)$ of the TLSs in terms of $\Dox$ and $\Doz$. A
simplified expression for the tunneling amplitude as a function of
the potential barrier height $V$ and width $d$ is
\begin{equation}\label{Tunn_amplitude}
    \Dox=\hbar\Omega e^{-\lambda}=\hbar\Omega e^{-d(2mV/\hbar^2)^{1/2}}
\end{equation}
where $\hbar\Omega$ is approximately equal to the average of the
ground states of both isolated wells, and $m$ is the particle
mass.\\
Typical values for $\lambda$ can be estimated demanding that $\Dox$
must be approximately equal to $kT$ if the tunneling state is to
contribute to thermal properties at a temperature $T$. At 1 K this
requires a tunnel splitting of $10^{-4}$ eV, which with
$\hbar\Omega\sim10^{-2}$ eV gives approximately $\lambda\sim5$. This
is equivalent to a bare proton tunneling across a barrier of 0.1 eV
with $d=0.7$ {\AA}.\\
For $\Doz$ it is argued that the distribution function must be
symmetric because both positive and negative values of $\Doz$ are
equally likely. The scale of energy variation is determined by the
thermal energy available at the glass-transition temperature $T_g$
where the fluctuating local potentials of the liquid are frozen in
the structure. Since $T_g$ is between 200 and 1000 K for most
glasses, this energy is about 0.05 eV, much larger than the thermal
energy available at 1 K. The low temperature properties are
therefore sensitive to the center of a broad symmetric distribution,
so that $P(\Dox,\Doz)$ can be taken as independent of $\Doz$.\\
In the case of the dependence on $\Dox$, due to the exponential
dependence of $\Dox$ on $\lambda$, only a relatively small range of
$\lambda$'s is sampled for a large range of $\Dox$ and over this
limited range the distribution of $\lambda$ can be assumed constant.
This leads immediately to $P(\Dox,\Doz)=P_0/\Dox$. This result is
slightly modified by a logarithmic factor if the distribution of
$\lambda$ varies slowly with the energy, but we will ignore this
dependence. The limits of the possible values for $\Dox$ and $\Doz$,
and the value of $P_0$, can be deduced from experiments (see next
subsection).

\subsection{Determination of values for $P_0$ and $\Delta^*$} \label{apP0}
The value $P_0\sim10^{44}$J$^{-1}$m$^{-3}$ is obtained by the
experimentalists through experiments of heat release in amorphous
solids, as follows (the derivations in this appendix follow very
closely the explanations appearing in ref.\cite{E98}): An amorphous
solid is taken which is at a temperature $T_1$, and cooled down very
fast to a temperature $T_0$. The slower TLSs, which were, as all the
rest, in thermal equilibrium at a temperature $T_1$, do not have
time to follow the cooling of the system, so they equilibrate with
the surroundings releasing part of their energy and heating up the
rest of the solid, until it reaches another temperature $T_2>T_0$.
One can calculate the time dependence of the heat release as
$\dot{Q}(t)=\dot{N}_\uparrow(t)\varepsilon=(1/2)\dot{\tilde{N}}(t)\varepsilon$,
where $N_\uparrow$ is the number of TLSs in the excited state,
$\varepsilon$ is the energy of such a state, and
$\tilde{N}=N_\downarrow-N_\uparrow$. Then, assuming that the
dynamical behavior of $\tilde{N}(t)$ can be described by the
relaxation time approximation
\begin{equation}\label{relaxapprox}
    \frac{d[\tilde{N}(t)-\tilde{N}_0(T)]}{dt}=-\frac{\tilde{N}(t)-\tilde{N}_0(T)}{\tau(T)}
\end{equation}
where $\tau(T)$ is the relaxation time of a TLS with energy
difference $\varepsilon$ at a temperature $T$. Taking $T=$const.,
one arrives at
\begin{equation}\label{populevol}
\dot{\tilde{N}}(t)=N\Bigl(\tanh\frac{\varepsilon}{2k_BT_0}-\tanh\frac{\varepsilon}{2k_BT_1}\Bigr)\frac{e^{-\frac{t}{\tau(T_0)}}}{\tau(T_0)}
\end{equation}
so that the heat release is given by
\begin{equation}\label{heat1}
\dot{Q}(t)=\frac{1}{2}\varepsilon
N\Bigl(\tanh\frac{\varepsilon}{2k_BT_0}-\tanh\frac{\varepsilon}{2k_BT_1}\Bigr)\frac{e^{-\frac{t}{\tau(T_0)}}}{\tau(T_0)}
\end{equation}
Replacing the total number N of TLSs by the distribution in terms of
$\varepsilon$ and $u$ ($\varepsilon=\sqrt{\Dx^2+\Delta_{z}^{2}}$ and
$u=\Dx/\varepsilon$) and the volume \textsl{V} of the sample, the
result is
\begin{equation}\label{heat2}
\dot{Q}(t)=\frac{P_0\textsl{V}}{2}\int_{0}^{\varepsilon_{max}}d\varepsilon
\varepsilon
N\Bigl(\tanh\frac{\varepsilon}{2k_BT_0}-\tanh\frac{\varepsilon}{2k_BT_1}\Bigr)\cdot\int_{u_{min}}^{1}du\frac{1}{u\sqrt{1-u^2}}
\frac{e^{-\frac{t}{\tau(T_0)}}}{\tau(T_0)}
\end{equation}
Notice that $\tau(T_0)$ is a function of $\varepsilon$ and $u$. For
low temperatures $\tau(T_0)$ can be approximated by the FGR result
\begin{equation}\label{tauFGR}
    \tau_{FGR}^{-1}(\varepsilon,\Dx,T) = \frac{4}{\pi}\frac{\gamma^2}{\rho\hbar^4}\frac{\Dx^2\varepsilon}{v^5}
  \coth\Bigl[\frac{\varepsilon}{2kT}\Bigr]
\end{equation}
so that
\begin{equation}\label{heat3}
\dot{Q}(t)=\frac{P_0\textsl{V}}{2}\int_{0}^{\varepsilon_{max}}d\varepsilon
\varepsilon
N\Bigl(\tanh\frac{\varepsilon}{2k_BT_0}-\tanh\frac{\varepsilon}{2k_BT_1}\Bigr)\cdot\int_{\tau_{min}}^{\tau_{max}}d\tau_{FGR}
\frac{P(\varepsilon,\tau_{FGR})e^{-t/\tau(T_0)}}{\tau_{FGR}(T_0)}
\end{equation}
For usual experimental conditions and most glasses the last integral
is insensitive to the limits of integration, and moreover, it can be
shown that the main contribution to the heat release comes from TLSs
with $\varepsilon\approx2k_BT_1$ and $\tau\approx t$. For the case
$\varepsilon\ll \varepsilon_{max}$, (\ref{heat3}) leads to
\begin{equation}\label{heat4}
    \dot{Q}(t)=\frac{\pi^2k_B^2}{24}P_0\textsl{V}(T_1^2-T_0^2)\frac{1}{t}e^{-t/\tau_{max}}
\end{equation}
For $t<0.1\,\tau_{max}$ the result is independent of $\tau_{max}$
and we obtain
\begin{equation}\label{heat5}
    \dot{Q}(t)=\frac{\pi^2k_B^2}{24}P_0\textsl{V}(T_1^2-T_0^2)\frac{1}{t}
\end{equation}
Fitting the heat release experiments to this formula determines
$P_0$. There are other types of measurements that can be used for
this purpose, but this is the most accurate one, as the
coupling constant $\gamma$ characterizing the coupling of the TLSs to the acoustic waves does not appear.\\

Regarding the upper cutoff of the distribution of TLSs, $\Delta^*$,
again the experiments, in this case sound velocity ones, fix a
minimum of about 5 K for oxide glasses, reasoning as follows: The
interaction of elastic waves with TLSs results in a relative change
of the sound velocity
\begin{equation}\label{sound1}
    \frac{\delta v}{v}=-\frac{1}{V}\sum_{TLS}\frac{\gamma^2}{2\rho
    v^2\hbar}\chi'(\om)\,,
\end{equation}
where $\chi(\om)=\chi'(\om)+i\chi''(\om)$ is the susceptibility of a
single defect. At low temperatures the relaxation rates of thermal
TLSs are much smaller than the external frequency $\om$,
$\om\tau\gg1$; then only the 'resonant' part of the susceptibility,
$\chi'=(\Dx^2/\varepsilon^3) \tanh\bigl[\varepsilon/2kT\bigr]$, is
relevant. Summing according to the distribution
$P(\Dx,\Delta_z)=P_0/\Dx$, the result is
\begin{equation}\label{sound2}
    \frac{\delta v}{v}=\frac{\gamma^2P_0}{\rho
    v^2}\log(T/T_{ref})\,\,\,\,\,\,(\Gamma\ll\om\ll k_BT/\hbar)
\end{equation}
where $T_{ref}$ is a reference temperature. This result is used to
fit the experiments on the low-temperature sound velocity of various
glasses. The logarithmic law arises from thermal TLSs with small
bias $\Delta_z$, i.e. with $\Dx$ close to temperature, and due to
the fact that this behavior has been observed up to several K, it
has been concluded that $\Dx^*/k_B$ is at least about 5 K.

\subsection{Predominant coupling of the strain to the asymmetry} \label{apCouplasymm}
Given the energy of a TLS as
$\varepsilon=(\Delta_x^2+\Delta_z^2)^{1/2}$, the coupling to strain
fields $e$ may be represented by the deformation potential
\begin{equation}\label{Defpotential}
    D=\frac{d\varepsilon}{de}=\frac{\Delta_z}{\varepsilon}\frac{d\Delta_z}{de}+\frac{\Delta_x}{\varepsilon}\frac{d\Delta_x}{de}\,.
\end{equation}
In the standard tunneling model the second term is neglected, with a
resulting coupling
\begin{equation}\label{Defpotential2}
    D=\frac{d\varepsilon}{de}\approx\frac{\Delta_z}{\varepsilon}\frac{d\Delta_z}{de}.
\end{equation}
Experimental support is given in ref.\cite{A86}, where the following
considerations are made: As there is a distribution of $\Delta_x$
and $\Delta_z$, there will be a distribution $n(\varepsilon,D)$ in
$D$. Due to the coupling $D$ the TLSs influence the low-temperature
behavior of the volume thermal-expansion coefficient $\beta$,
$\beta\propto n_0\langle D\rangle$, and ultrasonic attenuation
$\alpha$, $\alpha\propto n_0\langle D^2\rangle$ (the averages are
over the distribution of TLSs). From attenuation measurements
$\langle D^2\rangle^{1/2}\approx1$ eV, and from thermal expansion
measurements $\langle D\rangle\approx 10^{-3}$ eV, so $n(D)$ is a
very broad and almost symmetric distribution. Now, to justify
eq.(\ref{Defpotential2}) the author first proves that $\langle
D^2\rangle$ and $n_0$ are almost independent of pressure P, for
values of P up to 1000 atm and more. This, together with the
observed linear variation of $\alpha$ with temperature, indicates
that the distribution $n(\varepsilon,D)$ is essentially independent
of pressure. This is needed in order to get information from
experiments of ultrasonic attenuation where a pressure is applied at
low temperature (0.5 K). The TLSs affecting the propagation of sound
are mainly those with energies close to 0.5 K, and as increasing
pressure is applied, TLSs which were in this region get out from it,
and others, which were originally out, replace them. Given the high
pressures exerted (up to 1300 atm), and the value of $\langle
D^2\rangle$, the lack of change observed in the population of TLSs
around 0.5 K can only be explained if, for a TLS, when a pressure is
applied (equivalent to a variation of local strain), the energy as a
function of pressure first decreases and then increases, \textit{id
est}, D has to change sign at some point as P increases. The most
natural explanation is to assume that the main effect of pressure is
to change the relative positions of the two wells of the double-well
potential (\ref{Defpotential2}), so that the product
$\frac{\Delta_z}{\varepsilon}\frac{d\Delta_z}{de}$ changes sign at a
certain moment (due to the change of sign of $\Delta_z$), keeping
the absolute value approximately constant. If $D$ depended mainly on
the second term of eq.(\ref{Defpotential}), to explain the
experiments the size of the tunneling barrier would have first to
increase with P and then to decrease, which seems more unlikely.

\section{Path integral description of the dissipative two-state system.}\label{appath}
We will here remind the reader about the fundamental concepts
underlying the derivation of some of the results stated in section
\ref{secspinboson}. For a more detailed presentation of the basics
see \cite{FV63,BS05,PIW07}, and for a detailed study of the
spin-boson problem using a path integral approach see
\cite{Letal87,W99}.\\

A formulation of Quantum Mechanics different from the canonical one
developed by Schr\"odinger and Heisenberg but equivalent to it was
developed by Richard Feynman in the fourties \cite{F48}, the Path
Integral formulation, based on the following postulates
\cite{PIW07}:
\begin{enumerate}
  \item The probability for any fundamental event is given by the square modulus of a complex amplitude.
  \item The amplitude for some event is given by adding together all the histories which include that event.
  \item The amplitude a certain history contributes is proportional to $e^{i S/\hbar}$, where $S$ is the action
  of that history, or time integral of the Lagrangian.
\end{enumerate}
This approach is often used in semiclassical physics, where the
transition from classical to quantum behavior of a system is
studied. In terms of path integrals the classical behavior is
obtained imposing $\hbar\rightarrow0$, because in this way the
interferences between the contributions $e^{i S[x(t)]/\hbar}$ of
neighboring paths $x(t)$ will cancel out except for those close to
the path for which $\delta S/\delta x(t)=0$ (imposing of course the
proper boundary conditions at the initial and final time), which is
by definition the classical trajectory. The first quantum
corrections will be obtained by considering also the contributions
of the paths close to the classical solution, something usually done
using stationary phase approximation calculations \cite{BS05}, where
the otherwise unmanageable integrals reduce to Gaussian integrals.\\

Feynman's ideas constitute also a very convenient framework to study
the dynamics of open quantum systems, where a subsystem focus of our
interest is not isolated but coupled to many external degrees of
freedom (its "bath" or "environment") \cite{FH65}. In the path
integral approach, one starts with a generic path integral
containing all the possible histories followed by all the variables
of the composite "subsystem + environment" system. As we are only
interested in the variables describing the subsystem, the second
step is to integrate out the rest of variables as best as we can,
analogously to what is done in studies involving reduced density
matrices. Typically, the resulting integral, now containing only the
variables of the subsystem, contains an action $S$ composed of two
parts, the action $S_{\rm{isol}}$ of the variables of the subsystem
in absence of other degrees of freedom, plus a second term
containing the information about the effects due to the bath, the
so-called influence-functional $S_{\rm{infl}}$ \cite{FV63,FH65},
where memory effects mediated by the bath are manifested in terms of
"interactions" between the subsystem variables at different times.
Fortunately a broad class of open systems can be considered to be
coupled very weakly to each of the numerous degrees of freedom of
their environment, allowing for a description of these environmental
variables in terms of harmonic oscillators. The related path
integrals are Gaussian and can be performed exactly \cite{FV63}.
Thus the approximations will start only when dealing with
$S_{\rm{infl}}$, not before.\\

The spin-boson model or dissipative two-state system
\cite{Letal87,W99}, eq.(\ref{H2}), is perhaps the most successful
and studied example using the previous ideas, due to its generality
and applicability to many different systems, as well as due to its
star role in the study of environment-induced phase transitions.

\begin{figure}
\begin{center}
\includegraphics[width=14cm]{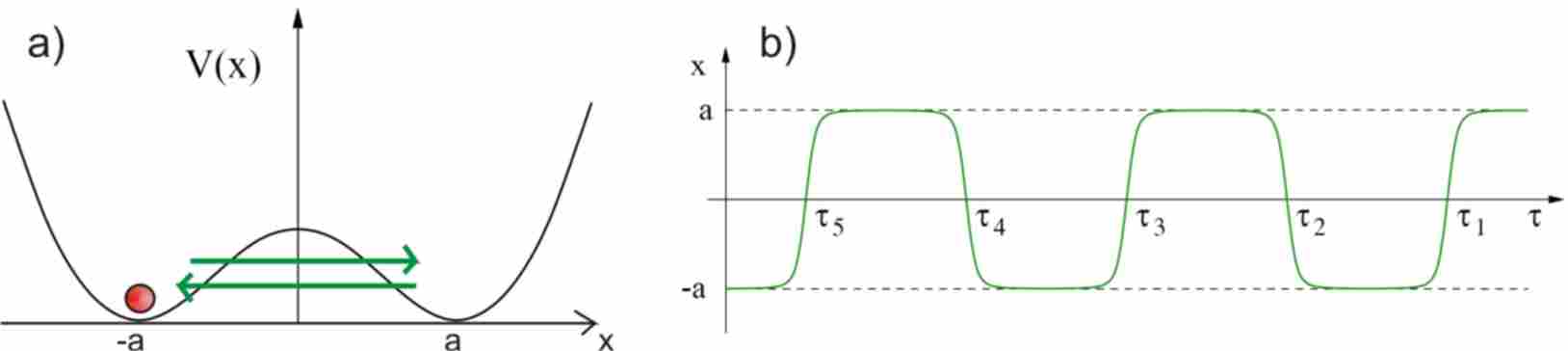}\\
\caption[Particle in a double well and multiple instanton
trajectory]{\label{Soliton} a) Particle in a double well potential.
The influence of purely quantum mechanical processes such as
tunneling is included in the path integral semiclassical approach by
considering the time as a general complex variable $\tau$, and
searching for stationary solutions $\delta S/\delta x(\tau)=0$ whose
contributions to the path integral are summed; after the solution in
terms of $\tau$ is found, the real-time dynamics is obtained by
performing an analytical continuation to real time. In a given
history of the particle dynamics, each tunneling process is called
"instanton", and in b) a prototypical multi-instanton stationary
solution is represented. Image (b) from \cite{BS05}.}
\end{center}
\end{figure}

In absence of dissipation, the path integral method gives a good
solution for the dynamics of the "tunneling particle",
fig.(\ref{Soliton}a), by adding up the contributions of all the
stationary solutions $\delta S/\delta x(\tau)=0$ where $\tau$ is
taken as a general complex variable, and performing in the final
result an analytical continuation to real times. This allows to take
into account quantum paths classically forbidden, including one or
several tunneling processes. In the jargon of the field each
tunneling process is called an "instanton" or "blip", and the
solution is the result of summing all possible multi-instanton
configurations, like the one in fig.(\ref{Soliton}b). The basic
approximation involved in the calculation is to consider the
tunneling processes as completely independent, with the total action
of a multi-tunneling path with $N$ tunneling events $S_{\rm{N}}$
being $N$ times the action of a single-tunneling path,
$S_{\rm{N}}=N\times S_{1}$, or expressed in terms of blips, there is
no interaction among them.

When a dissipative environment of harmonic oscillators is coupled to
this system, the path integral resulting from integrating out the
oscillators' degrees of freedom introduces, as previously told, an
influence-functional $S_{\rm{infl}}[|x(\tau)-x(\tau')|]$ that
provides memory to the dynamics. Now the result can be again
expressed in terms of sums over multi-blip paths, but interactions
play a role and complicate enormously the calculation. Leggett
\textit{et al.} introduced the so-called non-interacting blip
approximation (NIBA), which basically limits the memory effects to
the inclusion of the contribution to the action $S$ corresponding to
interaction between two consecutive periods separated by a tunneling
process.

The success of this approximation lies in its numerous virtues, two
out of which we highlight here for our purposes: i) It is a
controlled approximation, with a clearly defined range of validity
(basically low temperatures and weak tunneling amplitude as compared
to the characteristic frequency of the ground states of the wells)
and well defined expansion parameters, ii) It turns out to be a good
description in a very broad range of cases, from baths leading to a
weak damping to baths leading to a suppression of tunneling and
localization of the particle.

Remember that the Hilbert space of the system described in
fig.(\ref{Soliton}a) is reduced in the spin-boson model to the two
ground states of the wells, represented by a spin up and spin down
($\langle\sigma_z\rangle=1$ for the left well ground state and
$\langle\sigma_z\rangle=-1$ for the right well ground state, for
example). These two states become mixed by the tunneling amplitude
$\Dox$, introducing spin-flip processes in the dynamics. The basic
result obtained within NIBA for the time evolution
$P(t)=\langle\sigma_z(t)\rangle$ starting at $t=0$ in a given well,
$\langle\sigma_z(t=0)\rangle=1$, is
\begin{equation}\label{apPt1}
    P(t)=\frac{1}{2\pi i}\int_C e^{\lambda
    t}[\lambda+f(\lambda)]^{-1}d\lambda\,,
\end{equation}
where $C$ is the standard Bromwich contour, and
\begin{equation}\label{apPt2}
    f(\lambda)\equiv\Delta_x^2\int_0^\infty\cos\Bigl[Q_1(t)\Bigr]\exp-\Bigl[\lambda t+Q_2(t)\Bigr]\,,
\end{equation}
with $Q_i(t)$ defined by
\begin{eqnarray}
 \nonumber Q_1(t)&\equiv& \int_0^\infty \frac{J(\om)}{\om^2}\sin[\om t]d\om \\
  Q_2(t) &\equiv& \int_0^\infty \frac{J(\om)}{\om^2}(1-\cos[\om t])\coth[\hbar\om/2k_BT]d\om
\end{eqnarray}
The function $J(\om)$ appearing in the integrals is the spectral
density of the bath,
\begin{equation}\label{apJgen2}
    J(\om)= \int_{-\infty}^{\infty}dt\,e^{i\om t}\langle H_{int}(t)H_{int}(0)\rangle\,\,.
\end{equation}
At $T=0$, up to second order in perturbation theory, it can be
rewritten as
\begin{equation}\label{apJgen3}
    J(\om)= \sum_k\left| \lambda_k \right|^2 \delta (\omega - \omega_k )
\end{equation}
$J(\om)$ reflects the evolution as a function of $\om$ of the amount
of possible transitions (fluctuations) from a starting initial state
(at $T=0$ the ground state) to all final states for whom an energy
$\hbar\om$ has been exchanged between our subsystem and the
oscillators' bath.

Many of the oscillators' baths found in nature, like phonons in
solids or e-h excitations in metals, display a spectral density
$J(\om)\sim\om^s$ up to a certain cutoff (e.g. the Debye frequency
in the case of phonons). A very important result states that
whenever NIBA is applicable, the power $s$ of $J(\om)$ \emph{is the
crucial factor} determining the effect that the environment exerts
on the TLS ($\langle\sigma_z(t)\rangle$) dynamics. As explained with
an example in Appendices \ref{apTdepohm} and \ref{apTdepohm2}, at
finite temperatures $T\gg\om$ the amount of accessible transitions
thanks to the thermal energy present in the system determines
$J(\om,T>0)$, leading to $J(\om,T)\sim T^s$.

\subsection{Derivation of the spectral function for the
case of the modes of a quasi-1D nanoresonator} \label{DerivJ} The
starting point is the hamiltonian (\ref{Hamiltonian2}), where
$\partial_i u_j$ is a component of the deformation gradient matrix.
In the case of the bending modes of a rod of dimensions $L$, $t$ and
$w$, and mass density $\rho$, there are two variables $X(z),Y(z)$
obeying eqs.(\ref{apbending}). One can thus express $X(z),Y(z)$ in
terms of bosonic operators, for example $X(0)=\sum_k
\sqrt{\hbar/(2\rho twL\om_k)}(a_{k}^{\dag}+a_k)$. We can relate this
variables to the strain field $\partial_i u_j$ through the free
energy F:
\begin{eqnarray}
  \nonumber F_{rod}&=& \frac{1}{2}\int dz EI_y \Bigl ( \frac{\partial^2 X}{\partial z^2}  \Bigr )^2 +
   EI_x \Bigl ( \frac{\partial^2 Y}{\partial z^2}  \Bigr )^2=\frac{1}{2}\int dz \int dS \frac{1}{2}\lambda \sum_iu_{ii}^{2}+\mu \sum_{i,k}u_{ik}^{2} \\
  &\approx& \frac{1}{2}\int dz \int dS (\frac{3}{2}\lambda +9\mu)u_{ij}^{2}
\end{eqnarray}
extracting an average equivalence for one component $u_{ij}$,
$u_{ij}\approx2\sqrt{EI_y/(3\lambda + 18\mu)tw}\partial^2 X/\partial
z^2$. The interaction term in the hamiltonian is then
\begin{eqnarray}
  \nonumber H_{\rm{int}} &=& \hbar \sigma_z \sum_k\lambda\frac{k^2}{\sqrt{\omega_k}}(\adk +a_k) \\
   &=& \hbar \sigma_z \sum_{ij}\sum_k \Bigl [2\gamma \frac{\Dox}{\varepsilon}\sqrt{\frac{EI_y}{(3\lambda
 + 18\mu)tw}}\sqrt{\frac{1}{2\rho tw \hbar L}} \Bigr]\frac{(k^{ij})^2}{\sqrt{\omega_k^{ij}}}(a_{k}^{ij\dag}+a_{k}^{ij})
\end{eqnarray}
So we have approximately 9 times the same hamiltonian, once for each
$u_{ij}$, and the corresponding spectral function $J(\omega)$ will
be nine times the one calculated for
\begin{eqnarray}\label{deflambda}
  \nonumber H_{\rm{int}} &=& \hbar \sigma_z \sum_k\lambda\frac{k^2}{\sqrt{\omega_k}}(\adk +a_k) \\
   &\simeq& \hbar \sigma_z\sum_k \Bigl [2\gamma \frac{\Dox}{\varepsilon}\sqrt{\frac{EI_y}{(3\lambda+ 18\mu)tw}}
   \sqrt{\frac{1}{2\rho tw \hbar L}}\Bigr]\frac{k^2}{\sqrt{\omega_k}}(\adk +a_{k})
\end{eqnarray}
For this hamiltonian $J(\omega)$ is given by
\begin{equation}
    J(\omega)=\frac{1}{2\pi}\sum_k \Bigl[2\gamma
 \frac{\Dox}{\varepsilon}\sqrt{\frac{EI_y}{(3\lambda+ 18\mu)tw}}\sqrt{\frac{1}{2\rho tw \hbar L}}
\frac{k^2}{\sqrt{\omega_k}} \Bigr]^2 \delta (\omega - \omega_k)
\end{equation}
Taking the continuum limit ($\frac{1}{L}\sum_k \rightarrow
\frac{1}{\pi}\int dk$):
\begin{equation}
    J(\omega)=\frac{2L}{(2\pi)^2}\int_{k_{min}}^{k_{max}} dk \Bigl[2\gamma
 \frac{\Dox}{\varepsilon}\sqrt{\frac{EI_y}{(3\lambda+ 18\mu)tw}}\sqrt{\frac{1}{2\rho tw \hbar L}}
\Bigr]^2\frac{k^4}{\omega_k} \delta (\omega - \omega_k)
\end{equation}
Using the dispersion relation $\omega_{j}(k) = \sqrt{EI_j/(\rho
ab)}\times k^2=c\times k^2$ we express the integral in terms of the
frequency:
\begin{eqnarray}
  \nonumber J(\omega) &=& \frac{L}{(2\pi)^2}\int_{\omega_{min}}^{\wco} \frac{d\omega_k}{\sqrt{c\,\omega_k}} \Bigl[2\gamma
 \frac{\Dox}{\varepsilon}\sqrt{\frac{EI_y}{(3\lambda+ 18\mu)tw}}\sqrt{\frac{1}{2\rho tw \hbar L}}
\Bigr]^2\frac{k^4}{\omega_k} \delta (\omega -\omega_k) \\
   &=& \frac{L}{(2\pi)^2}\Bigl[2\gamma \frac{\Dox}{\varepsilon}\sqrt{\frac{EI_y}{(3\lambda+ 18\mu)tw}}\sqrt{\frac{1}{2\rho tw \hbar L}}
\Bigr]^2\frac{\sqrt{\omega}}{c^{5/2}}
\end{eqnarray}
$J_{{\rm flex}}(\omega)$ is just 9 times this,
eq.(\ref{Jsubohmic}).\\

For the compression and twisting modes the derivation follows
analogous steps. Longitudinal or compression modes correspond to the
displacement vector $u_i$ obeying the wave equation
(\ref{aplongitudinal}). Hence, there are plane waves with linear
dispersion relation $\om_k = \sqrt{\frac{E}{\rho}}k$. The strain
field $\partial u$ appears in the hamiltonian coupled to a TLS
located at $z = 0$. The quantized $u(0)$ and $\partial u \mid_{z=0}$
are given by
\begin{equation}
    u(0)=\sum_k \sqrt{\frac{\hbar}{2\rho twL
    \om_k}}(a_{k}^{\dag}+a_k) \phantom{X},\phantom{X}\partial u
    \mid_{z=0}= \sqrt{\frac{\hbar}{2\rho twL}}\sum_k \frac{k}{\sqrt{\om_k}}(a_{k}^{\dag}+a_k)
\end{equation}
 The hamiltonian is:
\begin{eqnarray}
  \nonumber \textsl{H} &=& \varepsilon\sigma_x +
    \hbar \sigma_z \sum_k \Bigl [\gamma \frac{\Dox}{\varepsilon}
    \sqrt{\frac{1}{2\rho twL\hbar }} \frac{k}{\sqrt{\om_k}}\Bigr ](a_{k}^{\dag}+a_k)+\sum_{k}\hbar\om_{k}(a_{k}^\dag a_{k}+\rm{h.c.}) \\
   &=& \varepsilon\sigma_x +
    \hbar \sigma_z \sum_k \lambda_k(a_{k}^{\dag}+a_k)+\sum_{k}\hbar\om_{k}(a_{k}^\dag a_{k}+\rm{h.c.})
\end{eqnarray}
With such a form for the coupling the spectral function is given by
eq.(\ref{alflongitudinal}).

The twisting or torsional modes obey the wave equation
(\ref{aptorsional}). In analogy with the previous case we can
quantize $\phi$ and get the same results for $\phi(0)$ and $\partial
\phi \mid_{z=0}$ as for u(0) and $\partial u \mid_{z=0}$, just
substituting $\rho$ by $\rho I$ and $E$ by $C$. But in our
hamiltonian the coupling term involves $\partial u \mid_{z=0}$ and
not $\partial \phi \mid_{z=0}$, so the latter has to be expressed in
terms of the former. Proceeding as with the bending modes:
\begin{eqnarray}
  \nonumber F_{rod} &=& \frac{1}{2}\int dz C \left( \frac{\partial \phi}{\partial z}  \right)^2 = \frac{1}{2}\int dz \int dS 4 \mu
  \left[ \left( \frac{\partial u_x}{\partial z} \right)^2 +
   \left( \frac{\partial u_y}{\partial z} \right)^2 \right]\\
   &\approx&  \frac{1}{2}\int dz \int dS 8 \mu \left( \frac{\partial u_x}{\partial z} \right)^2
\end{eqnarray}
Approximating further
\begin{equation}
\int dS 8 \mu\left(\frac{\partial u_x}{\partial z}\right)^2 \approx
8 \mu a b \left(\frac{\partial u_x}{\partial z}\right)^2=C \left(
\frac{\partial \phi}{\partial z}  \right)^2
\end{equation}
the equivalence $\partial u \mid_{z=0}=\partial u_x/\partial
z=\sqrt{C/(8\mu tw)}\partial \phi \mid_{z=0}$ is found. Therefore in
this case one has an interaction term in the hamiltonian
\begin{equation}\label{Hinttors}
H_{int}=\hbar \sigma_z \sum_k \gamma
\frac{\Dox}{\varepsilon}\sqrt{\frac{C}{8\mu tw}}
    \sqrt{\frac{1}{2\rho I \hbar L}} \frac{k}{\sqrt{\om_k}}(a_{k}^{\dag}+a_k)
\end{equation}
leading to an ohmic spectral function, eq.(\ref{alflongitudinal}),
as in the previous case.

\section{Dissipation from symmetric non-resonant
TLS$\rm{s}$}\label{Aoffresonance}

\subsection{Spectral function of a single TLS coupled to the subohmic bending modes}
We follow the method of ref.\cite{G85}. The form of $A(\om)$, the
spectral function of a single TLS, for frequencies $\om \ll
\varepsilon  $ and $\om \gg \varepsilon $ can be estimated using
perturbation theory. Without the interaction, the ground state
$|a\rangle$ of the TLS is the antisymmetric combination of the
ground states of the two wells, and the excited state is the
symmetric one, $|s\rangle$. We will use Fermi's Golden Rule applied
to the subohmic spin-boson hamiltonian, $H=\varepsilon \sigma_x +
    \hbar\lambda \sigma_z \sum_k \Bigl [ \frac{k^2}{\sqrt{\om_k}}\Bigr ](\adk +a_k)
    +\sum_k\hbar\om(k) a_{k}^{\dag}a_{k}$, where $a_k$ is the annihilation operator of a bending mode $k$.
Considering only the low energy modes $\om(k) \ll \varepsilon  $, to
first order the ground state and a state with energy $\om(k)$ are
given by
\begin{eqnarray}
  \nonumber |g\rangle &\simeq& |a\rangle - \frac{\frac{\lambda k^2}{\sqrt{\om_k}}}{2\varepsilon  }
  a_{k}^{\dag}|s\rangle + ...\hphantom{xx},\\
  |k\rangle &\simeq& a_{k}^{\dag}|a\rangle - \frac{\frac{\lambda k^2}{\sqrt{\om_k}}}{2\varepsilon  }
  |s\rangle + ...\hphantom{xx}.
\end{eqnarray}
We estimate the behavior of $A(\om)$ by taking the matrix element of
$\sigma_z$ between these two states, obtaining (remember that
$\om(k)\propto k^2$)
\begin{equation}\label{deltabig}
    A(\om)\sim \frac{\hbar \alpha_b
    \sqrt{\wco}\sqrt{\om}}{\varepsilon  ^{2}}+ ...\hphantom{xx},\hphantom{xx}\om(k) \ll \varepsilon  .
\end{equation}
The expression in the numerator is proportional to the spectral
function of the coupling,  $J(\om)=\alpha_b \sqrt{\wco}\sqrt{\om}$.
Now we turn our attention to the case $\om(k) \gg \varepsilon  $,
where the ground state $|g \rangle$ and an excited state $|k\rangle$
can be written as
\begin{eqnarray}
  \nonumber |g\rangle &\simeq& |a\rangle - \frac{\lambda k^2 /\sqrt{\om_k}}
  {\hbar \om_k + 2\varepsilon  }a_{k}^{\dag}|s\rangle + ...\hphantom{xx},\\
  |k\rangle &\simeq& a_{k}^{\dag}|a\rangle + \frac{\lambda k^2/\sqrt{\om_k}}
  {\hbar \om_k - 2\varepsilon  }|s\rangle + ...\hphantom{xx}.
\end{eqnarray}
The matrix element $\langle 0 |\sigma_z|k \rangle$ is $\sim
\frac{\lambda k^2 }{\sqrt{\om_k}}\frac{4\varepsilon  }{(\hbar
\om_k)^2}$, leading to
\begin{equation}\label{deltasmall}
    A(\om)\sim \frac{\alpha_b
    \sqrt{\wco}\varepsilon  ^{2}}{\hbar^3\om^{7/2}}+ ...\hphantom{xx},\hphantom{xx}\om(k) \gg \varepsilon  .
\end{equation}
\subsection{Value of $A_{\rm{off-res}}^{\rm{tot}}(\wo)$}
Now we will add the contributions of all the non-resonant TLSs using
the probability distribution $P(\Dox,\Doz)=P_0/\Dox$
\cite{AHV72,P72}. For the case of weak coupling, $\alf<1/2$ and $\wo
\geq (2\alpha_b)^2\wco$, which is the one found in experiments, one
has
\begin{eqnarray}\label{withresonance}
      \nonumber A_{\rm{off}}^{\rm{tot}}(\wo) \sim \int_{\hbar[\wo+\Gamma(\wo)]}^{\e_{max}}&&d\Dox\int_{-\Dox}^{\Dox}
      d\Dz\frac{P}{\Dox}\frac{\hbar \alpha_b \sqrt{\wco}\sqrt{\wo}}{(\Dox)^{2}}\\
       \nonumber +\int_{\hbar(2\alpha_b)^2\wco}^{\hbar[\wo-\Gamma(\wo)]}
    &&d\Dox \int_{-\Dox}^{\Dox}d\Dz\frac{P}{\Dox}\frac{\alpha_b\sqrt{\wco}(\Dox)^{2}}{\hbar^3\wo^{7/2}}
    \end{eqnarray}
obtaining the result $A_{\rm{off-res}}^{\rm{tot}}(\wo)\approx 2 P
\alpha_b \sqrt{\wco / \wo}$.
\subsection{The off-resonance contribution for T $>$ 0}
Using the same scheme, the modifications due to the temperature will
appear in the density of states of absorption and emission of energy
corresponding to a "dressed" TLS, $A(\varepsilon,\om,T)$. Now there
will be a probability for the TLS to be initially in the excited
symmetric state, $|s\rangle$, proportional to
$\exp[-\varepsilon/kT]$, and to emit energy $\hbar\om$ giving it to
our externally excited mode, $|k,n\rangle$, thus compensating the
absorption of energy corresponding to the opposite case (transition
from $|a\rangle|k,n\rangle$ to $|s\rangle|k,n-1\rangle$), but
contributing in an additive manner to the total amount of
fluctuations, which are the ones defining the linewidth of the
vibrational mode observed in experiments, fixing the value of
$Q^{-1}(\om,T)$. The expression for $A(\varepsilon,\om,T)$ is given
by
\begin{equation}\label{AofT}
    A(\om,T)=\frac{1}{Z}\sum_{i}\sum_{f}|\langle
    i|\sigma_{z}|f\rangle|^2 e^{-\frac{E_i}{kT}}\delta[\hbar\om-(E_f-E_i)]
\end{equation}
We consider a generic state $|i_a\rangle=|a\rangle|k_1 n_1,...,k_j
n_j,...\rangle$ or $|i_s\rangle=|s\rangle|k_1 n_1,...,k_j
n_j,...\rangle$, and states that differ from it in $\hbar\om_j$,
\begin{equation}
    |f_{a\pm}\rangle=|a\rangle|k_1 n_1,...,k_j
    n_j\pm1,...\rangle\phantom{XX},\phantom{XX}|f_{s\pm}\rangle|s\rangle|k_1 n_1,...,k_j n_j\pm1,...\rangle
\end{equation}
As for $T = 0$, we will correct them to first order in the
interaction hamiltonian $H_{int}= \hbar\lambda \sigma_z \sum_k
\sqrt{\om_k}(\adk +a_k)$, and then calculate the square of the
matrix element of $\sigma_z$, $|\langle i|\sigma_{z}|f\rangle|^2$.

Elements $|\langle i_{a,s}|\sigma_{z}|f_{a+,s+}\rangle|^2$
correspond to absorption by the "dressed" TLS of an energy
$\hbar\om_j$ from the mode $k_j$, while elements $|\langle
i_{a,s}|\sigma_{z}|f_{a-,s-}\rangle|^2$ correspond to emission and
"feeding" of the mode with a phonon $\hbar\om_j$. The expressions
for the initial states are
\begin{eqnarray}
  \nonumber |i_a\rangle=|a\rangle|k_1 n_1,...,k_j n_j,...\rangle &\rightarrow& |a\rangle|k_1 n_1,...,k_j n_j,...\rangle +
  \sum_{k_i\ni n_i>0}\frac{\hbar\lambda\sqrt{n_i\om_i}}{\hbar\om_i-2\varepsilon  }|s\rangle|...k_i n_i-1...\rangle\\
  \nonumber &-& \sum_{\forall k_i}\frac{\hbar\lambda\sqrt{(n_i+1)\om_i}}{\hbar\om_i+2\varepsilon  }|s\rangle|...k_i n_i+1...\rangle\\
    \nonumber |i_s\rangle=|s\rangle|k_1 n_1,...,k_j n_j,...\rangle &\rightarrow& |s\rangle|k_1 n_1,...,k_j n_j,...\rangle +
  \sum_{k_i\ni n_i>0}\frac{\hbar\lambda\sqrt{n_i\om_i}}{\hbar\om_i+2\varepsilon  }|a\rangle|...k_i n_i-1...\rangle\\
  &-& \sum_{\forall k_i}\frac{\hbar\lambda\sqrt{(n_i+1)\om_i}}{\hbar\om_i-2\varepsilon  }|a\rangle|...k_i n_i+1...\rangle
\end{eqnarray}
and, for example, the state $|f_{a+}\rangle$ is given by
\begin{eqnarray}
  \nonumber |f_{a+}\rangle&=&|a\rangle|k_1 n_1,...,k_j n_j+1,...\rangle\rightarrow\\
  \nonumber && |a\rangle|k_1 n_1,...,k_j n_j+1,...\rangle +
  \sum_{k_i\ni n_i>0,i\neq j}\frac{\hbar\lambda\sqrt{n_i\om_i}}{\hbar\om_i-2\varepsilon  }|s\rangle|...k_i n_i-1...k_j n_j+1...\rangle \\
   \nonumber&+& \frac{\hbar\lambda\sqrt{(n_j+1)\om_j}}{\hbar\om_j-2\varepsilon  }|s\rangle|...k_j
   n_j...\rangle - \sum_{\forall k_i,i\neq j}\frac{\hbar\lambda\sqrt{(n_i+1)\om_i}}{\hbar\om_i+2\varepsilon  }|s\rangle|...k_i n_i+1...k_j
   n_j+1...\rangle\\
   &-&\frac{\hbar\lambda\sqrt{(n_j+2)\om_j}}{\hbar\om_j+2\varepsilon  }|s\rangle|...k_j n_j+2...\rangle
\end{eqnarray}
with similar expression for the rest of states. The value of
$|\langle i_{a,s}|\sigma_{z}|f_{a+,s+}\rangle|$ (absorption) is
$|\langle
i_{a,s}|\sigma_{z}|f_{a+,s+}\rangle|=\Bigl|\frac{\lambda\sqrt{(n_j+1)\om_j}4\varepsilon
}{(\hbar\om_j)^2-4\varepsilon  ^2}\Bigr|$ , with $n_j=0,1,...$,
while for emission the result is $|\langle
i_{a,s}|\sigma_{z}|f_{a-,s-}\rangle|=\Bigl|\frac{\lambda\sqrt{n_j\om_j}4\varepsilon
}{(\hbar\om_j)^2-4\varepsilon  ^2}\Bigr|$,
 with $n_j=1,...$. Taking the limits we considered at T = 0
($\hbar\om_j\ll\varepsilon  $ and $\hbar\om_j\gg\varepsilon  $) the
results are the same as for T = 0 for absorption, except for a
factor $(n_j+1)$, and we also have now the possibility of emission,
with the same matrix element but with the factor $n_j$:
\begin{eqnarray}\label{matrixelT}
    \nonumber &\rm{Absorption}&\left\{
                       \begin{array}{ll}
                         \sim \frac{(n_j+1)\hbar\alf\sqrt{\wco}\sqrt{\om}}{\varepsilon  ^2} & \hbar\om_j\ll\varepsilon   \\
                         \sim \frac{(n_j+1)\alf\sqrt{\wco}\varepsilon  ^2}{\hbar^3\om^{7/2}}, & \hbar\om_j\gg\varepsilon
                       \end{array}
                     \right.\\
     &\rm{Emission}&\left\{
                       \begin{array}{ll}
                         \sim \frac{n_j\hbar\alf\sqrt{\wco}\sqrt{\om}}{\varepsilon  ^2} & \hbar\om_j\ll\varepsilon   \\
                         \sim \frac{n_j\alf\sqrt{\wco}\varepsilon  ^2}{\hbar^3\om^{7/2}}, & \hbar\om_j\gg\varepsilon
                       \end{array}
                     \right.
\end{eqnarray}
Now we have to sum over all initial states, and noting that the
first order correction to the energy of any eigenstate is 0, the
partition function, Z, is easy to calculate, everything factorizes,
and the result is, for example in the case $\hbar\om_j\ll\varepsilon
$:
\begin{eqnarray}
  \nonumber A_{abs}(\varepsilon  ,\om_j,T) &=& \frac{1}{Z}\sum_{i\neq j,n_i=0}^{\infty}\sum_{n_j=0}^{\infty}
  \frac{\hbar\alf\sqrt{\wco}\sqrt{\om_j}}{\varepsilon  ^2}(n_j+1)e^{-\frac{\hbar\om_jn_j}{kT}}
  e^{-\frac{\hbar\sum_in_i\om_i}{kT}}\\
  \nonumber&=&\frac{\hbar\alf\sqrt{\wco}}{\varepsilon  ^2}\frac{\sqrt{\om_j}}
  {1-e^{-\frac{\hbar\om_j}{kT}}}  \\
  \nonumber A_{em}(\varepsilon  ,-\om_j,T) &=& \frac{1}{Z}\sum_{i\neq j,n_i=0}^{\infty}\sum_{n_j=1}^{\infty}
  \frac{\hbar\alf\sqrt{\wco}\sqrt{\om_j}}{\varepsilon  ^2}n_je^{-\frac{\hbar\om_jn_j}{kT}}
  e^{-\frac{\hbar\sum_in_i\om_i}{kT}}\\
  &=&\frac{\hbar\alf\sqrt{\wco}}{\varepsilon  ^2}
  \frac{\sqrt{\om_j}e^{-\frac{\hbar\om_j}{kT}}}{1-e^{-\frac{\hbar\om_j}{kT}}}
\end{eqnarray}
In $A_{abs}(\varepsilon  ,\om_j,T)$ we have added the contributions
from the matrix elements
\begin{equation}
    |\langle i_a|\sigma_{z}|f_{a+}\rangle|^2e^{\frac{\varepsilon }{kT}}+|\langle
i_s|\sigma_{z}|f_{s+}\rangle|^2e^{\frac{-\varepsilon }{kT}}=|\langle
i_a|\sigma_{z}|f_{a+}\rangle|^2[e^{\frac{\varepsilon
}{kT}}+e^{\frac{-\varepsilon }{kT}}]
\end{equation}
The sum of exponentials cancels with the partition function of the
TLS (appearing as a factor in the total Z), leading in this way to
the expression above (the same applies for $A_{em}(\varepsilon
,-\om_j,T)$). The total fluctuations will be proportional to their
sum, which thus turns out to be at this level of approximation
$\propto\rm{cotanh}[\hbar\om_j/kT]$:
\begin{equation}
A_{diss}(\varepsilon ,\om_j,T)=A_{abs}(\varepsilon
,\om_j,T)+A_{em}(\varepsilon ,-\om_j,T)=A_{diss}(\varepsilon
,\om_j,T=0)\rm{cotanh}\Bigl[\frac{\hbar\om_j}{kT}\Bigr]
\end{equation}
In fact, this result applies for any other type of modes,
independently of its dispersion relation, provided the coupling
hamiltonian is linear in $\sigma_z$ and $(a_{-k}^{\dagger}+a_{k})$
and everything is treated at this level of perturbation theory.
Moreover, it can be proven that if one has an externally excited
mode with an average population $\langle n_j\rangle$ , and
fluctuations around that value are thermal-like, with a probability
$\propto \exp[-|n_j-\langle n_j\rangle|/kT)]$, one recovers again
the same temperature dependence,
$\propto\rm{cotanh}[\hbar\om_j/kT]$.

\section{$Q^{-1}$ due to the delayed response of biased TLSs (relaxation mechanism)}\label{aprelax}

The derivation presented here is based mainly on pages 396-406 of
ref.\cite{E98}. We want to deduce the formula for the relaxation
contribution to the attenuation of acoustic waves:
\begin{equation}\label{QEsq}
    Q^{-1}(\omega,T)=\frac{P_0\gamma^2}{\rho
    v^2}\int_{0}^{\varepsilon_{max}}d\varepsilon\int_{u_{min}}^{1}du\frac{\sqrt{1-u^2}}{u}\frac{1}{kT}\frac{1}
    {\cosh^2\bigl[\varepsilon/2kT\bigr]}\frac{\omega\tau}{1+(\omega\tau)^2}
\end{equation}
Consider a set of TLSs with a difference of energy between their two
eigenstates $\varepsilon=\varepsilon_2-\varepsilon_1$ and with
instantaneous occupation fractions $n_1$ and $n_2$, that obey
\begin{equation}\label{rateeq}
    \left\{
        \begin{array}{ll}
          1=n_1+n_2 &  \\
          \dot{n}_1=-\nu_{21}n_1+\nu_{12}n_2 &
        \end{array}
      \right.
\end{equation}
where $\nu_{12}$ and $\nu_{21}$ are the transition rates due to the
interaction of the TLS with the phonons, which reflect the finite
response time of the TLS to a perturbation. This delayed response is
responsible, together with the finite bias $\Doz>0$, of the
dissipation (\ref{QEsq}), as we shall demonstrate. We will simplify
the elasticity tensors and treat them as scalars. If we apply a
periodic stress $\s(t)=\s_0e^{-i\omega t}$ we perturb the energy
levels, in the linear response approximation, according to
\begin{equation}\label{levelshift}
    \varepsilon_{\alpha}(\s)=\varepsilon_{\alpha}(0)-v_0\lambda_{ij}^{\alpha}\s_{ij}\,\,\,\,\,\,\,,\,\,\,\,\,\,\,\,
\varepsilon_{\alpha}(u)=\varepsilon_{\alpha}(0)-v_0p_{ij}^{\alpha}u_{ij}
\end{equation}
where
\begin{itemize}
  \item $v_0$ is the unit cell volume,
  \item $\lambda_{ij}^{\alpha}=-(1/v_0)\partial
\varepsilon_{\alpha}/\partial\s_{ij}$ is the elastic dipole of state
$\alpha$ (also defined as the strain $u_{ij}$ created by a
homogeneous distribution of $c$ defects per mole,
$\lambda_{ij}^{\alpha}=\partial u_{ij}/\partial c^{\alpha}$),
  \item $p_{ij}^{\alpha}=-(1/v_0)\partial
\varepsilon_{\alpha}/\partial u_{ij}$ is the so-called double force
tensor (defined too as the stress $\s_{ij}$ created by a homogeneous
distribution of $c$ defects per mole, $p_{ij}^{\alpha}=\partial
\s_{ij}/\partial c^{\alpha}$)
\end{itemize}
Note that from the first definition of $p_{ij}^{\alpha}$ one can see
that it is approximately proportional to the deformation potential
$\gamma_{ij}^{\alpha}$ ,
$\gamma_{ij}^{\alpha}\simeq(1/2)\partial\Delta_{z}^{\alpha}/\partial
u_{ij}=-v_0p_{ij}^{\alpha}$ . This last quantity is related to our
coupling constant $\gamma_{ij}\sim\gamma$ of the interaction
hamiltonian $H_{\rm{int}}$ through
$\gamma_{ij}=2(\gamma_{ij}^{2}-\gamma_{ij}^{1})$. So returning again
to the applied periodic stress, it perturbs the energy levels,
causing a variation in the TLS populations and the corresponding
strain, according to $u_{ij}^{TLS}=c_{\alpha}\lambda_{ij}^{\alpha}$
. We want to find an expression relating $\Delta u_{ij}^{TLS}$ and
$\s(t)$, from which we can deduce the associated dynamic compliance
$\Delta S(\omega)=\Delta u/\s$. $S(\omega)$ is the inverse of the
relative change of the elastic stiffness, $\delta
C(\omega)=\s/\Delta u$, the magnitude giving us the frequency shift
and quality factor (eq.(\ref{QEsq})) due to
relaxational processes of TLSs.\\
We start dividing the instantaneous occupation fraction $n_1(t)$ as
\begin{equation}\label{n1}
    n_1(t)=\bar{n}_1^0+\Delta\bar{n}_1(t)+\delta n_1(t),
\end{equation}
where $\bar{n}_1^0$ is the equilibrium value in the absence of
stress, $\Delta\bar{n}_1(t)$ the perturbation to the equilibrium
value due to the applied stress,
\begin{equation}\label{deltan1}
    \Delta\bar{n}_1(t)=\frac{d\bar{n}_1}{d\varepsilon}\frac{\partial \varepsilon}{\partial
\s}\s_0e^{-i\omega t}=\Delta\bar{n}_1e^{-i\omega t},
\end{equation}
$\bar{n}_1(t)=\bar{n}_1^0+\Delta\bar{n}_1(t)$ is the instantaneous
equilibrium value, $\delta n_1(t)$ is the deviation from the
instantaneous equilibrium value due to the delayed response of the
defects, and $\Delta n_1(t)=\Delta\bar{n}_1(t)+\delta n_1(t)$ is the
deviation from the static equilibrium in the absence of applied
stress,
\begin{equation}\label{Deltan1}
    \Delta n_1(t)=\Delta n_1e^{-i\omega t}\,.
\end{equation}
The definitions of $\Delta n_1(t)$ and $\delta n_1(t)$ imply that
their amplitudes are complex, because they include a phase lag
between the excitation and the response, due to the finite time
needed for reaching the instantaneous equilibrium according to
(\ref{rateeq}). The strain $\Delta u_{ij}^{TLS}$ we need to compute
$\Delta S(\omega)$ is that due to $\Delta n_1$, which is
\begin{equation}\label{strain1}
    \Delta u=\lambda_1\Delta n_1 + \lambda_2\Delta n_2=\frac{c}{v_0}\frac{\partial \varepsilon}{\partial
\s}\Delta n_1
\end{equation}
Now we want to extract the value of the complex phase appearing in
eq.(\ref{Deltan1}), and for that sake we have to manipulate a bit
the rate equation (\ref{rateeq}): we begin using the condition for
the equilibrium values of the populations (detailed balance
condition),
\begin{equation}\label{detbal}
    \frac{d\bar{n}_1(t)}{dt}=0\,\,\,\,\rightarrow\,\,\,\bar{n}_1(t)\nu_{21}=\bar{n}_2(t)\nu_{12}
\end{equation}
With this relation we transform (\ref{rateeq}) into
\begin{equation}\label{rateeq2}
    \frac{dn_1(t)}{dt}=-\frac{\delta n_1(t)}{\tau}
\end{equation}
where $\tau^{-1}=\nu_{12}+\nu_{21}$. (\ref{rateeq2}) indicates that
the rate of change of the population is proportional to the
deviation from the instantaneous equilibrium. Noting that
\begin{equation}\label{rateeq3}
    \frac{dn_1(t)}{dt}=\frac{d\Delta n_1(t)}{dt}=-i\omega\Delta n_1(t)
\end{equation}
and
\begin{equation}\label{rateeq4}
    \delta n_1(t)=\Delta n_1(t)-\Delta \bar{n}_1(t)
\end{equation}
and using (\ref{rateeq2}), one obtains
\begin{equation}\label{rateeqfinal}
    \Delta n_1=\frac{\Delta \bar{n}_1}{1-i\omega\tau}=\Delta
\bar{n}_1\Bigl[\frac{1}{1+(\omega\tau)^2}+i\frac{\omega\tau}{1+(\omega\tau)^2}\Bigr]
\end{equation}
So the dynamic compliance can be written as
\begin{equation}\label{Compliance}
    \Delta S(\omega)=\frac{\Delta u}{\s}=\frac{c}{v_0}\frac{d\bar{n}_1}{d\varepsilon}\Bigl(\frac{\partial
\varepsilon}{\partial \s}\Bigr)^2\frac{1}{1-i\omega\tau}
\end{equation}
When one compares this result to the static compliance, it appears
that the dynamic response function is obtained multiplying the
static one by $(1-i\omega\tau)^{-1}$. With the dynamic modulus
$\delta C(\omega)$ an equivalent expression can be derived, with the
roles of $u$ and $\s$ exchanged,
\begin{equation}\label{Modulus}
    \delta C_{ijhk}(\omega)=\frac{\s}{\Delta u}=\frac{c}{v_0}\sum_{\alpha\beta}\frac{\partial n_{\alpha}}
{\partial \varepsilon_{\beta}} \frac{\partial
\varepsilon^{\alpha}}{\partial u_{ij}}\frac{\partial
\varepsilon^{\beta}}{\partial u_{hk}}\frac{1}{1-i\omega\tau}
\end{equation}
Now, this can be rewritten observing that
\begin{equation}\label{Corderorel}
    \frac{\partial n_{\alpha}}{\partial \varepsilon_{\beta}}=\beta n_{\alpha}(n_{\beta}-\delta_{\alpha\beta})\,\,\,\,\,\,\,\,
\rm{and}\,\,\,\,\,\,\,\sum_{\alpha}n_{\alpha}=1
\end{equation}
leading to
\begin{equation}\label{Modulus2}
    \delta C_{ijhk}(\omega)=\frac{c}{v_0}\beta n_{\alpha}n_{\beta}\Bigl(\frac{\partial \varepsilon^{\beta}}{\partial u_{ij}}-
\frac{\partial \varepsilon^{\alpha}}{\partial
u_{ij}}\Bigr)\Bigl(\frac{\partial \varepsilon^{\beta}}{\partial
u_{hk}}-\frac{\partial \varepsilon^{\alpha}}{\partial
u_{hk}}\Bigr)\frac{1}{1-i\omega\tau}
\end{equation}
To get to (\ref{QEsq}) we need a couple more transformations of this
expression. First, if the energies of the eigenstates under strain
$u$ are given by
\begin{equation}\label{apasymmetry}
    \varepsilon_{1,2}(u)=\mp\frac{1}{2}\sqrt{(\Delta_x)^2+(\Delta_z+\sum_{ij}\gamma_{ij}u_{ij})^2}\,,
\end{equation}
then eq.(\ref{Eofgamma}) is obtained, so that, together with the
definition $\varepsilon=\varepsilon_2-\varepsilon_1$, and the
equivalence $n_1n_2=(1/4)\rm{sech}^2(\varepsilon/2k_BT)$, results in
\begin{equation}\label{Modulus3}
    \delta C_{ijhk}(\omega)=\frac{c}{v_0}\Bigl(\frac{\Delta_z}{\varepsilon}\Bigr)^2\gamma_{ij}\gamma_{hk}
\frac{\beta}{4}\rm{sech}^2(\varepsilon/2k_BT)\frac{1}{1-i\omega\tau}
\end{equation}
This is the result for a concentration of $c$ defects per mole of
TLSs with fixed $\varepsilon$ and $\Delta_z$. In the case of an
amorphous solid one has instead a probability distribution in terms
of the parameters $\Delta_x$ and $\Delta_z$,
$g(\Delta_x,\Delta_z)d\Delta_xd\Delta_z=P_0/\Delta_xd\Delta_xd\Delta_z$,
which can be expressed with the variables
$\varepsilon=\sqrt{(\Delta_x)^2+(\Delta_z)^2}$ and
$u=\Delta_z/\varepsilon$ as $g(\varepsilon,u)d\varepsilon
du=P_0/(u\sqrt{1-u^2})d\varepsilon du$ , with $\varepsilon$ running
from 0 to $\varepsilon_{max}$ and $u$ from $u_{min}$ to 1. Writing
$\delta C_{ijhk}(\omega)$ in terms of $\varepsilon$ and $u$, the
inverse quality factor associated to the variations of the Young
Modulus $E$, the component of $C_{ijhk}$ entering the equations of
motion of the vibration, is, as anticipated,
\begin{equation}\label{apQEsq2}
    Q^{-1}(\omega,T)=\rm{Im}(\delta E)/E=\frac{P_0\gamma^2}{\rho
    v^2}\int_{0}^{\varepsilon_{max}}d\varepsilon\int_{u_{min}}^{1}du\frac{\sqrt{1-u^2}}{u}\frac{1}{kT}\frac{1}{\cosh^2\bigl[\varepsilon/2kT\bigr]}
    \frac{\omega\tau}{1+(\omega\tau)^2}
\end{equation}

\section{Derivation of $Q^{-1}_{\rm{rel}}(\wo,T)$, eq.(\ref{QEsq2})}\label{Relanalysis}
As discussed after eq.(\ref{QEsq}), we have to sum over underdamped
TLSs, $\varepsilon\geq [30 \alf\sqrt{ \wco} T]^{2/3}$, using the
approximation for $\Gamma$,
$\Gamma(\varepsilon,T)\sim30\alf\sqrt{\wco}T/\sqrt{\varepsilon}$.
Moreover, if $\wo\geq \Gamma(\e=[30 \alf\sqrt{ \wco} T]^{2/3},T)$
then in the whole integration range
$\wo\gg\Gamma(\varepsilon,T)\Leftrightarrow\wo\tau(\varepsilon,T)\gg1$,
so that $\wo\tau/[1+(\wo\tau)^2]\approx1/(\wo\tau)$, see
fig.(\ref{regimesQ}). $Q^{-1}_{\rm{rel}}(\wo,T)$ follows:
\begin{figure}
\begin{center}
\includegraphics[width=7cm]{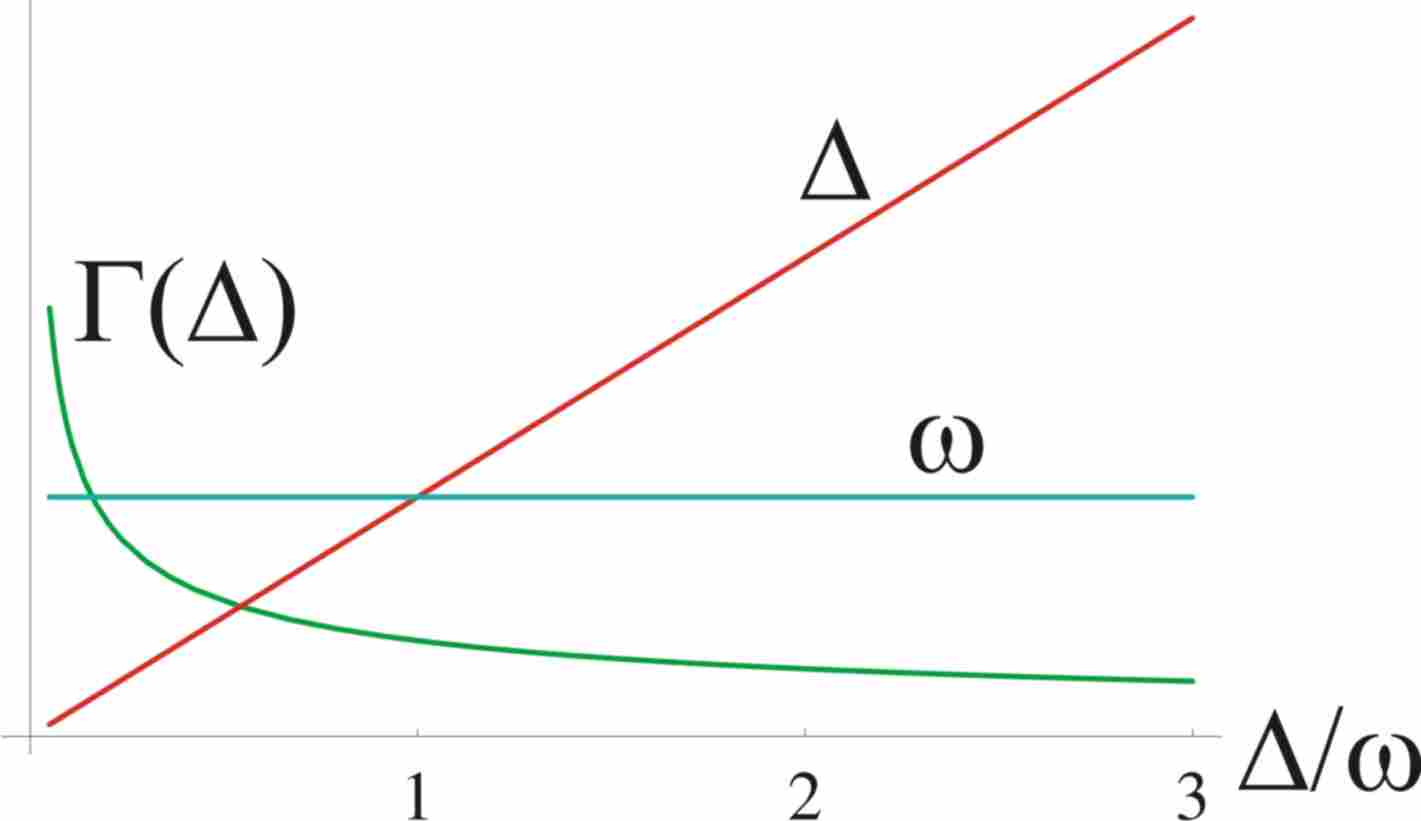}\\
    \caption[Evolution with $\Dox$ of several relevant parameters]{Evolution with $\Dox$ of the different
    quantities determining the approximations to be taken in the integrand
 $\wo\tau/[1+(\wo\tau)^2]$ of eq.(\ref{QEsq}).}
    \label{regimesQ}
\end{center}
\end{figure}
\begin{equation}\label{apQEsq}
    Q^{-1}_{\rm{rel}}(\wo,T)\approx\frac{P_0\gamma^2}{ET}\int_{[30 \alf\sqrt{ \wco} T]^{2/3}}^{T}d\varepsilon\int_{u_{min}}^{1}du
    \frac{\sqrt{1-u^2}}{u}\frac{\Gamma(\varepsilon,T)}{\wo}
\end{equation}
For temperatures $T\gg[30 \alf\sqrt{ \wco} T]^{2/3}$, which holds
for reasonable T and sizes, the integral, which renders a result of
the kind $Q^{-1}_{\rm{rel}}(\wo,T)\approx
(1/\wo)(A\sqrt{T}-BT^{1/3})$, can be approximated by just the first
term, obtaining eq.(\ref{QEsq2}). In any case for completeness we
give the expression for B:
\begin{equation}
 B\approx\frac{500P_0\gamma^{14/3}}{t^2w^{4/3}}\frac{(1+\nu)^{4/3}(1-2\nu)^{4/3}}{E^{11/3}(3-5\nu)}\Bigl(\frac{\rho}{E}\Bigr)^{1/3}
\end{equation}
Also for completeness we give the result for higher temperatures,
although for current sizes the condition $\Gamma(\varepsilon=[30
\alf\sqrt{ \wco} T]^{2/3},T)>\wo$ implies values of T above the
range of applicability of the Standard Tunneling Model. Now for some
range of energies $\Gamma(\varepsilon,T)>\wo$, and the range of
integration is divided into two regions, one where $\wo\tau\gg1$ and
one where the opposite holds:
\begin{eqnarray}\label{QforbigT}
     Q^{-1}_{\rm{rel}}(\wo,T)&=&\frac{P_0\gamma^2}{ET}\int_{u_{min}}^{1}du\frac{\sqrt{1-u^2}}{u}\times\\
\nonumber&&\phantom{X}\Bigl\{\int_{[30 \alf\sqrt{ \wco} T]^{2/3}}^
{[16\alf\sqrt{\wco}2T/\sqrt{\wo}]^2}d\varepsilon\,\,\wo\tau(\varepsilon,T)+\int_{[16\alf\sqrt{\wco}2T/\sqrt{\wo}]^2}^{T}d\varepsilon
\frac{1}{\wo\tau(\varepsilon,T)}\Bigr\}
\end{eqnarray}
The final result is $Q^{-1}_{\rm{rel}}(\wo,T)\approx
-7P_0\gamma^2\wo/T+A\sqrt{T}/\wo-CT/\wo^2$, with $A$ defined by
eq.(\ref{QEsq2}) and C by:
\begin{equation}
 C\approx\frac{1500P_0\gamma^{6}}{t^3w^{2}}\frac{(1+\nu)^{2}(1-2\nu)^{2}}{E^{3}(3-5\nu)^2}\Bigl(\frac{\rho}{E}\Bigr)^{1/2}
\end{equation}
All the results for $Q^{-1}_{\rm{rel}}(\wo,T)$ have to be multiplied
by the fraction of volume of the resonator presenting amorphous
features, $V_{amorph}/(twL)$.

\section{Derivation of eq.(\ref{E0tors})}\label{Tors}
To calculate classically the energy stored of a torsional mode
$\phi_j(z,t)=A\sin[(2j-1)\pi z/(2L)]\sin(\om_j t)$ we just calculate
the kinetic energy in a moment where the elastic energy is zero, for
example at time $t=0$. If an element of mass is originally at
position $(x,y,z)$ ($x,y$ transversal coordinates), with a torsion
$\phi(z,t)$ it moves to
\begin{equation}\label{vector}
    \bold{r}(t)=\Bigl ( \sqrt{x^2+y^2}\cdot\cos\Bigl[\arccos\frac{x}{\sqrt{x^2+y^2}}+\phi\Bigr],
    \sqrt{x^2+y^2}\cdot\sin\Bigl[\arccos\frac{x}{\sqrt{x^2+y^2}}+\phi\Bigr],z\Bigr )
\end{equation}
The kinetic energy at time $t=0$ is
\begin{equation}\label{kin}
    E_{0}=\int_{0}^{L}dz\int_{-t/2}^{t/2}dx\int_{-w/2}^{w/2}dy\cdot\frac{1}{2}\rho
    \Bigl|\frac{d\bold{r}}{dt}\Bigr|^2_{t=0}
\end{equation}
Substituting the expression for $\vec{r}(t)$ in the integrand, one
arrives at
\begin{eqnarray}\label{Erod}
    \nonumber E_{0}&=&\int_{0}^{L}dz\int_{-t/2}^{t/2}dx\int_{-w/2}^{w/2}dy\cdot\frac{1}{2}\rho
    A^2\om_j^2\sin^2\Bigl [\frac{(2j-1)\pi}{2L}z\Bigr ](x^2+y^2)\\
    &=&\frac{1}{48}A^2\om_j^2\rho L(t^3w+w^3t)
\end{eqnarray}
In terms of the creation and annihilation operators
\begin{equation}
    \phi_j(z,t)=\frac{\hbar}{2L\rho I\om_j}(a_{j}^{\dag}+a_j)e^{i(k_jz-\om t)}\,,
\end{equation}
so the mean square of its amplitude is $\langle
\phi_j^2\rangle=A^2/2=\hbar(2n+1)/[2L\rho I\om_j]$. Substituting
this in eq.(\ref{Erod}) the eq.(\ref{E0tors}) for $E_0$ is obtained.



\chapter{Appendix to Chapter \ref{chg2}}

\section{Electron-phonon coupling: friction in terms of the susceptibility $\chi$}\label{apChi}
\subsection{Damping of a phonon mode due to Coulomb interactions between charges in a device}\label{apChi1}
We have seen in subsection \ref{linear_response} how the response of
system to a given external perturbing field can be expressed, for
weak perturbations, in terms of a susceptibility or response
function. To be more concrete, we started from an interaction
hamiltonian $H_{int}$ acting as a perturbation on the initial one,
$H_0$, and its eigenfunctions $\phi_i$,
\begin{equation}\label{chipertham}
    H_{int}=\int d{\bf \vec{r}}V({\bf \vec{r}},t)\rho({\bf \vec{r}})
\end{equation}
where $\rho({\bf \vec{r}})=\sum_{{\bf \vec{k}},{\bf
\vec{k}'}}[c^\dagger_{{\bf \vec{k}}+{\bf \vec{k}'}}c_{{\bf
\vec{k}}}e^{i{\bf \vec{k}}'{\bf \vec{r}}}+$h.c.$]$ is the charge
operator, and the potential is an oscillating function of frequency
$\om$, $V({\bf \vec{r}},t)=V({\bf \vec{r}})e^{i\om t}$. We saw in
\ref{linear_response} that to linear order in $H_{int}$ the
variation in the particle density $n({\bf
\vec{r}},t)=\sum_i^{occ}|\phi_i'({\bf \vec{r}},t)|^2$ with respect
to the case in absence of perturbation can be expressed as
\begin{equation}\label{chipartdens2}
    \delta n({\bf \vec{r}})=\int d{\bf \vec{r}}'\chi({\bf \vec{r}},{\bf \vec{r}}',\om)V({\bf \vec{r}}').
\end{equation}
This can be taken as the definition of the \emph{retarded
density-density response function} $\chi$ for the charge density
change at ${\bf \vec{r}}$ induced by a potential field at ${\bf
\vec{r}}'$. In terms of a given approximation for the
single-particle eigenstates $|i\rangle$ and eigenenergies $e_i$ of
the total hamiltonian $H_{tot}=H_0+H_{int}$, it is expressed as
\begin{equation}\label{chiresponse1}
    \chi ({\bf \vec{r}},{\bf \vec{r}}',\om)=\sum_i^{occ}\sum_j\langle j|n({\bf \vec{r}})|i\rangle\langle
    i|n({\bf \vec{r}}')|j\rangle\Bigl(\frac{1}{e_i-e_j-\hbar\om}+
    \frac{1}{e_i-e_j+\hbar\om}\Bigr)\,.
\end{equation}
The response frequencies of the system are given by the poles of the
response function, which depend in turn on the approximation of the
basis. For convenience we will generalize eq.(\ref{chiresponse1}) in
terms of the $N$-body states $|\alpha\rangle$ of our system composed
of $N$ fermions. $|\alpha\rangle$ can be given, for example, by a
Slater determinant of one-body states $|i\rangle$. We assume that
the system starts in its ground state $|0\rangle$ (T=0 case). We
define the fermionic creation operators $c_i^\dagger$ creating
one-body states $|i\rangle$, so that
\begin{equation}\label{chiresponse2}
    \chi (i,j,\om)=\sum_{\alpha}\langle \alpha|c_i^\dagger c_j|0\rangle\langle
    0|c_j^\dagger c_i|\alpha\rangle\Bigl(\frac{1}{\varepsilon_0-\varepsilon_\alpha-\hbar\om}+
    \frac{1}{\varepsilon_0-\varepsilon_\alpha+\hbar\om}\Bigr)\,.
\end{equation}
Now we are interested in the probability per unit time that the
field $V({\bf \vec{r}},t)$ transfers energy $\hbar\om$ to the
system. According to Fermi's Golden Rule of second-order
perturbation theory the probability we are looking for is given by
\begin{equation}\label{transprob1}
    \Gamma(V,\om)=\frac{2\pi}{\hbar}\sum_\alpha|\langle \alpha|\hat{V}|0\rangle|^2 \delta(\hbar\om-\varepsilon_\alpha+\varepsilon_0)
\end{equation}
Here $\hat{V}=\sum_{i,j}\langle j|V({\bf \vec{r}},t)|i\rangle
c_i^\dagger c_j$ is the operator corresponding to $V({\bf
\vec{r}},t)$ in the second-quantized basis $c_i^\dagger$.
Eq.(\ref{transprob1}) can be expressed in terms of $\chi$ by
providing the latter with an imaginary part through the addition of
a small term $i\eta$ to $\om$, and then using the relation
Im$[1/(x-i\eta)]=\pi\delta(x)$. In this way we obtain the expression
for $\Gamma$ used in the thesis,
\begin{equation}\label{transprob2}
    \Gamma(V,\om)=\int d{\bf \vec{r}}d{\bf \vec{r}}'V({\bf \vec{r}})V({\bf \vec{r}}')\rm{Im}\chi({\bf \vec{r}},{\bf \vec{r}}',\om)\,.
\end{equation}
Now the goal is to use a good approximation for the susceptibility
function of our system.

\subsection{Clean metal susceptibility}\label{apclean_metal}
Suppose that our system is a metal. When the presence of impurities,
electron-phonon collisions and electron-electron interactions is
such that the collision frequency $\nu=1/\tau$ of an electron is
much less than the frequencies $\om$ we are interested in,
$\om\tau\ll1$, and/or the mean free path is much longer than the
length scales $L$ we study, $v_F\tau\gg L$, we are in the clean
metal limit\cite{NP99}, where the susceptibility can be expressed in
terms of elementary excitations. If one neglects electron-electron
interactions (free electron gas), the retarded susceptibility is
given by the bare bubble diagram:
\begin{equation}\label{chi0}
    \chi_0^R({\bf\vec{q}},\om)=\frac{1}{\textit{V}}\sum_{{\bf\vec{k}},\sigma}\frac{n_F(\e_{\bf\vec{k}})-n_F(\e_{{\bf\vec{k}}+{\bf\vec{q}}})}
    {\e_{\bf\vec{k}}-\e_{{\bf\vec{k}}+{\bf\vec{q}}}+\hbar\om+i\eta},
\end{equation}
where $\textit{V}$ is the volume of the system, $n_F(\e)$ is the
Fermi function, $\e_{\bf\vec{k}}-\e_{{\bf\vec{k}}+{\bf\vec{q}}}$ are
the excitation energies of the quasiparticles and
$\eta\rightarrow0$. The response of the system is determined by the
excitation of these one-body quasiparticle states, corresponding to
zeroes of the real part of the denominator. The main effect of the
inclusion of Coulomb interactions among electrons is the appearance
of collective excitations, formed by a superposition of one-body
states, called plasmons in the case of the electron gas. Their
presence is manifested for example in an RPA calculation. RPA is the
result of summing, for the interacting electron gas, the most
important diagrams of each order \cite{BF04}, which turn out to be
the ones of the bubbles, and these diagrams increase their
importance as the density increases, so RPA is the high density
limit of the interacting electron gas. The RPA result is the
following:
\begin{equation}\label{chiRPA}
    \chi_{RPA}^R({\bf\vec{q}},\om)=\frac{\chi_0^R({\bf\vec{q}},\om)}{1-V_{Coul}({\bf\vec{q}})\chi_0^R({\bf\vec{q}},\om)},
\end{equation}
where $V_{Coul}$ represents the Coulomb potential. The response will
show peaks again in the zeroes of the real part of the denominator
of $\chi_0^R({\bf\vec{q}},\om)$, but now also when
$1-V_{Coul}({\bf\vec{q}})\chi_0^R({\bf\vec{q}},\om)=0$, which
corresponds to collective excitations.

\subsection{Dirty metal susceptibility}\label{apdirty_metal}
When due to the presence of impurities, electron-phonon collisions
and electron-electron interactions the collision frequency
$\nu=1/\tau$ of an electron is higher than the frequencies $\om$ we
are interested in, $\om\tau\gg1$, and the mean free path is much
shorter than the length scales $L$ we study, $v_F\tau\ll L$, one
enters the so-called hydrodynamic regime\cite{NP99}. A given
quasiparticle is subject to many collisions during one period of the
exciting external field. It is thus completely "thermalized" well
before it can sample the periodicity of the field. Therefore the
best way to proceed is to assume that the collisions act to bring
about everywhere a state of local thermodynamic equilibrium, and use
just macroscopic "local" quantities (density, current, pressure,
etc) to describe the state of the system; the response to the
external field may be obtained by using the usual laws of
thermodynamics and hydrodynamics. For high enough temperatures one
arrives at an expression for the response function $\chi({\bf
\vec{q}},\om)$ which contains a peak corresponding to the plasmon
and a continuous background arising from thermal diffusion. This
background is the one described by eq.(\ref{susclayers}). Taking
into account the plasmon contribution, eq.(\ref{susclayers}) becomes
\begin{equation}\label{chidirtyplasmon}
    \chi({\bf\vec{q}},\om)=\frac{\nu D q^2}{Dq^2-\om^2\tau - i\om},
\end{equation}
The plasmon contribution to the dissipation of the vibrational modes
we study in this thesis can be safely neglected. Basically one has
to see if $Dq^2\gg \om^2\tau$ in each of the layers (graphene and
inversion layer in the Si-SiO$_2$ interface). In both cases one can
take $D\nu\sim10^3$, $\tau=l/v_F\sim10^{-7}/10^6\sim10^{-13}$ s
(assuming $l\sim100$ nm), $q\sim1/L\sim10^6$ m$^{-1}$,
$D=v_Fl\sim10^{11}$m$^2/$s. The density of states (DOS) of a stack
of N graphene layers is $\nu^C=N\gamma/\hbar^2v_F^2$, being
$\gamma\sim0.3$ eV the interlayer hopping element. The DOS of the
approximately two-dimensional electron gas lying in the inversion
layer Si-SiO$_2$ is $\nu_{2DEG}\sim m*/\pi\hbar^2$, with
$m*\sim0.3m_{e^-}$ the effective mass of the electrons in Si. Using
these values, $N\sim10$, and for the frequencies of the oscillators
analyzed, $\om\sim100$ MHz, one immediately verifies that indeed
\begin{equation}\label{chidirtyplasmon2}
    Dq^2\gg\om,\,\om^2\tau\,\,\,\,\rightarrow\,\,\,\,\rm{Im}\chi({\bf\vec{q}},\om)\approx\frac{\nu\om}{Dq^2}
\end{equation}
The inequalities are quite large, so there is place for substantial
variations of $l$ or $N$ and they will still hold, justifying our
use of eq.(\ref{susclayers}) and the limit $Dq^2\gg\om,\,\om^2\tau$.

\subsection{Microscopic derivation of eq.(\ref{susclayers})}\label{apdirty_metal2}
We sketch the analysis presented in \cite{AS06}. This form of the
susceptibility can be derived assuming a distribution of short-range
scattering centers represented by potentials, creating an $V({\bf
\vec{r}})=\sum_iV_{imp}({\bf \vec{r}}-{\bf \vec{r}}_i)$. To
calculate propagators and physical magnitudes one has to perform in
a final step a disorder average $\langle ...\rangle_{dis}=\int
DV\,\,P[V](...)$, where the probability measure $P[V]$ describes the
statistical properties of the potential $V$. In most applications it
is sufficient to implement a Gaussian distribution,
$P[V]=\exp\Bigl[-\frac{1}{2\gamma^2}\int d^drd^dr'V({\bf
\vec{r}})K^{-1}({\bf \vec{r}}-{\bf \vec{r}}')V({\bf
\vec{r}}')\Bigr]$, where $\gamma$ measures the strength of the
potential and $K$ describes its spatial correlation profile,
$\langle V({\bf \vec{r}})V({\bf
\vec{r}}')\rangle_{dis}=\gamma^2K({\bf \vec{r}}-{\bf \vec{r}}')$.
Very often one finds that the finite spatial correlation of V does
not matter, in which case one may set $K({\bf \vec{r}})=\delta({\bf
\vec{r}})$, so that
\begin{equation}\label{imppot}
    \langle ...\rangle_{dis}=\int DV\,\,\exp\Bigl[-\frac{1}{2\gamma^2}\int d^drV^2({\bf\vec{r}})\Bigr]
\end{equation}
For length scales longer than the mean free path, so that the
electrons experience multiple scattering, details of the form chosen
for the potential $V({\bf \vec{r}})$ are quickly erased.

In practice, this average is very difficult to perform, and a series
of techniques have been developed for mapping it into a different
form. For example, in the replica field theory, the expectation
value of an operator $O$ is written in terms of a source $J$
introduced conveniently in the action of the path integral
representing the replicated partition function, $Z^R[J]=\int
D(\Psi,\bar{\Psi})\exp\Bigl[-\sum_{a=1}^RS[\Psi^a,\bar{\Psi}^a,J]\Bigr]$,
\begin{equation}\label{replica2}
    \langle O\rangle=-(\partial/\partial J)\ln Z[J]=\lim_{R\rightarrow0}\frac{1}{R}\frac{\partial Z^R[J]}{\partial
 J}=\lim_{R\rightarrow0}\frac{1}{R}\sum_{a=1}^R\langle O(\Psi^a,\bar{\Psi}^a)\rangle_\Psi
\end{equation}
When performing now the disorder average of the functional $Z^R[J]$
over the distribution $P[V]$, the result is the presence of a new
effective interaction term in the action of the system, with respect
to the case without disorder:
\begin{equation}\label{replica}
    \langle Z[J]\rangle_{dis}=\int D(\Psi,\bar{\Psi})\exp\Bigl[-\sum_{a=1}^RS|_{V=0}[\Psi^a,\bar{\Psi}^a\Bigr]+
    \sum_{a=1}^R]\frac{\gamma^2}{2}\sum_{mn}\int d^dr
    \bar{\Psi}^a_m({\bf \vec{r}})\Psi^a_m({\bf \vec{r}})\bar{\Psi}^b_n({\bf \vec{r}})\Psi^b_n({\bf \vec{r}})
\end{equation}
Then, one can apply perturbation theory to this interaction term to
know what are the disorder averaged values of observables, or in our
case of the susceptibility, using diagrammatic theory and an
\textit{ansatz} for the imaginary part of the self-energy correction
to the fermion propagator, Im$\Sigma(\om_n)=-$sgn$(\om_n)/2\tau$,
where $\tau$ is the so-called elastic scattering time, related to
the elastic mean free path $l$ by $\tau\equiv l/v_F$. One can see
that this parameter is related to the potential strength $\gamma^2$
of the disorder potential by $\tau^{-1}=2\pi\nu\gamma^2$, where
$\nu$ is the density of states. Usually $\gamma^2$ is expressed
through $\tau$ from the outset. The main results are

\begin{eqnarray}\label{disorderresults}
  \nonumber \langle G(\textbf{x},\textbf{y};\tau)\rangle_{dis} &=& G(\textbf{x},\textbf{y};\tau)|_{V=0}e^{-|\textbf{x}-\textbf{y}|/2l} \\
  \langle\chi(\textbf{r},\tau)\rangle_{dis}\,\,\,\,&=& \frac{T}{L^d}\sum_{\textbf{q},\om_m}e^{i\textbf{q}\textbf{r}+i\om_m\tau}\frac{\nu Dq^2}
  {|\om_m|+Dq^2}
\end{eqnarray}
where $\chi(\textbf{r},\tau)$ is the density-density correlation
function (the susceptibility of a dirty metal).

\section{Im$\chi$, $Q^{-1}$ and temperature dependence of friction}
In order to make a connection between theory and experimental data a
first point that has to be kept in mind is the following: $Q^{-1}$,
the linewidth observed in the frequency domain of the response to an
external driving force, does not correspond exclusively to an
irreversible loss of energy of the mode (damping), but to all the
processes which alter the phonon population of an excited mode of a
certain frequency, including both increases and decreases of
population. That is, Im$\chi$, $Q^{-1}$ correspond to the
fluctuations of the given mode due to its coupling to other degrees
of freedom of the system, giving a finite lifetime of the mode
$(\textbf{k}_0,n_{\textbf{k}_0})$ subject of study due to any kind
of process, including those increasing $n_{\textbf{k}_0}$. Friction,
instead, corresponds just to processes involving a decrease of
$n_{\textbf{k}_0}$.

\subsection{Temperature dependence of Im$ \chi$, $Q^{-1}$ due to excitation and relaxation of e-h
pairs}\label{apTdepohm}

Friction experienced by, for example, traveling charged particles
through a host metal due to the excitation of e-h pairs has been
subject of intensive research since a long time \cite{G84}. In the
graphene resonator's case a similar effect produces absorption and
fluctuations of energy, but there is an important difference with
the previous example: we are studying the linewidth of a single
bending mode of a well-defined wavevector ${\bf \vec{q}}$ and
frequency $\wq$, so we will not sum over frequencies to calculate
the total absorption and fluctuations of the resonator's vibrational
modes due to this mechanism. More specifically, we want to calculate
\begin{equation}
\Gamma(\wq,T) = \int d ^3{\bf \vec{r}}  \int d^3{\bf \vec{r}}'
\rm{Re}V_{\rm  scr} ( {\bf \vec{r}} , \wq ) \rm{Re} V_{\rm scr} (
{\bf \vec{r}}' , \wq )  \rm{Im} \chi [ {\bf \vec{r}} - {\bf
    \vec{r}}' , \wq ,T]
\label{gammamode12}
\end{equation}
And for this we need Im$ \chi [ {\bf \vec{r}} - {\bf \vec{r}}' , \wq
,T]$. To obtain the dependence with T we can recall the fact that,
embedded in Im$ \chi$, what we are looking for is the sum of e-h
transitions that absorb an energy $\wq$ from the vibrational mode
$({\bf \vec{q}},\wq)$ through $H_{int}$ plus desexcitations that
transfer an energy $\wq$ to that mode. We will begin by analyzing
the $T=0$ case, where only the former can take place, so that
dissipation and fluctuations are indistinguishable. We will see that
for a given frequency $\om\ll E_F/\hbar$, the e-h transitions
available are proportional to $\om$, independently of the dimension
of the fermionic bath. It is a direct consequence of the exclusion
principle through the restrictions imposed by the Fermi distribution
functions $n_F(E)=1/[1+\exp[(E-E_F)/k_BT]]$ appearing. Taking for
example a 3D Fermi gas, with dispersion relation
$E=\sum_i\frac{\hbar^2k_i^2}{2m}$, the probability of exciting an
e-h pair of frequency $\om$ will be proportional to the number of
such pairs available, $P(\om)$,
\begin{equation}\label{phexc1}
    P(\om)=\int d^3{\bf \vec{k}}\int d^3{\bf \vec{k'}}n_F(E({\bf
    \vec{k}}))\Bigl[1-n_F(E({\bf \vec{k'}}))\Bigr]\delta\Bigl(\om-\frac{E({\bf \vec{k}})-E({\bf \vec{k'}})}{\hbar}\Bigr)
\end{equation}
Expressing the integrals in terms of energies and making the
approximation $\hbar\om\ll E_F$,
\begin{eqnarray}\label{phexc2}
  \nonumber P(\om) &\propto& \int_0^\infty dE\sqrt{E(E+\hbar\om)}n_F(E)[1-n_F(E+\hbar\om)] \\
   &=& \int_{E_F-\hbar\om}^{E_F}dE\sqrt{E(E+\hbar\om)} \simeq E_F\times \hbar\om\sim \om
\end{eqnarray}
In the case we took a fermionic bath of a different dimension, the
only difference in the calculation would arise in the factor which
in the 3D case is $\sqrt{E(E+\hbar\om)}$, which will be in general a
function $f(E,\hbar\om)$. But as long as this is a smooth function
of $E$ and $\hbar\om$, for  $\hbar\om\ll E_F$ we can approximate
$f(E,\hbar\om)\approx f(E_F,0)$ and arrive again at $P(\om)\sim\om$.
At $T=0$ only excitations of e-h pairs from the ground state can
occur, so the net absorption rate involving excitations of frequency
$\om$ will be indeed proportional to
$\om$.\\
At finite temperatures there will be absorption and emission
processes. First we will calculate the number of e-h excitations of
a certain frequency $\om$ which will cause energy absorption, as in
the $T=0$ case:
\begin{equation}\label{phexc3}
    P(\om,T)=\int d^n{\bf \vec{k}}\int d^n{\bf \vec{k'}}n_F(E({\bf
    \vec{k}})/k_BT)\Bigl[1-n_F(E({\bf \vec{k'}})/k_BT)\Bigr]\delta\Bigl(\om-\frac{E({\bf \vec{k}})-E({\bf \vec{k'}})}{\hbar}\Bigr)
\end{equation}
For frequencies and temperatures such that $\hbar\om\ll k_BT\ll E_F$
one can approximate
\begin{eqnarray}\label{phexc4}
  \nonumber P(\om,T) &\propto& \int_0^\infty dE f(E,\hbar\om)n_F(E/k_BT)[1-n_F([E+\hbar\om]/k_BT)]\\
  \nonumber &\approx& f(E_F,0)\int_0^\infty dE  n_F(E/k_BT)[1-n_F(E/k_BT)] \\
   &=& f(E_F,0)\times k_BT \times \int_0^\infty dx\, n_F(x)[1-n_F(x)]\sim T
\end{eqnarray}
So for excitation frequencies much smaller than the temperature the
number of available excitations absorbing energy is proportional to
$T$, and roughly independent of $\om$. The number of desexcitations
compensating this absorption processes has a similar expression:
\begin{equation}\label{phexc5}
    P'(\om,T)\propto\int_0^\infty dE f(E,\hbar\om)n_F([E+\hbar\om]/k_BT)[1-n_F(E/k_BT)],
\end{equation}
and is also proportional to $T$, leading to Im$ \chi
(T),Q^{-1}(T)\sim T$.

\subsection{Temperature dependence of energy loss due to excitation and relaxation of e-h pairs}
The net amount of energy absorbed by the fermionic environment of
e-h pairs of frequency $\om$ at a finite temperature, $A(\om,T)$,
for $\hbar\om\ll k_BT\ll E_F$, is proportional to
\begin{equation}\label{phexc5}
    A(\om,T)\propto P(\om,T)-P'(\om,T)\propto\int_0^\infty dE f(E,\hbar\om)\Bigl\{n_F(E/k_BT)-n_F([E+\hbar\om]/k_BT)\Bigr\},
\end{equation}
and it can be approximated as
\begin{eqnarray}\label{phexc6}
    \nonumber A(\om,T)&\propto& f(E_F,0)\int_0^\infty dE \Bigl\{n_F(E/k_BT)-n_F([E+\hbar\om]/k_BT)\Bigr\}\\ \nonumber &=&
    f(E_F,0)\times k_BT\times\int_{-E_F/k_BT}^{-(E_F-\hbar\om)/k_BT} dx\, n_F(x)\\
    &\approx& f(E_F,0)\times \frac{1}{1+e^{-E_F/k_BT}}\times \hbar\om\sim \om
\end{eqnarray}
The net absorption of energy due to pairs of frequency $\om$ is
therefore to a first approximation proportional to $\om$ and
independent of $T$.

\subsection{Extension to a generic system + bath}\label{apTdepohm2}
Suppose that the subsystem subject of study, characterized by the
index ${\bf \vec{k}}$, is coupled linearly to a bath of excitations,
\begin{equation}\label{Jgen1}
    H_{int}\propto (c^\dagger_{{\bf \vec{k}}}+c_{{\bf \vec{k}}})\times\sum_{{\bf \vec{q}}}\lambda_{{\bf \vec{q}}}
    (b^\dagger_{{\bf \vec{q}}}+b_{{\bf \vec{q}}}),
\end{equation}
where $c^\dagger_{{\bf \vec{k}}}$ and $b^\dagger_{{\bf \vec{q}}}$
represent subsystem and bath creation operators, respectively. We
can calculate at $T=0$
\begin{equation}\label{Jgen2}
    J_{gen}(\om)= \int_{-\infty}^{\infty}dt\,e^{i\om t}\langle H_{int}(t)H_{int}(0)\rangle\sim\om^s\,\,,
\end{equation}
which as we saw in Appendix \ref{appath} reflects the evolution as a
function of $\om$, up to second order in perturbation theory, of the
amount of possible transitions (fluctuations) from a starting
initial state (at $T=0$ the ground state) to all final states for
whom an energy $\hbar\om$ has been exchanged between our subsystem
and the bath. As temperature grows, for $T\gg\om$, the amount of
accessible transitions thanks to the thermal energy present in the
system will consequently grow as $T^s$, so that Im$ \chi
(T),Q^{-1}(T)\sim T^s$, at least when the perturbation $H_{int}$ is
not too strong.

\section{Charge impurity density in SiO$_2$ and the SiO$_2$-Si interface}\label{apcharge_imp}
From refs.\cite{S81,N82}. In ref.\cite{S81} they speak of four types
of interface trapped charges:
\begin{enumerate}
  \item Just in the interface, due to the
interruption of the periodic lattice structure, there is the
so-called \textit{interface-trapped charge} $Q_{it}$. Densities as
low as $Q_{it}\sim10^{10}$ cm$^{-2}$ are achieved by low-temperature
(450ºC) hydrogen annealing. These charges are associated to traps,
called interface traps, which change occupancy with gate bias
changes and have energy levels distributed throughout the bandgap.
Interface traps characteristic of thermal oxidation are
\textit{donor} type in the upper half of the bandgap.
  \item Close to the Si-SiO$_2$ interface there are also charge centers
called \textit{oxide fixed charge}, $Q_{fc}$, predominantly
positive, immobile and with fixed charge when a gate voltage is
applied. Experiments locate it within 30 {\AA} of the interface.
  \item There are \textit{positive mobile ionic charges}, $Q_m$, related to
  trace contamination by alkali metal ions, causing reliability
  problems in semiconductor devices operated at high temperatures
  and voltages. Films impervious to mobile ions, such as Si$_3$N$_4$
  can prevent mobile ionic charge contamination of the oxide during
  device life, but there is no such film in our case. Effective
  densities $Q_{m}\sim10^{10}$ cm$^{-2}$ can be achieved without
  films.
  \item There are oxide traps associated with defects in SiO$_2$,
  which are usually electrically neutral, and are charged by
  introducing electrons and holes into the oxide, originating \textit{oxide
  trapped charge}.
  \item During the diffusion of oxigen creating the oxide layer, if
  the process is fast enough, a considerable amount of dopant
  impurities (B, P, As or Sb) may be left in the oxide, although their diffusion
  coefficients in silicon dioxide are very small compared to the
  ones in silicon. In \cite{YOTT79} experiments were made where up
  to 20\% of the phosphorous was trapped inside the oxide during the
  oxidation in wet oxygen process at temperatures between 750 and 900ºC.
\end{enumerate}
These charges give rise to a charge density in the oxide layer,
$\rho_{ox}$. Unfortunately, the published literature lacks the
information on this quantity for oxides used in MEMS applications.
What has been extracted from high-frequency capacitance-voltage
curves of MOS capacitors is the \textit{effective interface charge
density} $Q_f$, which corresponds to
\begin{equation}\label{interfeffcharge}
    Q_f=\int_0^{t_{ox}}\frac{x}{t_{ox}}\rho_{ox}dx + Q_{it}
\end{equation}
In \cite{FKDB02} they mention that the breakdown field value for
SiO$_2$ is $\sim1$ V/nm. This means that in typical graphene
experiments, where the potentials are about 100 V applied through an
oxide layer 300 nm thick, the field is 1/3 of the breakdown field,
so most probably all the potentially chargeable impurities within
the oxide will be charged. Indeed, in \cite{KHDS02} a field of about
0.03 V/nm was sufficient to polarize the SiO$_2$ dielectric, and in
\cite{RFTJ02} they mention that the bulk oxide traps become charged
with injection of electrons from the nanotube at gate fields above
0.03 V/nm.

\subsection{Estimate of the charge concentration, and comparison with numbers given in other references to fit experiments}
From the experimental $Q_f$ values in \cite{AP87}, where a MOS setup
with a 200 nm oxide layer, and p-type silicon (the most common
choice for mass-produced chips) with doping $N_A=4\cdot10^{15}$
cm$^{-3}$ was the subject of the measurements, the authors in
\cite{WS04} extract the value of the quantity
\begin{equation}\label{oxidecharge}
    Q_{ox}=\int_0^{t_{ox}}\rho_{ox}dx + Q_{it},
\end{equation}
where $t_{ox}$ is the thickness of the oxide layer, assuming several
distributions of charge in the oxide. If one chooses the charges to
be concentrated in a layer of 20 nm starting from the interface and
$Q_{it}=0$, the result is
\begin{equation}\label{oxchargedens}
    \rho_{ox}\sim1.5\cdot10^{16}\,\,\,\rm{charges/cm}^3
\end{equation}
If instead of that one assumes a uniform distribution of charge
throughout the oxide, the result is
\begin{equation}\label{oxchargedens2}
    \rho_{ox}\sim3\cdot10^{15}\,\,\,\rm{charges/cm}^3
\end{equation}

\subsection{Thickness of the charged layer in the doped Si
gate}\label{apthicknessSi} From \cite{S81}. When one applies a
voltage between the graphene and the p-doped Si electrode, the first
effect at low voltages ($\sim1$ V) is to populate the intergap
states associated with the impurities, creating a region of a
certain thickness where all impurities are charged, so there the
charge density is just the doping $N_A$, which for a typical
low-doped Si wafer is $N_A\sim10^{15}-10^{16}$cm$^{-3}$. For that
scale of voltages the thickness soon reaches a maximum value $W_m$
before the complete depletion regime is reached. This value can be
calculated as
\begin{equation}\label{Wm}
    W_m=\sqrt{\frac{4\epsilon_sk_BT\ln(N_A/n_i)}{q^2N_A}},
\end{equation}
where $\epsilon_s\sim11.9\e_0$ is the dielectric constant of Si,
$n_i\sim10^{10}$ cm$^{-3}$ its intrinsic doping (everything at room
temperature), and $q$ the electron charge. For a low-doped Si wafer
one has $W_m\sim450$ nm. The complete depletion regime is reached
when the voltages applied are enough to curve the bands close to the
Si-SiO$_2$ interface until the Fermi level reaches the bottom of the
conduction band. Due to the high density of states of the band the
charge density will increase dramatically in a narrow region
$\sim10$ nm thick as the voltage is further raised, overwhelming the
amount of charge corresponding to the intergap states, so for high
voltages (for example typical values $V\sim100$ V) the approximation
of a 2D charged Si layer is reasonable for typical oxide thicknesses
$t_{SiO_2}\sim300$ nm.

The voltage that really matters is the one at the Si-SiO$_2$
interface, which can be safely taken as of the order of the
potential applied between the electrodes.

\subsection{Information about the SiO$_2$ surface and structure. Conclusions from experiments with carbon
nanotubes}\label{apsilica_surface} The strong adherence of the
graphene layer to the SiO$_2$ substrate is probably due to the
presence of Hydroxyl groups in the surface of the latter, which form
hydrogen bonds with the carbon layer. When setting the graphene
sheet into periodic motion, a repeated breaking and healing of some
these bonds (like a Velcro) can be a source of short-wavelength
phonons which carry away part of the energy initially stored in the
resonator, thus damping its motion. How many OH's are there? In
ref.\cite{SG95} they analyze Hydroxyl groups on a well-defined
silica surface, and mention that isolated Si-OH will survive anneals
at over 1000 $^o$C, presumably because there is nowhere for the
hydrogen to go: it needs to find another hydrogen to be removed as
H$_2$ or even H$_2$O. As the concentration of H at the surface is
reduced this becomes increasingly improbable. Adding hydrogen atoms
from e.g. a hydrogen plasma helps desorb the remaining isolated
Si-OH, but in any case the surface density of a previously fully
hydroxylated surface after high-temperature annealing is $\sim1$
OH/nm$^2=10^{14}$OH/cm$^2$. This is in agreement with another
experiment \cite{DPX98}, which estimates a surface density of
isolated silanol groups on a flat silica surface of about
$9\cdot10^{13}$silanol/cm$^2$, thus supporting the idea of either a
simultaneous breaking of a great number hydrogen bonds, for a
resonator $\sim1\mu$m wide, or no breaking of bonds at all.\\

The presence of oxide trapped charges within the SiO$_2$ and of
other surface impurities has attracted the attention in several
studies where its influence and possible use in carbon nanotube
field-effect transistors has been contemplated
\cite{FKDB02,RFTJ02,Ketal03,Fetal06}. The most interesting
conclusion for us is the one of \cite{Ketal03}, where they show
after systematic studies that there is a monolayer or submonolayer
of water molecules hydrogen bonded to the Si-OH silanol groups all
over the SiO$_2$ surface, which cannot be removed by pumping in
vacuum over extended periods of time at room temperature. These
molecules can store charge when a voltage is applied between the CNT
and the underlying p-doped Si backgate, and are the main cause of
hysteresis in I-V cycles, much more than the underlying oxide
trapped charges within the SiO$_2$ substrate. These bulk charges
were invoked incorrectly in \cite{FKDB02,RFTJ02} as the main cause
of hysteresis.

Returning to the surface water molecules, they manage in
\cite{Ketal03} to estimate the charging per H$_2$O to be roughly
$0.1|e|$ per molecule.

They also suggest ways to get rid of the molecules by combining i)
thermal annealing in a dry atmosphere with simultaneous passivation
of the surface with PMMA, which links strongly to the silanol groups
and is also hydrophobic, ii) the permanent immersion of the
passivated surfaces in dry oxygen atmosphere. In \cite{Fetal06} they
also point out that a PMMA coating of the oxide surface changes
locally the dielectric constant (it is 3.3-3.9 at 60 KHz, and
2.2-3.2 at 1 MHz, a bit smaller than the one of the oxide, 3.9).

In \cite{Fetal06} they say that logically the regions in the oxide
which will be first charged will be those subject to the highest
fields, which in the case of nanotubes will be close to the
nanotube. They also estimate that trapped charges close to the CNT
are expected to produce potential fluctuations on the length scale
of the screening length
$\lambda_{scr}\sim\sqrt{\frac{\e_{CNT}}{\e_{ox}}t_{CNT}t_{ox}}\sim40$
nm ($\e_{CNT}\sim20$, $t_{CNT}\sim3$ nm is the diameter of the CNT
and $t_{ox}=100$ nm is the oxide thickness in their case). This
formula for estimating the screening length is derived in
\cite{YOL92}.

\section{Screening of the potentials at the graphene sheet and Si gate}\label{metallic_layers}
The equations for the selfconsistent potentials $v_{scr}(z,{\bf
\vec{r}}-{\bf \vec{r}}',\om)$ as a function of the bare potentials
$v^j_0(z,{\bf \vec{r}}-{\bf \vec{r}}',\om)$ are given by

\begin{eqnarray}
   v_{scr}(d,{\bf \vec{r}}-{\bf \vec{r}}',\om) &=& v^C_0(d,{\bf \vec{r}}-{\bf \vec{r}}',\om)+v^G_0(d,{\bf \vec{r}}-{\bf
  \vec{r}}',\om)\\ \nonumber&+&\int_Cd{\bf \vec{r}}_1\int_Cd{\bf \vec{r}}_2
   v_{\text{Coul}}(d,{\bf \vec{r}}-{\bf \vec{r}}_1,\om)\cg({\bf \vec{r}}_1-{\bf \vec{r}}_2,\om)v_{scr}(d,{\bf \vec{r}}_2-{\bf \vec{r}}',\om)\\
  \nonumber &+& \int_Gd{\bf \vec{r}}_3\int_Gd{\bf \vec{r}}_4
   v_{\text{Coul}}(d,{\bf \vec{r}}-{\bf \vec{r}}_3,\om)\csi({\bf \vec{r}}_3-{\bf \vec{r}}_4,\om)v_{scr}(0,{\bf \vec{r}}_4-{\bf \vec{r}}',\om) \\
   \nonumber v_{scr}(0,{\bf \vec{r}}-{\bf \vec{r}}',\om) &=& v^G_0(0,{\bf \vec{r}}-{\bf \vec{r}}',\om)+v^C_0(0,{\bf \vec{r}}-{\bf
  \vec{r}}',\om)\\ \nonumber&+& \int_Gd{\bf \vec{r}}_1\int_Gd{\bf \vec{r}}_2
   v_{\text{Coul}}(0,{\bf \vec{r}}-{\bf \vec{r}}_1,\om)\csi({\bf \vec{r}}_1-{\bf \vec{r}}_2,\om)v_{scr}(0,{\bf \vec{r}}_2-{\bf \vec{r}}',\om)\\
   \nonumber&+& \int_Cd{\bf \vec{r}}_3\int_Cd{\bf \vec{r}}_4
   v_{\text{Coul}}(0,{\bf \vec{r}}-{\bf \vec{r}}_3,\om)\cg({\bf \vec{r}}_3-{\bf \vec{r}}_4,\om)v_{scr}(d,{\bf \vec{r}}_4-{\bf
   \vec{r}}',\om)\,\,,
\end{eqnarray}
where for example in the first equation $v^C_0(d,{\bf \vec{r}}-{\bf
\vec{r}}',\om)$ represents the bare potential experienced by a point
charge $e$ in the graphene layer due to the presence of charges in
that same layer, while $v^G_0(d,{\bf \vec{r}}-{\bf \vec{r}}',\om)$
is the bare potential experienced by a point charge $e$ in the
graphene layer due to the presence of charges in the Si plane.
$v_{\text{Coul}}$ is the two-dimensional bare Coulomb potential.
These equations simplify considerably in the ${\bf \vec{q}}$ space,
assuming 2D translational invariance:

\begin{equation}\label{scr1}
    \left\{
      \begin{array}{l}
         v_{scr}(d,{\bf \vec{q}},\om)=v^C_0(d,{\bf \vec{q}},\om)e^{qd}+ v^G_0(d,{\bf \vec{q}},\om)+\\
         \vphantom{consruc}\\
         \phantom{xxxxxxxxxxxxxxxx}+v_q\cgq v_{scr}(d,{\bf \vec{q}},\om)+  v_qe^{-qd}\csiq v_{scr}(0,{\bf \vec{q}},\om)\\
\\
         v_{scr}(0,{\bf \vec{q}},\om)=v^C_0(0,{\bf \vec{q}},\om)+ v^G_0(d,{\bf \vec{q}},\om)e^{qd}+\\
         \vphantom{consruc}\\
         \phantom{xxxxxxxxxxxxxxxx}+v_qe^{-qd}\cgq v_{scr}(d,{\bf \vec{q}},\om)+  v_q\csiq v_{scr}(0,{\bf \vec{q}},\om)
      \end{array}
    \right.
\end{equation}
, where $v_q=2\pi e^2/|{\bf \vec{q}}|$ is the Fourier transform of
the Coulomb potential in two dimensions, and where $v^G_0(0,{\bf
\vec{q}},\om)$ and $v^C_0(d,{\bf \vec{q}},\om)$ have been expressed
in terms of $v^G_0(d,{\bf \vec{q}},\om)$ and $v^C_0(0,{\bf
\vec{q}},\om)$. Now we can calculate $v_{scr}(d,{\bf \vec{q}},\om)$
and $v_{scr}(0,{\bf \vec{q}},\om)$ in terms of the rest of the
variables,
\begin{equation}
  \left(
      \begin{array}{c}
         v_{scr}(d)\\
        v_{scr}(0) \\
      \end{array}
    \right)= \left(
              \begin{array}{cc}
                1-v_q\cg & -v_qe^{-qd}\csi \\
                -v_qe^{-qd}\cg & 1-v_q\csi \\
              \end{array}
            \right)^{-1}\times \left(
                             \begin{array}{cc}
                                e^{qd} & 1 \\
                               1 & e^{qd} \\
                               \end{array}
                           \right)\left(
                     \begin{array}{c}
                       v_0^C(0) \\
                       v_0^G(d) \\
                     \end{array}
                   \right) \\
\end{equation}
The dependence on ${\bf \vec{q}}$ and $\om$ has been omitted for the
sake of clarity. Now, if we are interested only in the long
wavelenght limit $v_q\cg,v_q\csi\gg 1$, the last equation simplifies
to
\begin{eqnarray}\label{vscr2}
  \left(
      \begin{array}{c}
         v_{scr}(d)\\
        v_{scr}(0) \\
      \end{array}
    \right) &=& \frac{1}{v_q^2\cg\csi\Bigl(1-e^{-2qd}\Bigr)}\times \\
  \nonumber &\times& \left(
    \begin{array}{cc}   v_q\Bigl(\cg e^{-qd}-\csi e^{qd}\Bigr) &
      v_q\Bigl(-\csi+\cg\Bigr) \\  v_q\Bigl(-\cg+\csi\Bigr) &
      v_q\Bigl(\csi e^{-qd}-\cg e^{qd}\Bigr) \\   \end{array}  \right)\left(
                     \begin{array}{c}
                       v_0^C(0) \\
                       v_0^G(d) \\
                     \end{array}
                   \right)
\end{eqnarray}

\subsubsection{Values of $v_0^C(0,{\bf \vec{q}},\om)$ and $v_0^G(d,{\bf \vec{q}},\om)$}
Now we will calculate the parts of these terms which will give rise
to a coupling to the vibration. When the graphene layer is set into
motion with a bending mode of wavevector ${\bf \vec{q}}$ and
amplitude $A_{\bf \vec{q}}$, the potential $v_0^C(0,{\bf
\vec{r}},t)$ of a point charge $e$ in the Si plane due to the charge
in the graphene layer is

\begin{eqnarray}\label{voz}
    \nonumber v_0^C(0,{\bf \vec{r}},t)&=& \frac{1}{2}\int_Cd{\bf \vec{r}}'v_{\text{Coul}}({\bf \vec{r}}-{\bf \vec{r}}',z')\rho({\bf
 \vec{r}}',z',t)\\
 \nonumber &=&\frac{1}{2}\int_Cd{\bf \vec{r}}'\frac{2\pi e^2\rho_0}{\sqrt{({\bf \vec{r}}-{\bf \vec{r}}')^2 +(d+A_{\bf \vec{q}}e^{i({\bf \vec{q}}
 {\bf \vec{r}}'-\wq t)})^2}}\\ \nonumber&\approx&\frac{1}{2}\int_Cd{\bf \vec{r}}'\frac{2\pi e^2\rho_0}{\sqrt{({\bf \vec{r}}-{\bf
    \vec{r}}')^2+d^2}}+\frac{1}{2}\int_Cd{\bf \vec{r}}'\frac{2\pi e^2\rho_0A_{\bf \vec{q}}e^{i({\bf \vec{q}}{\bf \vec{r}}'-\wq t)}d}
    {\Bigl(({\bf \vec{r}}-{\bf \vec{r}}')^2+d^2\Bigr)^{3/2}}\\ &\approx&
    f({\bf \vec{r}})+\pi e^2\rho_0A_{\bf \vec{q}}e^{-dq}e^{i({\bf \vec{q}}{\bf \vec{r}}-\wq t)}
\end{eqnarray}
where in the third line an expansion for small $A_{\bf \vec{q}}$ has
been performed. The Fourier transform for $\om\neq0$ is
\begin{equation}\label{voz2}
    v_0^C(0,{\bf \vec{k}},\om')=\pi e^2\rho_0A_{\bf
    \vec{q}}e^{-dq}\delta({\bf \vec{k}}-{\bf
    \vec{q}})\delta(\om'-\wq)\,\,\,\,,\,\,\,|{\bf \vec{q}}|=1/L
\end{equation}
Similarly, the potential of a point charge in the oscillating
graphene sheet due to the charge in the Si plane $v_0^G(d)$, is

\begin{equation}\label{voz3b}
    v_0^G(d,{\bf \vec{r}},t)= \frac{1}{2}\int_Gd{\bf \vec{r}}'\frac{2\pi e^2\rho_0}
 {\sqrt{({\bf \vec{r}}-{\bf \vec{r}}')^2 +(d+A_{\bf \vec{q}}e^{i({\bf \vec{q}}{\bf \vec{r}}-\wq t)})^2}}\approx
  f({\bf \vec{r}})+\pi e^2\rho_0A_{\bf \vec{q}}e^{i({\bf \vec{q}}{\bf \vec{r}}-\wq t)}
\end{equation}
leading to the same expression as eq.(\ref{voz2}) but without the
factor $e^{-qd}$
\begin{equation}\label{voz4}
    v_0^G(d,{\bf \vec{k}},\om')=v_0^C(0,{\bf \vec{k}},\om')e^{qd}
\end{equation}
Substituting (\ref{voz2},\ref{voz4}) in eq.(\ref{vscr2}), one
obtains eq.(\ref{vscr0}).

\section{Dissipation due to two-level systems in graphene
resonators}\label{apTLSgraphene}
\subsection{Model and vibrating modes of a 2D sheet}
The hamiltonian describing the coupling of the effective TLS's and
the oscillating graphene sheet is given by \cite{SGN07}
\begin{equation}\label{apHam_TLS}
    H=\e\,\sigma_x+\gamma\frac{\Dox}{\e}\sigma_z\sum_{\textbf{k}}\lambda_{\textbf{k}}(b_{\textbf{k}}+b_{\textbf{-k}}^{\dagger})+
\sum_{\textbf{k}}\hbar\om_{\textbf{k}}b_{\textbf{k}}^{\dagger}b_{\textbf{k}}
\end{equation}
where $\e=\sqrt{(\Dox)^2+(\Doz)^2}$, $b_{\textbf{k}}^{\dagger}$
represent the phonon creation operators associated to the different
vibrational modes of a sheet, and
$\sum_{\textbf{k}}\lambda_{\textbf{k}}(b_{\textbf{k}}+b_{\textbf{-k}}^{\dagger})$
represents the coupling to the strain tensor $u_{ik}$. There are two
types, compression modes (longitudinal waves) and bending modes. The
equations governing the former are \cite{LL59}
\begin{equation}\label{compression_sheet}
    \left\{
      \begin{array}{l}
        \frac{\rho}{E}\frac{\partial^2 u_x}{\partial t^2}=\frac{1}{1-\nu^2}\frac{\partial^2 u_x}{\partial x^2}+\frac{1}{2(1+\nu)}
\frac{\partial^2 u_x}{\partial y^2}+\frac{1}{2(1-\nu)}\frac{\partial^2 u_y}{\partial x\partial y} \\
        \frac{\rho}{E}\frac{\partial^2 u_y}{\partial t^2}=\frac{1}{1-\nu^2}\frac{\partial^2 u_y}{\partial y^2}+\frac{1}{2(1+\nu)}
 \frac{\partial^2 u_y}{\partial  x^2}+\frac{1}{2(1-\nu)}\frac{\partial^2 u_x}{\partial x\partial y}
      \end{array}
    \right.
\end{equation}
Choosing a solution $e^{ik_xx-\om t}$ the equations simplify to wave
equations:
\begin{equation}\label{compression_sheet2}
  \frac{\partial^2 u_x}{\partial t^2}-\frac{E}{\rho(1-\nu^2)}\frac{\partial^2 u_x}{\partial x^2}=0\,\,\,\,,\,\,\,\frac{\partial^2 u_y}{\partial
t^2}-\frac{E}{2\rho(1+\nu)}\frac{\partial^2 u_y}{\partial x^2}=0
\end{equation}
so that the speed of the longitudinal ($v_l=\sqrt{E/\rho(1-\nu^2)}$)
and transversal ($v_t=\sqrt{E/2\rho(1+\nu)}$) waves are different.
Therefore we will have two types of compression modes, longitudinal
and transversal, each parametrized with a vector $\textbf{k}$ in the
plane. The expression for the local deformations associated to the
longitudinal component in second quantization is
\begin{equation}\label{compression_sheet3}
  u^l(0)=\sum_{\textbf{k}}\sqrt{\frac{\hbar}{2\rho twL\om_{\textbf{k}}}}(b_{\textbf{k}}^l+b_{\textbf{-k}}^{\dagger,l})\,\,\,\,,\,\,\,\partial u^l
 =\sqrt{\frac{\hbar}{2\rho twL}}\sum_{\textbf{k}}\frac{k^{l}}{\sqrt{\om_{\textbf{k}}}}(b_{\textbf{k}}^l+b_{\textbf{-k}}^{\dagger,l})
\end{equation}
and similarly for the transversal one. They give rise to an
interaction hamiltonian
\begin{eqnarray}\label{compression_sheet4}
  \nonumber H_{int} &=& \hbar\sigma_z\sum_{j=x,y}\sum_{i=l,t}\sum_{k}\Bigl[\gamma\frac{\Dox}{\e}\sqrt{\frac{1}{2\hbar\rho twL}}\frac{k^i_j}
  {\sqrt{\om^i_{k_j}}}\Bigr](a_{-k_j}^{\dagger,i}+a_{k_j}^i) \\
   &=& \hbar\sigma_z\sum_{j=x,y}\sum_{i=l,t}\sum_{k}\gamma_{eff}\frac{k^i_j}{\sqrt{\om^i_{k_j}}}(a_{-k_j}^{\dagger,i}+a_{k_j}^i)
\end{eqnarray}
which has an associated \textbf{superohmic} spectral function
\begin{equation}\label{compression_sheet5}
  J(\om)=\sum_{j=x,y}\sum_{i=l,t}\sum_{k}\gamma_{eff}^2\frac{(k^i_j)^2}{\om^i_{k_j}}\delta(\om-\om^i_{k_j})=\alpha_c\,\om^2\,\,,
\end{equation}
with $\alpha_c$ given by
\begin{equation}\label{compression_sheet6}
  \alpha_c=\frac{Lw}{(2\pi)^2}\gamma_{eff}^2\Bigl[\frac{1}{v_l^4}+\frac{1}{v_t^4}\Bigr]=\frac{1}{(2\pi)^2}\Bigl(\gamma\frac{\Dox}{\e}\Bigr)^2
 \frac{1}{2\hbar\rho t}\Bigl[\Bigl(\frac{\rho(1-\nu^2)}{E}\Bigr)^2+\Bigl(\frac{2\rho(1+\nu)}{E}\Bigr)^2\Bigr]\,\,.
\end{equation}
In the case of the bending modes the governing equation is
\begin{equation}\label{bending_sheet}
        \rho tw\frac{\partial^2 z}{\partial t^2}+\frac{Et^2}{12(1-\nu^2)}\triangle^2z=0
\end{equation}
so there is a non-linear dispersion relation,
$\om=k^2\sqrt{Et^2/12\rho(1-\nu^2)}$. To obtain the approximate
relation between the coordinate $z$ which will have the bosonic
operators associated, and the strain tensor $u_{ik}$ appearing in
the interaction part of the hamiltonian, we equate two expressions
for the elastic free energy,
\begin{eqnarray}\label{bending_sheet2}
  \nonumber F_{sheet}&=& \frac{Et^3}{24(1-\nu^2)}\int dxdy\Bigl[\Bigl(\frac{\partial^2 z}{\partial x^2}+\frac{\partial^2 z}{\partial y^2}\Bigr)^2+
 2(1-\nu)\Bigl\{\Bigl(\frac{\partial^2 z}{\partial x\partial y}\Bigr)^2-\frac{\partial^2 z}{\partial x^2}\frac{\partial^2 z}{\partial y^2}\Bigr\}\Bigr]\\
   &=& \frac{Et}{2(1+\nu)}\int dxdy\Bigl[u_{ik}^2+\frac{\nu}{1-2\nu}u_{ll}\Bigr]
\end{eqnarray}
and obtain an average relation
\begin{equation}\label{bending_sheet3}
    \langle u_{ij}\rangle\approx t\sqrt{\frac{1}{3(1-\nu)(9+\frac{3\nu}{1-2\nu})}}\langle\frac{\partial^2 z}{\partial x^2}\rangle\,\,\,.
\end{equation}
With this information $H_{int}$ reads
\begin{equation}\label{bending_sheet4}
    H_{int}=\hbar\sigma_z\sum_{j=x,y}\sum_{l,m=1,2,3}\sum_{k}\Bigl[\gamma\frac{\Dox}{\e}\frac{t}{\sqrt{3(1-\nu)(9+\frac{3\nu}{1-2\nu})}}
 \frac{1}{\sqrt{2\hbar twL\rho}}\frac{(k^{lm}_j)^2}{\sqrt{\om^{lm}_{k_j}}}\Bigr](a_{-k_j}^{\dagger,lm}+a_{k_j}^{lm})\,\,.
\end{equation}
Proceeding as with the previous modes, the spectral function
associated turns out to be \textbf{ohmic}:
\begin{equation}\label{bending_sheet5}
    J(\om)=\alpha\,\om\,\,\,\,\,\,\,\,\,,\,\,\,\,\,\,\,\,\,\alpha=\frac{9\sqrt{3}}{\pi}\Bigl(\gamma\frac{\Dox}{\e}\Bigr)^2\frac{\rho^{1/2}(1+\nu)^{3/2}
 (1-\nu)^{1/2}}{\hbar t^2E^{3/2}(9+\frac{3\nu}{1-2\nu})}
\end{equation}
We will restrict ourselves to the study of dissipation caused by the
ohmic bath of bending modes, which will prevail at low temperatures
over the superohmic bath of compression modes.

\subsection{Losses due to the ohmic bath. Temperature dependence} An estimate is given
of the properties of the bath, namely $\alpha$, $\winf$ and $\wco$,
for two cases, the one of a single layer graphene sheet (whose
thickness we take to be $t\sim3$ {\AA}), and a stack of layers 10 nm
thick, using the value $\gamma\sim1$ eV, in table
(\ref{Parameters_ohmicbath}). This value of $\gamma$ is a big
overestimate, since it corresponds to the value of the coupling of
TLS's embedded in a vibrating amorphous structure to the
oscillations, and as stated previously this is not our case, and the
coupling will be reduced by some power of the factor $( a / d ) \sim
10^{-4}$. We thus conclude that $\alpha$ is extremely small, and
there is no appreciable renormalization of the tunneling amplitude,
$\Dr=\Dox\cdot(\Dox/\wco)^{\alpha/(1-\alpha)}\simeq\Dox$
\cite{Letal87}.
\begin{table}
  \centering
\begin{tabular}{|c|c|c|}
  \hline
   & 1 layer & $t=10$ nm \\\hline
  $\winf$ & $7\cdot10^{7}$ & $2.5\cdot10^{9}$ \\
  $\wco$ & $8\cdot10^{14}$ & $3\cdot10^{13}$ \\
  $\alpha$ & $\sim0.03$ & $\sim3\cdot10^{-5}$ \\
  \hline
\end{tabular}
  \caption{Estimate of the parameters of the bath, for $\gamma\sim1$ eV.}\label{Parameters_ohmicbath}
\end{table}
In ref.\cite{W99} an extensive analysis of the correlation function
$C(t)=\langle\sigma_z(t)\sigma_z(0)\rangle$ of a TLS in presence of
an ohmic bath is carried out, and for the cases we consider
($\alpha\ll1$, biased TLS) the following results are obtained for
the quasielastic (relaxational) peak around $\om=0$:\\
\begin{enumerate}
   \item For $\e<\pi\alpha kT$:
\begin{equation}\label{relax1}
    C_{rel}(\om)=\frac{2\gamma_r}{\om^2+\gamma_r^2}\,\,\,\,,\,\,\,\,\gamma_r=\frac{\hbar\Dr^2}{2\pi\alpha kT}\frac{1}
   {1+(\Doz/2\pi\alpha kT)^2}\approx\frac{\hbar\Dr^2}{2\pi\alpha kT}
\end{equation}
  \item For $\pi\alpha kT<\e\leq kT$:
\begin{equation}\label{relax2}
    C_{rel}(\om)=2\Bigl(\frac{(\Doz)^2}{\e^2}-\tanh^2\Bigl(\frac{\e}{2kT}\Bigr)\Bigr)\frac{\gamma_r}{\om^2+\gamma_r^2}\approx
2\frac{(\Doz)^2}{\e^2}\frac{\gamma_r}{\om^2+\gamma_r^2}\,\,\,\,,
\end{equation}
with
\begin{equation}\label{relax1b}
    \gamma_r=2\pi\alpha\frac{kT}{\hbar}\frac{\Dr^2}{(\Doz)^2+\Dr^2}
\end{equation}
  \item For $\e\geq kT$:
\begin{equation}\label{relax3}
    C_{rel}(\om)=2\frac{(\Doz)^2/\e^2}{\cosh^2\Bigl(\frac{\e}{2kT}\Bigr)}\frac{\gamma_r}{\om^2+\gamma_r^2}\approx
4e^{-\e/kT}(\Doz)^2/\e^2\frac{\gamma_r}{\om^2+\gamma_r^2}\,\,\,\,,
\end{equation}
with
\begin{equation}\label{relax1b}
    \gamma_r=\pi\alpha\coth\Bigl(\frac{\e}{2kT}\Bigr)\Dr^2/\hbar\e\approx\pi\alpha\Dr^2/\hbar\e
\end{equation}
\end{enumerate}
These expressions for $C_{rel}(\om)$ substitute the one found
usually in the context of sound attenuation in amorphous solids due
to 3D acoustic modes,
$C_{rel}^{3D}(\om)=\frac{(\Doz)^2/\e^2}{\cosh^2(\e/2kT)}\frac{\Gamma}{\om^2+\Gamma^2}$,
and they will enter into the expression for the dissipation as
follows:
\begin{equation}\label{Qgeneral}
    Q^{-1}(\om,T)=\frac{P\gamma^2}{EkT}\int_0^{\e_{max}} d\e\int_{u_{min}}^{1}
du\frac{1}{u\sqrt{1-u^2}}\om C(\om)
\end{equation}
where $u=\Dr/\e$, $\e_{max}\sim5$ K, and $(u\sqrt{1-u^2})^{-1}$
comes from the
probability density of TLS's in an amorphous solid, like SiO$_2$.\\
When $kT>\e_{max}$ there will be TLS's only in the regimes of eqs.
(\ref{relax1}) and (\ref{relax2}), and moreover, thanks to the very
low value of $\alpha$ and the range of $\om$ typical of experiments
we will have $\gamma_r<\om$ for all values of $u$, so that
$\gamma_r/(\om^2+\gamma_r^2)\approx\gamma_r/\om^2$, and
\begin{eqnarray}\label{QrelaxhighTgraphene}
  \nonumber Q^{-1}(\om,T)  &=& \frac{P\gamma^2}{EkT}\Bigl\{\int_0^{\pi\alpha kT} d\e\int_{u_{min}}^{1}
 du\frac{1}{u\sqrt{1-u^2}}\frac{2\e^2u^2}{2\pi\hbar\om\alpha kT}\\
 \nonumber&&\phantom{XXXXX}+\int_{\pi\alpha kT}^{\e_{max}} d\e\int_{u_{min}}^{1}
 du\frac{1}{u\sqrt{1-u^2}}2(1-u^2)\frac{2\pi\alpha kT
u^2}{\hbar\om}\Bigr\} \\
   &\approx& \frac{P\gamma^2}{E\hbar\om}\Bigl\{\frac{4\pi}{3}\alpha\e_{max}+\frac{\pi^2}{3}\alpha^2kT\Bigr\}
\end{eqnarray}
So for high temperatures it will give a constant contribution to
dissipation, plus a weak (due to the prefactor $\alpha^2$) linear T dependence.\\
In the case of lower temperatures, $kT<\e_{max}$, there are some
TLS's in the regime corresponding to eq. (\ref{relax3}), and again
one can approximate
$\gamma_r/(\om^2+\gamma_r^2)\approx\gamma_r/\om^2$ thanks to the low
$\alpha$, leading to
\begin{eqnarray}\label{QrelaxlowTgraphene}
  \nonumber Q^{-1}(\om,T)  &=& \frac{P\gamma^2}{EkT}\Bigl\{\int_0^{\pi\alpha kT} d\e\int_{u_{min}}^{1}
 du\frac{1}{u\sqrt{1-u^2}}\frac{2\e^2u^2}{2\pi\hbar\om\alpha kT}\\
 \nonumber&&\phantom{XX}+\int_{\pi\alpha kT}^{kT} d\e\int_{u_{min}}^{1}
 du\frac{1}{u\sqrt{1-u^2}}2(1-u^2)\frac{2\pi\alpha kT u^2}{\hbar\om} \\
 \nonumber &&\phantom{XX}+ \int_{kT}^{\e_{max}} d\e\int_{u_{min}}^{1}du\frac{1}{u\sqrt{1-u^2}}4(1-u^2)e^{-\e/kT}\frac{\pi\alpha u^2\e}{\hbar\om}\Bigr\}\\
   \nonumber&\approx& \frac{P\gamma^2}{E\hbar\om}\Bigl\{\frac{4\pi}{3}\alpha kT+\frac{\pi^2}{3}\alpha^2kT+\frac{8\pi}{\sqrt{e}}\alpha kT\Bigr\}\\
   &\approx&\frac{P\gamma^2\alpha}{E\hbar\om}\Bigl\{\frac{4\pi}{3}+\frac{8\pi}{\sqrt{e}}\Bigr\}kT
\end{eqnarray}
Thus, at low temperatures the contribution from TLS's to attenuation
is linear in T. These tendencies come basically from the ohmic
character of the vibrational bath and the probability density
function of the TLS's. In case the TLS's resided in a crystalline
material, or for some other reason their distribution function
differed from the one of amorphous solids, changes in the behavior
of $Q^{-1}(T,\om)$ are to be expected.


\newpage
\cleardoublepage

\chapter{Appendix to Chapter \ref{ch3}}

\section{Dipole matrix element from single-particle mean-field states}\label{apDipole}
The derivation given in this section is borrowed from \cite{W06}. We
want to calculate the dipole matrix element $d_{\alpha\beta}$
defined in (\ref{dab})
\begin{equation}\label{dab2}
    d_{\alpha\beta}=\langle\alpha|\Bigl[z\Theta(a-r)+\frac{za^3}{r^3}\Theta(r-a)\Bigr]|\beta\rangle
\end{equation}
For that sake we need a description of the eigenstates
$|\alpha\rangle$ of the mean-field Hamiltonian (\ref{Hrelative3}).
The TDLDA numerical calculations briefly presented in Sec.
\ref{secMeanfield}inform us on the shape of the self-consistent mean
field potential $V (r)$, as shown in Fig. \ref{TDLDA}. It suggests
that for analytical calculations, $V (r)$ can be approximated by a
spherical well of radius $a$ and finite height $V_0$,
$V(r)=V_0\Theta(r-a)$, where $V_0=\e_F+W$, $\e_F$ and $W$ being the
Fermi energy and the work function of the considered nanoparticle,
respectively.

Because of the spherical symmetry of the problem, the one-particle
wave functions
\begin{equation}\label{1pwavef}
    \psi_{\e lm}(\textbf{r})=\frac{u_{\e l}(r)}{r}Y_l^m(\theta,\phi)
\end{equation}
decompose into radial and angular parts given by the spherical
harmonics $Y_l^m(\theta,\phi)$, where $l$ and $m$ are the angular
momentum quantum numbers. The radial wave functions $u_{\e l}(r)$
satisfy the reduced Schrödinger equation
\begin{equation}\label{Schrodinger}
    \Bigl[-\frac{\hbar^2}{2m_e}\frac{d^2}{dr^2}+\frac{\hbar^2l(l+1)}{2m_er^2}+V(r)\Bigr]u_{\e
    l}(r)=\e u_{\e l}(r)
\end{equation}
with the conditions $u_{\e l}(0)=0$ and
lim$_{r\rightarrow\infty}[u_{\e l}(r)/r]=0$. This yields the
single-particle eigenenergies $\e$ in the mean-field potential $V
(r)$. Thus, the matrix elements $d_{\alpha\beta}$ can be separated
into an angular part ${\cal A}_{l_\alpha l_\beta}^{m_\alpha m_\beta}
$ and a radial part ${\cal R}_{l_\alpha l_\beta}(\varepsilon_\alpha
, \varepsilon_\beta)$,
\begin{equation}
\label{d_ph2} d_{ph} = {\cal A}_{l_\alpha l_\beta}^{m_\alpha
m_\beta} {\cal R}_{l_\alpha l_\beta}(\varepsilon_\alpha ,
\varepsilon_\beta)\,.
\end{equation}
With (\ref{dab2}), we have
\begin{equation}
\label{d_ph3} {\cal A}_{l_\alpha l_\beta}^{m_\alpha m_\beta}
=\int_0^\pi d\theta\sin\theta\int_0^{2\pi}d\phi
Y_{l_\alpha}^{m_\alpha}\,^{*}(\theta,\phi)\cos\theta
Y_{l_\beta}^{m_\beta}(\theta,\phi)
\end{equation}
which can be expressed in terms of Wigner-3j symbols \cite{E60} as
\begin{equation}\label{sp_angular2}
    {\cal A}_{l_\alpha l_\beta}^{m_\alpha m_\beta} =(-1)^{m_\alpha} \sqrt{(2l_\alpha+1)(2l_\beta+1)}\times
\begin{pmatrix}
l_\alpha & l_\beta & 1 \\
0 & 0 & 0
\end{pmatrix}
\begin{pmatrix}
l_\alpha & l_\beta & 1 \\
-m_\alpha & m_\beta & 0
\end{pmatrix}\,.
\end{equation}
The Wigner-3j symbols contain the dipole selection rules
$l_\alpha=l_\beta\pm 1$ and $m_\alpha=m_\beta$.

Concerning ${\cal R}_{l_\alpha l_\beta}(\varepsilon_\alpha ,
\varepsilon_\beta)$, in the limit of strong electronic confinement
$V_0\gg\e_F$, the single-particle wave function vanishes outside the
nanoparticle, enabling us to take the following approximation
\begin{equation}\label{sp_radial1}
    {\cal R}_{l_\alpha l_\beta}(\varepsilon_\alpha
    ,\varepsilon_\beta)\simeq\int_0^a dru_{\varepsilon_\alpha l_\alpha}^{*}(r)ru_{\varepsilon_\beta l_\beta}(r)
\end{equation}
The commutator of $r$ with the self-consistent Hamiltonian
$H_{sc}=\textbf{p}^2/2m_e+V(r)$ reads
\begin{equation}\label{commutatorr}
    [r,H_{sc}]=\frac{i\hbar}{m_e}p_r\,,
\end{equation}
where $p_r$ is the conjugated momentum to the variable $r$.
Calculating the matrix element of this commutator between two
single-particle eigenstates $|\alpha\rangle$ and $|\beta\rangle$ of
$H_{sc}$ then allows to write
\begin{equation}\label{commutatorr2}
    \langle\alpha|r|\beta\rangle=\frac{i\hbar}{m_e(\varepsilon_\beta-\varepsilon_\alpha)}\langle\alpha|p_r|\beta\rangle\,.
\end{equation}
Now we use $[p_r,H_{sc}]=-i\hbar\,dV/dr$ to obtain
\begin{equation}\label{commutatorr3}
    \langle\alpha|p_r|\beta\rangle=-\frac{i\hbar}{\varepsilon_\beta-\varepsilon_\alpha}\langle\alpha|\frac{dV}{dr}|\beta\rangle\,,
\end{equation}
and substituting in (\ref{commutatorr2}) one gets
\begin{equation}\label{commutatorr4}
    \langle\alpha|r|\beta\rangle=\frac{\hbar^2}{m_e(\varepsilon_\beta-\varepsilon_\alpha)^2}\langle\alpha|\frac{dV}{dr}|\beta\rangle\,,
\end{equation}
so that the radial part of the dipole matrix element can be
reexpressed as a function of $dV/dr$ as
\begin{equation}\label{sp_radial2}
    {\cal R}_{l_\alpha l_\beta}(\varepsilon_\alpha
    ,\varepsilon_\beta)\simeq\frac{\hbar^2}{m_e(\varepsilon_\beta-\varepsilon_\alpha)^2}\int_0^a dru_{\varepsilon_\alpha l_\alpha}^{*}(r)\frac{dV}{dr}u_{\varepsilon_\beta
    l_\beta}(r)\,,
\end{equation}
which for an approximate step potential such as ours,
$V(r)=V_0\Theta(r-a)$, leads to
\begin{equation}\label{sp_radial3}
    {\cal R}_{l_\alpha l_\beta}(\varepsilon_\alpha
    ,\varepsilon_\beta)\simeq\frac{\hbar^2}{m_e(\varepsilon_\beta-\varepsilon_\alpha)^2}V_0u_{\varepsilon_\alpha l_\alpha}^{*}(a)u_{\varepsilon_\beta
    l_\beta}(a)\,.
\end{equation}
For $V_0\rightarrow\infty$ and $r\leq a$, the regular solutions of
(\ref{Schrodinger}) satisfying $u_{\e l}(a)=0$ are given by
\begin{equation}\label{bessel1}
    u_{\e l}(r)=\frac{\sqrt{2}}{a^{3/2}}\frac{rj_l(kr)}{j_{l+1}(ka)}\,,
\end{equation}
where $j_\nu(z)$ are spherical Bessel functions of the first kind
\cite{AS70}, and $k=\sqrt{2m_e\e}/\hbar$. The condition $j_l(ka)=0$
yields the quantization of the single-particle states in the
infinitely deep spherical well. Inserting now the radial wave
function (\ref{bessel1}) into (\ref{sp_radial3}),
\begin{equation}\label{sp_radial4}
    {\cal R}_{l_\alpha l_\beta}(\varepsilon_\alpha
    ,\varepsilon_\beta)=\frac{\hbar^2}{m_ea(\varepsilon_\beta-\varepsilon_\alpha)^2}V_0\frac{j_{l_\alpha}(k_\alpha a)}{j_{l_\alpha+1}(k_\alpha a)}
    \frac{j_{l_\beta}(k_\beta a)}{j_{l_\beta+1}(k_\beta a)}\,.
\end{equation}
Now, for $r\rightarrow a^+$, $u_{\e l}(r)\rightarrow 0$ and
$V(r)=V_0\rightarrow\infty$, so that the radial Schrödinger equation
(\ref{Schrodinger}) takes the form
\begin{equation}\label{Schrodinger2}
    \Bigl[-\frac{\hbar^2}{2m_e}\frac{d^2}{dr^2}+V_0\Bigr]u_{\e
    l}(r)=0\,.
\end{equation}
With $k_0=\sqrt{2m_e V_0}/\hbar$, the physical solution of this
equation is
\begin{equation}\label{uel}
    u_{\e l}(r\rightarrow a^+)\sim e^{-k_0r}\,.
\end{equation}
Then
\begin{equation}\label{uel2}
    \frac{du_{\e l}}{dr}\Bigl |_{r\rightarrow a^+}= -k_0u_{\e l}(r\rightarrow a^+)\,,
\end{equation}
and imposing the continuity of the radial wave function and of its
derivative with respect to $r$ at $r=a$, we obtain
\begin{equation}\label{uel3}
    \frac{du_{\e l}}{dr}\Bigl |_{r\rightarrow a^-}= -k_0u_{\e l}(r\rightarrow a^-)\,.
\end{equation}
Therefore, using (\ref{bessel1})
\begin{equation}\label{bessel2}
    \lim_{V_0\rightarrow\infty}\Bigl[\frac{\sqrt{2m_eV_0}}{\hbar}j_l(ka)\Bigr]=-\frac{dj_l(kr)}{dr}\Bigr|_{r=a}=
    -k\frac{dj_l(\eta)}{d\eta}\Bigr|_{\eta=ka}\,.
\end{equation}
Using the recurrence relation \cite{AS70}
\begin{equation}\label{bessel3}
    \frac{l}{\eta}j_l(\eta)-\frac{dj_l}{d\eta}=j_{l+1}(\eta)
\end{equation}
and the fact that $j_l(\eta)=j_l(ka)=0$, we reach the final result
for the radial matrix element \cite{YB92}
\begin{equation}
\label{R_phap} {\cal R}_{l_\alpha l_\beta}(\varepsilon_\alpha ,
\varepsilon_\beta)=\frac{2\hbar^2}{m_{\rm e} a}
\frac{\sqrt{\varepsilon_\alpha
\varepsilon_\beta}}{(\varepsilon_\alpha -\varepsilon_\beta )^2}\,.
\end{equation}
Note that the dipole selection rule $l_\alpha=l_\beta\pm1$ appearing
in the angluar matrix element (\ref{sp_angular2}) implies that
$\varepsilon_\alpha\neq\varepsilon_\beta$, and thus (\ref{R_phap})
does not diverge. Furthermore, it decreases with increasing energy
$|\varepsilon_\alpha -\varepsilon_\beta|$ of the dipole transition
$|\alpha\rangle\rightarrow|\beta\rangle$.

\section{Lowest energy of the particle-hole spectrum}
\label{app_Emin} Using the large $ka$ expansion of (\ref{bessel1})
(semiclassical high-energy limit), the quantisation condition reads
\begin{equation}
\label{qcondition} ka=\pi\left(\frac{l}{2}+n\right)
\end{equation}
with $l$ and $n$ non-negative integers. The energy of a
single-particle (hole) state is related to its wavevector
$k_{p(h)}$, its total angular momentum $l_{p(h)}$, and its radial
quantum number $n_{p(h)}$ as
\begin{equation}\label{qcondition2}
    \varepsilon_{p(h)}=\frac{\hbar^2k_{p(h)}^2}{2m_{\rm e}}=\varepsilon_\textrm{F}\left(\frac{\pi}{k_{\rm F}a}\right)^2
\left(\frac{l_{p(h)}}{2}+n_{p(h)}\right)^2\,.
\end{equation}
Thus, the energy of a p-h excitation entering the RPA sum
\eqref{RPAsum_inter} is
\begin{equation}\label{Ediff}
    \Delta\varepsilon_{ph}=\varepsilon_\textrm{F}\left(\frac{\pi}{k_{\rm
F}a}\right)^2\left(\frac{l_p-l_h}{2}+n_p-n_h\right)\times
\left(\frac{l_p+l_h}{2}+n_p+n_h\right)\,.
\end{equation}
Notice that using the exact quantum mechanical spectrum in
\eqref{RPAsum_inter} would not change significantly the result
depicted in Figure \ref{sum}, since the approximation
\eqref{qcondition} is very reliable for states close to the Fermi
energy. We have also checked that generating the p-h excitation
energies randomly in the RPA sum \eqref{RPAsum} does not affect the
physical picture of Figure \ref{sum}. Indeed, the main ingredient to
understand such a picture is the fast decay of the dipole matrix
element with the p-h energy.

The expressions for the dipole matrix element \eqref{R_ph} as well
as for the typical dipole matrix element \eqref{D1} diverge in the
limit of a small p-h energy. It is therefore crucial for our
analysis to determine the appropriate minimal p-h energy
$\Delta\varepsilon_{\rm min}$ that renders this divergence
unphysical.

This can be achieved by imposing the dipole selection rules in
\eqref{Ediff} and that the energy difference is minimal. The first
condition dictates that $l_h=l_p±1$ and $m_h=m_p$. Therefore there
are two ways of obtaining the minimal energy difference: $n_p=n_h$
with $l_p=l_h+1$ and $n_p=n_h+1$ with $l_p=l_h-1$. In both cases we
have
\begin{equation}
\label{dleminGuillaume} \Delta\varepsilon_\textrm{min}\simeq
\frac{\varepsilon_{\rm F}}{k_{\rm F}a/\pi}\frac{k_h}{k_{\rm F}}\,.
\end{equation}
Since we are interested in states close to the Fermi level, we can
simplify \eqref{dleminGuillaume} to expression
\eqref{SpacingGuillaume}.

If we consider sodium clusters with $k_\textrm{F}a=30$
($a=\unit[3.3]{nm}$ and $N\simeq4000$ conduction electrons per spin
direction), we have
$\Delta\varepsilon_\textrm{min}\approx\varepsilon_{\rm F}/10$. This
is a much larger energy than the lowest one we can observe in the
numerically generated excitation spectrum (see Fig.~1 in
Ref.~\cite{MWJ02}). However, the two results are reconciled once we
take into account the large degeneracy yielded by our approximate
quantisation condition (\ref{qcondition}).

The degeneracy of p-h excitations with minimal energy is given by
twice the number of pairs ($l_h,n_h$) compatible with $k_h=k_{\rm
F}$ and $\Delta\varepsilon_{ph}=\Delta\varepsilon_{\rm min}$.
Indeed, we have seen that there are two possible particle states $p$
starting from $h$ and verifying the above-mentioned conditions. For
each $n$ between 1 and $k_{\rm F}a/\pi$, there is a value of
$l=2(k_{\rm F}a/\pi-n)$ and therefore the number of degenerate p-h
excitations with energy $\Delta\varepsilon_{\rm min}$ is
\begin{equation}\label{deg1}
    {\cal N}=2\sum_{n=0}^{k_{\rm F}a/\pi}(2l+1)=2\sum_{n=0}^{k_{\rm F}a/\pi}\left[4\left(\frac{k_{\rm
F}a}{\pi}-n\right)+1\right]\simeq 4\left(\frac{k_{\rm
F}a}{\pi}\right)^2\,.
\end{equation}
This degeneracy factor has to be included in Figure \ref{sum}, and
it is crucial for the determination of the collective excitation.

\section{Density of states at a fixed angular momentum $l$, $\rho_l(\varepsilon)$}
For our derivations we need to have an estimate for the $l$-fixed
density of states, which we will call $\rho_l(\varepsilon)$. It
corresponds to the number of eigenstates per energy unit whose
radial part $u_{kl}(r)$ satisfy (\ref{Schrodinger}) with the
conditions $u_{\e l}(0)=0$ and lim$_{r\rightarrow\infty}[u_{\e
l}(r)/r]=0$. It is important to notice that the variable $r$ is
limited to positive values and that the centrifugal potential
possesses a singularity at $r=0$. This prevents us from a naïve
application of the WKB approximation \cite{M70} to treat this radial
problem, as would be done for a standard 1D Schrödinger equation.
First we have to rewrite (\ref{Schrodinger}) in terms of a variable
where such a singularity is absent, and then we will apply WKB. The
change of variables is
\begin{equation}\label{changevar}
    x=\ln r,\,\,\,\,\,\,,\,\,\,\,\,\chi_{kl}(x)=e^{x/2}u_{kl}(r)\,,
\end{equation}
and (\ref{Schrodinger}) becomes non-singular
\begin{equation}\label{Schrodinger2}
    \Bigl[\frac{d^2}{dx^2}+\Pi_l^2(x)\Bigr]\chi_{kl}(x)=0\,,
\end{equation}
with
\begin{equation}\label{Schrodinger3}
    \Pi_l^2(x)=\frac{2m_e}{\hbar^2}e^{2x}[\varepsilon_{kl}V(e^x)]-\Bigl(l+\frac{1}{2}\Bigr)^2\,.
\end{equation}
Using now the WKB approximation for $\chi_{kl}(x)$ amounts to change
the centrifugal potential in (\ref{Schrodinger}) according to the
Langer modification \cite{L37,BM72}
\begin{equation}\label{Langermod}
    l(l+1)\Rightarrow \Bigl(l+\frac{1}{2}\Bigr)^2\,.
\end{equation}
Therefore the modified effective radial potential becomes
\begin{equation}\label{Radpotl}
    V_l^{eff}(r)=\frac{\hbar^2(l+1/2)^2}{2m_er^2}+V(r)\,,
\end{equation}
where in our case $V(r)$ is the spherical billiard potential
\begin{equation}\label{Radpotl2}
    V(r)=\left\{
           \begin{array}{ll}
             0, & r<a\,, \\
             \infty, & r\geq a\,,
           \end{array}
         \right.
\end{equation}
\begin{figure}[t]
\begin{center}
\includegraphics[width=13cm]{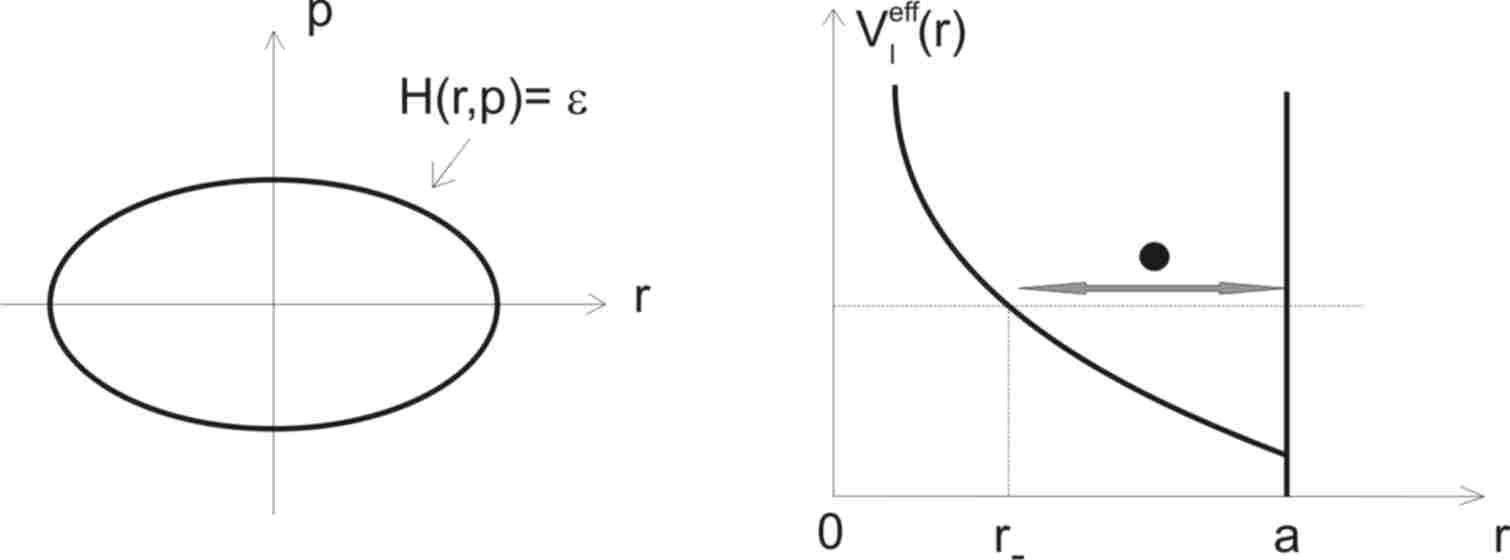}
\caption[Effective radial potential $V_l^{eff}$]{\label{Effpotential}%
Left: Phase-space trajectory of a periodic motion of energy
$\varepsilon$ in a confining potential. Right: Radial potential
$V_l^{eff}$ for fixed $l$, of the spherical cavity as a function of
the radial coordinate $r$.}
\end{center}
\end{figure}
A plot of $V_l^{eff}(r)$ is provided in fig.(\ref{Effpotential}). A
semiclassical theory can be developed to calculate
$\rho_l(\varepsilon)$, obtaining an expression whose first
corrections lead to an extra oscillatory component
$\rho_l^{osc}(\varepsilon)$ added to the Thomas-Fermi result for
$\rho_l(\varepsilon)$ \cite{WMWJ05,W06}, $\rho_l^0(\varepsilon)$:
\begin{equation}\label{rhothomas}
   \rho_l^0(\varepsilon)=\frac{1}{h}\int dp\,dr\delta(\varepsilon-H_l(r,p))
\end{equation}
where $H_l(r,p)$ is the hamiltonian with potential (\ref{Radpotl}).
But for our purposes in this thesis the Thomas-Fermi result
suffices, because the quantities we want to calculate will be
integrals whose integrands include products of the kind
$\rho_l(\varepsilon)\rho_{l\pm1}(\varepsilon)$, where the effects of
the oscillating corrections are strongly suppressed.

To calculate (\ref{rhothomas}) we start showing that the phase space
integral on the right hand side equals the period of the classical
trajectory of energy $\varepsilon$ of the particle confined in the
$V_l^{eff}(r)$ shown in fig.(\ref{Effpotential})
\begin{equation}\label{rhothomas2}
   \int dp\,dr\delta(\varepsilon-H_l(r,p))=\tau_l(\varepsilon)
\end{equation}
The delta function can be expressed as
\begin{equation}\label{rhothomas3}
   \delta(\varepsilon-H_l(r,p))=\sum_{\substack {(r',p')\text{ such that} \\ H(r',p')=\varepsilon}}\frac{\delta(r-r',p-p')}{|\nabla
H(r',p')|}\,,
\end{equation}
where $\delta(r-r',p-p')$ fulfills $\int
dp\,dr\delta(r-r',p-p')f(r,p)=f(r',p')$, for a generic function
$f(r,p)$. Therefore the surface integral (\ref{rhothomas2}) is
converted by the sum over delta functions into a line integral over
the trajectory in phase space whose energy is fixed at the value
$\varepsilon$. The denominator can be reexpressed with the aid of
the Hamilton equations satisfied by the classical trajectory:
\begin{equation}\label{Hamiltoneq}
   \frac{dp}{dt}=-\frac{\partial H}{\partial r}\,\,\,\,\,\,,\,\,\,\,\,\,\frac{dr}{dt}=\frac{\partial H}{\partial p}\,,
\end{equation}
so that $|\nabla H(r',p')|=\sqrt{[\partial H/\partial r]^2+[\partial
H/\partial p]^2}$ becomes $\sqrt{dr^2+dp^2}/dt=dl/dt$. Substituting
in (\ref{rhothomas2}):
\begin{equation}\label{rhothomas4}
   \int dp\,dr\sum_{\substack {(r',p')\text{ such that} \\ H(r',p')=\varepsilon}}\frac{\delta(r-r',p-p')}{|\nabla
H(r',p')|}=\oint dl \frac{1}{dl/dt}=\oint dt=\tau_l(\varepsilon)\,.
\end{equation}
The next step is to calculate $\tau_l(\varepsilon)$. For that sake
we use the classical action functional
\begin{equation}\label{actionS}
    S[\varepsilon,\tau_l]=\int_0^{\tau_l}\Bigl[\frac{p^2}{2m}-V_l^{eff}(r)\Bigr]dt+\varepsilon\tau_l=\int_{r(t=0)}^{r(t=\tau_l)}dr\cdot
p\,,
\end{equation}
which satisfies $\partial S/\partial\varepsilon = \tau_l$. We will
calculate $S$ using its last form in terms of $r$ and $p$, recalling
the fact that for a trajectory of fixed energy
\begin{equation}\label{momentum}
   \frac{p^2}{2m}+V_l^{eff}(r)=\varepsilon\,\,\,\,\,\rightarrow\,\,\,\,\, p=\sqrt{2m(\varepsilon-V_l^{eff}(r))}
\end{equation}
The value of $S$ can then be calculated as
\begin{equation}\label{actionS2}
   S[\varepsilon,\tau_l]=\int_{r(t=0)}^{r(t=\tau)}dr\cdot
p=2\int_{r_{-}(\varepsilon)}^{r=a}dr\sqrt{2m(\varepsilon-V_l^{eff}(r))}\,.
\end{equation}
The value of the turning point $r_{-}(\varepsilon)$ on the left of
the classically allowed region is given by the condition
$V_l^{eff}(r_{-}(\varepsilon))=\varepsilon$, i.e.,
$r_{-}(\varepsilon)=\hbar(l+1/2)/\sqrt{2m_e\varepsilon}$, see
fig.(\ref{Effpotential}). Substituting the form of the potential
$V_l^{eff}(r)$ in the integral (\ref{actionS2}), after a
straighforward calculation we get
\begin{equation}\label{actionS3}
   S[\varepsilon,\tau_l]=2\hbar\Bigl[\sqrt{(2m\varepsilon a^2/\hbar^2)-\Bigl(l+\frac{1}{2}\Bigr)^2}-\Bigl(l+\frac{1}{2}\Bigr)
 \arccos\Bigl(\frac{l+1/2}{\sqrt{(2m\varepsilon a^2/\hbar^2)}}\Bigr)\Bigr]\,,
\end{equation}
and the period $\tau_l$ is obtained performing the partial
derivative with respect to $\varepsilon$, obtaining in this way the
desired Thomas-Fermi estimate for the partial density of states:
\begin{equation}\label{rhole}
    \tau_l(\varepsilon)=\frac{\hbar\sqrt{(2m\varepsilon
a^2/\hbar^2)-\Bigl(l+\frac{1}{2}\Bigr)^2}}{\varepsilon}\,\,\,\,\,\rightarrow\,\,\,\,\,
\rho_l(\varepsilon)\simeq\rho_l^0(\varepsilon)=\frac{\sqrt{(2m\varepsilon
a^2/\hbar^2)-\Bigl(l+\frac{1}{2}\Bigr)^2}}{2\pi\varepsilon}
\end{equation}

\section{Local density of the dipole matrix element}
\label{app_A} Equation \eqref{lddme} defines the local density of
dipole matrix elements connecting states at energies $\varepsilon$
and $\varepsilon+\Delta\varepsilon$. We used particle and hole
states in our definition, since this is the main interest of our
work. But note that the calculation presented in this appendix is
not restricted to that case and can be easily extended to any
states. The result would be of course unchanged.

Local densities of matrix elements of arbitrary operators have been
thoroughly studied as they can be easily connected with physical
properties, ranging from far-infrared absorption in small particles
\cite{MR97} to electronic lifetimes of quantum dots \cite{GJS04}. A
semiclassical theory for the local density of matrix elements has
been developed \cite{FP86,W87,EFMW92}, where \eqref{lddme} can be
expressed as a smooth part given by correlations along classical
trajectories plus a periodic orbit expansion. We will not follow
here this general procedure, but continue to use the simple form of
the dipole matrix elements \eqref{d_ph} for states confined in a
hard-wall sphere and the semiclassical approximation applied to the
radial (fixed $l$) problem \cite{MWJ02,MWJ03,WMWJ05}.

Using the $l$-fixed density of states (\ref{rhole}) we can write
with the help of equations (\ref{d_ph2}, \ref{sp_angular2},
\ref{R_phap})
\begin{eqnarray}
  \nonumber &&{\cal C}(\varepsilon,\Delta\varepsilon)= \sum_{\al\beta}\Bigl(\textsl{A}_{l_\al l_\beta}^{m_\al m_\beta}\Bigr)^2 \Bigl(\textsl{R}_{l_\al l_\beta}
(\e_\al , \e_\beta)\Bigr)^2  \delta(\e-\e_\al)\delta(\e+\dle-\e_\beta)\\
  \nonumber &=& \int_0^{\infty}d\e_\al\sum_{l_\al m_\al}\rho_{l_\al}(\e_\al)\int_0^{\infty}d\e_\beta\sum_{l_\beta m_\beta}\rho_{l_\beta}(\e_\beta)
  \Bigl(\textsl{A}_{l_\al l_\beta}^{m_\al m_\beta}\Bigr)^2 \Bigl(\textsl{R}_{l_\al l_\beta}(\e_\al , \e_\beta)\Bigr)^2\delta(\e-\e_\al)\delta(\e+\dle-\e_\beta) \\
   &=&\sum_{l_\al m_\al,l_\beta m_\beta}\rho_{l_\al}(\e)\rho_{l_\beta}(\e+\dle)\Bigl(\textsl{A}_{l_\al l_\beta}^{m_\al m_\beta}\Bigr)^2
   \Bigl(\textsl{R}_{l_\al l_\beta}(\e_\al , \e_\beta)\Bigr)^2
\end{eqnarray}
$\textsl{A}_{l_\al l_\beta}^{m_\al m_\beta}\neq0$ only if
$m_\al=m_\beta$ and $l_\al=l_\beta\pm1$, therefore
\begin{equation}\label{CofEA0}
  C(\e,\dle)= \Bigl(\frac{2\hbar^2}{m a}\Bigr)^2\frac{\e(\e+\dle)}{\dle^4}\sum_{l_\al,m_\al,l_\beta=l_\al\pm1}\rho_{l_\al}(\e)
  \rho_{l_\beta}(\e+\dle)\Bigl(\textsl{A}_{l_\al l_\beta}^{m_\al m_\al}\Bigr)^2
\end{equation}
We first perform the sum over $l_\beta$: \begin{eqnarray}
  \nonumber &&\sum_{l_\beta=l_\al\pm1}\Bigl(\textsl{A}_{l_\al l_\beta}^{m_\al
m_\al}\Bigr)^2\rho_{l_\beta}(\e+\dle)=
  \Bigl(\textsl{A}_{l_\al l_\al+1}^{m_\al m_\al}\Bigr)^2\rho_{l_\al+1}(\e+\dle)+\Bigl(\textsl{A}_{l_\al l_\al-1}^{m_\al m_\al}\Bigr)^2
  \rho_{l_\al-1}(\e+\dle) \\
   &=&\frac{(l_\al+1-m_\al)(l_\al+1+m_\al)}{(2l_\al+1)(2l_\al+3)}\rho_{l_\al+1}(\e+\dle)+\frac{(l_\al-m_\al)(l_\al+m_\al)}{(2l_\al-1)(2l_\al+1)}
   \rho_{l_\al-1}(\e+\dle)
\end{eqnarray}
and then over $m_\al$: \begin{eqnarray}
  \nonumber \sum_{m_\al=-l_\al}^{l_\al}(l_\al+1-m_\al)(l_\al+1+m_\al)&=& \frac{1}{3}(l_\al+1)(2l_\al+1)(2l_\al+3), \\
     \sum_{m_\al=-l_\al}^{l_\al}(l_\al-m_\al)(l_\al+m_\al)&=&\frac{1}{3}l_\al(2l_\al-1)(2l_\al+1),
\end{eqnarray}
yielding \begin{equation}\label{CofEA1}
    {\cal C}(\varepsilon,\Delta\varepsilon)=\left(\frac{2\hbar^2}{m_{\rm
  e}a}\right)^2\frac{\varepsilon\varepsilon'}{3\Delta\varepsilon^4}\times\sum_{l_h=0}^{l_{\rm
  max}}\varrho_{l_h}(\varepsilon)\left[(l_h+1)
  \varrho_{l_h+1}(\varepsilon')+l_h\varrho_{l_h-1}(\varepsilon')\right]
\end{equation}
where $l_{\rm max}$ is the maximum allowed $l_h$ for an energy
$\varepsilon$, while $\varepsilon'=\varepsilon+\Delta\varepsilon$.
In the semiclassical limit we can take $l_h\simeq l_h+1\gg1$ and
convert the sum into an integral. Thus, \begin{eqnarray}
  \nonumber C(\e,\dle) &=& \frac{2}{3}\Bigl(\frac{2\hbar^2}{m a}\Bigr)^2\frac{\e(\e+\dle)}{\Delta
  \e^4}\int_{0}^{l_{max}}dl\,l\frac{\sqrt{2m\e a^2/\hbar^2-l^2}}{2\pi\e}\frac{\sqrt{2m(\e+\dle) a^2/\hbar^2-l^2}}{2\pi(\e+\dle)} \\
   &=&\frac{1}{6\pi^2}\Bigl(\frac{2\hbar^2}{m a}\Bigr)^2\frac{1}{\dle^4}\int_{0}^{y}dl\,l\sqrt{y^2-l^2}\sqrt{y'^2-l^2}
\end{eqnarray}
where $y^2=2m\e\, a^2/\hbar^2$ and $y'^2=y^2+2m\dle\,
a^2/\hbar^2$.Changing variables now to $u=l^2$, $du=2ldl$, we get
\begin{equation}\label{CofEA3}
  C(\e,\dle)= \frac{1}{12\pi^2}\Bigl(\frac{2\hbar^2}{m a}\Bigr)^2\frac{1}{\dle^4}\int_{0}^{y^2}du\sqrt{y^2-u}\sqrt{y'^2-u}
\end{equation}
Using the results of ref.\cite{GR00} we obtain
\begin{equation}\label{CofEA3}
  \int_{0}^{y^2}du\sqrt{y^2-u}\sqrt{y'^2-u}=\frac{(y'^2-y^2)^2}{8}\ln\Bigl[\Bigl|\frac{y'-y}{y'+y}\Bigr|\Bigr]+\frac{(y'^2+y^2)yy'}{4}
\end{equation}
And substituting the values of $y$ and $y'$ one arrives at the
desired expression \eqref{C}.

\section{Density of particle-hole excitations}
\label{app_B} The density of p-h excitations with energy
$\Delta\varepsilon$ is defined in \eqref{rph1} and can be written as
\begin{equation}\label{rph2}
    \rho^\textrm{p-h}(\Delta\varepsilon)=\int_{\varepsilon_\textrm{F}-\Delta\varepsilon}^{\varepsilon_\textrm{F}}
 {\rm d}\varepsilon_h\sum_{l_h}(2l_h+1)\varrho_{l_h}(\varepsilon_h)\times
\left[\varrho_{l_h+1}(\varepsilon_h+\Delta\varepsilon)
+\varrho_{l_h-1}(\varepsilon_h+\Delta\varepsilon)\right]\,.
\end{equation}
Using the semiclassical density of states \eqref{rhole} and
performing the sum in the limit $l_h\gg1$, we obtain
\begin{equation} \rho^\textrm{p-h}(\Delta\varepsilon)\simeq
\frac{\Delta\varepsilon^2}{8\pi^2}\left(\frac{2m_{\rm
e}a^2}{\hbar^2}\right)^2
\int_{\varepsilon_\textrm{F}-\Delta\varepsilon}^{\varepsilon_\textrm{F}}
{\rm d}\varepsilon_h\frac{F(\varepsilon_h/\Delta\varepsilon)}
{\varepsilon_h (\varepsilon_h+\Delta\varepsilon)}\,,
\end{equation}
where the function $F$ has been defined in \eqref{F}. Performing the
remaining integral over the hole energy in the limit
$\Delta\varepsilon\ll\varepsilon_{\rm F}$ is straightforward and
leads to the result \eqref{rph3}.


\nocite{}

{
\ssp 

\bibliography{NEMs1}
}


\end{document}